%% file: main.tex
\documentclass[reqno,a4paper,9pt]{article}
\input{macros}

\title{\LARGE\textsf{\textbf{Quantitative and Metric Rewriting: \\ 
\Large{Abstract, Non-Expansive, and Graded Systems}}} 
}
\author{\normalsize
\textsc{Francesco Gavazzo}\thanks{University of Bologna and INRIA Sophia Antipolis}\; \&\; \textsc{Cecilia Di Florio}\footnotemark[1]}
\date{}

\begin{document}

\maketitle

\epigraph{\emph{On May 2, the first author became a father. 
The theory presented in this work is dedicated to his son, Giulio Febo.
}}{}

\section{INTRODUCTION}
%\section{Introduction}

Modern mathematics begins with \emph{symbolic manipulation}. 
The central role of signs and symbols \emph{per se} 
is one of the main achievement of the Medieval culture \cite{medieval-semiotics} leading, among others, to the development of
elementary or \emph{symbolic algebra}. 
Starting from the latter, the syntactic manipulation of symbols 
more or less independently of their meaning --- i.e. to what symbols stand for --- has become 
an essential part of mathematical reasoning, not to say of reasoning \emph{in general}. 
Today, symbolic manipulation is not just a pillar of mathematics, 
but it is at the very hearth of \emph{computation}. Indeed, the symbolic manipulations 
of elementary algebra carry a computational content and, vice versa, 
computational processes can be fully described symbolically. 

Rewriting theory \cite{newman,terese} is the discipline that studies 
(the computational content of) symbolic manipulation in general. 
As such, rewriting has its origin both in symbolic algebra as the study of 
the algorithmic properties of equational reasoning,
and in computability and programming language theory,
where rewriting systems have been used to define symbolic models of computation --- such as the 
$\lambda$-calculus~\cite{Barendregt/Book/1984} and combinatory logic~\cite{curry-combinatory-logic-1985,lambda-calculus-and-combinators-hindley-seldin} --- as well as 
the (operational) semantics and implementation of programming languages~\cite{asperti-optimal}. 
In both cases, rewriting is motivated by the need to define 
\emph{operational} notions of equality revealing the computational content 
of equational deductions. Remarkably, operationality is ultimately achieved by  
making equality asymmetric, so that the aforementioned computational content  
can be fully uncovered by orienting equations. Nowadays, these oriented equations (and the evolution thereof)
are known as \emph{rewriting} --- or \emph{reduction} --- relations.
% this 
% making the latter asymmetric. By orienting equations, one obtains \emph{operational} notions of 
% equality which  the notion of
% rewriting as an oriented equality came up
% carry a computational content and that such a content is obtained by orienting equality, this 
% making the latter asymmetric. By orienting equations, one obtains \emph{operational} notions of 
% equality which  the notion of rewriting as an oriented equality came up
%independent 
%and complete model of computation themselves. Besides its foundational role,  
%
%the theory of rewriting systems
%is a well-established field of theoretical computer science  is a main branch of theoretical 
%computer science fields 
%such as programming language theory and automated reasoning, just to mention but a few. 
%For instance, rewriting systems are widely used to study the operational 
%semantics of programming languages, as well as the semantics of computational effects. 
All of that highlights a crucial trait of rewriting theory, namely its deep connection 
with equational reasoning. In fact, rewriting 
does not actually focus on arbitrary symbolic transformations, 
but with \emph{equality-preserving} ones: a rewriting relation \emph{refines} 
equality by making the latter operational, and it is thus contained in it. 
% In the theory of computation and programming languages, neither program 
% evaluation (which can be ultimately described as a rewriting process) nor program transformations should 
% change program semantics, i.e. they have to preserve semantic equality. 
%Abstracting from these examples, we can thus seen rewriting systems 
%as describing the computational content of semantic equality. 

%Rewriting systems 
%as describing the computational content of semantic equality. 
%
%
%
%In spite of the heterogeneity of its applications, the theory of rewriting systems builds upon some 
%background principles. 
%
%
%In particular, rewriting processes are required to preserve \emph{semantic equality}, 
%meaning that when rewriting an expression $t$ into an expression $s$, 
%the semantic of $t$ and $s$ must be the same. That is, $t$ and $s$ must be semantically \emph{equal}.  
%The reason behind such an equality principle is clear: in algebra, a rewriting relation describes 
%the computational content of equational reasoning; consequently, rewriting \emph{refines} 
%equational reasoning, and it is thus contained in it. 
%Analysing symbolic computation and program execution are rewriting processes, we face a similar 
%situation: executing a program should not change its semantics. 
%Abstracting from these examples, we can thus seen rewriting systems 
%as describing the computational content of semantic equality. 

% All of that highlights the central role played by equality in rewriting, 
% up to the point that rewriting systems can be seen 
% as describing the \emph{computational} content of semantic equality. 
Recent advances in theoretical computer science, however, have questioned the central 
role played by equality in semantics, arguing for more quantitative and approximated 
forms of equivalence. For instance, equality is a too strong notion for reasoning about 
probabilistic computations, where even small perturbations  
break the equivalence between probabilistic processes. To overcome this problem, 
researchers have thus refined equality to \emph{distances} between probabilistic processes, this 
way replacing equivalences with \emph{metrics}.  
Similarly, metric-based and approximated equivalences have been used to 
to reason about privacy and security of systems \cite{Pierce/DistanceMakesTypesGrowStronger/2010,GaboardiEtAl/POPL/2017}, 
not to mention reasoning about intensional 
aspects of computation, such as resource consumption~\cite{DBLP:journals/pacmpl/LagoG22a,modal-reasoning-equal-metric-reasoning}.

Prompted by that, several theories of semantic equality have been refined giving rise to \emph{quantitative 
theories of semantic differences}, prime examples being general theories of program~\cite{Arnold/Metric-interpretations/1980,DeBakker/Semantics-concurrency/1982,Escardo/Metric-model-PCF/1999,GaboardiEtAl/POPL/2017,Pierce/DistanceMakesTypesGrowStronger/2010,Gavazzo/LICS/2018,CrubilleDalLago/LICS/2015,CrubilleDalLago/ESOP/2017,DBLP:phd/basesearch/Gavazzo19,DBLP:journals/pacmpl/LagoG22a,DBLP:conf/ictcs/LagoG20,DLR2022,DBLP:conf/icalp/LagoGY19,DALLAGO2021} and system 
distances~\cite{Deng-Gebler/Behavioural-pseudometrics/2016,Bonchi/Behavioral-metrics-via-functor-lifting/FSTTCS/2014,Bonchi/Towards-trace-metrics-via-functor-fifting/CALCO/2015,Panangaden/Metric-markov-finite/2004,Panangaden/Metric-markov-infinite/2005,Gebler/Compositional-biismulation/2016}
and the theory of \quant{} algebras and \quant{} equational reasoning~\cite{plotkin-quantitative-algebras-2016,plotkin-quantitative-algebras-2017,plotkin-quantitative-algebras-2018,plotkin-quantitative-algebras-2018-bis,plotkin-quantitative-algebras-2021,plotkin-quantitative-algebras-2021-bis,DBLP:conf/lics/MioSV21}. 
The latter, in particular, aim to provide a common foundation for general 
quantitative reasoning by refining traditional, set-based algebraic structures to metric-like ones 
and by replacing traditional equations with \emph{\quant{} equations} bounding the 
difference (or distance) between the equated elements. Accordingly, classic equations $t = s$ 
are replaced by expressions of the form  $t \qequal{\qone} s$, with the 
informal reading that $t$ and $s$ are at most $\qone$ apart, or that 
they are equal up to an error $\qone$. Thus, quantitative algebraic theories 
are not theories about \emph{equality} between objects, but about \emph{distances} between them, 
and can thus be seen as the quintessence of \quant{} and metric reasoning. 

But what about the \emph{computational content} of \quant{} equational reasoning? 
What is an \emph{operational} notion of \quant{} equality or distance allowing us 
to effectively compute distances by means of \quant{} equations? And, more generally, what is the theory of 
\emph{\quant{} symbolic manipulation}, where symbolic transformations can break semantic equivalence? 
The development of such a theory, which is the main topic of this work, 
is of paramount importance not only to make \quant{} equational 
deduction effective, but also to develop a general \quant{} theory of programming language semantics. 

In this paper, we introduce the theory of \emph{\quant{} and metric rewriting systems}
as a first step towards a general theory of the computational content of \quant{} symbolic manipulations. 
Such a theory is rich and subsumes and largely (and nontrivially) extends traditional rewriting. 
The goal of this paper is to lay the foundation of \quant{} rewriting systems, 
this way opening the door to a larger research program. 
More specifically, in this work we introduce the notion of a \quant{} abstract 
rewriting system and its general theory. We define \quant{} notions of confluence and termination, 
refining cornerstone results such as the Newman~\cite{newman} and Hindley-Rosen~\cite{hindley-1964,rosen-70} 
Lemma to a \quant{} and metric setting. 
Such notions are crucial for the \emph{operational} study of what we shall call \emph{metric word problems}, 
the \quant{} refinement of traditional word problems. 
We then introduce \emph{\quant{} linear and non-expansive term rewriting systems} 
and apply the general theory 
previously developed to such systems. Linearity and the related notion of non-expansiveness 
will be crucial to avoid distance triviliasation phenomena \cite{CrubilleDalLago/ESOP/2017,DBLP:phd/basesearch/Gavazzo19} and the failure 
of major confluence theorems. Concerning the latter, in fact,
we shall prove general \quant{} critical pair-like lemmas~\cite{Huet80} 
ensuring confluence of large families of linear and non-expansive systems. 
Finally, we go beyond linearity and non-expansiveness by introducing
\emph{graded \quant{} term rewriting systems}. In such systems, rewriting is not only 
\quant{} but also \emph{modal} and \emph{context-sensitive}, meaning that contexts 
are not required to non-expansively propagate distances --- as in linear systems --- 
but they directly act on them, this way behaving as generalised Lipschitz continuous 
functions. We will extend the confluence results proved for non-expansive systems to 
graded ones, as well as prove an additional confluence result for orthogonal systems.

All our theory is developed following the abstract relational theory of distances initiated by 
\citet{Lawvere/GeneralizedMetricSpaces/1973}, whereby we work with relations taking values 
in arbitrary quantales \cite{Rosenthal/Quantales/1990}. Such an approach has been successfully applied 
to the study of general theories of program and process distances \cite{Worrell-omega-categories,Gavazzo/LICS/2018,DBLP:phd/basesearch/Gavazzo19,paul-wild-2022}.
Moreover, since abstract metric and modal reasoning are 
essentially equivalent~\cite{DBLP:journals/pacmpl/LagoG22a,modal-reasoning-equal-metric-reasoning}, 
our theory can be seen both as a general theory of metric and quantitative rewriting systems 
and as a theory of
modal and substructural rewriting, this way suggesting possible connections with 
modal and coeffectful systems~\cite{Orchard-icfp-2019,DBLP:conf/esop/GhicaS14,Mycroft-et-al/ICFP/2014,Gaboradi-et-al/ICFP/2016,DBLP:journals/pacmpl/AbelB20}.

In addition to the just outlined theoretical development, in this paper we deal with several examples 
of \quant{} (and modal) systems. Such examples come from the field of algorithms (notably edit distances on strings), 
\quant{} algebras and algebraic effects (e.g. \quant{} barycentric algebras), 
programming language theory (\quant{} and graded combinatory logic), and combinations thereof.

\paragraph{From Equality to Distances: A Gap}
Before outlining the main contents and contributions of the present work in more detail, it is 
instructive to shortly stress the gap between traditional, equality-based reasoning and 
\quant{} one (this gap will be the main theme of the first example in \autoref{section:long-intro}). 
When it comes to reason about equality between objects, we usually have at our disposal a 
\emph{heterogenous} arsenal of techniques, ranging from semantic and denotational characterisations of equality 
to symbolic and operational ones. 
Think about equality between (natural) numbers: 
we can approach it foundationally using set-theory or Peano arithmetic --- 
depending on whether we prefer a semantic or syntactic approach --- 
but we can also study it using algebra, category theory, or type theory, this way relying  
on its inductive nature. And that is not the end of the story, 
as we can also use plain number theory, this way building upon 
numerical and analytical techniques, rather than symbolic ones. 

When we move to
\quant{} equality and metric reasoning, the situation vastly changes and only a few 
of the aforementioned techniques are available, with a strong orientation 
towards numerical and analytical ones. On natural numbers, we can consider the Euclidean 
distance, which is ultimately defined numerically. And when it comes to reason about 
it, numerical and analytical techniques are largely used, other techniques 
being simply not available. However, the Euclidean distance (between natural numbers)
has an embarrassingly simple inductive definition (and thus an associated induction principle) 
in terms \quant{} equations, as well as a well-behaved associated 
notion of operational equality (i.e. rewriting). 
The same story can be told for many other (more challenging) distances, ranging from edit to 
transportation distances \cite{encyclopedia-of-distances}, in all cases obtaining elegant 
\quant{} equational characterisations and well-behaved \quant{} rewriting systems. 
As already mentioned, in the last decade researchers have started to realise that the mathematical
heterogeneity characterising equality 
%--- which makes it, at the same time, 
%an algebraic, logical, numerical, symbolical, etc notion --- 
pertains to \quant{} equality and 
distances too, although this new awareness it still in its infancy. 
This paper has the ambitious goal to contribute to all of that by beginning the exploration 
of the computational content of \quant{} equality. 

\paragraph{Structure of the Paper}
We dedicate \autoref{section:long-intro} to gently introduce the reader 
to the theory of \quant{} and metric rewriting systems by means of 
concrete examples. After that, we move to the technical development of our theory, 
which is divided in three sections. Once recalled the necessary mathematical 
preliminaries (\autoref{sect:quantales}), in \autoref{section:qars} 
we outline a theory of \emph{abstract} \quant{} rewriting systems, focusing 
on \quant{} notions of confluence and termination. The main results proved are \quant{} 
refinements of 
Newman's Lemma, Church-Rosser Theorem, and Hindley-Rosen Lemma.
In \autoref{sect:qtrs}, we specialise the theory of \autoref{section:qars}  
to \quant{} \emph{term} rewriting systems. We prove several \quant{} critical pair-like lemmas 
for \emph{linear} and \emph{non-expansive} systems, and 
use them to infer nontrivial properties of the systems introduced in 
\autoref{section:long-intro}. Finally, in \autoref{sect:beyond-non-expansive-systems}, 
we go beyond linearity and introduce the theory of \emph{graded} (modal) \quant{} 
rewriting systems. We extend \quant{} critical pair lemmas to graded systems and 
prove that (graded) orthogonal systems are always confluent. Using such results, 
we obtain a confluence result for a system of graded combinators extending bounded 
combinatory logic.

\section{%QUANTITATIVE AND METRIC REWRITING: SHAPING A THEORY 
BEYOND TRADITIONAL REWRITING: SHAPING A THEORY
}
\label{section:long-intro}
\renewcommand{\termone}{t}
\renewcommand{\termtwo}{s}
\renewcommand{\termthree}{u}

% \section{Quantitative and Metric Rewriting: Shaping a Theory}
In this section, we gently introduce the reader to quantitative rewriting systems
by looking at some simple examples of quantitative systems coming from diverse fields 
(e.g. algebra, programming language theory, and biology). 
For pedagogical reasons, we shall focus on \emph{non-expansive} 
systems  (i.e. systems where rewriting inside terms 
non-expansively propagates distances) only, postponing 
graded systems to \autoref{sect:beyond-non-expansive-systems}.

\subsection{From Equality to Differences: Warming-Up} 
\label{sect:natural-numbers}
Let us begin with one of the simplest possible example: the system of natural numbers 
with the addition operation. 
Such a system can be define in several
ways (algebraically, set-theoretically, numerically, etc) each giving
a specific perspectives on equality between numbers. 
In this example, we model natural numbers \emph{symbolically} using the signature 
$\signature_{\mathcal{N}} \defeq \{\nzero, \nsucc, \nadd\}$ containing a constant $\nzero$ for zero, 
a unary function symbol $\nsucc$
for the successor function, and a binary function
symbol $\nadd$ for addition. Fixed a set $\variables$ of variables, we 
use the set\footnote{Given a signature $\signature$ and a set 
$\variables$ of variables, we denote by 
$\terms{\signature}{\variables}$ the collection of $\signature$-terms over 
$\variables$.}
$\terms{\signature_{\mathcal{N}}}{\variables}$ to study natural numbers 
syntactically. 
Equality between terms in $\terms{\signature_{\mathcal{N}}}{\variables}$ 
is given by the relation 
$=_{\natrelone}$ inductively defined
by the following rules.\footnote{Given an expression $\termone$ and a substituion 
$\sone$ (i.e. a maps from variables to terms), we write $\subst{\termone}{\sone}$ for the application 
of the substitution $\sone$ to $\termone$.}
%\footnote{
%For conciseness, we have omitted the usual reflexivity, symmetry, and transitivity 
%rules for $=_{\mathcal{N}}$.
%\[
%\infer{x =_{\mathcal{N}} x}{} 
%\qquad
%\infer{y =_{\mathcal{N}} x}{x =_{\mathcal{N}} y}
%\qquad
%\infer{x =_{\mathcal{N}} z}{x =_{\mathcal{N}} y & y =_{\mathcal{N}} z}
%\]
%}
%We call $\mathcal{N}_{=}$ the resulting system $(\signature, =_{\mathcal{N}})$ 

{
\centering
\begin{tcolorbox}[boxrule=0.5pt,width=(\linewidth*2)/3,colframe=black,colback=black!0!white,arc=0mm]
\[
\infer{\nzero =_{\natrelone} \nzero}{\phantom{\Phi}}
\qquad
\infer{\nadd(x, \nzero) =_{\natrelone} x}{}
\qquad
\infer{\nadd(x, \nsucc(y)) =_{\natrelone} \nsucc(\nadd(x,y))}{}
\]

\[
\infer{x =_{\natrelone} x}{} 
\qquad
\infer{y =_{\natrelone} x}{x =_{\natrelone} y}
\qquad
\infer{x =_{\natrelone} z}{x =_{\natrelone} y & y =_{\natrelone} z}
\]

\[
\infer{\nsucc(x) =_{\natrelone} \nsucc(y)}{x =_{\natrelone} y}
\qquad 
\infer{\nadd(x,y) =_{\natrelone} \nadd(x',y')}{x =_{\natrelone} x' & y =_{\natrelone} y'}
\qquad
\infer{\subst{\termone}{\sone} =_{\natrelone} \subst{\termtwo}{\sone}}{\termone =_{\natrelone} \termtwo}
\]
\end{tcolorbox}
}
The first three rules are the defining axioms of $=_{\natrelone}$, whereas the last three rules 
close $=_{\natrelone}$ under substitutions and function symbols in $\signature_{\natrelone}$. 
The remaining three rules, finally, makes $=_{\natrelone}$ reflexive, symmetric, and transitive, 
and thus an equivalence.

% \begin{notation}
%     When unambiguous, we use the lighter notation $\termone = \termtwo$ in place of 
%     $\termone =_{\mathcal{N}} \termtwo$. 
% \end{notation}

The equational system $(\signature_{\mathcal{N}}, =_{\natrelone})$ gives a symbolic approach to numerical equality. 
In fact, given two numbers (or numerical expressions defined as sums of natural numbers) 
$m$ and $n$, we can check whether $m$ and $n$ are equal in (at least) two ways:
either we compute $m$ and $n$ numerically (assuming to have ways to perform calculations) 
\emph{or} we look at $m$, $n$ as expressions $\termone$, $\termtwo$ in 
$\terms{\signature_{\mathcal{N}}}{\variables}$ and 
manipulate them symbolically to produce a formal derivation of $\termone =_{\natrelone} \termtwo$. 
For instance, we see that $1+2$ is equal to 
$2+1$ because we numerically compute them --- obtaining $3$ in both cases --- or because we observe 
that $\nadd(x,y) =_{\natrelone} \nadd(y,x)$ is provable in $(\signature_{\mathcal{N}}, =_{\natrelone})$, and thus 
$\nadd(\nsucc(\nzero), \nsucc(\nsucc(\nzero))) =_{\natrelone} \nadd(\nsucc(\nsucc(\nzero)), \nsucc(\nzero))$ 
is derivable.
Furthermore --- and most importantly --- we can uncover the computational content of $=_{\natrelone}$ by 
orienting its defining equational axioms, this way obtaining a rewriting (or reduction) relation 
$\reduce_{\natrelone}$ defined as follows:

{
\centering
\begin{tcolorbox}[boxrule=0.5pt,width=(\linewidth*2)/3,colframe=black,colback=black!0!white,arc=0mm]
\vspace{-0.3cm}
\[
    \nadd(x, \nzero) \stepto_{\natrelone} x
    \qquad
    \nadd(x, \nsucc(y)) \stepto_{\natrelone} \nsucc(\nadd(x,y))
    \qquad
    \infer{C[\subst{\termone}{\sone}] \reduce_{\natrelone} C[\subst{\termtwo}{\sone}]}
    {\termone \stepto_{\natrelone} \termtwo}
\]
\end{tcolorbox}
}
The first two axioms define the relation $\stepto_{\natrelone}$, whereas the last rule 
extends to $\stepto_{\natrelone}$ to the (full) rewriting relation $\reduce_{\natrelone}$. 
To define the latter, we have denoted by $C[\cdot]$ a context in $\terms{\signature_{\mathcal{N}}}{\variables}$ 
--- i.e. a term with a single occurrence of a hole $\Box$ --- and by $C[\termone]$ the term obtained by 
replacing the hole $\Box$ with $\termone$ in $C[\cdot]$. 
Accordingly, we see that $\reduce_{\natrelone}$ is obtained by applying (substitution) instances of $\stepto_{\natrelone}$ inside 
arbitrary terms. 

% \begin{notation}
% As before, when unambiguous, we write $\stepto$ and $\reduce$ 
% in place $\stepto_{\mathcal{N}}$ and $\reduce_{\mathcal{N}}$, respectively.
% \end{notation}

The rewriting system $(\signature_{\mathcal{N}}, \stepto_{\natrelone})$ fully describes 
equality in $(\signature_{\mathcal{N}}, =_{\natrelone})$ \emph{operationally},
in the sense that an equality $\termone =_{\natrelone} \termtwo$ is provable if and only if 
$\termone$ and $\termtwo$ are $\reduce_{\natrelone}$-convertible, meaning that there is a rewriting path from 
$t$ to $s$ obtained by performing a finite number of bidirectional rewriting steps (i.e. from 
left to right as well as from right to left). Additionally, $(\signature_{\mathcal{N}}, \stepto_{\natrelone})$
enjoys several nice properties. In particular, it is confluent and 
terminating, meaning that convertibility (and thus $=_{\natrelone}$) is decidable 
and coincides with having the same normal form. 

\paragraph{The Computational Content of a Distance}

What we have seen so far shows that equality between natural numbers can not only be defined symbolically 
as $=_{\natrelone}$,
but also \emph{operationally} via $\reduce_{\natrelone}$, 
this way making explicit its computational content. All of that is no more 
than a classic introductory example to rewriting theory. 
Let us make a step further and ask the following question:
what happens if we move from \emph{equality} to \emph{distances} between numbers?
That is, what happens if instead of determining whether two numbers are equal or not, 
we ask the finer question about how much \emph{different} they are? 
Answering these questions numerically is not a problem at all: we consider the Euclidean metric 
and define the distance between 
two numbers $n$ and $m$ as
%\footnote{Notice that 
%to talk about the distance between numbers we need to have (possibly truncated) subtraction.} 
$|n - m|$. 
But what about symbolic approaches? And what is the \emph{computational content}, if any, of 
the Euclidean distance? 

Answering these questions in the affirmative ultimately means finding  
systems like $(\signature_{\mathcal{N}}, =_{\natrelone})$ 
and $(\signature_{\mathcal{N}}, \stepto_{\natrelone})$ describing, however, 
the Euclidean distance between numbers, rather than their equality. 
To define such systems, we 
refine $=_{\natrelone}$ 
and $\stepto_{\natrelone}$ \emph{quantitatively}. 
Let us begin with $=_{\natrelone}$. Following the methodology of \quant{} equational 
theories \cite{plotkin-quantitative-algebras-2016,plotkin-quantitative-algebras-2017}, 
we move from traditional equations $\termone = \termtwo$ to 
\emph{\quant{} equations},
%\footnote{To emphasise the shift from Boolean to 
%\quant{} equations, we move from Latin to Greek letters (which we will also use later 
%to denote distances). 
%We will thus denote by $=_{\natrelone}$ the \quant{} 
%refinement of $=_{\mathcal{N}}$.} 
that is \emph{ternary relations}
$\mrel{\qone}{\termone}{=}{\termtwo}$
relating pairs of terms $\termone, \termtwo$ with non-negative numbers\footnote{For the moment, 
whether we work with natural, rational, or real numbers is not relevant.} $\qone$,
the informal reading of a quantitative equation $\mrel{\qone}{\termone}{=}{\termtwo}$ 
being that $\termone$ and $\termtwo$ are at most $\qone$-apart.\footnote{
Other possible readings come from the world of metric spaces 
(\emph{the distance between $\termone$ and $\termtwo$ is at most $\qone$}), 
intensional and resource analysis (\emph{given resource $\qone$, the terms $\termone$ and $\termtwo$ can 
be proved equal}), and fuzzy and graded logic(s) (\emph{$\termone$ is equal to $\termtwo$ with degree $\qone$}).
}

\begin{notation}
To improve readability, we oftentimes abbreviate $\mrel{\qone}{\termone}{=}{\termtwo}$  
as $\termone \qequal{\qone} \termtwo$.
\end{notation}

This shift from traditional to \quant{} equality leads to a change in the classic rules 
of equational deduction which now have a \quant{} flavour: transitivity,
for instance, now describes the usual triangular inequality axiom of metric spaces.
\[
\infer{\mrel{\qone + \qtwo}{\termone}{\equal}{\termtwo}}
{\mrel{\qone}{\termone}{\equal}{\termthree} & \mrel{\qtwo}{\termthree}{\equal}{\termtwo}}
\]
We will see this kind of rules in detail throughout this paper, but for the moment 
we can leave them aside. Traditional equality now corresponds to the null (zero) 
distance, whereas congruence rules give \emph{non-expansiveness} of syntactic constructs. 
Non-expansiveness is a crucial feature of \quant{} systems and we will say 
more about that in \autoref{sect:combinators-intro} and \autoref{sect:beyond-non-expansive-systems}.

Finally, in order to deal with natural numbers, we add a single distance-producing, \quant{} equation: 
$\mrel{1}{\nsucc(x)}{\equal_{\natrelone}}{x}$. This equation simply states that a number and its successor 
are at distance one.
Overall, we obtain the following \quant{} refinement of system
$(\signature_{\mathcal{N}}, =_{\natrelone})$ which, overloading the notation, we still denote 
by $(\signature_{\mathcal{N}}, =_{\natrelone})$ (this will not create confusion, since 
from now on we shall deal
\quant{} systems only).\footnote{
The complete definition of $=_{\natrelone}$ actually requires the addition of all rules of 
\quant{} equational deduction 
previously mentioned. Such rules (which are formally described in \autoref{sect:qtrs}) 
include the \quant{} refinements of reflexivity, symmetry, and transitivity --- 
which essentially correspond to the usual identity of indiscernibles, symmetry, and triangle inequality 
axioms of metric spaces, respectively --- as well as structural rules for $=_{\natrelone}$ 
(for instance, there is a weakening rule stating that whenever 
$\mrel{\qone}{\termone}{\equal_{\natrelone}}{\termtwo}$ is 
derivable, then so is $\mrel{\qtwo}{\termone}{\equal_{\natrelone}}{\termtwo}$, for any $\qtwo \geq \qone$.
}

{
\centering
\begin{tcolorbox}[boxrule=0.5pt,width=(\linewidth*2)/3,colframe=black,colback=black!0!white,arc=0mm]
\vspace{-0.3cm}
\[
\infer{\nsucc(x) \qequal{1}_{\natrelone} x}{\phantom{F}}
\qquad
\infer{\nzero \qequal{0}_{\natrelone} \nzero}{}
\qquad
\infer{\nadd(x, \nzero) \qequal{0}_{\natrelone} x}{}
\qquad
\infer{\nadd(x, \nsucc(y)) \qequal{0}_{\natrelone} \nsucc(\nadd(x,y))}{}
\]

\[
\infer{\nsucc(x) \qequal{\qone}_{\natrelone} \nsucc(y)}{x \qequal{\qone}_{\natrelone} y}
\qquad 
\infer{\nadd(x,y) \qequal{\qone + \qtwo}_{\natrelone} \nadd(x',y')}{x \qequal{\qone}_{\natrelone} x' & y \qequal{\qtwo}_{\natrelone} y'}
\qquad
\infer{\subst{\termone}{\sone} \qequal{\qone}_{\natrelone} \subst{\termtwo}{\sone}}{\termone 
\qequal{\qone}_{\natrelone} \termtwo}
\]
\end{tcolorbox}
}
% to indicate that $\stepto_{\vrelone}$ rewrites $t$ into $s$ with distance (or error) $a$, 
% and use the lighter 
% notation $t \qstepto{a} s$ when the label $\vrelone$ can be safely omitted.

% {
% \centering
% \begin{tcolorbox}[boxrule=0.5pt,width=\linewidth,colframe=black,colback=black!0!white,arc=0mm]
% \vspace{-0.3cm}
% \[
% \infer{
% \mrel{1}{\nsucc(x)}{\equal_{\natrelone}}{x}}
% {\phantom{F}}
% \qquad
% \infer{
% \mrel{0}{\nzero}{\equal_{\natrelone}}{\nzero}
% }{}
% \qquad
% \infer{
% \mrel{0}{\nadd(x, \nzero)}{\equal_{\natrelone}}{x}
% }{}
% \qquad
% \infer{
% \mrel{0}{\nadd(x, \nsucc(y))}{\equal_{\natrelone}}{\nsucc(\nadd(x,y))}
% }{}
% \]

% \[
% \infer{
% \mrel{\qone}{\nsucc(x)}{\equal_{\natrelone}}{\nsucc(y)}}
% {\mrel{\qone}{x}{\equal_{\natrelone}}{y}}
% \qquad 
% \infer{
% \mrel{\qone + \qtwo}{\nadd(x,y)}{\equal_{\natrelone}}{\nadd(z,w)}
% }
% {\mrel{\qone}{x}{\equal_{\natrelone}}{z} 
% & 
% \mrel{\qtwo}{y}{\equal_{\natrelone}}{w}
% }
% \qquad
% \infer{
% \mrel{\qone}
% {\subst{\termone}{\sone}}
% {\equal_{\natrelone}} 
% {\subst{\termtwo}{\sone}}
% }
% {
% \mrel{\qone}
% {\termone}
% {\equal_{\natrelone}}
% {\termtwo}
% }
% \]
% \end{tcolorbox}
% }

Notice that $\equal_{\natrelone}$ is an \emph{inductive} notion
and that it defines a distance $E$ on 
$\terms{\signature_{\mathcal{N}}}{\variables}$ 
as
$$
E(\termone,\termtwo) \defeq \inf \{\qone \mid \termone \qequal{\qone}_{\natrelone} 
\termtwo\}.
$$
Such a distance is a pseudometric and when it is applied to terms representing natural numbers 
it indeed gives the Euclidean distance between such numbers, hence showing that 
the latter distance has an inductively-defined algebraic characterisation. 

Let us now uncover the computational content of the Euclidean distance 
by giving an \emph{operational} account of $=_{\natrelone}$.
We do so by refining the rewriting relation previously introduced  
to the (ternary) \emph{\quant{} rewriting relation} $\reduce_{\natrelone}$ 
giving information on the distance produced when rewriting terms. 
We thus read $\mrel{\qone}{\termone}{\reduce_{\natrelone}}{\termtwo}$ as stating that 
reducing $\termone$ to $\termtwo$ produces a difference $\qone$ between the former and the latter.

\begin{notation}
As before, we often abbreviate $\mrel{\qone}{t}{\reduce_{\natrelone}}{s}$  
as $t \qreduce{\qone}_{\natrelone} s$ (and similarly for the other rewriting 
relations we are going to introduce).
\end{notation}

We first define $\stepto_{\natrelone}$ by stipulating that actual distances between terms 
are produced by deleting successor functions and extends it to 
the full \quant{} rewriting relations $\reduce_{\natrelone}$ by 
\emph{non-expansively} propagating distances produced by substitution instances of $\stepto_{\natrelone}$ 
throughout arbitrary contexts of the language.

{
\centering
\begin{tcolorbox}[boxrule=0.5pt,width=(\linewidth*2)/3,colframe=black,colback=black!0!white,arc=0mm]
\vspace{-0.3cm}
\[
    \nadd(x, \nzero) \qstepto{0}_{\natrelone} x
    \qquad
    \nadd(x, \nsucc(y)) \qstepto{0}_{\natrelone} \nsucc(\nadd(x,y))
    \qquad
    \nsucc(x) \qstepto{1}_{\natrelone} x
    \]
    \vspace{-0.2cm}
    \[
    \infer{C[\subst{\termone}{\sone}] \qreduce{\qone}_{\natrelone} C[\subst{\termtwo}{\sigma}]}
    {\termone \qstepto{\qone}_{\natrelone} \termtwo}
\]
\end{tcolorbox}
}

The relation $\to_{\natrelone}$ induces a (rewriting) distance 
$\natrelone$ between terms defined by
$$
\natrelone(\termone, \termtwo) \defeq \inf \{\qone \mid \termone \qreduce{\qone}_{\natrelone} \termtwo\}
$$
so that we obtain a \quant{} relational system 
$\NATS = (\terms{\signature_{\mathcal{N}}}{\variables}, \natrelone)$,
which is our first example of an \emph{abstract \quant{} rewriting system}. 
We will study such systems in \autoref{section:qars}. For the moment, we simply 
say that as an abstract rewriting system consists of a set $A$ of objects together with 
a relation $\relone \subseteq A \times A$ on it, a 
\emph{\quant{} abstract rewriting system} is given by a set $A$ together with 
a \emph{\quant{} relation} $\vrelone: A \times A \to [0,\infty]$ on it.\footnote{
Actually, we will consider a more general form of \quant{} relations (see \autoref{section:qars}), 
but for the moment it is more convenient to restrict to 
$[0,\infty]$-valued relations.} \emph{Quantitative term rewriting systems} 
are a special class of \quant{} abstract rewriting systems where objects are terms 
and the \quant{} relation $\vrelone$ is 
canonically defined as $\vrelone(\termone,\termtwo) \defeq 
\inf\{\qone \mid \termone \qreduce{\qone}_{\vrelone} \termtwo\}$ 
starting from a ternary relation $\stepto_{\vrelone}$ (then extended to $\reduce_{\vrelone}$)
like those we have seen so far as.

\begin{remark}
Notice that in a \quant{} abstract rewriting system $(A, \vrelone)$, 
the \quant{} relation $\vrelone$ gives the rewriting distance between elements 
of $A$. Such a distance, however, is \emph{not} required to obey 
the usual (psuedo)metric axioms, nor a subset thereof. Such a requirement, in fact, 
would be morally the same as requiring a traditional rewriting relation to be an 
equivalence, which is clearly undesirable. 
\end{remark}

Let us expand on quantitative relations. 
As pointed out by Lawvere \cite{Lawvere/GeneralizedMetricSpaces/1973}, 
quantitative relations (or distances) are governed by 
an algebra close to the one of ordinary relations\footnote{We could think about 
such an algebra as a monoidal algebra of relations.}~\cite{relational-mathematics}
so that a large part of the calculus of relations~\cite{tarski-1941,relational-mathematics,algebra-of-programming} 
can be refined to give rise to 
a calculus of
\quant{} relations. In fact, by viewing binary relations as maps 
$\relone: A \times B \to \{\false, \true\}$, we see that a quantitative 
relation simply refines the structure $(\{\false, \true\}, \leq, \wedge)$ by replacing it with 
$([0,\infty], \geq, +)$, so that we can use this similarity to generalise many relational constructions 
and their properties to a \quant{} setting. For instance, by refining 
the existential quantifier $\exists$ as the infimum $\inf$ 
and the Boolean meet $\wedge$ as 
addition $+$, we can define the composition between \quant{} relations 
$\vrelone$, $\vreltwo$ by 
$$
(\vrelone;\vreltwo)(\tone,\tthree) \defeq \inf_\ttwo \vrelone(\tone, \ttwo) + 
\vreltwo(\ttwo,\tthree).
$$
Consequently, we will say that a \quant{} relation $\vrelone$ is transitive 
if $\vrelone;\vrelone \geq \vrelone$, i.e. if 
$$
\inf_\ttwo \vrelone(\tone, \ttwo) + \vrelone(\ttwo,\tthree) \geq \vrelone(\tone,\tthree),
$$
which is nothing but 
the usual triangle inequality law. 
In the same way, we can refine the notions of reflexivity, symmetry, and transitivity 
to \quant{} relations, this way obtaining exactly the defining axioms of a pseudometric. 
In particular, as any rewriting relation induces --- 
by taking its reflexive, symmetric, and transitive closure --- 
an equality between terms as convertibility,
any rewriting distance $\vrelone$ defines a \emph{convertibility distance} (pseudometric, acutally)
$\makedistance{\vrelone}$
by means of its reflexive, symmetric, and transitive closure. 
% , we will see in detail the general theory of 
% abstract \quant{} relations \emph{\`a la} Lawvere. What is relevant, for the moment, 
% is that by modelling traditional rewriting relationally, 
% we can then rely on such a general theory for quantitatively refining them. 
We shall see in detail the general theory of 
abstract \quant{} relations \emph{\`a la} Lawvere 
in \autoref{sect:quantales}. What is relevant, for the moment, 
 is that by modelling traditional rewriting relationally~\cite{backshouse-calculational-approach-to-mathematical-induction}, 
 we can then rely on such a general theory for quantitatively refining them.
 
Let us apply the ideas seen so far to the rewriting distance $\natrelone$. 
The pseudometric $\makedistance{\natrelone}$ gives the convertibility distance 
between terms, which is nothing but the distance $E$ induced by $=_{\natrelone}$, i.e. 
the Euclidean distance. 
Consequently, the Euclidean distance is not only obtained symbolically via $=_{\natrelone}$, 
but it is also completely described operationally as the convertibility distance induced by the 
\quant{} rewriting system $\NATS$ (and thus by $\reduce_{\natrelone}$). 

At this point, it is natural 
to ask whether $\makedistance{\natrelone}$ (and thus $E$) has nice computational properties. 
Without much of a surprise, 
the `nice computational properties' we have in mind are the \quant{} refinements of 
well-known rewriting notions, such as confluence and termination. We postpone 
precise definitions of these notions until \autoref{section:qars} 
and content ourselves with some intuitions behind them for now. 
Suppose we are approximating the distance $\makedistance{\natrelone}(\termone,\termtwo)$ 
with a bidirectional reduction path of the form 
$$ 
\termone \qreduce{\qone_1} \cdot \stackrel{\qone_2}{\leftarrow} \cdot \qreduce{\qone_3}  \cc 
\stackrel{\qone_{n-1}}{\leftarrow} 
\cdot \qreduce{\qone_n} \termtwo
$$
so that the convertibility distance between $\termone$ and $\termtwo$ given by this path is 
$\sum_{i=1}^n \qone_i$. When asked to compute or approximate such a distance, 
it is desirable to have a term $\termthree$ such that
$$
\termone \qreduce{\qtwo_1} \cc \qreduce{\qtwo_m} \termthree \qreduceleft{\qthree_p} \cc \qreduceleft{\qthree_1} 
\termtwo
\quad \text{ and } \quad
%s \qreduce{c_1} \cc \qreduce{c_m} u
\sum_{i=1}^n \qone_i \geq \sum_{j=1}^m \qtwo_i + \sum_{k=1}^p \qthree_k.
$$
Moving to distances, that means that to approximate $\makedistance{\natrelone}(\termone,\termtwo)$ 
(and thus $E$), we can restrict ourselves to 
proper rewriting rather than convertibility. Formally: 
$$
\makedistance{\natrelone}(\termone,\termtwo) = \inf_{\termthree} \natrelone^*(\termone,\termthree) + 
\natrelone^*(\termtwo,\termthree),
$$
where $\natrelone^*$ denotes the reflexive and transitive closure of $\natrelone$ 
(which is precisely the distance induced by the reflexive and transitive closure of $\reduce_{\natrelone}$).  
This is nothing but the \quant{} refinement of the well-known Church-Rosser property. 
In a similar fashion, we obtain the \quant{} refinement of confluence; and since
$\NATS$ is confluent --- as we will see in \autoref{sect:qtrs-confluence} ---
it also has the aforementioned (\quant{)} Church-Rosser property. 
Additionally, $\NATS$ is terminating 
(in a suitable sense that we will see in \autoref{section:qars}), so that not only we can approximate 
$\makedistance{\natrelone}(\termone,\termtwo)$ by measuring the distance between 
the common reducts of $\termone$ and $\termtwo$, but we can also reduce the search space to 
normal forms. 

Now that the reader is warmed-up, we can move to slightly more involved (and interesting) 
examples: those will also give us the chance to introduce, still at an informal level, 
a few more notions related to \quant{} rewriting systems.

\subsection{Quantitative String Rewriting Systems} 
\label{sect:string-rewriting}

Historically, rewriting systems have first appeared in the form of 
\emph{string rewriting systems}~\cite{string-rewriting-systems,thue}: 
we thus find appropriate to include examples of \emph{\quant{} string rewriting system} 
in this motivational section.  
Recall that given an alphabet $\signature$, a string rewriting system is given by 
a relation $\stepto_{\relone}$ on strings $\Sigma^*$ over $\signature$. The relation 
$\stepto_{\relone}$ induces a rewriting relation $\reduce_{\relone}$ 
that rewrites substrings according $\stepto_{\relone}$. That is, whenever
we have $t \stepto_{\relone} s$, then we have $u t v \reduce_{\relone} u s v$ too, where 
$u,v$ are strings and we write string concatenation as juxtaposition. 

In this example, we consider a family of classic examples of string rewriting systems: \emph{DNA-based systems}. 
Let us fix the alphabet $\signature_{\DNAone} \defeq \{\A, \C, \G, \T\}$ of DNA bases 
(the latter $\DNAone$ stands form \emph{molecule}). 
We view strings 
over this alphabet as representing DNA molecules or DNA sequences, so that, for instance, 
we view a string such as 
$\T\A\G\C\T\A\G\C\T\A\G\C\T$ as describing a DNA molecule. 
A string rewriting system over $\signature$ specifies how DNA molecules can be 
transformed into one another, and thus it is a 
%Here is an example of such a system~\cite{DBLP:journals/corr/abs-2002-12554}:\footnote{When unambiguous, we write 
% $\stepto$ and $\reduce$ in place of $\stepto_{\relone}$ and $\reduce_{\relone}$, respectively.}
% \begin{align*}
% \C\T &\stepto \T  
% & 
% \G\A &\stepto \A
% \\ 
% \T\A\T &\stepto \T
% & 
% \A\T\A &\stepto \A 
% \\
% \A\G\T &\stepto \A\T
% & 
% \T\C\A &\stepto \T\A
% %\T\C\A\T &\stepto \T 
% %\\
% %\G\A\G &\stepto \A\G
% %\\
% %\C\T\C & \stepto \T\C 
% %\\
% %\A\G\T\A &\stepto \A 
% %\\
% %\T\A\T &\stepto \C\T
% \end{align*}
 crucial tool to deal with \emph{word-problems}, i.e. 
problems asking whether two DNA molecules 
are equal. 
% Instead of explicitly defining equality (as we did in the previous section) between 
% DNA sequences, here we directly work with convertibility relations, this 
% way regarding two molecules as equal if they are convertible. 
% We write $\equiv_{\relone}$ for the convertibility relation 
% (i.e. the reflexive, symmetric, and transitive closure of $\reduce_{\relone}$) 
% induced by 
% $\reduce_{\relone}$ (and thus by $\stepto_{\relone}$)
% adopting the same notational conventions introduced in the previous section (abd omitting the 
% subscript $\relone$ when unambiguous). 
% \footnote{Any string rewriting system $(\signature, \stepto)$ 
% induces an equivalence relation 
% $\approx$ as the reflexive, symmetric, and transitive closure of $\reduce$. 
% The word problem asks whether $t \approx s$ 
% is decidable. Notice that one can also proceed in the opposite direction: to study computational properties of an 
% equality relation $=$ on strings, we define a string rewriting systems $\stepto$ such that $=$ coincides with $\approx$. 
% Solving the word problem for $\approx$ then entails decidability of equality.} 
In fact, once we know that equality coincides with convertibility in a string rewriting 
system, it is sufficient to prove
confluence of the latter to obtain semi-decidability of equality (and thus of its associated word problem); 
and if, additionally, 
the system is terminating, then equality is decidable. 
% since the above system is confluent 
% and terminating~\cite{DBLP:journals/corr/abs-2002-12554}, 
% the word problem\footnote{To be precise, the word problem ask for decidability 
% of a notion of equality $=$ over $\signature^*$, and a standard way to solve it 
% is to extract from $=$ a rewriting relation $\stepto_{\relone}$ such that 
% $=$ coincides with $\equiv_{\relone}$, and to prove decidability of the latter 
% by showing that $\to_{\relone}$ is confluent and terminating.}
% definition of the 
% for it is decidable and one can show that, e.g., $\T\A\G\C\T\A\G\C\T\A\G\C\T$ and
% $\C\T\G\C\T\A\C\T\G\A\C\T$ are equivalent.

\paragraph{Quantitative String Rewriting Systems}
When we transform a DNA molecule into another one, however, we usually obtain \emph{different} 
molecules, so that reasoning about DNA sequences in terms of equality or convertibility is often too restrictive.
And in fact, researchers are more interested in measuring distances between molecules 
rather than studying their equivalence. 
For instance, if we modify a DNA molecule to cure or prevent a disease, we obviously do \emph{not} want our 
modification to make the involved molecules equivalent. 
Similarly, to measure DNA compatibility and similarity
it is not realistic to look at exact equivalence between molecules: 
instead, one should look for metrics and distances between them.
%Similarly, to mix the DNA of different species it would nonsense to look for equivalence: 
%what matters is to ensure the difference between the DNA of such species to be close enough. 

To cope with these problems, we move from traditional string rewriting systems to 
\emph{\quant{} string rewriting systems}. 
Following the same ideas of \autoref{sect:natural-numbers}, 
we refine string rewriting relations %$\stepto_{\relone}$ 
to ternary \quant{} (rewriting) relations 
relating pairs of strings with non-negative extended real numbers.\footnote{
We apply the same notational conventions of previous section, 
hence using the notations $\mrel{\qone}{\termone}{\stepto_{\vrelone}}{\termtwo}$ and
$\termone \qstepto{\qone}_{\vrelone} \termtwo$ interchangeably. 
}
Here is an example of 
\quant{} string rewriting system --- called $\DNAone$ --- over the DNA alphabet, where $\emptystring$ denotes the empty string, 
$\moleculeone, \moleculetwo \in \{\A, \C, \G, \T\}$ and $\moleculeone \neq \moleculetwo$ in the last rule.

{
\centering
\begin{tcolorbox}[boxrule=0.5pt,width=\linewidth/2,colframe=black,colback=black!0!white,arc=0mm]
%\vspace{-0.3cm}
% \begin{align*}
% 	M &\qstepto{1} \varepsilon  \qquad M \in \{\A, \C, \G, \T\}
% 	\\
% 	\varepsilon &\qstepto{1} N \qquad N \in  \{\A, \C, \G, \T\} 
% 	\\
% 	M &\qstepto{1} N \qquad M,N \in  \{\A, \C, \G, \T\} \text{ and } M \neq N
% \end{align*}
$$
\moleculeone \qstepto{1}_{\dnarelone} \emptystring  
\qquad
\emptystring \qstepto{1}_{\dnarelone} \moleculeone
\qquad
\moleculeone \qstepto{1}_{\dnarelone} \moleculetwo
$$
\end{tcolorbox}
}
Ignoring its quantitative dimension, system $\DNAone = (\signature_{\DNAone}, \stepto_{\dnarelone})$ 
allows us to substitute bases with 
one another inside any molecule --- so that, for instance, we can always replace $\A$ with $\G$ ---  
as well as to arbitrarily erase and insert bases inside a molecule. 
This results in an inconsistent (equational) system, in the sense that any two molecules are convertible. 
The situation changes when we take the \quant{} information into account. A rewriting step 
$\termone \qstepto{\qone} \termtwo$ gives the distance between $\termone$ and $\termtwo$ and a rewriting sequence
$$
\termtwo_1 \qreduce{\qone_1} \termtwo_2 \qreduce{\qone_2} \cc \qreduce{\qone_{n-1}} \termtwo_n
$$
produces the distance $\qone_i$ when rewriting $\termtwo_i$ into $\termtwo_{i+1}$, so that 
the overall distance between $\termtwo_1$ and $\termtwo_n$ is bounded by $\sum_i \qone_i$. 
For this example, we 
have stipulated the distance produced by substitution, inserting, and deleting a base 
to be $1$, although we could have chosen any non-negative extended real number. 
For instance, the rewriting relation %$\stepto_{\dnareltwo}$ 
defined below 
measures mutations between a purine ($\A,\G$) and a pyrimidine $(\C$, $\T$) only.\footnote{
The distance induced by these mutations is used to study virus and
cancer proliferation under control of drugs or the immune system~\cite{encyclopedia-of-distances}.}

{
\centering
\begin{tcolorbox}[boxrule=0.5pt,width=\linewidth/2,colframe=black,colback=black!0!white,arc=0mm]
\vspace{-0.3cm}
\begin{align*}
    \A &\qstepto{1}_{\dnareltwo} \C 
    &
    \G &\qstepto{1}_{\dnareltwo} \T
    &
    \A &\qstepto{1}_{\dnareltwo} \T
    \\
    \A &\qstepto{0}_{\dnareltwo} \G 
    &
     \G &\qstepto{1}_{\dnareltwo} \C
     &
     \C &\qstepto{0}_{\dnareltwo} \T
\end{align*}
\end{tcolorbox}
}

But what is the meaning of such distances? As before,
the rewriting relation $\reduce_{\dnarelone}$ induces a distance 
$\dnarelone$ on molecules 
defined by
$
\dnarelone(\termone,\termtwo) \defeq \inf\{\qone \mid \termone \qreduce{\qone}_{\dnarelone} \termtwo\},
$
so that $(\Sigma^*_{\mathcal{M}}, \dnarelone)$ is a \quant{} abstract rewriting system. 
Consequently, we can consider the convertibility pseudometric $\makedistance{\dnarelone}$ 
induced by $\dnarelone$ and realise that the 
latter is nothing but the \emph{Levenshtein distance} \cite{string-algorithms,encyclopedia-of-distances}
between DNA molecules.
% Moreover, writing $\equal_{\gamma}$ for the \quant{} 
% notion of equality induced by $\reduce_{\gamma}$, 
% we obtain another distance $\makedistance{\vrelone}$ on molecules as
% $$
% \makedistance{\gamma}(t,s) \defeq \inf\{a \mid t \qequal{a}_{\gamma} s\}.
% $$
% And it does not take much to realise that such the latter distance is actually a metric and, precisely,
% the \emph{Levenshtein distance} [REF]
% (or edit distance) between molecules.
This means that system $\DNAone$ is a way to formalise the computational content of 
the Levenshtein distance, and that $\dnarelone$ is an operational definition of the latter. 
In particular, any bidirectional rewriting sequence between molecules $\termone$ and $\termtwo$ 
approximates the Levenshtein distance between them.

At this point, we can study properties of the Levenshtein distance and, most importantly, of its computation 
relying on the theory of \quant{} rewriting systems that we will introduce later in this paper.
As we shall see, $\dnarelone$ is confluent, so that we can approximate the convertibility 
distance (i.e. the Levenshtein distance) between molecules as the sum of the rewriting distances 
into their common reducts:
% $$
% t \qreduce{b_1} \cc \qreduce{b_n} u \qreduceleft{c_m} \cc \qreduceleft{c_1} s
% \quad \text{ and } \quad
% %s \qreduce{c_1} \cc \qreduce{c_m} u
% a \geq \sum_i b_i + \sum_j c_j.
% $$
% That means that to approximate $\makedistance{\vrelone}(t,s)$, we can restrict ourselves to 
% proper rewriting rather than bidirectional one: 
$$
\makedistance{\dnarelone}(\termone,\termtwo) = \inf_\termthree \dnarelone^*(\termone,\termthree) +
\dnarelone^*(\termtwo,\termthree).
$$
Additionally, even if system $\DNAone$ is not terminating, we can 
extract a terminating system out of it. 
All of that holds not only for $\DNAone$, but also for its variations. 
For instance, allowing $\stepto_{\dnarelone}$ to perform substitutions only (so that we simply have 
the rule $\moleculeone \qstepto{1}_{\dnarelone} \moleculetwo$, for $\moleculeone,\moleculetwo$ different bases), 
we see that $\makedistance{\dnarelone}$ 
measures the number of mutations between DNA sequences, and thus
gives the \emph{Hamming distance} between  molecules~\cite{string-algorithms}.
Similarly, the distance induced by the 
previously defined \quant{} relation measuring 
mutations between purines ($\A,\G$) and pyrimidines $(\C$, $\T$)
gives the so-called Eigen–McCaskill–Schuster distance\footnote{One 
obtains the Watson–Crick distance in a similar way~\cite{encyclopedia-of-distances}.} 
between molecules~\cite{encyclopedia-of-distances}.

\paragraph{Metric Word Problems}
Other interesting properties of the Levenshtein distance, as well as of the other aforementioned distances, 
can be described operationally in terms of \emph{metric word problems}. 
Contrary to traditional (string) rewriting systems, in the \quant{} world a word problem can take several forms 
to which we shall generically refer to as \emph{metric word problems}. 
Here are some of those.

\begin{description}
    \item[\textbf{Reachability}] The \emph{reachability problem} is
        the \quant{} refinement of the traditional word problem. 
        Given a \quant{} (abstract) rewriting system $\qarsone = (A, \vrelone)$, 
        the reachability problem asks whether $\makedistance{\vrelone}(\tone,\ttwo) < \infty$, 
        for elements $\tone, \ttwo \in A$. 
        If $\vrelone$ is confluent, then the reachability problem is 
        semi-decidable; if, additionally, $\vrelone$ is terminating, 
        then we obtain decidability of the reachability problem. 
        The reachability problem for $\DNAone$ --- i.e. 
        for its associated abstract system $(\signature^*_{\DNAone}, \dnarelone)$ --- 
        as well as for variations 
        thereof, is indeed decidable. In fact, in this paper we will introduce 
        several techniques to prove confluence of $\dnarelone$. 
        % and extract an equivalent terminating system out of it. 
        % there exists $a \neq \infty$ such that $\makedistance{\vrelone}(t,s) < a$. 
        % The strong version, instead, asks whether 
        % there exists $a \neq \infty$ such that $t \qequal{a} s$. 
        % As for traditional rewriting, confluence alone is not 
        % enough to decide reachability: we need termination.  
        Moreover, even if $\dnarelone$ is not terminating, we 
        can easily extract a terminating rewriting relation out of it 
        (for instance, 
        we force substitution rules to be asymmetric and stipulate that molecules 
        can be deleted, but not inserted).  
        %This way, we can trivially solve the reachability problem, as all molecules have common reducts.
        % \footnote{ 
        % The reachability problem for our system is not very interesting, as the edit distance 
        % between strings is always defined. Nonetheless, the problem becomes interesting 
        % when dealing with partially defined metrics. For instance, if we omit the insertion and erasing rules, 
        % we obtain a new \quant{} string rewriting system whose corresponding metric is the Hamming distance 
        % between strings. In that case, strings with different lengths have distance $\infty$, so 
        % that not all strings are reachable with finite distance. }
    \item[\textbf{$\qone$-Reachability}] 
        More interesting metric word problems are obtained by strengthening the reachability problem 
        to what we shall call \emph{$\qone$-reachability} problems.
        Fixing a number 
        $\qone$, the $\qone$-reachability problem asks 
        whether $\makedistance{\vrelone}(\tone,\ttwo) < \qone$ holds. 
        Equivalently, the $\qone$-reachability problem asks whether 
        there exists a bidirectional
        $\vrelone$-rewriting sequence between $\tone$ and $\ttwo$ 
        producing distance $\qone$. 
        %, i.e. whether $ \stackrel{\qone}{\equiv}_{\vrelone} s$ holds.
        % The strong version, instead, 
        % asks whether $t \qequal{a} s$ holds.
        Contrary to reachability (which is nothing but $\infty$-reachability), 
        confluence and termination are in general not enough to solve the 
        $\qone$-reachability problem. 
        In fact, looking at the rewriting paths leading to the common normal form (if any) 
        of two objects $\tone, \ttwo$ can give too coarse (over)approximations of 
        $\makedistance{\vrelone}(\tone,\ttwo)$ only, 
        as illustrated by the following rewriting diagram:
        
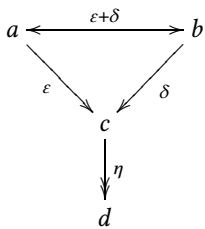
\[
        \xymatrix{
        \tone \ar@{<->}[rr]^{\qone + \qtwo} \ar[rd]_{\qone} & & \ttwo \ar[ld]^{\qtwo}
        \\
        & \tthree \ar@{->>}[d]^{\qthree} &
        \\
        & \tfour &
        }
        \]
        Assuming the system to be confluent, one can try to obtain better approximations 
        by enlarging the state space and looking at arbitrary common reducts 
        (and this strategy is indeed sound and complete, since confluence of 
        $\vrelone$ entails $\makedistance{\vrelone}(\tone,\ttwo) = \inf_{\tthree} \vrelone^*(\tone,\tthree) + 
        \vrelone^*(\ttwo,\tthree)$). That, however, does not solve the 
        $\qone$-reachability problem either, as 
        there may be infinitely many such reducts (see, for instance, \autoref{ex:example-4}). 
    \item[\textbf{Shortest Path}] The \emph{shortest path} is a problem specific to 
        \quant{} string and term rewriting systems. Given such a system 
        $\qtrsone = (\signature, \stepto_{\vrelone})$, the shortest path problem for $\qtrsone$ 
        asks to determine 
        whether 
        % $$
        % \makedistance{\vrelone}(\termone,\termtwo) = \min\{\qone \mid \termone 
        % \stackrel{\qone}{\equiv}_{\vrelone} \termtwo\},
        % $$
         $$
        \makedistance{\vrelone}(\termone,\termtwo) = \min\{\qone \mid 
        \mrel
        {\qone}
        {\termone}
        {\equiv_{\vrelone}} 
        {\termtwo}\},
        $$
        i.e. whether the infimum 
        $\inf\{\qone \mid \mrel
        {\qone}
        {\termone}
        {\equiv_{\vrelone}} 
        {\termtwo}\}$ 
        is achieved 
        by an actual conversion 
        $\mrel
        {\qone}
        {\termone}
        {\equiv_{\vrelone}} 
        {\termtwo}
        $
        (we write $\equiv_{\vrelone}$ for the conversion ternary relation induced by 
        $\reduce_{\vrelone}$). 
        % The optimal path problem is specific to \quant{} string and term rewriting 
        % systems, as such systems come with a ternary relation $\equiv_{\vrelone}$ 
        % for convertibility, $\makedistance{\vrelone}$ being the distance associated 
        % to it. 
        All the systems seen in this section have a shortest path, although 
        finding such a path is usually difficult. 
        Indeed, in all these cases, shortest paths are usually found relying on optimisation
        techniques and dynamic programming~\cite{string-algorithms}, 
        and it is thus an interesting question whether solutions to 
        this problem can be given in terms of \quant{} rewriting. 
         The shortest path problem, additionally, is particularly interesting from a rewriting perspective 
         because it opens the door 
         to another problem: the \emph{optimal strategy problem}. 
    \item[\textbf{Optimal Strategy}] Assuming a term or string rewriting system 
        $\qtrsone = (\signature, \stepto_{\vrelone})$ to have shortest paths, 
        the \emph{optimal strategy} problem asks whether there 
        exists a \quant{} rewriting strategy $\reduce_{\vrelone_{\mathtt{s}}}$ 
        such that:\footnote{Actually, several variations of this problem can be 
        given simply by replacing $\to_{\vrelone_{\mathtt{s}}}^*$ with other 
        relations related to $\reduce_{\vrelone_{\mathtt{s}}}$.}
        % $$
        % \termone \stackrel{\qone}{\equiv}_{\vrelone} \termtwo \text{ } \iff \text{ } 
        % %t \stackrel{a}{\reduce_{\vrelone_{\mathtt{s}}}^*} s.
        % \termone \mathrel{\stackrel{\qone}{\to_{\vrelone_{\mathtt{s}}}}\hspace{-0.7em}{^*}} \termtwo.
        % %t \qequal{a} s \text{ } \iff \text{ } t \qreducee{a}_{\mathtt{s}} s.
        % $$
        $$
        \mrel
        {\qone}
        {\termone} 
        {\equiv_{\vrelone}}
        {\termtwo} 
        \text{ } \iff \text{ } 
        \mrel
        {\qone}
        {\termone} 
        {\to_{\vrelone_{\mathtt{s}}}^* %\hspace{-0.7em}{^*}
        }
        {\termtwo}.
        %t \qequal{a} s \text{ } \iff \text{ } t \qreducee{a}_{\mathtt{s}} s.
        $$
        An optimal strategy or an approximation thereof for $\DNAone$ 
        can be then used to efficiently compute distances
        distance between DNA molecules. To the best of the authors' knowledge, optimal strategy 
        problems for the systems considered so far are still open. 
\end{description}

\subsection{Beyond Traditional Rewriting: Quantitative Term Rewriting Systems} 
\label{sect:combinators-intro} 
We now go beyond string rewriting systems and take a closer 
look at examples of \quant{} \emph{term} rewriting systems. 
Even if we have already seen an example of a \quant{} term rewriting system --- 
the system of natural numbers of \autoref{sect:natural-numbers} ---
we now consider more interesting examples of such systems and take the chance to 
illustrate further features of \quant{} rewriting. 
Among the many examples available, we focus on those coming from the field of (\quant) algebras and 
programming language theory.

\paragraph{Affine Combinatory Logic} 

As a first example, we consider a basic system of affine combinators that we shall 
enrich  with 
effectful and \quant{} primitives in subsequent sections:
 system $\bck$ of
\emph{affine combinatory logic}~\cite{Barendregt/Book/1984,hindley-basic-simple-type-theory,lambda-calculus-and-combinators-hindley-seldin}. System $\bck$ has three constants (known as basic combinators) --- $\Bcomb$, 
$\Ccomb$,
and $\Kcomb$ --- and a single binary operation symbol $\cdot$ for application. 
We denote by $\signature_{\bck}$ the signature thus obtained. As usual, 
we assume application to associate to the left and omit unnecessary parentheses. 
We refer to terms written by means of variables, basic combinators, and application as 
\emph{combinators}.
Even if $\bck$ is historically defined as an equational theory (from which a rewriting system 
is then extracted), we directly define $\bck$ by means of rewriting rules as follows, 
with $\stepto$ being the (ground) reduction relation and $\reduce$ being defined
by applying substitution instances 
of $\stepto$ inside arbitrary context.

{
\centering
\begin{tcolorbox}[boxrule=0.5pt,width=(\linewidth*3)/4,colframe=black,colback=black!0!white,arc=0mm]
\vspace{-0.3cm}
\[
     \Bcomb \cdot x \cdot y \cdot z \stepto x \cdot (y \cdot z)
    \qquad
     \Ccomb \cdot x \cdot y \cdot z \stepto x \cdot z \cdot y
    \qquad
    \Kcomb \cdot x \cdot y  \stepto x
    \]
    \vspace{-0.2cm}
    \[
    \infer{C[t^\sigma] \reduce C[s^\sigma]}{t \stepto s}
    \]
\end{tcolorbox}
}

% \begin{align*}
%     \Bcomb \cdot x \cdot y \cdot z &\stepto x \cdot (y \cdot z)
%     \\
%     \Ccomb \cdot x \cdot y \cdot z &\stepto x \cdot z \cdot y
%     \\
%     \Kcomb \cdot x \cdot y  &\stepto x
% \end{align*}
% The relation $\reduce$ is then defined, as usual, by applying substitution instances 
% of $\stepto$ inside arbitrary context
% \[
% \infer{\subst{t}{\substone} \reduce \subst{s}{\substone}}{t \stepto s}
% \qquad 
% \infer{t \cdot u \reduce s \cdot u}{t \stepto s}
% \qquad
% \infer{u \cdot t\reduce u \cdot s}{t \stepto s}
% \]
To obtain a \quant{} refinement of system $\bck$, 
we assign distances in $[0,\infty]$ to basic rewriting rules, this way obtaining 
the \quant{} rewriting relation $\stepto_{\combrelone}$ defined by\footnote{
We use the same notational conventions introduced for string rewriting systems.}
% \begin{align*}
%     0 \Vdash \Bcomb \cdot x \cdot y \cdot z &\stepto x \cdot (y \cdot z)
%     \\
%     0 \Vdash \Ccomb \cdot x \cdot y \cdot z &\stepto x \cdot z \cdot y
%     \\
%     0 \Vdash \Kcomb \cdot x \cdot y  &\stepto x,
% \end{align*}
\begin{align*}
   \Bcomb \cdot x \cdot y \cdot z &\qstepto{0}_{\combrelone} x \cdot (y \cdot z)
    \\
    \Ccomb \cdot x \cdot y \cdot z &\qstepto{0}_{\combrelone} x \cdot z \cdot y
    \\
    \Kcomb \cdot x \cdot y  &\qstepto{0}_{\combrelone} x,
\end{align*}
and then extending $\stepto_{\combrelone}$ to $\to_{\combrelone}$ by 
non-expansively propagating distances produced by substitution instances of $\stepto_{\combrelone}$ 
throughout arbitrary contexts of the language.
% \[
% \infer{\mrel{a}{\subst{t}{\substone}}{\reduce}{\subst{s}{\substone}}}{\mrel{a}{t}{\stepto}{s}}
% \qquad 
% \infer{\mrel{a}{t \cdot u}{\reduce}{s \cdot u}}{\mrel{a}{t}{\stepto}{s}}
% \qquad
% \infer{\mrel{a}{u \cdot t}{\reduce}{u \cdot s}}{\mrel{a}{t}{\stepto}{s}}.
% \]
\[
    \infer{C[t^\sigma] \qreduce{\qone}_{\combrelone} C[s^\sigma]}{t \qstepto{\qone}_{\combrelone} s}
\]

Although \quant, the system thus obtained can only produce trivial distances (i.e. either $0$ or 
$\infty$), 
since no rewriting rule creates non-zero 
distances. In the next section, we will introduce effectful \quant{} extensions of $\bck$. 
For the moment, we simply extend $\bck$ with (the combinatory 
counterpart of) system $\NATS = (\signature_{\NATS}, \stepto_{\natrelone})$
of \autoref{sect:natural-numbers}. That is, we add to $\bck$ natural numbers and addition.
Even if 
the resulting system is not particularly interesting from a programming language perspective, 
it gives us the chance to illustrate the role 
played by linearity in \quant{} and metric reasoning. 
We thus consider the three additional basic combinators: 
$\Zcomb$, $\Scomb$, and $\Acomb$ for zero, successor, and addition, respectively. 
The system $\BCKNATS = (\signature_{\BCKNATS}, \stepto_{\combnatrelone})$ 
of affine combinators with natural numbers is given by 
the signature $\signature_{\BCKNATS} \defeq \signature_{\BCK} \cup \{\Zcomb,\Scomb, \Acomb\}$ 
and the \quant{} rewriting relation defined thus:

{
\centering
\begin{tcolorbox}[boxrule=0.5pt,width=(\linewidth*3)/4,colframe=black,colback=black!0!white,arc=0mm]
\vspace{-0.3cm}
\[
     \Bcomb \cdot x \cdot y \cdot z \qstepto{0}_{\combnatrelone} x \cdot (y \cdot z)
     \qquad
    \Ccomb \cdot x \cdot y \cdot z \qstepto{0}_{\combnatrelone} x \cdot z \cdot y
    \qquad
    \Kcomb \cdot x \cdot y  \qstepto{0}_{\combnatrelone} x
    \]
    \vspace{-0.2cm}
    \[
    \Acomb \cdot x \cdot \Zcomb \qstepto{0}_{\combnatrelone} x
    \qquad
    \Acomb \cdot x \cdot (\Scomb \cdot y) \qstepto{0}_{\combnatrelone} \Scomb \cdot (\Acomb \cdot x \cdot y)
    \qquad
    \Scomb \cdot x \qstepto{1}_{\combnatrelone} x
    \]
    \vspace{-0.2cm}
    \[
    \infer{C[t^\sigma] \qreduce{\qone}_{\combnatrelone} C[s^\sigma]}{t \qstepto{\qone}_{\combnatrelone} s}
    \]
\end{tcolorbox}
}
As usual $\stepto_{\combnatrelone}$ induces a distance $\combnatrelone$ on combinators defined by
$
\combnatrelone(t,s) \defeq \inf\{\qone \mid t \qreduce{\qone}_{\combnatrelone} s\},
$
from which we obtain a pesudometric $\makedistance{\combnatrelone}$. 
Let us now turn our attention to the definition of $\reduce_{\combnatrelone}$. 
Given a \quant{} rewriting relation $\stepto_{\vrelone}$, all the systems 
considered so far define $\reduce_{\vrelone}$ by forcing \emph{non-expansiveness} 
of contexts and substitution (cf. \quant{} equational theories).
System $\BCKNATS$ is no exception. 
The defining rules of $\reduce_{\combnatrelone}$ ensures that the application operation is non-expansive 
with respect to $\makedistance{\combnatrelone}$. Formally:
% $$
% \vrelone(t,s) \geq \vrelone(u \cdot t, u \cdot s)
% \text{ and }
% \vrelone(t,s) \geq \vrelone(t \cdot u, s \cdot u).
% $$
% This means, in particular, that if $t \qreduce{\qone} t'$ and $s \qreduce{\vtwo} s'$, then 
% $t\cdot s \qreducee{\qone + \qtwo} t' \cdot s'$, which gives the following non-expansiveness result:
$$
\makedistance{\combnatrelone}(t,t') + \makedistance{\combnatrelone}(s,s') \geq 
\makedistance{\combnatrelone}(t\cdot s, t' \cdot s').
$$
In particular, 
%if $t \qreduce{\qone} t'$ and $s \qreduce{b} s'$, then 
%$t\cdot s \stackrel{\qone + \qtwo}{\reduce^*} t' \cdot s'$. 
if $\mrel{\qone}{\termone}{\reduce_{\combnatrelone}}{\termone'}$ and
$\mrel{\qtwo}{\termtwo}{\reduce_{\combnatrelone}}{\termtwo'}$, then 
$\mrel{\qone+\qtwo}{\termone \cdot \termtwo}{\reduce_{\combnatrelone}}{\termone' \cdot \termtwo'}$.

Non-expansiveness, however, does not come for free: it is a direct consequence of \emph{linearity}\footnote{
The word \emph{linearity} is used both in rewriting and in logic with different, although similar, 
meaning. For the moment, we use it informally to indicate the absence of variable duplication, 
leaving formal definitions to the technical part of this paper.} 
of $\BCKNATS$. In fact, the addition of a non-linear combinator such as $\Wcomb$ directly leads to 
breaking non-expansiveness (in the sense that forcing non-expansiveness 
leads to undesired results, such as distance trivialisation and non-confluence). 
To see that, let us add the basic combinator $\Wcomb$ together with the following rewriting rule
to our system.
$$
\Wcomb \cdot x \cdot y \qstepto{0}_{\combnatrelone} x \cdot y \cdot y
$$
We now show that the presence of $\Wcomb$ makes \quant{} reasoning trivial. 

\begin{notation}
Let us write $\code{n}$ for the combinator $\Scomb \cdot ( \cc \cdot (\Scomb \cdot \Zcomb))$, 
with $n$ applications of $\Scomb$,
so that $\makedistance{\combnatrelone}(\code{n}, \code{m}) = |n - m|$ and 
$\makedistance{\combnatrelone}(\Acomb \cdot \code{n} \cdot \code{m}, \code{n+m}) = 0$.
\end{notation}

\begin{proposition}[Distance Trivialisation]
\label{prop:distance-trivialisation}
In presence of the combinator $\Wcomb$, the convertibility distance $\makedistance{\combnatrelone}$ 
trivialises, meaning that the distance $\makedistance{\combnatrelone}(t,s)$ is either $0$ 
or $\infty$, for all combinators $t,s$.
\end{proposition}

\begin{proof}
Given combinators $t,s$, let $\makedistance{\combnatrelone}(t,s) = \qone$. 
If $\qone$ is $\infty$, we are done. Otherwise, $\qone$ is a natural number,
%We prove $\makedistance{\combrelone}(t,s) = \infty$ by contradiction. 
%So assume $\makedistance{\combrelone}(t,s) = \qone < \infty$, for some $\qone$. 
since the defining rule of $\stepto_{\combnatrelone}$
ensures the codomain of $\makedistance{\combnatrelone}$ to actually be 
$\mathbb{N}^{\infty}$.  
Consequently, we have combinators $\code{m}$, $\code{n}$ such that 
$\makedistance{\combnatrelone}(\code{m}, \code{n}) = \qone$.
Notice also that whenever we have combinators $t,t',s,s'$ such that
$\makedistance{\combnatrelone}(t,t') = 0$ and $\makedistance{\combnatrelone}(s,s') = 0$, then 
$\makedistance{\combnatrelone}(t,s) = \makedistance{\combnatrelone}(t',s')$. Thus, for instance, 
we see that 
$$
\makedistance{\combnatrelone}(\Wcomb \cdot t \cdot s, \Wcomb \cdot t' \cdot s') = 
\makedistance{\combnatrelone}(t \cdot s \cdot s, t' \cdot s' \cdot s').
$$
Non-expansiveness of application then gives (using $\makedistance{\combnatrelone}(\Wcomb,\Wcomb) = 0$ 
and $\makedistance{\combnatrelone}(\Acomb,\Acomb) = 0$):
\begin{align*}
    \makedistance{\combnatrelone}(\code{n}, \code{m}) 
    &\geq \makedistance{\combnatrelone}(\Wcomb \cdot \Acomb \cdot \code{n}, \Wcomb \cdot \Acomb \cdot \code{m}) 
    \\
    &= \makedistance{\combnatrelone}(\Acomb \cdot \code{n} \cdot \code{n}, \Acomb \cdot \code{m} \cdot \code{m})
    \\
    &= \makedistance{\combnatrelone}(\code{n+n}, \code{m+m}),
\end{align*}
meaning that we have $\qone \geq \qone + \qone$. In $\mathbb{N}^{\infty}$ this is possible only for 
$0$ and $\infty$, and thus we conclude $\makedistance{\combnatrelone}(t,s)=0$.
\end{proof}
% Non-expansiveness of application then gives (using $\makedistance{\vrelone}(\Wcomb,\Wcomb) = 0$ 
% and $\makedistance{\vrelone}(\mathbf{A},\mathbf{A}) = 0$):
% \begin{align*}
%     \makedistance{\vrelone}(\code{n}, \code{m}) 
%     &\geq \makedistance{\vrelone}(\Wcomb \cdot \mathbf{A} \cdot \code{n}, \Wcomb \cdot \mathbf{A} \cdot \code{m}) 
%     \\
%     &= \makedistance{\vrelone}(\mathbf{A} \cdot \code{n} \cdot \code{n}, \mathbf{A} \cdot \code{m} \cdot \code{m})
%     \\
%     &= \makedistance{\vrelone}(\code{2n}, \code{2m}).
% \end{align*}
% It then follows 
% $$
% |n - m| 
% = \makedistance{\vrelone}(\code{n}, \code{m}) 
% \geq \makedistance{\vrelone}(\code{2n}, \code{2m})
% = 2|n-m|
% $$
% which does not make sense for non-trivial (i.e. neither $0$ nor $\infty$) distances. 

\autoref{prop:distance-trivialisation} is known as \emph{distance trivialisation} or 
\emph{distance amplification}
\cite{DBLP:phd/basesearch/Gavazzo19,CrubilleDalLago/ESOP/2017} and it has been deeply 
investigated studying (effectful) program distancing. 
Linearity, and variations thereof,
are a way to avoid trivialisation of \quant{} reasoning. Additionally, we shall see 
in \autoref{sect:qtrs-confluence} that 
 linearity is also crucial to ensure \quant{} forms of confluence, 
the latter being, together with distance trivialisation, the reason why
in \autoref{sect:qtrs} we 
will focus on \emph{linear non-expansive} rewriting systems. 
In \autoref{sect:beyond-non-expansive-systems}, we will see how moving to 
\emph{graded (modal) systems} gives us a way to go beyond the linearity assumption 
and refine non-expansive systems to \emph{Lipschitz continuous} ones.
 
 \subsubsection{Effectful Combinatory Logic}

% \begin{align*}
% \makedistance{\vrelone}(t,t') + \makedistance{\vrelone}(s,s')
% &\geq 
% \makedistance{\vrelone}(\Wcomb \cdot t \cdot s, \Wcomb \cdot t' \cdot s') 
% &= \makedistance{\vrelone}(t, t') + \makedistance{\vrelone}(s, s')
% \end{align*}

% For, 
% \begin{align*}
%  \makedistance{\vrelone}(\Wcomb \cdot t \cdot s, \Wcomb \cdot t' \cdot s') 
%  &= 0 + \makedistance{\vrelone}(\Wcomb \cdot t \cdot s, \Wcomb \cdot t' \cdot s') + 0
%  \\
%  &= \makedistance{\vrelone}(t \cdot s \cdot s, \Wcomb \cdot t \cdot s) + 
%  \makedistance{\vrelone}(\Wcomb \cdot t \cdot s, \Wcomb \cdot t' \cdot s') + 
%  \makedistance{\vrelone}(\Wcomb \cdot t' \cdot s', t' \cdot s' \cdot s')
%  \\
%  & \geq \makedistance{\vrelone}(t \cdot s \cdot s, t' \cdot s' \cdot s').
% \end{align*}
% Non-expansiveness of application then gives
% $$
% \makedistance{\vrelone}(\Wcomb \cdot t \cdot s, \Wcomb \cdot t' \cdot s') 
% \leq \makedistance{\vrelone}(t, t') + \makedistance{\vrelone}(s, s')
% $$

System $\BCKNATS$ of affine combinators and arithmetic allowed us to highlight the role 
of linearity and non-expansiveness in \quant{} reasoning. Apart from that, system $\BCKNATS$ is not 
particularly interesting. Here, we extend affine combinators 
with \quant{} algebraic theories modelling computational effects~\cite{plotkin-quantitative-algebras-2016,plotkin-quantitative-algebras-2017,plotkin-quantitative-algebras-2018,plotkin-quantitative-algebras-2018-bis,plotkin-quantitative-algebras-2021,plotkin-quantitative-algebras-2021-bis,DBLP:conf/lics/MioSV21}. 

Probabilistic~\cite{foundations-probabilistic-programming} and, more generally, 
effectful programming languages 
have been extensively studied in the last decade using, among others, 
probabilistic~\cite{Bournez-2002,Bournez-2005,Faggian-2019,Faggian-Ronchi-2019,dal-lago-avanzini-yamada} 
and effectful~\cite{Gavazzo-Faggian-2021} rewriting systems. 
Such systems come in two flavours, depending on whether effects 
are considered internally or externally to the system. In the latter case, 
one obtains \emph{probabilistic}~\cite{Bournez-2002,Bournez-2005,dal-lago-avanzini-yamada} and 
\emph{monadic rewriting systems}~\cite{Gavazzo-Faggian-2021}, 
In the former case, instead, one models (equational theories 
defining) computational effects themselves as 
rewriting systems and then combines the latter with
the actual 
calculus or programming language at hand, which is modelled as a rewriting system itself. 
Here, we follow the later approach and look 
at computational effects as defined by \quant{} equational theories~\cite{plotkin-quantitative-algebras-2016,plotkin-quantitative-algebras-2017}.

% Before doing that, however, we shortly introduce another \quant{} variation of 
% $\bck$ showing that interesting \quant{} analyses can be obtained by looking 
% at \emph{intensional} features of combinators. 

% \paragraph{The Landauer Distance} 
% According to Landauer's principle [REF] the energetic cost of computation 
% can be defined in terms of logical entropy which in turn shows that energy 
% is consumed only when information is erased. Moving from that observation, 
% researchers have developed theories of reversible computing --- i.e. 
% computations where information is never erased --- which established 
% interesting connections with logical linearity. Accordingly, 
% since a linear program does not neither erase nor duplicate data, 
% it is reversible and thus energy efficient. In system $\bck$, 
% the combinators $\Bcomb$ and 
% $\Ccomb$ are linear, whereas $\Kcomb$ erases its first input, and thus its execution 
% is non-reversible and thus it consumes energy. However, instead of forbidding 
% its use, we can use a \quant{} rewriting system to track how $\Kcomb$ is used. 
% To do so, we simply refine our system as follows, where $\ell$ is a cost unit 
% for data erasing.
% \begin{align*}
%     0 \Vdash \Bcomb \cdot x \cdot y \cdot z &\stepto x \cdot (y \cdot z)
%     \\
%     0 \Vdash \Ccomb \cdot x \cdot y \cdot z &\stepto x \cdot z \cdot y
%     \\
%     \ell \Vdash \Kcomb \cdot x \cdot y  &\stepto x,
% \end{align*}

\paragraph{Barycentric Algebras}

Let us begin with one of the main examples of a quantitative equational 
theory: \emph{barycentric algebras}. 
Barycentric algebras have been introduced by \citet{stone49} as an equational axiomatisation 
of finite distributions, and they have recently refined as a \emph{\quant{} 
equational theory} by \citet{plotkin-quantitative-algebras-2016,plotkin-quantitative-algebras-2017}. 
Here, we present such a \quant{} refinement 
directly as a \quant{} term rewriting system.
Let us consider a signature $\signature_{\BA}$ 
containing a family of binary probabilistic choice 
operations $\barplus{\probone}$ indexed by rational numbers $\probone \in \mathbb{Q} \cap [0,1]$. 
The \quant{} term rewriting system $\BA = (\signature_{\BA}, \stepto_{\baryrelone})$ 
of Barycentric algebras is defined thus:
% \begin{align*}
%     x \barplus{1} y &\qequal{0} x
%     \\
%     x \barplus{\probone} x &\qequal{0} x
%     \\
%     x \barplus{\probone} y &\qequal{0} y \barplus{1 - \probone} x
%     \\
%     (x \barplus{\probone_1} y) \barplus{\probone_2} z &= 
%     x \barplus{\probone_1 \probone_2} (y \barplus{\frac{\probone_1 - \probone_1\probone_2}{1 - \probone_1\probone_2}} z) 
%     \qquad  where \probone_1, \probone_2 \in (0,1)
% \end{align*}
% \begin{align*}
%             0 \Vdash x \barplus{1} y &\stepto x & 
%             \\
%     %         0 \Vdash  x \barplus{\probone} x &\stepto x &
%     %        \\
%             0 \Vdash  x \barplus{\probone} y &\stepto y \barplus{1 - \probone} x &
%             \\
%              0 \Vdash  (x \barplus{\probone_1} y) \barplus{\probone_2} z &\stepto
%             x \barplus{\probone_1 \probone_2} (y \barplus{\frac{\probone_1 - \probone_1\probone_2}{1 - \probone_1\probone_2}} z) 
%             & \probone_1, \probone_2 \in (0,1)
%             \\
%              \qone \Vdash x \barplus{\probone} y &\stepto z \barplus{\probone} y
%             & \probone \leq \qone \in \mathbb{Q} \cap [0,1]
%  \end{align*}

{
\centering
\begin{tcolorbox}[boxrule=0.5pt,width=(\linewidth*3)/4,colframe=black,colback=black!0!white,arc=0mm]
\vspace{-0.3cm}
\begin{align*}
             x \barplus{1} y &\qstepto{0}_{\baryrelone} x & 
            \\
    %         0 \Vdash  x \barplus{\probone} x &\stepto x &
    %        \\
            x \barplus{\probone} y &\qstepto{0}_{\baryrelone} y \barplus{1 - \probone} x &
            \\
              (x \barplus{\probone_1} y) \barplus{\probone_2} z &\qstepto{0}_{\baryrelone}
            x \barplus{\probone_1 \probone_2} (y \barplus{\frac{\probone_1 - \probone_1\probone_2}{1 - \probone_1\probone_2}} z) 
            & \probone_1, \probone_2 \in (0,1)
            \\
             x \barplus{\probone} y &\qstepto{\qone}_{\baryrelone} z \barplus{\probone} y
            & \probone \leq \qone \in \mathbb{Q} \cap [0,1]
 \end{align*}
\end{tcolorbox}
}

 Notice that system $\BA$ does not have the idempotency rule 
 $ x \barplus{\probone} x \qstepto{0}_{\baryrelone} x$, meaning that we are actually modelling 
 \emph{multi-distributions}~\cite{dal-lago-avanzini-yamada} rather than distributions: 
 this guarantees linearity of $\BA$ 
 and thus agrees with the definition of a probabilistic rewriting system by \citet{dal-lago-avanzini-yamada}, 
 which is indeed based on multi-distributions.
 The operation $\barplus{\probone}$ behaves as an unfair (binary) probabilistic choice operation weighted by 
$\probone$, so that we can read $x \barplus{\probone} y$ as stating that we have $x$ 
with probability $\probone$ and $y$ with probability $1 - \probone$. 
Accordingly, for a set $\variables$ of variables, 
a term in $\terms{\signature_{\BA}}{\variables}$ 
 can be 
seen as a finite \emph{formal sum}, i.e. a syntactic representation of a finitely supported distribution. 

As usual, starting from $\stepto_{\baryrelone}$, we obtain the rewriting relation $\reduce_{\baryrelone}$,
the rewriting distance $\baryrelone$, and the convertibility distance $\makedistance{\baryrelone}$. 
Remarkably, the latter is precisely the 
\emph{total variation distance}~\cite{Villani/optimal-transport/2008} between multi-distributions 
(see \autoref{sect:beyond-non-expansive-systems} for another example of a probabilistic distance).
We can then combine systems $\BA$ and 
$\BCK$ (or even $\BCKNATS$), this way obtaining 
the \quant{} term rewriting system $\BCKprob = (\signature_{\bck} \cup \signature_{\BA}, 
\stepto_{\combrelone\baryrelone})$ for
probabilistic affine combinatory logic,\footnote{Another, more powerful, system is obtained 
by modelling operations $\barplus{\probone}$ as combinators.} as 
 summarised in \autoref{figure:probabilistic-bck}.
\begin{figure}
{
\centering
\begin{tcolorbox}[boxrule=0.5pt,width=\linewidth,colframe=black,colback=black!0!white,arc=0mm]
\vspace{-0.3cm}
   \begin{align*}
   \Bcomb \cdot x \cdot y \cdot z &\qstepto{0}_{\combrelone\baryrelone} x \cdot (y \cdot z) 
    &   (x \barplus{\probone_1} y) \barplus{\probone_2} z &\qstepto{0}_{\combrelone\baryrelone}
            x \barplus{\probone_1 \probone_2} (y \barplus{\frac{\probone_1 - \probone_1\probone_2}{1 - \probone_1\probone_2}} z) 
            & \probone_1, \probone_2 \in (0,1) 
    \\
    \Ccomb \cdot x \cdot y \cdot z &\qstepto{0}_{\combrelone\baryrelone} x \cdot z \cdot y 
    &   x \barplus{\probone} y &\qstepto{\qone}_{\combrelone\baryrelone} z \barplus{\probone} y
            & \probone \leq \qone \in \mathbb{Q} \cap [0,1]
    \\
     \Kcomb \cdot x \cdot y  &\qstepto{0}_{\combrelone\baryrelone} x 
    &  x \barplus{1} y &\qstepto{0}_{\combrelone\baryrelone} x & 
\\
    & &   x \barplus{\probone} y &\qstepto{0}_{\combrelone\baryrelone} y \barplus{1 - \probone} x &
 \end{align*}
 \end{tcolorbox}
% \[
% \infer{\mrel{a}{\subst{t}{\substone}}{\reduce}{\subst{s}{\substone}}}{\mrel{a}{t}{\stepto}{s}}
% \qquad 
% \infer{\mrel{a}{t \cdot u}{\reduce}{s \cdot u}}{\mrel{a}{t}{\stepto}{s}}
% \qquad
% \infer{\mrel{a}{u \cdot t}{\reduce}{u \cdot s}}{\mrel{a}{t}{\stepto}{s}}
% \qquad 
% \infer{\mrel{\qone}{t \barplus{\probone}u}
% {\reduce}{s \barplus{\probone} u}}{\mrel{\qone}{t}{\stepto}{s}}
% \qquad
% \infer{\mrel{\qone}{u \barplus{\probone} t}{\reduce}{u \barplus{\probone}s}}{\mrel{\qone}{t}{\stepto}{s}}.
% \]
}
\caption{The \quant{} rewriting relation $\stepto_{\combrelone\baryrelone}$}
 \label{figure:probabilistic-bck}
\end{figure}
In particular, the convertibility distance induced by $\stepto_{\combrelone\baryrelone}$ 
is essentially $\makedistance{(\combrelone \min \baryrelone)}$, where 
$(\combrelone \min \baryrelone)(t,s) \defeq \min(\combrelone(t,s), \baryrelone(t,s))$, 
which gives the total variation distance between probabilistic 
combinators. That puts together the usual equational theory of combinators 
with the \quant{} analysis of probabilistic choice, this way giving a \quant{} 
theory of probabilistic (affine) computation. 
In light of that, \quant{} rewriting properties and metric word problems 
become interesting both for the (\quant{}) equational theory 
of probabilistic affine computations (is the theory consistent? Is it decidable or 
semi-decidable?) and for its operational semantics (is reduction confluent? Do 
we have an optimal strategy?).

% At this point, several nontrivial theoretical questions arises: is the 
% rewriting relation confluent? Does that imply that the quantitative theory is 
% consistent? Do we have optimal strategies? 

In this paper, we shall prove confluence of $(\signature_{\bck} \cup \signature_{\mathbf{BA}}, 
\stepto_{\combrelone\baryrelone})$. Consequently, we will obtain consistency of its 
(\quant{}) equational theory and semi-decidability of the reachability problem. 
Achieving such a result is nontrivial and requires 
the introduction of several new results on \quant{} rewriting system. In particular, 
we will prove confluence in a modular fashion relying on a suitable \quant{} 
refinement of the Hindley-Rosen Lemma~\cite{hindley-1964,rosen-70} (\autoref{lemma:hindley-rosen}) 
and proving confluence of $(\signature_{\bck}, \stepto_{\combrelone})$ and 
$(\signature_{\mathbf{BA}}, \stepto_{\baryrelone})$ separately (\autoref{sect:qtrs-confluence-part-2}), 
the latter requiring the extension of 
critical pair-like lemmas~\cite{Huet80} to \quant{} rewriting systems (\autoref{sect:qtrs-confluence} and \autoref{sect:qtrs-confluence-part-2}).

\paragraph{Ticking}
Barycentric algebras are just \emph{one} example of a \quant{} algebraic theory used 
to model computational effects. Other examples include the theory of 
\quant{} semilattices~\cite{plotkin-quantitative-algebras-2016} (whose associated distance is the Hausdorff distance), \quant{} global states, and \quant{} output \cite{bacci-mardare-panangaden-plotkin-2020}. 
Here, we introduce the \quant{} theory of ticking, a specific instance of \quant{} output 
used in improvement theory and cost analysis \cite{Sands/Improvement-theory/1998} 
to study intensional aspects of programs. 

Let us consider the monoid $(\mathbb{N}, +, 0)$ of natural numbers with addition 
endowed with the Euclidean distance.\footnote{
A more general definition can be given by fixing a \quant{} output monoid, that is 
a monoid endowed with a generalised distance~\cite{Lawvere/GeneralizedMetricSpaces/1973} 
making monoid multiplication non-expansive. Besides the monoid of natural numbers with the Euclidean distance, 
another classic example of \quant{} output monoid is given by 
the monoid of words over an alphabet endowed with the least common prefix distance.} 
The (\quant{}) term rewriting system $\TICK = (\signature_{\TICK}, \stepto_{\tickrelone})$ 
of ticking is defined by the signature 
$\signature_{\TICK}$ of unary 
operation symbols $\writeop{n}{(\cdot)}$ indexed by elements $n \in \mathbb{N}$ 
and the following rewriting rules:

{
\centering
\begin{tcolorbox}[boxrule=0.5pt,width=(\linewidth*2)/3,colframe=black,colback=black!0!white,arc=0mm]
\vspace{-0.3cm}
    \begin{align*}
        \writeop{0}{x} &\qstepto{0}_{\tickrelone} x &
        \\
         \writeop{n}{(\writeop{m}{x})} & \qstepto{0}_{\tickrelone} \writeop{(n+m)}{x} &
        \\
        \writeop{n}{x} & \qstepto{\qone}_{\tickrelone} \writeop{m}{x} & \qone \geq E(n,m)
    \end{align*}
\end{tcolorbox}
}
    The operation $\writeop{n}{\termone}$ can be informally read as \emph{count $n$ unit of cost, 
    then continue as $\termone$}. Oftentimes, one writes terms of the form $\writeop{1}{\termone}$ 
    as $\checkmark \termone$ and decorates programs with $\checkmark$ annotations 
    to count computation steps (for instance, in systems based on the 
    $\lambda$-calculus or combinatory logic, applications $\termone \cdot \termtwo$ 
    are decorated as $\checkmark(\termone \cdot \termtwo)$: this way, one measures the 
    cost of a computation as the numbers of applications performed).\footnote{
    The ticking operation can be seen as a particular instance of an output operation, 
    where the output produced is the cost of computation. 
    \citet{bacci-mardare-panangaden-plotkin-2020} 
    have shown 
    that the \quant{} equational theory associated to this reading of ticking is exactly the 
    theory of the \quant{} writer monad, which specialises to the one of the cost or ticking monad 
    \cite{DBLP:conf/esop/LagoG19,Sands/Improvement-theory/1998}.} 
    In this case, we actually obtain a simplified system $\TICK_{\checkmark}$ whose signature 
    contains the unary function symbol $\checkmark$ only 
    and whose (unique) rewriting rule is the following 
    \begin{align*}
        \checkmark x &\qstepto{1} x 
    \end{align*}
Let us now come back to system $\TICK$.
    The first two rewriting rules of system $\TICK$ model null cost 
    production and cost sequencing, whereas the last rule allows us to measure differences between cost traces. 
    Accordingly, variations of $\TICK$ are obtained by changing the way we measure cost differences. 
    For instance, having in mind program refinement, one may want to to replace the Euclidean distance 
    with its asymmetric counterpart. 
    Finally, we can combine systems $\TICK$ and $\BCK$ --- or even $\BCKprob$ --- together, 
    this way obtaining systems for the \quant{} cost analysis of affine and probabilistic 
    computations. 
    We can then (and we will) prove confluence of the resulting systems
    \emph{compositionally}
    relying on the 
    \quant{} Hindley-Rosen lemma (\autoref{lemma:hindley-rosen}) and  
    proving confluence of each system separately.
    
\subsection{Further Examples and Where to Find Them}
    The systems we have seen so far are just \emph{some} of the many examples of \quant{} rewriting 
    systems one can either find in the literature 
    or design independently. For instance, several \quant{} equational theories of computational effects 
    have been recently developed in addition to the ones introduced in this motivational section. 
    Examples of those include \quant{} nondeterminism (describing the 
    Hausdorff distance between sets) \cite{plotkin-quantitative-algebras-2016}, 
    global stores~\cite{bacci-mardare-panangaden-plotkin-2020},
    and combined pure-probabilistic nondeterminism~\cite{DBLP:conf/lics/MioSV21}. 
    All these theories can be analysed operationally as \quant{} rewriting systems and combined 
    with the systems introduced so far.
    
    A further source of examples follows the line of 
    \autoref{sect:string-rewriting}, where we have provided operational descriptions of 
    several edit distances (e.g. the Hamming and Levenshtein distance)
    on (DNA) strings. 
    Indeed, \quant{} rewriting systems are particularly well-suited to model 
    edit distances --- not necessarily on strings --- operationally.  
    The \emph{Encyclopedia of Distances} by \citet{encyclopedia-of-distances} 
    is a great source of potential examples of (edit) distances that could be approached operationally. 
    Potential applications of
    \quant{} rewriting systems are given by optimisation theory~\cite{optimization}, 
    where one naturally deals with weighted graphs and searches optimal paths. 
    By characterising such weighted graphs as the reduction graphs of \quant{} rewriting systems, 
    one may give a more symbolic account to optimisation. 
    
    Numeric and approximated computation provide interesting examples of \quant{} rewriting system 
    too. In fact, several computer algebra systems allow the user to combine symbolic 
    and numeric computation.\footnote{See, for instance, the library 
    \texttt{SymPy} (\url{https://www.sympy.org/en/index.html}).} 
    It then seems natural to consider exact rewriting to model the symbolic 
    part of a computation and \quant{} rewriting to model the numeric one, 
    as the latter naturally involves numerical approximations and precision errors. 
    For instance, the numerical evaluation of the symbolic constant $\pi$ 
    to, e.g., the numerical approximation $3.14$ could be modelled as 
    the reduction $\pi \qreduce{\qone} 3.14$, with $\qone$ the error produced 
    in the approximation.

    The reader should now be sufficiently familiar with examples of \quant{} 
    rewriting systems and basic ideas behind them. 
    The rest of the paper is devoted to 
    introduce the general theory of \quant{} and metric rewriting in full detail, 
    starting from abstract and then moving to \quant{} term rewriting systems. 
    In doing so, we also analyse the examples seen in this introductory section (as well 
    as new ones) formally.

\section{PRELIMINARIES: QUANTITATIVE RELATIONAL CALCULUS, \textnormal{\emph{\`a la}} LAWVERE} 
\label{sect:quantales}

We begin our analysis of \quant{} rewriting by developing a theory of \emph{\quant{} abstract rewriting systems}. 
To do so, we first recall some mathematical preliminaries.  

\paragraph{Quantales}
Traditional abstract rewriting systems can be naturally defined and studied 
\emph{relationally}. To define a theory of quantitative rewriting, it thus seems natural 
to rely on \emph{quantitative relational calculi}. Here, we follow the analysis 
of generalised metric spaces as enriched categories by \citet{Lawvere/GeneralizedMetricSpaces/1973} and work with relations taking values 
in a quantale \cite{Rosenthal/Quantales/1990}. Quantale-valued relations are 
extensively used in monoidal topolgy \cite{Hoffman-Seal-Tholem/monoidal-topology/2014} 
and they have been successfully applied to define metric and \quant{} semantics 
of higher-order languages~\cite{Gavazzo/LICS/2018,DBLP:phd/basesearch/Gavazzo19}, as well behavioural metrics~\cite{Worrell-omega-categories,paul-wild-2022}.
Let us begin recalling the definition of a quantale, which we view 
as modelling abstract quantities.

% first define a quantitative relational calculus that  generalises the classical, boolean-valued relational calculus. We have two possible ways of proceeding, either through the notion of modal relation or by resorting to the notion of quantal-valued relation. We introduce both these notions and verify that they are interchangeable. 
% \subsection{Modal Relations}
% \cecilia{qui devo scrivere per bene, ho solo appuntato alcune cose} Consider $(W,\leq, +, 0)$ consisting of a monoid $(W, +, 0)$ with unit $0$ and a pre-ordered possible word set $(W,\leq)$ (we can consider the preorder relation as the accesebility relation). Given two sets $X,Y$ a modal relation on $W?$ ,is defined a ternary relation $\relone\subseteq W\times X\times Y$. If $(a,x,y)\in \relone$ we will write $\mrel{\alpha}{x}{\relone}{y} $. The identity relation on $X$ is $I=\{(\alpha, a, a)\mid a\in X\}$.The composition between two modal relations is defined as: given  $\mrel{\alpha}{a}{\relone}{b}$ and $\mrel{\beta}{b}{\reltwo}{c}$ if $\gamma\geq \alpha+\beta$ then $\mrel{\alpha}{a}{(\relone;\reltwo)}{c}$. The tensor between  two relations is defined as: given $\mrel{\alpha}{a}{\relone}{b}$ and $\mrel{\beta}{a}{\reltwo}{b}$ if $\gamma\geq \alpha+\beta$ then $\mrel{\alpha}{a}{(\relone\tensor \reltwo)}{c}$.
% \subsection{Quantales and Quantale-valued Relations}

\begin{definition}
A (unital) quantale $\Quantale = (\quantale, \leq, \qunit, \tensor)$ consists of a monoid $(\quantale,  \qunit, \tensor)$ and a sup-semilattice $(\quantale,  \leq)$ satisfying the following distributivity laws:
\begin{align*}
    \qvaltwo \tensor \join_{i\in I} \qvalone_i  
    &= \join_{i \in I} (\qvaltwo \tensor \qvalone_i)
    \\
    (\join_{i \in I} \qvalone_i) \tensor \qvaltwo
    &= \join_{i \in I} (\qvalone_i \tensor \qvaltwo).
  \end{align*}
The element $\qunit$ is called unit of the quantale, whereas $\tensor$ is called 
the tensor (or multiplication) 
of the quantale. 
% Given quantales $\quantale$, $\quantaletwo$, a quantale lax morphism is a monotone map $h : \quantale \to \quantaletwo$ satisfying the following inequalities:
% $$l \leq h(k)$$
% $$h(\qvalone) \tensor h(\qvaltwo) \leq h(\qvalone \tensor \qvaltwo)$$ 
% where $l$ is the unit of $\quantaletwo$.
\end{definition}

 It is easy to see that $\tensor$ is monotone in both arguments. 
 We denote the top and bottom element of a quantale by and $\top$, $\qbot$ respectively. 
 Quantales having unit $\qunit$ coinciding with the top element are called \emph{integral} quantales. 
 Moreover, we say that a quantale is commutative if its underlying monoid is, 
 and that it is non-trivial if $\qunit \neq \qbot$. 
Integral quantales
are particularly well-behaved:
for instance, in an 
integral quantale $\qone_1 \tensor \qone_2$ is a lower bound 
of each $\qone_i$.\footnote{ 
By monotonicity of $\tensor$, we have: 
$
\qone_1 \tensor \qone_2 \leq \qone_i \tensor \top 
= \qone_i \tensor \qunit = \qone_i.
$
}
Additionally, in an integral quantale we have 
$\qone \tensor \qbot = \qbot$, for any $\qone \in \quantale$. 
If the opposite direction holds, i.e. whenever 
$\qone \tensor \qtwo = \qbot$, either $\qone = \qbot$ or 
 $\qtwo = \qbot$ holds, we say that the quantale is \emph{cointegral}.

From now on, we assume quantales to be 
 commutative, (co)integral, and non-trivial.
 We refer to such quantales as \emph{Lawvereian}. 
Finally, we say that a quantale is \emph{idempotent} if $\qone \tensor \qone = \qone$. 
 Notice that any quantale $(\quantale, \leq, \qunit, \tensor)$ induces an idempotent 
 quantale as $(\quantale, \leq, \top, \wedge)$ and that in any integral idempotent quantale 
$\wedge$ and $\tensor$ coincide.

\begin{example}
 \begin{enumerate}
  \item The \emph{boolean quantale}
  $\Two = (\two, \leq, \wedge, \true)$, where $\two = \{\true, \false\}$
  and $\false \leq \true$, is an idempotent Lawverian quantale. 
  %Actually, any complete Boolean algebra \cite{DaveyPriestley/Book/1990} 
  %is an idempotent quantale.
 \item Any frame\footnote{
    Recall that a frame \citep{Vickers/Topology-via-logic} 
   	consists of a sup lattice $(V, \leq, \join)$ satisfying the following 
   	distributivity laws:
   	\begin{align*}
    y\wedge \join_{i\in I} x_i  
    &= \join_{i \in I} (y \wedge x)
    &
    (\join_{i \in I} x_i) \wedge y
    &= \join_{i \in I} (x_i \wedge y).
  \end{align*}
  A main concrete example of a frame is the structure 
  $(\tau, \subseteq, \cap, X)$ given by the open sets $\tau$ of a topological space.} is an idempotent integral quantale. 
  If the frame is cointegral, then we obtain a Lawverian quantale. 
% \item The powerset $\powerset(M)$ of a commutative monoid 
%           $(M, \cdot, 1)$ with multiplication defined by 
%           $\mu \tensor \nu = \{m \cdot n \mid 
%           m \in \mu, n \in \nu\}$ is a Lawverian quantale. 
% \item The collection of binary relations on a set $X$ 
%   	forms a quantale with tensor product given by relation composition. 
%   	This quantale is not Lawverian (it is not even commutative).
  \item The \emph{Lawvere quantale} $\Lawvere = ([0, \infty], \geq, +,0)$ 
  	consisting of the extended real half-line 
    ordered by 
    the ``greater or equal'' relation $\geq$ and 
    extended\footnote{We extend 
    ordinary addition as follows: 
    $x + \infty \defeq \infty \defeq \infty + x$.} 
    addition as tensor product is a Lawverian quantale. 
    Notice that we use the opposite of the natural ordering, 
          so that, e.g., $0$ is the top element 
          of $\Lawvere$.
  \item The \emph{Strong Lawvere quantale}
    $\StrongLawvere = ([0,\infty], \geq, \max, 0)$ obtained by
    replacing addition with maximum in the Lawvere quantale is an idempotent Lawverian quantale/
    Notice that in the strong Lawvere quantale tensor and meet coincide, and thus the quantale is idempotent. 
  \item The unit interval $\UnitInterval = ([0,1], \leq, *)$ endowed with a left continuous 
  \emph{triangular norm}~\cite{fuzzy-metamathematics} ($t$-norm for short)\footnote{Recall that a $t$-norm 
    is a binary operator 
    $*: [0,1] \times [0,1] \to [0,1]$ that induces a quantale 
    structure over the complete lattice $([0,1], \leq)$ in 
    such a way that the quantale is commutative.} $*$ is an integral quantale.
    Examples of $t$-norms are:
    \begin{enumerate}
      \item The \emph{product $t$-norm}: $x *_p y \defeq x \cdot y$.
      \item The \emph{\L{}ukasiewicz $t$-norm}: $x *_l y \defeq \max\{x + y - 1, 0\}$.
      \item The \emph{G\"{o}del $t$-norm}: $x *_g y \defeq \min\{x,y\}$.
    \end{enumerate}
    If, additionally, $x * y = 0$ implies $x=0$ or $y=0$, then we obtain a Lawverian quantale. 
    In particular, both the product and G\"{o}del $t$-norms give Lawverian quantales.
    Such quantales are used to model Fuzzy reasoning, and thus we refer to
    $\UnitInterval = ([0,1], \leq, *,1)$ as the Fuzzy quantale(s).
\item 
    The set of \emph{monotone modal predicates} $2^W$ 
    on a preorder monoid with top element $(W, \leq, +, 0, \top)$ of possible worlds,
    endowed with the tensor product defined below, is a Lawverian quantale.
    $$(p \tensor q)(w) \iff \exists u,v.\ w \geq u + v \wedge 
    p(u) \wedge q(v).$$
    Such a quantale is used to study modal and coeffectful properties of 
    programs \cite{modal-reasoning-equal-metric-reasoning,DBLP:journals/pacmpl/LagoG22a}.
\item If in the previous example of modal predicates we replace $2$ with a 
    (Lawverian) fuzzy quantale $([0,1], \leq, *,1)$, then we obtain the Lawverian quantale 
    of \emph{Fuzzy modal predicates}.
\item The set $\mathbb{F} \defeq \{f \in [0,1]^{[0,\infty]} \mid f \text{ monotone and }
            f(a) = \join_{b < a} f(b)\}$ used by  \citet{HOFMANN20131} 
            to model probabilsitic metric spaces is a quantale. Notice that 
            elements of $\mathbb{F}$ can be seen as Fuzzy modal predciates over 
            the set of possible worlds $[0,\infty]$.
\end{enumerate}
\end{example}

To help the reader working with quantales, we summarise the
 correspondence between the Boolean ($\Two$), Lawvere ($\Lawvere$), and 
 Strong Lawvere ($\StrongLawvere$) quantale --- our main running examples --- as well as a generic quantale $\Quantale = (\quantale, \leq, \qunit, \tensor)$, in \autoref{table:correspondence-quantale}.
\begin{table*}[htbp]
\centering
\begin{tabular}{|c|c|c|c|c|}
	\hline
  & $\Two$ (Boolean) & $\Lawvere$ (Lawvere) & $\StrongLawvere$ (strong Lawvere)& 
  $\Quantale$ (quantale) \\
  \hline 
Carrier & $\two $	& $[0,\infty]$ & $[0,\infty]$ & $\quantale$ 
  \\
Order & $\leq$ 	& $\geq$ & $\geq$	& $\leq$
  \\
Join & $\exists$	& $\inf$ & $\inf$ 	& $\join$ 
  \\
Meet & $\forall$ 	&$\sup$ &$\sup$ 	& $\meet$
  \\
Tensor & $\wedge$ 	& $+$ & $\max$		& $\tensor$ 
  \\
Unit & $\true$	& $0$ & $0$	& $\qunit$				
  \\
  \hline
  %$\mathsf{false}$	& $\infty$ 		& $\qbot$.
\end{tabular}
%\vspace{-0.2cm}
\caption{Correspondence $\two$-$[0,\infty]$-$\quantale$.}
\label{table:correspondence-quantale}
\end{table*}

Since any quantale is, in particular, a complete lattice and tensor product 
is monotone in both arguments, the latter has both left and right adjoints 
(which coincide in our case, as we 
assume $\tensor$ to be commutative):
$$
\qone \tensor \qtwo \leq \qthree \iff \qtwo \leq \qone \mmap \qthree.
$$
Explicitly, we have $\qone \mmap \qtwo \defeq \join \{\qthree \mid \qone \tensor \qthree \leq \qtwo\}$.
For instance, in the Boolean quantale $\mmap$ is ordinary implication, whereas 
in the Lawvere quantale $\mmap$ is truncated subtraction.

This is all the reader has to know about quantales to understand 
\quant{} \emph{abstract} rewriting systems. When it comes to move to \quant{} 
\emph{term} rewriting systems, a little more notions about (and a little 
more conditions on) quantales are needed. In particular, 
the addition of structural rules akin to \quant{} equational theories 
requires us to work with \emph{continuous} quantales~\cite{continuous-lattices-and-domains} 
(but see \autoref{rem:structural-rules}).

\begin{definition}
        Given a quantale $\Quantale$ and elements $\qone, \qtwo \in \quantale$, 
        the \emph{way-below} relation $\ll$ is defined thus: 
        $\qtwo \ll \qone$ if and only if for every subset $A \subseteq \quantale$, 
        whenever $\qone \leq \join A$, there exists a finite
        subset $A_0 \subseteq A$ such that $\qtwo \leq \join A_0$. 
        We say that $\Quantale$ is \emph{continuous} 
        if and only if 
        $$
        \qone = \join_{\qtwo \ll \qone} \qtwo.
        $$
\end{definition}

\begin{example}
 The Boolean quantale $\Two$ being finite it is trivially continuous. 
 Both the Lawvere and the strong Lawvere quantales are continuous with $>$ (i.e. greater than) 
 as the way below relation. There, we extend $>$ to $[0,\infty]$ by stipulating  
 $\infty > \infty$. In the same way, one obtains continuity of Fuzzy quantales. 
 There, the way below relation is given by $<$ (i.e. less than) extended by 
 stipulating $0 < 0$.
\end{example}

% ▶ Example 11. The Boolean quantale (({0 ≤ 1}, ∨), ⊗ := ∧) is finite and thus continuous [14].
% Since it is continuous, {0, 1} itself is a basis for the quantale that satisfies the conditions above.
% For the Gödel t-norm [11] (([0, 1], ∨), ⊗ := ∧), the way-below relation is the strictly-less
% relation < with the exception that 0 < 0. A basis for the underlying lattice that satisfies
% the conditions above is the set Q ∩ [0, 1]. Note that, unlike real numbers, rationals numbers
% always have a finite representation. For the metric quantale (also known as Lawvere quantale)
% (([0, ∞], ∧), ⊗ := +), the way-below relation corresponds to the strictly greater relation with
% ∞ > ∞, and a basis for the underlying lattice that satisfies the conditions above is the set of
% extended non-negative rational numbers. The latter also serves as basis for the ultrametric
% quantale (([0, ∞], ∧), ⊗ := max).

\paragraph{Quantale-valued Relations}
We now move to quantale-valued relations, our main tool to model 
\quant{} rewriting. As quantales model abstract quantities, quantale-valued relations 
provide abstract notions of distances. 

\begin{definition}
Given a quantale $\Quantale = (\quantale, \leq, \qunit, \tensor)$, a $\Quantale$-relation $\vrelone: A \torel B$ between sets $A$ and $B$ is a function $\vrelone: A \times B \to \quantale$. For any set $A$, we 
define the identity (or diagonal) $\Quantale$-relation 
$\idvrel_{A} : A \torel A$ mapping diagonal elements $(\tone,\tone)$ to $\qunit$, 
and all other elements to $\qbot$.
Moreover, the composition $\vrelone; \vreltwo: A \torel C$ of $\Quantale$-relations 
$\vrelone: A \torel B$ and $\vreltwo: B \torel C$ is defined 
by the so-called matrix multiplication formula \cite{Hoffman-Seal-Tholem/monoidal-topology/2014}:
$$
(\vrelone;\vreltwo)(\tone,\tthree) \defeq \join_{\ttwo \in B} \vrelone (\tone,\ttwo) 
\tensor \vreltwo (\ttwo,\tthree).
$$
\end{definition}

In general, we think about a $\Quantale$-relation as giving the distance or the 
degree of relatedness of two elements~\cite{Flagg-1992,Flagg-1997,Flagg-1997-b,Hoffman-Seal-Tholem/monoidal-topology/2014}. 
For instance, when the quantale is 
Boolean, elements are either related or not, whereas for Fuzzy quantales 
$\Quantale$-relations coincide with Fuzzy relations~\cite{Fuzzy-relational-systems}, 
and thus they 
give the degree to which elements are related, as well as proximity and similarity relations. 
When we move to the Lawvere quantale (and 
quantales alike), $\Quantale$-relations give general notions of 
distances~\cite{Lawvere/GeneralizedMetricSpaces/1973}, and thus act 
as a foundation for metric reasoning~\cite{BonsangueBreguelRutten/GeneralisedMetricSpaces/1998}. 
Coarser forms of metric reasoning are obtained by 
considering interval-based~\cite{Geoffroy-Pistone-2021} and probabilistic quantales~\cite{HOFMANN20131}, where 
instead of establishing the distance between elements exactly, one obtains 
only an interval to which such a distance belongs, or a probability of the accuracy of 
its measurement. 
Finally, considering the quantale of (fuzzy) modal predicates, we obtain (fuzzy) modal and 
coeffectful 
relations~\cite{DBLP:journals/pacmpl/LagoG22a,Routley-1972-II,Routley-1972-III,Routley-1973,Urquhart-1972}, 
whereby (the degree of) relatedness of elements is given with respect a 
possible world (such as the available resources).  

\begin{example}
We summarise composition on the Boolean, Lawvere, 
and Strong Lawvere quantale in 
\autoref{figure:composition-boolean-lawevere-stronglawvere}.
\end{example}

\begin{table*}[htbp]
\centering
\begin{tabular}{|c|c|c|c|}
	\hline
& $\Two$ & $\Lawvere$ & $\StrongLawvere$  \\
\hline
$(\vrelone;  \vreltwo)(\tone,\tthree)$ 
&
$ \exists y.\ \vrelone (\tone,\ttwo) \wedge \vreltwo (\ttwo,\tthree)$
&
$\inf_y \vrelone (\tone,\ttwo) + \vreltwo (\ttwo,\tthree)$ 
&
$\inf_y \max(\vrelone (\tone,\ttwo), \vreltwo (\ttwo,\tthree))$
\\
\hline
\end{tabular}
%\vspace{-0.2cm}
\caption{Composition on $\Two$, $\Lawvere$, and $\StrongLawvere$}
\label{figure:composition-boolean-lawevere-stronglawvere}
\end{table*}

Since $\Quantale$-relation composition is associative and has 
$\idvrel$ as unit element, for any quantale $\Quantale$ we have a category, denoted by 
$\vrel$, with sets as objects and $\Quantale$-relations as arrows. 
 Moreover, the complete lattice structure of $\Quantale$ lifts to 
$\Quantale$-relations pointwise, so that
we can say that a $\Quantale$-relation $\vrelone: A \torel A$ is 
\emph{reflexive} if $\idvrel \leq \vrelone$; \emph{transitive} 
if $\vrelone; \vrelone \leq \vrelone$; and symmetric if 
$\dual{\vrelone} \leq \vrelone$, 
where the transpose of $\vrelone: A \torel B$
is the $\Quantale$-relation $\dual{\vrelone}: B \torel A$ defined by 
$\dual{\vrelone}(\ttwo,\tone) \defeq \vrelone(\tone,\ttwo)$. 
When read pointwise, reflexivity, transitivity, and symmetry give the 
following inequalities:
\begin{align*}
\qunit &\leq \vrelone(\tone,\tone)
\\
\vrelone(\tone,\ttwo) \tensor \vrelone(\ttwo,\tthree) &\leq \vrelone(\tone,\tthree)
\\
\vrelone(\tone,\ttwo) &\leq \vrelone(\ttwo,\tone).
\end{align*}
Altogether, we obtain the notion of a preorder (i.e. reflexive and transitive) and 
equivalence (i.e. reflexive, transitive, and symmetric) $\Quantale$-relation. 

\begin{notation}
Fixed a quantale $\mathbb{\Theta}$, we oftentimes refer to $\Quantale$-relation on 
$\mathbb{\Theta}$ as $\mathbb{\Theta}$-relations. Thus, for example, 
$\Lawvere$-relations are just $\Quantale$-relations on the Lawvere quantale 
$\Lawvere$.
\end{notation}

      \begin{example}
     % The following are examples of quantales.
      \begin{enumerate}
        \item On the Boolean quantale,  
          $\Two$-relations are ordinary (binary) relations, and 
          preorder and equivalence $\Two$-relations 
          coincide with traditional preorders and equivalences.
        \item On the Lawvere quantale, $\Lawvere$-relations are distances. 
          Instantiating transitivity on $\Lawvere$, we obtain the 
          usual \emph{triangle inequality} formula:
          $$
          \inf_{\ttwo} \vrelone(\tone,\ttwo) + \vrelone(\ttwo,\tthree) \geq \vrelone(\tone,\tthree)
        $$
        Similarly, reflexivity gives the identity of indiscernibles inequality: 
        $$
        0 \geq \vrelone(\tone, \tone).
        $$ 
        Altogether, we see that preorder $\Lawvere$-relations coincide with \emph{generalised metrics} 
          \cite{Lawvere/GeneralizedMetricSpaces/1973,BonsangueBreguelRutten/GeneralisedMetricSpaces/1998} and 
          equivalence $\Lawvere$-relations with \emph{pseudometrics} \cite{steen/CounterexamplesTopology/1995}.
        \item Moving from the Lawvere to the Strong Lawvere quantale, we 
        replace addition with binary maximum, so that transitivity now gives 
          the \emph{strong triangle inequality} formula:
          $$
          \inf_{\ttwo} \max(\vrelone(\tone,\ttwo), \vrelone(\ttwo,\tthree)) \geq \vrelone(\tone,\tthree)
          $$
          Consequently, 
          equivalence $\StrongLawvere$-relations coincide with \emph{ultra-pseudometrics}.
         \item  
         On the quantale $\mathbb{F}$, equivalence $\mathbb{F}$-relations 
            give \emph{probabilistic metric spaces} \cite{HOFMANN20131}. 
            The informal reading of a $\mathbb{F}$-relation $\vrelone$ is that 
            $\vrelone(\tone,\ttwo)(\qone)$ gives the probability that $\tone$ and $\two$ are 
            at most $\qone$-far.
        \item On the unit interval (fuzzy) quantale(s), $\UnitInterval$-relations 
        coincide with fuzzy relations \cite{Fuzzy-relational-systems}. Equivalence 
        $\UnitInterval$-relations are often called similarity or proximity relations.
      \end{enumerate}
      \end{example}
      
We summarise how reflexivity, symmetry, and transitivity instantiate 
on the Boolean, Lawvere, and Strong Lawevere quantale in
\autoref{figure:correspondence-reflexivity-symmetry-transitivity}.

\begin{table*}[htbp]
\centering
\begin{tabular}{|c|c|c|}
	\hline
$\Two$ & $\Lawvere$ & $\StrongLawvere$  \\
\hline
$\true \leq \vrelone(\tone,\tone) $	
&
$0 \geq \vrelone(\tone,\tone)$ 
&
$0 \geq \vrelone(\tone,\tone)$ 
\\
$\vrelone(\tone,\ttwo) \leq \vrelone(\ttwo,\tone) $	
&
$\vrelone(\tone,\ttwo) \geq \vrelone(\ttwo,\tone)$ 
&
$\vrelone(\tone,\ttwo) \geq \vrelone(\ttwo,\tone)$ 
\\ 
$\vrelone(\tone,\ttwo) \wedge \vrelone(\ttwo,\tthree) \leq \vrelone(\tone,\tthree)$	
&
$\vrelone(\tone,\ttwo) + \vrelone(\ttwo,\tthree) \geq \vrelone(\tone,\tthree)$ 
&
$\max(\vrelone(\tone,\ttwo), \vrelone(\ttwo,\tthree)) \geq \vrelone(\tone,\tthree)$ 
\\
\hline
\end{tabular}
%\vspace{-0.2cm}
\caption{Correspondences reflexivity-symmetry-transitivity.}
\label{figure:correspondence-reflexivity-symmetry-transitivity}
\end{table*}

% \begin{table*}[htbp]
% \centering
% \begin{tabular}{|c|c|c|c|}
% 	\hline
% $\Two$ & $\Lawvere$ & $\StrongLawvere$ & $\quantale$ \\
% \hline
% $\true \leq \vrelone(x,x) $	
% &
% $0 \geq \vrelone(x,x)$ 
% &
% $0 \geq \vrelone(x,x)$ 
% &
% $\qunit \leq \vrelone(x,x)$
% \\
% $\vrelone(x,y) \leq \vrelone(y,x) $	
% &
% $\vrelone(x,y) \geq \vrelone(y,x)$ 
% &
% $p(x,y) \geq \vrelone(y,x)$ 
% &
% $\vrelone(x,y) \leq \vrelone(y,x)$
% \\ 
% $\vrelone(x,y) \wedge \vrelone(y,z) \leq \vrelone(x,z)$	
% &
% $\vrelone(x,y) + \vrelone(y,z) \geq \vrelone(x,z)$ 
% &
% $\max(\vrelone(x,y), \vrelone(y,z)) \geq \vrelone(x,z)$ 
% &
% $\vrelone(x,y) \tensor \vrelone(y,z) \leq \vrelone(x,z)$
% \\
% \hline
% \end{tabular}
% %\vspace{-0.2cm}
% \caption{Correspondences reflexivity-symmetry-transitivity.}
% \label{figure:correspondence-reflexivity-symmetry-transitivity}
% \end{table*}

 Finally, we notice that the ``algebra'' of $\Quantale$-relations is close to the on 
 ordinary relations,\footnote{As the category of traditional relations, the 
 category $\vrel$ is a 
 \emph{quantaloid}~\cite{Hoffman-Seal-Tholem/monoidal-topology/2014,introduction-to-quantaloids}. 
 } so that we can refine a large part of calculi of relations \cite{relational-mathematics} 
 to a quantale-based setting. 
 In fact, we can even think about $\Quantale$-relations as ``monoidal relations''. 
 Since many notions of traditional rewriting can be given in purely relational terms, 
 we can take advantage of that and rephrase them in terms of $\Quantale$-relations. 
 For the moment, we simply recall the following useful closure operations.

\begin{definition} 
Let   $\vrelone: A \torel A$ be a $\Quantale$-relation. For $n \in \mathbb{N}$, we define 
the $n$-th iterate of $\vrelone$, notation $\vrelone^n$ by
 $\vrelone^{0} \defeq \idvrel$ and $\vrelone^{n+1} \defeq \vrelone;\vrelone^{n}$. We define:
\begin{enumerate}
    \item The \emph{reflexive closure} of $\vrelone$ as 
    	$\reflex{\vrelone} \defeq \vrelone \vee \idvrel$.
   %\item The \emph{symmetric closure} of $\vrelone$ as $\symclosure{\vrelone} \defeq \vrelone \vee \dual{\vrelone}$
   % \item The \emph{transitive closure} of $\vrelone$ as $\vrelone^+ \defeq \join_{n\geq 1} \vrelone^{n}$.
    \item The \emph{transitive and reflexive closure} of $\vrelone$ as
    ${\vrelone}^{*} \defeq \join_{n\geq 0} \vrelone^{n}$.
    \item The \emph{equivalence closure} of $\vrelone$ as $\makedistance{\vrelone} \defeq (\vrelone \vee \dual{\vrelone})^*$.
 \end{enumerate}
\end{definition}

As already remarked, our approach to \quant{} abstract rewriting systems will be algebraic and relational. 
Accordingly, we shall prove several nontrivial rewriting properties relying on the 
algebra of $\Quantale$-relations. To do so, it is useful to exploit fixed point 
characterisations of relational constructions, 
as well as their adjunction 
properties~\cite{Backhouse-fixed-point-and-galois-connection,algebra-of-programming}. 
Recall that $\vrel(A,B)$ carries a complete lattice structure, so that any 
monotone map $F: \vrel(A,B) \to \vrel(A,B)$ has least and greatest fixed points, denoted by
$\mu X.F(X)$ and $\nu X.F(X)$, respectively. Consequently, 
we can define $\Quantale$-relations both \emph{inductively} and \emph{coinductively}. 
That gives us the following (fixed point) induction and (fixed point) coinduction proof principles:
\[
\infer{\mu X. F(X) \leq \vrelone}{F(\vrelone) \leq \vrelone}
\qquad 
\infer{\vrelone \leq \nu X. F(X)}{\vrelone \leq F(\vrelone)}
\]

In particular, we notice that $\vrelone^*$ is the least solution 
to the equation 
$
X = \idvrel \vee \vrelone;X,
$
so that $\vrelone^*$ can be equivalently 
defined the least fixed point $\mu X.\idvrel \vee \vrelone;X$, and thus as the least pre-fixed point
of the map $F(X) \defeq \idvrel \vee \vrelone;X$. Consequently, we obtain the following 
least fixed point induction rule:
\[
\infer{\vrelone^* \leq \vreltwo}{\idvrel \vee \vrelone;\vreltwo \leq \vreltwo}
\]

\begin{notation}
We denote by $\relbot$ the $\Quantale$-relation $\mu X.X$ assigning distance $\qbot$ to all elements,
and by $\reltop$ the $\Quantale$ relation $\nu X.X$, i.e. the indiscrete $\Quantale$-relation assigning distance 
$\qunit$ to all elements. Explicitly, we have $\relbot(\tone, \ttwo) = \qbot$ and 
$\reltop(\tone, \ttwo) = \qunit$, for all $\tone, \ttwo$.
\end{notation}

Finally, we mention that $\vrel(A,A)$ being not only a complete lattice, but a quantale, 
$\Quantale$-relation composition has both left and right adjoints, often referred to 
as left and right division \cite{algebra-of-programming}:
$$
\vrelone; \vreltwo \leq \vrelthree \iff \vreltwo \leq \vrelone \setminus \vrelthree
\qquad 
\vrelone; \vreltwo \leq \vrelthree \iff \vrelone \leq \vrelthree / \vreltwo.
$$

\paragraph{Ternary Relations}
Even if we model \quant{} rewriting relations as $\Quantale$-relations, in \autoref{section:long-intro} 
we have defined 
rewriting systems by means of suitable ternary relations from which we have then extracted a $\Quantale$-relations. 
This process, known as \emph{strata extension} \cite{Hoffman-Seal-Tholem/monoidal-topology/2014}, 
is an instance of a more general correspondence~\cite{modal-reasoning-equal-metric-reasoning} 
between abstract distances and suitable ternary relations 
akin to substructural Kripke 
relations~\cite{Routley-1972-II,Routley-1972-III,Routley-1973,Urquhart-1972}
as used in the relational analysis of coeffects~\cite{DBLP:journals/pacmpl/LagoG22a, DBLP:journals/pacmpl/AbelB20}.
Since we will extensively switch between $\Quantale$-relations and ternary relations, we recall the notion of 
a $\Quantale$-ternary relation.

{
\renewcommand{\relone}{R}
\newcommand{\reltodist}[1]{#1^{\bullet}}
\newcommand{\disttorel}[1]{#1^{\circ}}

\begin{definition} 
Given a quantale $\Quantale$, a \emph{$\Quantale$-ternary relation} over $A \times B$ 
is a ternary relation $\relone \subseteq A \times \quantale \times B$ antitone in 
its second argument (meaning that $\relone(\tone,\qone,\ttwo)$ implies $\relone(\tone,\qtwo,\ttwo)$, for 
any $\qtwo \leq \qone$).
\end{definition} 
Any ternary $\Quantale$-relation $\relone$ induces a $\Quantale$-relation $\reltodist{\vrelone}$ thus:
$$
\reltodist{\vrelone}(\tone,\ttwo) \defeq \join_{\relone(\tone,\qone,\ttwo)} \qone.
$$
Vice versa, any  $\Quantale$-relation $\vrelone$ induces a $\Quantale$-ternary relation $\disttorel{\relone}$ 
defined by
$$
\disttorel{\relone}(\tone,\qone,\ttwo) \iff \qone \leq \vrelone(\tone,\ttwo).
$$
This two processes are each other inverses, meaning that $\vrelone^{\circ\bullet} = \vrelone$ and 
$\relone^{\bullet\circ} = \relone$, so that we can freely switch between $\Quantale$-ternary relations and 
$\Quantale$-relations. 

\begin{notation}
Oftentimes, we will use modal relations to define rewriting systems. In those cases, 
we will use notations of the form $\reduce_{\vrelone}$ and write $\mrel{\qone}{\tone}{\reduce_{\vrelone}}{\ttwo}$ 
in place of $\reduce_{\vrelone}(\tone,\qone,\ttwo)$. Moreover, we shall denote by 
$\vrelone$ the $\Quantale$-relation 
associated to $\reduce_{\vrelone}$. That is, 
$\vrelone(\tone,\ttwo) \defeq \join_{\mrel{\qone}{\tone}{\reduce_{\vrelone}}{\ttwo}}\qone$.
%As already mentioned, the notion of modal relation and the quantale-valued relation are mutually linked. Indeed $\mathcal{P}(W,X,Y)\simeq 2^{W\times X\times Y}\simeq X\times Y \to 2^{W}$ where $2^{W}$ is a quantale, thus we can translate a modal relation into a quantale-valued one. Vice versa, if we have $\Quantale$-relation $\vrelone: X\torel Y$ then we can define the modal relation $R$ such that $\mrel{v}{a}{R}{b}$ iff $v\in \vrelone(a,b)$. Since the two notions are interchangeable, we will switch from one to the other depending on which best suits the context. Let us begin by defining abstract quantitative rewriting systems on the basis of the quantal-relational calculus defined. 
\end{notation}

\begin{remark}
To make definitions computationally lighter, the literature on \quant{} algebraic
theories usually considers ternary relations over a \emph{base} of $\quantale$, 
the carrier of such 
a base usually being considerably smaller than $\quantale$. 
For instance, \quant{} equational theories are often defined using ternary relations 
over non-negative rationals, the latter being a base for $[0,\infty]$. 
Since our results are independent of working with bases or with full quantales, 
we keep the necessary mathematical preliminaries as minimal as possible, this way  
defining (in \autoref{sect:qtrs}) \quant{} term rewriting 
systems relying on quantales rather than on their bases 
(the interested reader can consult the recent work by \citet{An-Internal-Language-for-Categories-Enriched-over-Generalised-Metric-Spaces} 
to convince herself that our theory is invariant with respect to such a design choice). 
\end{remark}
}

\section{QUANTITATIVE ABSTRACT REWRITING SYSTEMS} 
\label{section:qars}
%\section{Quantitative Abstract Rewriting Systems}

In this section, we introduce \emph{\quant{} abstract rewriting systems} and their theory. 
These systems constitute the foundation of \quant{} rewriting, and all other notions 
of \quant{} rewriting system, such as string- and term-based systems, can be ultimately regarded 
as \quant{} abstract rewriting systems. Moreover, we shall use the latter to define crucial notions 
and properties of rewriting, such as confluence and termination, that we will later specialise
to term-based systems. Throughout this and later sections, we fix a (Lawverian) quantale 
$\Quantale = (\quantale, \leq, \qunit, \tensor)$.

\begin{definition}
\label{def:qars}
A  \emph{Quantitative Abstract Rewriting Systems} (\qars, for short) is a pair $(A, \vrelone: A \torel A)$.
\end{definition}
 
\autoref{def:qars} is extremely simple: as a traditional abstract rewriting system is defined as a set of objects 
together with a binary (rewriting) relation on it, a \qars{} is defined as a set of objects 
together with a binary (rewriting) $\Quantale$-relation on it. 
Given elements $\tone,\ttwo \in A$, we say that $\tone$ rewrites into (or reduces to) $\ttwo$ 
%--- or, more compactly, 
%that $\tone$ reduces to $\ttwo$ ---
if $\vrelone(\tone,\ttwo) \neq \qbot$: in that case, we say that the (rewriting) distance or difference between 
$\tone$ and $\ttwo$ is $\vrelone(\tone,\ttwo)$. 
Further possible informal reading (possibly depending on the quantale considered) 
refer to $\vrelone(\tone,\ttwo)$ as the \emph{degree} of the reduction, the \emph{cost} of the reduction, or 
as the \emph{resource} required for the reduction.\footnote{This is the case, in particular, for quantales 
of modal predicates, where rewriting is ultimately performed in a possible world describing intensional 
aspects of the rewriting process, such as the available resource.}
Rewriting paths are obtained by iterating $\vrelone$. In particular, we say that:
\begin{enumerate}
    \item $\tone$ reduces to $\ttwo$ 
%in $n$-steps if $\vrelone^{n}(\tone,\ttwo) \neq \qbot$ and that $\tone$ reduces to $\ttwo$ 
in finitely many-steps if $\vrelone^*(\tone,\ttwo) \neq \qbot$; 
%: in those cases, 
% the distance between $\tone$ and $\ttwo$ is 
%$\vrelone^{n}(\tone,\ttwo)$ and $\vrelone^*(\tone,\ttwo)$, respectively. 
%Similarly, we say 
    \item $\tone$ reduces to $\ttwo$ in one or zero steps if $\reflex{\vrelone}(\tone,\ttwo) \neq \qbot$;
%Finally, we say 
    \item $\tone$ is convertible with $\ttwo$ if 
    $\makedistance{\vrelone}(\tone,\ttwo) \neq \qbot$.
%in that case, the distance between $\tone$ and $\ttwo$ is $\makedistance{\vrelone}(\tone,\ttwo)$. 
\end{enumerate}
Notice that given a \qars{} $(A, \vrelone: A \torel A)$, the convertibility $\Quantale$-relation $\makedistance{\vrelone}$ 
generated by $\vrelone$ is a $\Quantale$-equivalence and thus endows $A$ with a metric-like structure. 

Sometimes, we will need to explicitly consider reduction sequences. 
We thus say that a finite sequence $(\tone_0, \hh, \tone_n)$ is a $\vrelone$-reduction sequence if 
%$\bigotimes_{i=0}^n \vrelone(\tone_i,\tone_{i+1}) \neq \qbot$, 
$$
\vrelone(\tone_0,\tone_1) \tensor \cc \tensor \vrelone(\tone_{n-1},\tone_n) \neq \qbot
$$
and that an infinite sequence $(\tone_0, \hh, \tone_n, \hh)$ is a $\vrelone$-reduction sequence if 
$\vrelone(\tone_0,\tone_1) \tensor \cc \tensor \vrelone(\tone_{n-1},\tone_n) \neq \qbot$, for any $n \geq 0$. 
 Every reduction sequence\footnote{If the underlying \qars{} $(A, \vrelone)$ is clear from the context, 
 we simply refer to \emph{reduction sequences} for $\vrelone$-reduction sequences.}
 has a first element: if a reduction sequence starts from $\tone$, then we refer to it as a 
 reduction sequence of $\tone$. Notice that since 
 $\Quantale$ is Lawverian, for any reduction sequence $(\tone_0, \hh, \tone_n)$, we have 
$\vrelone(\tone_i, \tone_{i+1}) \neq \qbot$, for any $i$.\footnote{In case the 
underlying quantale is not Lawverian, then one should take this condition as 
part of the definition of a reduction sequence.}

% Given two elements $\qvalone,\qvaltwo$ of $A$, if   $(\qvalone,\qvaltwo)\in \vrelone$ then $\qvaltwo$ is called a one-step reduct of $\qvalone$; in some cases we will simply write $\qvalone\to^\vrelone$. If   $(\qvalone,\qvaltwo)\in \vrelone^{*}$ then we will say that $\qvaltwo$ is  a reduct of $\qvalone$ or  alternatively, that there is a reduction from $\qvalone$ to $\qvaltwo$; in some cases we will simply write  $\qvalone\to^{\vrelone^{*}}\qvaltwo$ \cecilia{dubbio da chiedere}. 
 %$\vrelone(\qvalone,\qvaltwo)\neq \qbot$ then $\qvaltwo$ is called a one-step reduct of $\qvalone$; in some cases we will simply write $\qvalone\to^\vrelone$. A reduction path or a simply a reduction from $\qvalone$ to $\qvaltwo$ is a sequence $(x_{1},...,x_{n})$ such that $\vrelone(x_{1},x_{2})\tensor\vrelone(x_{2},x_{3})\tensor \cdot\cdot\cdot\tensor\vrelone(x_{n-1},x_{n})\neq \qbot$, with $\qvalone=x_{1}$ and $\qvaltwo=x_{n}$; in this case we will write $\qvalone\to^{\vrelone^{*}}\qvaltwo$.

\subsection{Confluence}
Given a set $A$ of objects together with an equivalence 
$\equiv$ on it, traditional rewriting systems are often introduced as ways to give computational content to $\equiv$. 
Accordingly, one considers a rewriting relation ${\to} \subseteq A \times A$ on $A$ 
such that $\to$-convertibility 
coincides with $\equiv$. At this point, properties of $\to$ are proved so to ensure 
$\equiv$ to be computationally well-behaved. Among those, the so-called
Church-Rosser property states that whenever $\tone \equiv \ttwo$, there exists an object $\tthree$ 
such that both $\tone$ and $\ttwo$ can be reduced to $\tthree$ in a finite number of steps. 
Formally, ${\equiv}$ coincides with $\to^*; \prescript{*}{}{\leftarrow}$ 
(where $\leftarrow$ stands for $\dual{\to}$). 
The Church-Rosser property thus implies that to study $\equiv$ it is  enough to 
study directional rewriting. Moreover, if $\equiv$ is defined axiomatically, 
then the Church-Rosser property gives a powerful tool to test consistency of $\equiv$:
if $\tone$ and $\ttwo$ have no common reduct, then they cannot be equivalent.

In a \quant{} setting, the relation $\equiv$ is replaced by a $\Quantale$-equivalence
$\vdistone$, and the rewriting relation $\to$ is replaced by a $\Quantale$-rewriting relation 
$\vrelone$ such that $\makedistance{\vrelone} = \vdistone$. 
On then looks for properties $\vrelone$ ensuring $\vdistone$ to be computationally well-behaved. 
In this section, we explore some of these properties, viz. \emph{(\quant) confluence}, 
\emph{(metric) Church-Rosser}, and \emph{termination}.
We begin with confluence, which states that two reductions 
originating from the same element can be joined into a common element, as in the classical case, 
but with the additional property that the merging reduction is achieved without increasing distances.

 \begin{definition} 
 \label{def:commutation}
 Let  $\vrelone, \vreltwo: A \torel B$ be $\Quantale$-relations.
 \begin{enumerate}
 \item We say that $\vrelone$ \emph{commutes} with $\vreltwo$ if 
 	$\dual{\vrelone}; \vreltwo \leq \vreltwo; \dual{\vrelone}$. 
 \item We say that $\vrelone$ satisfies the \emph{diamond property}  if 
 	$\vrelone$ commutes with itself, i.e. 
	$\dual{\vrelone}; \vrelone \leq \vrelone; \dual{\vrelone}$.
 \end{enumerate}
 \end{definition}  
 
Let us comment on \autoref{def:commutation} by analysing the diamond property. 
On the Boolean quantale, we recover the usual diamond property as defined for 
traditional rewriting systems. More interesting is the case of the Lawevere quantale, 
which we use as vehicle to move to the general case. 
Pointwise, the diamond property reads as follows:
$$
\inf_\tthree \vrelone(\tthree,\tone) + \vrelone(\tthree,\ttwo) \geq \inf_\tfour \vrelone(\tone,\tfour) + 
\vrelone(\ttwo,\tfour).
$$
The left-hand-side of the inequality, 
namely $\inf_\tthree \vrelone(\tthree,\tone) + \vrelone(\tthree,\ttwo)$, 
gives %the minimal convertibility distance between $x$ and $y$ or, equivalently, 
the minimal \emph{peak distance} between $\tone$ and $\ttwo$, that is the shortest connection 
between $\tone$ and $\ttwo$ obtained through a pick $\tthree$ reducing to both $\tone$ and $\ttwo$.
The right-hand-side, instead, 
gives the minimal \emph{valley distance} between $\tone$ and $\ttwo$.
Let us say that $\tthree$ is a peak over $\tone$ and $\ttwo$ if $\vrelone(\tthree,\tone) + 
\vrelone(\tthree,\ttwo) \neq \infty$, 
so that none of $\vrelone(\tthree,\tone)$ and $\vrelone(\tthree,\ttwo)$ is $\infty$.\footnote{In the general case, 
we say that $\tthree$ is a peak over $\tone$ and $\ttwo$ if $\vrelone(\tthree,\tone) \tensor \vrelone(\tthree,\ttwo) \neq \qbot$, 
so that $\vrelone(\tthree,\tone) \neq \qbot$ and $\vrelone(\tthree,\ttwo) \neq \qbot$ follow since the 
quantale is Lawverian.}
 The diamond property ensures that whenever we have a peak $\tthree$ over $\tone$ and $\ttwo$, then 
 we also have a collection of valleys under $\tone$ and $\ttwo$, i.e. elements $\tfour$ such that 
 $\vrelone(\tone,\tfour) + \vrelone(\ttwo,\tfour) \neq \infty$, such that the infimum of 
 such valleys is smaller or equal than $\vrelone(\tthree,\tone) + \vrelone(\tthree,\ttwo)$. 
 In fact, the diamond property gives:
  $$
 \infty >  \vrelone(\tthree,\tone) + \vrelone(\tthree,\ttwo)  
 \geq \inf_{\tthree} \vrelone(\tthree,\tone) + \vrelone(\tthree,\ttwo) 
 \geq \inf_\tfour \vrelone(\tone,\tfour) + \vrelone(\ttwo,\tfour).
 $$
 And if there is no element $\tfour$ such that $\vrelone(\tone,\tfour) + \vrelone(\ttwo,\tfour) \neq \infty$, then 
 $ \inf_{\tfour} \vrelone(\tone,\tfour) + \vrelone(\ttwo,\tfour) = \infty$, 
 which gives a contradiction. Notice, however, 
 that there is no guarantee that there is an actual valley $\tfour$ such that 
 $
  \vrelone(\tthree,\tone) + \vrelone(\tthree,\ttwo)  \geq \vrelone(\tone,\tfour) + \vrelone(\ttwo,\tfour).
 $

\begin{example}
\label{ex:example-4}
Consider the Lawvere quantale and the \makears{$\Lawvere$} over the set 
$A \defeq \mathbb{R}^+ \cup \{\tone, \ttwo_1, \ttwo_2\}$ 
with $\vrelone(\tone, \ttwo_1) \defeq \vrelone(\tone, \ttwo_2) \defeq 0$, 
$\vrelone(\ttwo_1, \qone) \defeq \vrelone(\ttwo_2, \qone) \defeq \frac{\qone}{2}$, for each 
$\qone \in \mathbb{R}^+$, 
and $\vrelone(x,y) \defeq \infty$ otherwise. 
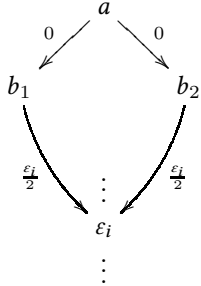
\[
\xymatrix@-0.5pc{
& \tone \ar[ld]_0 \ar[rd]^0 & 
\\
\ttwo_1 \ar@/_/[rd]_{\frac{\qone_i}{2}}   &   & \ttwo_2\ar@/^/[ld]^{\frac{\qone_i}{2}} 
\\
& \text{$\begin{matrix} \vdots \\ \qone_i \\ \vdots \end{matrix}$} & 
}
\] 
Then, there is no $\tthree$ such that $\vrelone(\ttwo_1, \tthree) + \vrelone(\ttwo_2, \tthree) = 0$, 
although $\inf_{\qone}  \vrelone(\ttwo_1, \qone) + \vrelone(\ttwo_2, \qone) =  
\inf_{\qone > 0} \frac{\qone}{2} + \frac{\qone}{2} = 0$.
 \end{example}

In the general setting of an arbitrary quantale $\Quantale$, we see that the diamond 
property has the following pointwise reading:
$$
\join_\tthree \vrelone(\tthree,\tone) \tensor \vrelone(\tthree,\ttwo) 
\leq \join_\tfour \vrelone(\tone,\tfour) \tensor \vrelone(\ttwo,\tfour).
$$
The abstract formulation suggests further non-distance-based readings of the diamond property (and properties alike); 
and among those, noticeable ones are obtained in terms of \emph{graded properties} 
and \emph{degree of reductions}. 
Accordingly, we read $\join_\tthree \vrelone(\tthree,\tone) \tensor \vrelone(\tthree,\ttwo)$ as the 
\emph{divergence degree} of $\tone$ and $\ttwo$, 
and $\join_\tfour \vrelone(\tone,\tfour) \tensor \vrelone(\ttwo,\tfour)$ as 
the \emph{convergence degree} of $\tone$ and $\ttwo$. 
The diamond property then states that the divergence degree between any two elements 
is always smaller or equal than their convergence degree; that is, the system tends more to 
converge than to diverge. 
Instantiating $\Quantale$ with the Boolean element (and thus recovering the traditional diamond 
property and properties alike), we stipulate degrees of convergence and divergence to be absolute 
values. On the other hand, taking the unit interval quantale, 
we let convergence and divergence be \emph{fuzzy} notions. 

Finally, we mention that we also have a \emph{modal} and \emph{coeffectful} reading of 
the diamond property along the lines coeffectful relational 
calculi~\cite{modal-reasoning-equal-metric-reasoning,DBLP:journals/pacmpl/LagoG22a}. Accordingly, 
we parametrise the latter property with respect to possible worlds 
(such as information states, security levels, available resources, etc.), 
this way obtaining a local and (more) intensional view of rewriting. 
% The left-hand-side of the inequality, namely $\join_u \vrelone(u,x) \tensor \vrelone(u,y)$, 
% gives the \emph{peak distance} between $x$ and $y$ and can be seen as the degree of 
% \emph{divergence} of $x$ and $y$. Dually,  the right-hand-side of the diamond property inequality, 
% namely $\join_z \vrelone(x,z) \tensor \vrelone(y,z)$, 
% gives the \emph{valley distance} between $x$ and $y$ and can be seen as the degree of 
% \emph{convergence} of $x$ and $y$. Accordingly, if $\vrelone$ has the diamond property, 
% then the degree of divergence of each pair of elements bounds their degree of 
% convergence. 
We summarise the pointwise reading of commutativity and of the diamond property in 
\autoref{figure:commutativity-and-diamond-property}.

\begin{table*}[htbp]
\begin{tcolorbox}[boxrule=0.5pt,width=\linewidth,colframe=black,colback=black!0!white,arc=0mm]
\centering
\begin{tabular}{cc}
%\hline
Commutation & Diamond Property 
\\
$$
\xymatrix@ -0.8pc{
& \tthree \ar[ld]_{\vrelone} \ar[rd]^{\vreltwo} & 
\\
\tone \ar@{.>}[rd]_{\vreltwo}   &   & \ttwo \ar@{.>}[ld]^{\vrelone} 
\\
& \tfour  & 
}
$$
%%%%%%%%%%%%%
&
%%%%%%%%%%%%% 
$$
\xymatrix@ -0.8pc{
& \tthree \ar[ld]_{\vrelone} \ar[rd]^{\vrelone} & 
\\
\tone \ar@{.>}[rd]_{\vrelone}   &   & \ttwo \ar@{.>}[ld]^{\vrelone} 
\\
&  \tfour & 
}
$$
\\
{$\displaystyle \join_\tthree \vrelone(\tthree,\tone) \tensor \vreltwo(\tthree,\ttwo) 
\leq \join_\tfour \vreltwo(\tone,\tfour) \tensor \vrelone(\ttwo,\tfour)$}
& 
{$\displaystyle \join_\tthree \vrelone(\tthree,\tone) \tensor \vrelone(\tthree,\ttwo) 
\leq \join_\tfour \vrelone(\tone,\tfour) \tensor \vrelone(\ttwo,\tfour)$}
\\
%\hline
\end{tabular}
\end{tcolorbox}
%\vspace{-0.2cm}
\caption{Commutativity and the Diamond Property}
\label{figure:commutativity-and-diamond-property}
\end{table*}

% Another possible reading of
% the diamond property (and properties alike)
% is in terms of \emph{graded properties} and \emph{degree of reductions}. 
% Accordingly, we read $\inf_u \vrelone(u,x) + \vrelone(u,y)$ as the 
% \emph{divergence distance} or \emph{divergence degree} of $x$ and $y$, 
% and $\inf_z \vrelone(x,z) + \vrelone(y,z)$ as 
% the \emph{convergence distance} or \emph{convergence degree} of $x$ and $y$. 
% The diamond property then states that the divergence degree between any two elements 
% is always smaller or equal than their convergence degree.

\begin{remark}
\label{remark:graded-properties}
We have seen that both confluence and the diamond property involve \emph{graded properties} --- 
namely degrees of divergence and convergence --- i.e. 
non-Boolean properties taking values in a quantale. Nonetheless, 
both confluence and the diamond property are \emph{Boolean} properties 
of $\Quantale$-relations, as they are essentially of the form $\qone \leq \qtwo$. 
It is natural to push the \quant{} perspective one step further and consider 
a \emph{graded} version of, e.g., commutation. 
In fact, by exploiting the adjunction property of $\Quantale$-relation 
composition, we see that requiring $\vrelone$ to commute with $\vreltwo$, i.e. 
$\dual{\vrelone}; \vreltwo \leq \vreltwo; \dual{\vrelone}$, means 
requiring 
$$
\idvrel \leq \dual{\vrelone}{\setminus}(\vreltwo; \dual{\vrelone}) / \vreltwo.
$$
Forgetting about $\idvrel$, we can think about the $\Quantale$-relation 
$\dual{\vrelone}{\setminus}(\vreltwo; \dual{\vrelone}) / \vreltwo$ 
as assigning to elements the degree of commutation of $\vrelone$ and $\vreltwo$ 
on them, i.e. their divergence-convergence distance. Pointwise, we thus obtain:
$$
(\dual{\vrelone}{\setminus}(\vreltwo; \dual{\vrelone}) / \vreltwo)(\tone, \ttwo) = 
\join_\tthree \vrelone(\tthree,\tone) \tensor \vreltwo(\tthree,\ttwo) 
\mmap \join_\tfour \vreltwo(\tone,\tfour) \tensor \vrelone(\ttwo,\tfour).
$$
Notice that the latter is an element of $\quantale$ rather than a (Boolean) truth value, and thus it 
indicates \emph{how much} $\vrelone$ commutes with $\vreltwo$. 
For instance, on the Lawvere quantale, 
$(\dual{\vrelone}{\setminus}(\vrelone; \dual{\vrelone}) / \vrelone)(\tone, \ttwo)$ gives the 
difference between the divergence and convergence distance on $\tone$ and $\ttwo$, and thus 
a measure of \emph{how much} $\vrelone$ has the diamond property on $\tone$ and $\ttwo$.
Using the vocabulary of (enriched) category theory \cite{Kelly/EnrichedCats,Lawvere/GeneralizedMetricSpaces/1973}, 
one may say that enrichment is given not only at the level of relations, but also at the level of 
equality and refinement of relations (that is, not only relations $\relone$ take values 
in $\quantale$, but also statements such as $\vrelone = \vreltwo$ and $\vrelone \leq \vreltwo$ do).
We leave the exploration of this further form of enrichment for future investigation. 
\end{remark}

As for the traditional case, we are interesting in rewriting paths rather than in single 
rewriting steps.  

\begin{definition}
\label{def:confluence}
Let $(A, \vrelone)$ be a \qars{}.
\begin{enumerate}
\item We say that $\vrelone$ is \emph{confluent} if 
	$\vrelone^*$ has the diamond property. 
\item We say that $\vrelone$ is \emph{locally confluent}\footnote{Notice that 
    $\stardual{\vrelone} = \dualstar{\vrelone}$.} if
	$\dual{\vrelone}; \vrelone \leq \vrelone^*; \dualstar{\vrelone}$.
\item We say that $\vrelone$ is \emph{Church-Rosser} (CR, for short) if 
	$\makedistance{\vrelone} = \vrelone^*; \dualstar{\vrelone}$
\end{enumerate}
\end{definition}

If a rewriting $\Quantale$-relation is confluent, then we can characterise the convertibility distance 
$\makedistance{\vrelone}$ in terms convergent sequences of rewriting steps. 

\begin{proposition}
\label{prop:church-rosser}
Let $(A, \vrelone)$ be a \qars{}. Then $\vrelone$ is confluent if and only if it 
is CR.  
\end{proposition}

\begin{proof}
Clearly, if $\vrelone$ is CR, then it is confluent. 
Suppose now $\vrelone$ to be confluent and recall that $\makedistance{\vrelone} =(\vrelone \vee \dual{\vrelone})^*$.
First, we notice that since $\dualstar{\vrelone} = \vrelone^{{\scriptstyle-}*}$, we have:
$$
\vrelone^*; \dualstar{\vrelone} 
\leq \vrelone^*; \vrelone^{{\scriptstyle-}*} 
\leq (\vrelone \vee \dual{\vrelone})^*; (\vrelone \vee \dual{\vrelone})^*
= (\vrelone \vee \dual{\vrelone})^*.
$$
It thus remains to show $(\vrelone \vee \dual{\vrelone})^* \leq \vrelone^*; \dualstar{\vrelone}$. 
We proceed by fixed point induction,
% \footnote{
% Recall that we have defined $\vrelone^*$ as $\join_n \vrelone^n$. That is the least solution 
% to the equation $X = \idvrel \vee \vrelone;X$, so that $\vrelone^*$ can be equivalently 
% defined the least fixed point $\mu X.\idvrel \vee \vrelone;X$, and thus as the least pre-fixed point
% of the map $X \mapsto \idvrel \vee \vrelone;X$ --- 
% i.e. the least $X$ such that $\idvrel \vee \vrelone;X \leq X$. Consequently, we obtain the following 
% least fixed point induction rule:
% \[
% \idvrel \vee \vrelone;\vreltwo \leq \vreltwo \implies \vrelone^* \leq \vreltwo.
% \]
% } 
showing that 
$$\idvrel \vee ((\vrelone \vee \dual{\vrelone}); \vrelone^*; \dualstar{\vrelone}) \leq \vrelone^*; \dualstar{\vrelone}.$$
Clearly, $\idvrel \leq \vrelone^*; \dualstar{\vrelone}$, so that it remains to show
$(\vrelone \vee \dual{\vrelone}); \vrelone^*; \dualstar{\vrelone} \leq \vrelone^*; \dualstar{\vrelone}$. 
We have:\footnote{Recall that for all $\vrelone_i: A \to B$ and 
$\vreltwo: B \to C$, we have $(\join_i \vrelone_i); \vreltwo =  \join_i \vrelone_i; \vreltwo$.} 
\begin{align*}
(\vrelone \vee \dual{\vrelone}); \vrelone^*; \dualstar{\vrelone}  
&= \vrelone;\vrelone^*; \dualstar{\vrelone}  \vee \dual{\vrelone}; \vrelone^*; \dualstar{\vrelone} &
\\
&\leq \vrelone^*; \dualstar{\vrelone}  \vee \dual{\vrelone}; \vrelone^*; \dualstar{\vrelone} &
\\
&\leq \vrelone^*; \dualstar{\vrelone}  \vee \vrelone^{{\scriptstyle -}*}; \vrelone^*; \dualstar{\vrelone} & 
\\
&= \vrelone^*; \dualstar{\vrelone}  \vee \dualstar{\vrelone}; \vrelone^*; \dualstar{\vrelone} & 
\\
&\leq \vrelone^*; \dualstar{\vrelone}  \vee  \vrelone^*; \dualstar{\vrelone}; \dualstar{\vrelone} & \text{(by CR)}
\\
&\leq \vrelone^*; \dualstar{\vrelone}  \vee  \vrelone^*; \dualstar{\vrelone} & 
\\
&= \vrelone^*; \dualstar{\vrelone} & 
\end{align*}
\end{proof}

\begin{notation}
Given a \qars{} $\mathcal{A} = (A,\vrelone)$ and a property $\varphi$ on $\Quantale$-relations, such as being 
confluent, we say that $\mathcal{A}$ has property $\varphi$ if $\vrelone$ has $\varphi$. 
Thus, for instance, we say that $\mathcal{A}$ is confluent if $\vrelone$ is.
\end{notation}
 
 Thanks to \autoref{prop:church-rosser}, we see that confluence is a crucial property in 
 \quant{} and metric reasoning. Proving confluence of quantitative systems, however, 
 can be cumbersome: indeed, \quant{} systems are often built \emph{compositionally} 
 by joining systems together. \autoref{section:long-intro} has already shown us several examples 
 of systems obtained that way. Consequently, it 
 is desirable to design \emph{modular techniques} to prove confluence of such systems 
 compositionally, i.e. relying on confluence of their component subsystems, 
 rather than proceed monolithically from scratches. 
 Among such modular techniques, Hindley-Rosen Lemma~\cite{hindley-1964,rosen-70} is arguably the most 
 well-known one in traditional rewriting. 
 \autoref{lemma:hindley-rosen} generalises such a result to quantitative systems. 
 Before proving it, we recall a few basic properties of $\Quantale$-relations.
 
 \begin{lemma}
 \label{lemma:auxiliary-lemma-hindley-rosen}
 Given $\Quantale$-relations $\vrelone$, $\vreltwo$, and $\vrelthree$, we have: 
 \begin{enumerate}
     \item If $\vrelone; \vreltwo \leq \vreltwo; \vrelone$  and  $\vrelone; \vrelthree \leq \vrelthree; \vrelone$, then 
        $\vrelone; (\vreltwo \vee \vrelthree)^* \leq (\vreltwo \vee \vrelthree)^*; \vrelone$.
    \item $(\vrelone^* \vee \vreltwo^*)^* = (\vrelone \vee \vreltwo)^*$.
 \end{enumerate}
 \end{lemma}
\begin{proof}
For the first item, we observe that proving the thesis 
amounts to prove $(\vreltwo \vee \vrelthree)^* \leq \vrelone \setminus ((\vreltwo \vee \vrelthree)^*; \vrelone)$, 
so that we can use fixed point induction. Proving $\idvrel \leq \vrelone \setminus ((\vreltwo \vee \vrelthree)^*; \vrelone)$ 
is straightforward. It remains to prove 
$$
(\vreltwo \vee \vrelthree); \vrelone \setminus ((\vreltwo \vee \vrelthree)^*; \vrelone) \leq \vrelone \setminus ((\vreltwo \vee \vrelthree)^*; \vrelone).
$$
Since $(\vreltwo \vee \vrelthree); \vrelone \setminus ((\vreltwo \vee \vrelthree)^*; \vrelone)  = 
(\vreltwo; \vrelone \setminus ((\vreltwo \vee \vrelthree)^*; \vrelone)) \vee
(\vrelthree; \vrelone \setminus ((\vreltwo \vee \vrelthree)^*; \vrelone))
$, it is sufficient to prove 
$\vreltwo; \vrelone \setminus ((\vreltwo \vee \vrelthree)^*; \vrelone) \leq \vrelone \setminus ((\vreltwo \vee \vrelthree)^*; \vrelone)$ 
and 
$\vrelthree; \vrelone \setminus ((\vreltwo \vee \vrelthree)^*; \vrelone)  \leq \vrelone \setminus ((\vreltwo \vee \vrelthree)^*; \vrelone)$. 
We prove the former which, by adjunction, is equivalent to 
$$
\vrelone; \vreltwo; \vrelone \setminus ((\vreltwo \vee \vrelthree)^*; \vrelone) \leq (\vreltwo \vee \vrelthree)^*; \vrelone.
$$
By commutation of $\vrelone$ with $\vreltwo$, we obtain:
\begin{align*}
\vrelone; \vreltwo; \vrelone \setminus ((\vreltwo \vee \vrelthree)^*; \vrelone)
&\leq \vreltwo; \vrelone; \vrelone \setminus ((\vreltwo \vee \vrelthree)^*; \vrelone) 
\\
&\leq \vreltwo; (\vreltwo \vee \vrelthree)^*; \vrelone 
\\
&\leq (\vreltwo \vee \vrelthree) ; (\vreltwo \vee \vrelthree)^*; \vrelone 
\\
&\leq (\vreltwo \vee \vrelthree)^*; \vrelone.
\end{align*}
Let us now move to the second item, which essentially amounts to prove 
$(\vrelone^* \vee \vreltwo^*)^* \leq (\vrelone \vee \vreltwo)^*$. 
We use fixed point induction and show $(\vrelone^* \vee \vreltwo^*);(\vrelone \vee \vreltwo)^* \leq (\vrelone \vee \vreltwo)^*$. 
We have:
\begin{align*}
    (\vrelone^* \vee \vreltwo^*);(\vrelone \vee \vreltwo)^* 
    &= \vrelone^*;(\vrelone \vee \vreltwo)^* \vee \vreltwo^*;(\vrelone \vee \vreltwo)^*
    \\
    &\leq (\vrelone \vee \vreltwo)^*;(\vrelone \vee \vreltwo)^* \vee (\vrelone \vee \vreltwo)^*;(\vrelone \vee \vreltwo)^*
    \\
    &= (\vrelone \vee \vreltwo)^* \vee (\vrelone \vee \vreltwo)^*
    \\
    &= (\vrelone \vee \vreltwo)^*.
\end{align*}
\end{proof}

 \begin{proposition}[Hindley-Rosen Lemma]
 \label{lemma:hindley-rosen}
% Let $\vrelone, \vreltwo: A \torel A$ be $\Quantale$-relations.
%  \begin{enumerate}
%  \item If $\vrelone$ commutes with $\vreltwo$ and both $\vrelone$ and $\vreltwo$ have the diamond property, then 
%  	$\vrelone\vee\vreltwo$ is confluent.
 If $\vrelone^{*}$ commutes with $\vreltwo^{*}$ and $\vrelone$, $\vreltwo$ are confluent, 
    then $\vrelone\vee\vreltwo$ is confluent.  
%\end{enumerate}
 \end{proposition}
 
 \begin{proof}
 By the second item of \autoref{lemma:auxiliary-lemma-hindley-rosen}, it is enough to 
 prove that if $\vrelone$ commutes with $\vreltwo$ and both $\vrelone$ and $\vreltwo$ have the diamond property, then 
 $\vrelone\vee\vreltwo$ is confluent. 
 Let us write $\vrelthree$ for $\vrelone \vee \vreltwo$. Using the adjoints of $\Quantale$-relation 
 composition, we obtain:
 $$\dualstar{\vrelthree};\vrelthree^* \leq \vrelthree^*; \dualstar{\vrelthree} 
 \iff 
 \dualstar{\vrelthree} \leq (\vrelthree^*; \dualstar{\vrelthree}) / \vrelthree^* 
 \iff 
 \stardual{\vrelthree} \leq (\vrelthree^*; \dualstar{\vrelthree}) / \vrelthree^* 
$$
 Therefore, it is sufficient to prove $\stardual{\vrelthree} \leq (\vrelthree^*; \dualstar{\vrelthree}) / \vrelthree^*$, 
 which we do using fixed point induction. That amounts to prove 
 $\idvrel \leq (\vrelthree^*; \dualstar{\vrelthree}) / \vrelthree^*$ 
 and $\dual{\vrelthree}; (\vrelthree^*; \dualstar{\vrelthree}) / \vrelthree^* \leq (\vrelthree^*; \dualstar{\vrelthree}) / \vrelthree^*$. 
The former holds since
 $$
 \idvrel \leq (\vrelthree^*; \dualstar{\vrelthree}) / \vrelthree^* 
 \iff \idvrel; \vrelthree^*  \leq \vrelthree^*; \dualstar{\vrelthree}
$$
and  $ \idvrel; \vrelthree^* =  \vrelthree^* = \vrelthree^*; \idvrel \leq \vrelthree^*; \dualstar{\vrelthree}$. 
For the latter, we have 
$$
\dual{\vrelthree}; (\vrelthree^*; \dualstar{\vrelthree}) / \vrelthree^* \leq (\vrelthree^*; \dualstar{\vrelthree}) / \vrelthree^* 
\iff
\dual{\vrelthree}; ((\vrelthree^*; \dualstar{\vrelthree}) / \vrelthree^*); \vrelthree^*  \leq \vrelthree^*; \dualstar{\vrelthree}
\impliedby \dual{\vrelthree}; \vrelthree^*; \dualstar{\vrelthree} \leq \vrelthree^*; \dualstar{\vrelthree}
$$
since 
$((\vrelthree^*; \dualstar{\vrelthree}) / \vrelthree^*); \vrelthree^*  \leq \vrelthree^*; \dualstar{\vrelthree}$. 
To prove 
$\dual{\vrelthree}; \vrelthree^*; \dualstar{\vrelthree} \leq \vrelthree^*; \dualstar{\vrelthree}$, we notice that
$$
\dual{\vrelthree}; \vrelthree^*; \dualstar{\vrelthree} 
= \dual{(\vrelone \vee \vreltwo)};  \vrelthree^*; \dualstar{\vrelthree} 
= (\dual{\vrelone} \vee \dual{\vreltwo});  \vrelthree^*; \dualstar{\vrelthree} 
= (\dual{\vrelone};  \vrelthree^*; \dualstar{\vrelthree}) \vee (\dual{\vreltwo};  \vrelthree^*; \dualstar{\vrelthree})
$$
so that it is sufficient to prove 
$\dual{\vrelone};  \vrelthree^*; \dualstar{\vrelthree} \leq \vrelthree^*; \dualstar{\vrelthree}$
and $\dual{\vreltwo};  \vrelthree^*; \dualstar{\vrelthree} \leq \vrelthree^*; \dualstar{\vrelthree}$. 
We prove the first inequality, as 
the second one is similar. 
Since $\vrelone$ commute both with itself and with $\vreltwo$, by \autoref{lemma:auxiliary-lemma-hindley-rosen}, we have:
\begin{align*}
\dual{\vrelone};  (\vrelone \vee \vreltwo)^*; \dualstar{(\vrelone \vee \vreltwo)}
&\leq (\vrelone \vee \vreltwo)^*; \dual{\vrelone}; \dualstar{(\vrelone \vee \vreltwo)} 
\\
&=  (\vrelone \vee \vreltwo)^*; \dual{\vrelone}; \stardual{(\vrelone \vee \vreltwo)}
\\
&\leq  (\vrelone \vee \vreltwo)^*; (\dual{\vrelone} \vee \dual{\vreltwo}); \stardual{(\vrelone \vee \vreltwo)}
\\
&\leq  (\vrelone \vee \vreltwo)^*; \dual{(\vrelone\vee \vreltwo)}; \stardual{(\vrelone \vee \vreltwo)}
\\
&\leq  (\vrelone \vee \vreltwo)^*; \stardual{(\vrelone \vee \vreltwo)}
\\
&=\vrelthree^*; \dualstar{\vrelthree}
\end{align*}
\end{proof}

\subsection{Locality and Termination}

By \autoref{prop:church-rosser}, we know that nice operational properties of 
a $\Quantale$-equivalence $\vdistone$ can be obtained by characterising 
 $\vdistone$ as the convertibility $\Quantale$-relation of a 
 \emph{confluent} rewriting $\Quantale$-relation $\vrelone$. 
 Even if \autoref{lemma:hindley-rosen} gives a technique to prove confluence 
 of composed systems 
 compositionally, proving confluence of
 ``atomic'' systems may still be not easy. In fact, by its very definition, 
 confluence is a \emph{global} property of a system, in the sense that it refers to 
 rewriting sequences, rather than to single rewriting steps. 
 Newman's Lemma \cite{newman} is a well-known result in the theory of abstract rewriting 
 stating that if a system is \emph{terminating}, then confluence follows from \emph{local confluence}. 
The rest of this section is dedicated to refining Newman's Lemma to a \quant{} setting. 
To do so, we first define the notion of a terminating $\Quantale$-relation and 
prove that terminating $\Quantale$-relations satisfy a suitable induction principle.
We then use the latter to extend Newman's Lemma to $\Quantale$-relations. 

Before proceeding any further, we observe that up to this point our analysis of 
\qars{s} has been relational, proceeding in an algebraic and pointfree fashion. 
To make this paper as accessible as possible, we now take a (temporary) break from 
that methodology and
first give a
\emph{pointwise} analysis of \quant{} termination and a 
\emph{pointwise} proof of (the \quant{} refinement of) Newman's Lemma 
similar to the one by \citet{fuzzy-rewriting-1} 
(see \autoref{sect:conclusion} for a precise comparison).
After that, we go back on our choice and review termination and Newman's Lemma in a novel way, 
following the relational and algebraic 
paradigm and extending the relational theory of induction and abstract rewriting by \citet{backshouse-calculational-approach-to-mathematical-induction} to 
a \quant{}, $\Quantale$-enriched setting. Such an extension is nontrivial and requires 
the introduction of suitable relational modalities akin to corelators \cite{DBLP:journals/pacmpl/LagoG22a}. 
The outcome (which the authors believe is worth the effort) is interesting not only because 
it gives a clean analysis of induction in a $\Quantale$-enriched setting --- as well as a (slightly) more general 
version of (\quant) Newman's Lemma --- but also for the methodology employed, which constitutes 
a nice example of \quant{} relational methods.

\subsubsection{Quantitative Termination, Induction, and Newman's Lemma}

As a first step towards a \quant{} refinement of Newman's Lemma, we extend the 
notion of termination to $\Quantale$-relations.
Contrary to traditional rewriting, in a \quant{} setting the notion of 
termination may be defined in many, non-equivalent ways. We could 
define, for instance, a terminal element as one having no \emph{nontrival} and \emph{non-null} reductions, 
so 
that we allow a terminal element $\tone$ to be reduced to another one $\ttwo$, only if the 
distance between $\tone$ and $\ttwo$ is either $\qunit$ or $\qbot$. 
This notion, which is meaningful in a genuine \quant{} setting (especially 
if interested in `metric reasoning modulo equality'), trivialises 
when instantiated to the Boolean quantale, as elements are always reducible. 
To stay closer to traditional rewriting, we may exclude null distances too, 
so that terminal elements can be reduced only to those element that are $\qbot$-apart from them.
Finally, we may also go beyond finitary notions of reduction \cite{infinitary-rewriting-1,infinitary-rewriting-2,infnitary-rewriting-3} and think about termination 
in the limit \cite{Faggian-2019}, i.e. as possibly infinite reductions converging 
to $\qunit$, in the limit. 

All these proposals are legitimate, and they all deserve 
to be investigated. A complete analysis of notions of \quant{} termination, however, is beyond the scope 
of this paper and we should thus fix a conceptually minimal notion of termination and focus on that one 
only. 
To do that, we take an operational approach and stipulate that a (rewriting) $\Quantale$-relation 
is terminating if it supports an induction principle. But what could the latter possibly be? 
%To answer that question, we follow the abstract approach by \citet{backshouse-calculational-approach-to-mathematical-induction}
%and define induction $\Quantale$-\emph{relationally}. 

Let us recall \cite{backshouse-calculational-approach-to-mathematical-induction} that, given a binary relation 
$\relone \subseteq A \times A$, a property $p$ on $A$ is 
$\relone$-\emph{inductive} if it satisfies the law
$$(\forall x.\ x \relone y \to p(x)) \to p(y),$$
for any $y \in A$. We then say that the relation $\relone$ 
\emph{admits induction} if for any $\relone$-inductive property $p$, 
$p(\tone)$ holds for any $\tone \in A$, 
%for any $\relone$-inductive property $p$. 
Consequently, if 
$\relone$ admits induction and we want to prove that each element of 
$A$ has a given property $p$, it is enough to prove $p$ to be 
$\relone$-inductive. We thus recover the familiar formulation of 
well-founded induction:
%\footnote{
% \citet{backshouse-calculational-approach-to-mathematical-induction} define induction in a purely relational way. 
% First, the (Boolean) calculus of classes is embedded into the calculus of relations, this way defining 
% properties as relations satisfying suitable laws (for instance, one may view a property as a coreflexive relation, 
% i.e. as a relation $p$ such that $p \subseteq \idvrel$). 
% Given a relation $\relone$ and a property $p$, the (semantics of the) property ${\relone}{\searrow}{p}$ 
% is then defined as
% $\{x \in A \mid \forall y.\ y \relone x \to p(y)\}$ 
% (${\relone}{\searrow}{p}$ is actually defined so relying on the 
% axioms of the calculus of relations only, rather than on their set-theoretic semantics).
% We thus see that $p$ is $\relone$-inductive if ${\relone}{\searrow}{p} \subseteq p$ 
% and that $\relone$ admits induction if 
%  $$
%  {\relone}{\searrow} p \subseteq p \implies \idvrel \subseteq p.
%  $$
%  }
\[
\infer{\forall y. p(y)}
{\forall y. (\forall x.\ x \relone y \to p(x)) \to p(y)}
\]
We now generalise this idea to $\Quantale$-relations and $\Quantale$-properties.
%First, let us define a $\Quantale$-predicate on a set $A$ as a map 
%$p: A \to \quantale$.

\begin{definition}
\label{def:inductive}
 Let $(A, \vrelone)$ be a \qars{}. 
 \begin{enumerate}
     \item A $\Quantale$-property 
     $p: A \to \quantale$ is \emph{$\vrelone$-inductive} if 
     the following holds for any $\ttwo \in A$:
         $$
         \meet_\tone \vrelone(\tone,\ttwo) \mmap p(\tone) \leq p(\ttwo).
         $$
     \item  We say that $\vrelone$ \emph{admits induction} if for any $\tone \in A$,
        $p(\tone) = \qunit$, for any 
         $\vrelone$-inductive predicate $p$.
 \end{enumerate}
\end{definition}

Notice that if $\vrelone$ admits induction and $p$ is a $\vrelone$-inductive predicate, 
then by \autoref{def:inductive} we obtain $p(\tone) = \qunit$, for any $\tone \in A$.\footnote{
Another (more liberal) option that we do not explore in this work is to require 
$\meet_\tone p(\tone) = \qunit$ in place of $(\forall \tone)\ p(\tone) = \qunit$.} 

\begin{remark}
Being $\vrelone$-inductive 
(as well as admitting induction) is a Boolean property. 
As already seen in \autoref{remark:graded-properties} for confluence, 
we could obtain a finer, 
\quant{} analysis of induction by $\Quantale$-enriching (i.e. grading) 
the properties of \autoref{def:inductive} in  
$\Quantale$, this way replacing, e.g., 
$\meet_\tone \vrelone(\tone,\ttwo) \mmap p(\tone) \leq p(\tone)$ with 
$\meet_\tone (\vrelone(\tone,\ttwo) \mmap p(\tone)) \mmap p(\tone)$. 
The latter formula, intuitively, gives the \emph{degree of inductiveness} of $\vrelone$.
\end{remark}

% \begin{definition}
% \label{def:well-founded}
%  We say that a $\Quantale$-relation $\vrelone$ is \emph{well-founded} if it admits induction, and 
%  that $\vrelone$ is \emph{terminating} if its $\dual{\vrelone}$ is well-founded.
% \end{definition}

Armed with \autoref{def:inductive}, we can now operationally 
identify terminating $\Quantale$-relations with those admitting induction. 
%by \autoref{def:well-founded}, we have essentially identified well-founded 
%$\Quantale$-relations with those admitting induction. 
Nonetheless, the reader may wonder whether 
there is an explicit characterisation of terminating relations 
in terms of familiar conditions akin to the equivalence between inductive and 
well-founded relations in the traditional case. 
The answer is in the affirmative 
and shows that $\Quantale$-relations admitting induction are 
precisely those that terminates in the strongest sense among those discussed 
at the beginning of this section.\footnote{It is an interesting question 
to determine if weaker and \quant{} refinements of \autoref{def:inductive} 
correspond to weaker and \quant{} notions of termination.}

% Even if this is operationally satisfactory, it is still natural to ask 
% for an explicit definition of terminating $\Quantale$-relations. 
% It turns out that $\Quantale$-relations admitting induction are 
% precisely those that terminates in the strongest sense among those discussed 
% at the beginning of this section.\footnote{It is an interesting questions 
% whether weaker and \quant{} refinements of \autoref{def:inductive} 
% correspond to weaker and \quant{} notions of termination.}

\begin{definition}
Let $(A, \vrelone)$ be a \qars{}. 
\begin{enumerate}
    % \item We say that a sequence 
    % $(x_{0},...,x_{n})$ of elements of $A$ is a \emph{($\vrelone$-)reduction} or \emph{rewriting 
    % sequence} if 
    % $$
    % \vrelone(x_{0},x_{1})\tensor  \cdots \tensor \vrelone(x_{n-1},x_{n}) \neq \qbot.
    % $$
    \item We say that $\tone \in A$ is a \emph{normal form} if $\join_\ttwo \vrelone(\tone,\ttwo) = \qbot$.
    \item We say that a reduction sequence
    $(\tone_{0},...,\tone_{n})$ \emph{terminates} if $\tone_n$ is a normal form.
    \item We say that $\vrelone$ is \emph{weakly normalizing} (WN) if for any $\tone \in A$ 
    there exists a normal form $\ttwo$ such that $\vrelone^*(\tone,\ttwo) \neq \qbot$. 
    \item We say that $\vrelone$ is \emph{strongly normalizing} (SN) if for each $\tone \in A$, 
    all reduction sequences starting from $\tone$ terminate.
\end{enumerate}
% \begin{align*}
%   &\forall (k\leq n).\vrelone(x_{0},x_{1})\tensor\vrelone(x_{1},x_{2})\tensor \cdot\cdot\cdot\tensor\vrelone(x_{k-1},x_{k})\neq \qbot  \\
% &\vrelone(x_{0},x_{1})\tensor\vrelone(x_{1},x_{2})\tensor \cdot\cdot\cdot\tensor\vrelone(x_{n-1},x_{n})\tensor\join_{y}\vrelone(x_{n},y)=\qbot
% \end{align*}
% We say that a sequence $(x_{1},...,x_{n})$ of elements of $A$ terminate or is strong normalizing iff 
% \begin{align*}
% &\vrelone(x_{0},x_{1})\tensor\vrelone(x_{1},x_{2})\tensor \cdot\cdot\cdot\tensor\vrelone(x_{n-1},x_{n})\neq \qbot \\
% &\join_{y}\vrelone(x_{n},y)=\qbot
% \end{align*}
% We say that $\vrelone$ is \emph{weakly normalizing} (WN) if each element 
% of $A$ has a terminating rewriting sequence starting from it. 
% We say that $\vrelone$ is strong normalizing (SN) if for each element $x \in A$, 
% all rewriting sequences starting from it terminate).
\end{definition}

% We say that a \qars{} $(A, \vrelone)$ is \emph{weakly normalizing} (WN, for short) if each element 
% of $A$ has a terminating rewriting sequence starting from it. Similarly, 
% we say that $(A, \vrelone)$ is strong normalizing (SN, for short) if each element of $A$ 
% has no non-terminating rewriting sequence starting from it (i.e. all rewriting sequences 
% starting from it terminate).

Notice that if a reduction sequence terminates, the sequence must be finite. 
Moreover, if $\vrelone$ is WN and $\tone \in A$, then there must exists an element $\ttwo$ 
such that $\vrelone^*(\tone,\ttwo) \neq \qbot$. That means $\join_n \vrelone^n(\tone,\ttwo) \neq \qbot$, 
which in turn means that there exists an actual index $n$ and elements 
$\tone = \tone_0, \tone_1, \hh, \tone_{n-1}= \ttwo$ such that $(\tone_0,\hh, \tone_{n-1})$ 
is a reduction sequence from $\tone$ 
(in particular, $\vrelone(\tone_i, \tone_{i+1}) \neq \qbot$, for any index $i$).

The next result shows that terminating and inductive $\Quantale$-relations 
are indeed one and the same. 

\begin{proposition}
\label{prop:induction-iff-SN}
 Let $(A, \vrelone)$ be a \qars{}. Then $\dual{\vrelone}$ admits induction if
 and only if $\vrelone$ is SN.
\end{proposition}

\begin{proof} We prove the two implications separately. 
\begin{itemize}
    \item[(${\implies}$)] Suppose $\dual{\vrelone}$ admits induction. We prove that 
    $\vrelone$ is SN. Let $p(\tone) = \qunit$ if all reduction sequences from $\tone$ 
    terminates, and $p(\tone) = \qbot$, otherwise. 
    We prove that $p$ is inductive, from which the thesis follows. 
    We have to show $\meet_\ttwo \vrelone(\tone,\ttwo) \mmap p(\ttwo) \leq p(\tone)$. 
    If $p(\tone) = \qunit$, then we are trivially done. Otherwise, $p(\tone) = \qbot$ 
    and we have a sequence $\tone = \tone_0, \tone_1, \hh$ such that $\vrelone(\tone_n, \tone_{n+1}) \neq \qbot$, 
    for any $n$. To prove $\meet_\ttwo \vrelone(\tone,\ttwo) \mmap p(\ttwo) \leq \qbot$, we 
    show $\vrelone(\tone,\tone_1) \mmap p(\tone_1) \leq \qbot$. By very definition of $\mmap$, 
    we have $\vrelone(\tone,\tone_1) \mmap p(\tone_1) = 
    \join \{\qone \mid \qone \tensor \vrelone(\tone, \tone_1) \leq p(\tone_1)\}$, 
    so that it is sufficient to show that for any $\qone$ such that 
    $\qone \tensor \vrelone(\tone, \tone_1) \leq p(\tone_1)$, we have $\qone \leq \qbot$. 
    Now, obviously $p(\tone_1) = \qbot$, so that $\qone \tensor \vrelone(\tone, \tone_1) = \qbot$ too. 
    Since $\Quantale$ is Lawverian, we then have that either $\qone = \qbot$ or
    $\vrelone(\tone, \tone_1) = \qbot$. Since $\vrelone(\tone, \tone_1) \neq \qbot$, we thus conclude 
    $\qone = \qbot$, and we are done.
    \item[(${\impliedby}$)] Suppose $\vrelone$ is SN and let $p$ be $\dual{\vrelone}$-inductive. 
    We prove $p(\tone) = \qunit$, for any $\tone \in A$. 
    We proceed by contradiction showing that if there exists $\tone \in A$ 
    such that $p(\tone) < \qunit$, then there also exists $\ttwo \in A$ 
    such that $\vrelone(\tone,\ttwo) \neq \qbot$ and $p(\ttwo) < \qunit$. 
    Therefore, if there is an $\tone$ such that $p(\tone) < \qunit$, we also 
    have a non-terminating reduction sequence from $\tone$, this way contradicting $SN$.
    So suppose to have an element $\tone$ such that $p(\tone) < \qunit$. 
    Suppose also, 
    for the sake of a contradiction, that for any $\ttwo$ either $\vrelone(\tone,\ttwo) = \qbot$ 
    or $p(\ttwo) < \qunit$. In both cases we obtain $\vrelone(\tone,\ttwo) \mmap p(\ttwo) = \qunit$, 
    and thus $\meet_\ttwo \vrelone(\tone,\ttwo) \mmap p(\tone) = \qunit$. Since $p$ is inductive, 
    we also have $\meet_\ttwo \vrelone(\tone,\ttwo) \mmap p(\ttwo) \leq p(\tone)$ and thus 
    $\qunit = \meet_\ttwo \vrelone(\tone,\ttwo) \mmap p(\ttwo) \leq p(\tone) < \qunit$. Contradiction. 
    % So, if $\vrelone(x,y) = \qbot$, then trivially $\vrelone(x,y) \mmap p(y) = \qunit$ 
    % (we have $\qbot \mmap a = \qunit$, for any $a$). If, instead 
    % $\vrelone(x,y) \neq \qbot$ but $p(y) < k$, then 
\end{itemize}
\end{proof}

We now have all the ingredients to quantitatively refining Newman's Lemma.

\begin{proposition}
\label{prop:newmans-lemma-pointwise}
 Let $(A, \vrelone)$ be a \emph{strongly normalising} \qars. 
 Then, $\vrelone$ is confluent if and only if it is locally confluent.
\end{proposition}

\begin{proof}
Obviously, if $\vrelone$ is confluent, then it is locally confluent too. 
We prove the converse. Suppose that $\vrelone$ is locally confluent and let 
us define the $\Quantale$-property $p$ as follows: 
$$
p(\tone)\defeq
\begin{cases}
\qunit  & \text{if } \forall b_1, b_2 \in A.\ \vrelone^*(a,b_1) \tensor \vrelone^*(a,b_2) \leq 
    \join_b \vrelone(b_1,b) \tensor \vrelone(b_2,b)
    \\
\qbot & \text{otherwise.}
\end{cases}
$$
Therefore, $p$ is a Boolean property, in the sense that for any $a \in A$, $p(a)$ is either $\qunit$ or 
$\qbot$. Moreover, we see that  
$p(a) = \qunit$ if and only if $\vrelone$ is confluent on $a$. 
Since $\vrelone$ is SN, it admits induction, and thus to prove the thesis it is sufficient to show that 
$p$ is inductive. Let us first notice that since $p$ is Boolean, 
we can simplify the proof of its inductiveness. 
\begin{claim}
To prove that $p$ is inductive it is enough to show:
\begin{align}
    \forall a.\ (\forall b.\ \vrelone(a,b) \neq \qbot \implies p(b) = \qunit) 
    &\implies p(a) = \qunit.
    \label{eq:newman-aux}
\end{align}
\end{claim}
\begin{proofoftheclaim}{}
To prove that $p$ is inductive, we have to show that for any $a$,  
$\meet_b \vrelone(a,b) \mmap p(b) \leq p(a)$. 
Now, since $p$ is Boolean, either $p(a) = \qunit$ of $p(a) = \qbot$. 
In the former case, we trivially have 
$\meet_b \vrelone(a,b) \mmap p(b) \leq \qunit = p(a)$
In the latter case (i.e. $p(a) = \qbot$), 
by \eqref{eq:newman-aux} there exists an element in $A$, call it $c$, such that 
$\vrelone(a,c) \neq \qbot$ and $p(c) = \qunit$. 
We then have
$$\meet_b \vrelone(a,b) \mmap p(b) \leq  \vrelone(a,c) \mmap p(c) = \qbot = p(a),
$$
since $\Quantale$ is Lawverian. 
\end{proofoftheclaim}

\noindent Coming back to the main proof, we have seen that we can conclude the main thesis by 
proving \eqref{eq:newman-aux}. So let us fix $a \in A$ and assume 
\begin{align}
    \forall b.\ \vrelone(a,b) \neq \qbot \implies p(b) = \qunit.
    \label{eq:newman-hyp}
\end{align}
We prove 
$p(a) = \qunit$, i.e. $\vrelone^*(a,b_1) \tensor \vrelone^*(a,b_2) \leq 
\join_b \vrelone^*(b_1,b) \tensor \vrelone^*(b_2,b)$, for arbitrary $b_1,b_2$. 
Let $\qone \defeq \join_b \vrelone^*(b_1,b) \tensor \vrelone^*(b_2,b)$.
Since $\vrelone = \idvrel \vee \vrelone;\vrelone^*$, it is sufficient to
prove 
$\idvrel(a,b_1) \vee (\vrelone;\vrelone^*)(a,b_1) \leq 
\vrelone^*(a,b_2) \mmap \qone$, which is itself implied by 
$\idvrel(a,b_1) \leq 
\vrelone^*(a,b_2) \mmap \qone$ and $(\vrelone;\vrelone^*)(a,b_1) \leq 
\vrelone^*(a,b_2) \mmap \qone$. For the former, we assume $a=b_1$ (the case for 
$a \neq b_1$ is trivial) and notice that 
$\vrelone^*(a,b_2) \leq \qone$ follows by taking $b = b_2$. 
For the former, we see that 
$(\vrelone;\vrelone^*)(a,b_1) \leq 
\vrelone^*(a,b_2) \mmap \qone$ is 
equivalent to $(\vrelone;\vrelone^*)(a,b_2) \leq 
\vrelone^*(a,b_1) \mmap \qone$.
We repeat the above argument, this time on $(\vrelone;\vrelone^*)(a,b_2)$, 
so that 
it is sufficient to prove $\idvrel(a,b_2) \leq (\vrelone;\vrelone^*)(a,b_1) \mmap \qone$ 
and $(\vrelone;\vrelone^*)(a,b_2) \leq (\vrelone;\vrelone^*)(a,b_1) \mmap \qone$. 
For the former, we proceed as usual. The real interesting case is the latter, 
which is equivalent to 
$$
(\vrelone;\vrelone^*)(a,b_1) \tensor (\vrelone;\vrelone^*)(a,b_2) \leq \qone.
$$
Since we have
$$
(\vrelone;\vrelone^*)(a,b_1) \tensor(\vrelone;\vrelone^*)(a,b_2) = 
\join_c \vrelone(a,c) \tensor \vrelone^*(c,b_1) \tensor \join_d \vrelone(a,d) \tensor \vrelone^*(d,b_2),
$$
it is enough to prove that for all $c,d$, we have 
$\vrelone(a,c) \tensor \vrelone^*(c,b_1) \tensor \vrelone(a,d) \tensor \vrelone^*(d,b_2) 
\leq \qone$. 
Now, if either $\vrelone(a,c) = \qbot$ or $\vrelone(a,d) = \qbot$, we are trivially done, since 
the quantale is integral. Suppose then they are all different from $\qbot$, so that we can apply 
\eqref{eq:newman-hyp} on both of them, this way obtaining $p(d) = p(c) = \qunit$. 
Since $\vrelone$ is locally confluent, we obtain:
\begin{align*}
    \vrelone(a,c) \tensor \vrelone^*(c,b_1) \tensor \vrelone(a,d) \tensor \vrelone^*(d,b_2)  
    &= \vrelone(a,c) \tensor \vrelone(a,d) \tensor \vrelone^*(c,b_1) \tensor \vrelone^*(d,b_2) 
    \\
    &\leq \join_e \vrelone^*(c,e) \tensor \vrelone^*(d,e) \tensor \vrelone^*(c,b_1) \tensor \vrelone^*(d,b_2)
\end{align*}
It is then sufficient to prove 
$\vrelone^*(c,e) \tensor \vrelone^*(d,e) \tensor \vrelone^*(c,b_1) \tensor \vrelone^*(d,b_2) \leq \qone$, 
for any $e \in A$. From $p(d) = p(c) = \qunit$, we obtain:
\begin{align*}
    \vrelone^*(c,e) \tensor \vrelone^*(d,e) \tensor \vrelone^*(c,b_1) \tensor \vrelone^*(d,b_2)
    &=  \vrelone^*(c,e) \tensor \vrelone^*(c,b_1)  \tensor \vrelone^*(d,e) \tensor \vrelone^*(d,b_2)
    \\
    &\leq \join_f \vrelone^*(e,f) \tensor \vrelone^*(b_1,f) \tensor \vrelone^*(d,e) \tensor \vrelone^*(d,b_2) 
    \\
    &= \join_f \vrelone^*(d,e) \tensor \vrelone^*(e,f) \tensor \vrelone^*(b_1,f)  \tensor \vrelone^*(d,b_2)
    \\
    &= \join_f \vrelone^*(d,f) \tensor \vrelone^*(b_1,f)  \tensor \vrelone^*(d,b_2)
    \\
    &= \join_f \vrelone^*(d,f) \tensor \vrelone^*(d,b_2) \tensor \vrelone^*(b_1,f)
    \\
    &\leq \join_f \join_g \vrelone^*(f,g) \tensor \vrelone^*(b_2,g) \tensor \vrelone^*(b_1,f)
    \\
    &= \join_f \join_g \vrelone^*(b_1,f) \tensor \vrelone^*(f,g) \tensor \vrelone^*(b_2,g)
    \\
    &\leq  \join_g \vrelone^*(b_1,g) \tensor \vrelone^*(b_2,g)
    \\
    &= \qone
\end{align*}
\end{proof}

\section{Quantitative Induction and Newman's Lemma, $\Quantale$-Relationally}

%To answer that question, we follow the abstract approach by \citet{backshouse-calculational-approach-to-mathematical-induction}
%and define induction $\Quantale$-\emph{relationally}. 

Even if mathematically fine, the previous section 
does not follow the relational style we have 
used to define \quant{} rewriting so far. 
\citet{backshouse-calculational-approach-to-mathematical-induction} 
have developed a relational theory of induction that allowed them to 
give an elegant, algebraic proof of Newman's Lemma. In this section, 
we extend their proof to \qars{s}. As already remarked, such an extension is 
nontrivial and builds upon the crucial notion of a relational modality 
(also known as a corelator \cite{DBLP:journals/pacmpl/LagoG22a,modal-reasoning-equal-metric-reasoning})
to define Boolean properties relationally. 

Let us begin by reviewing how \citet{backshouse-calculational-approach-to-mathematical-induction}  
deal with induction, algebraically.
In a nutshell, 
the (Boolean) calculus of classes is first embedded into the calculus of relations, this way defining 
(Boolean) predicates as relations satisfying suitable laws. 
Secondly, given a relation $\relone$ and a predicate $p$, 
the (semantics of the) predicate ${\relone}{\searrow}{p}$ 
is defined as\footnote{${\relone}{\searrow}{p}$ is actually defined relying on the 
axioms of the calculus of relations only, rather than on their set-theoretic semantics.}
$\{x \in A \mid \forall y.\ y \relone x \to p(y)\}$.
We then say that a predicate $p$ is $\relone$-inductive if ${\relone}{\searrow}{p} \subseteq p$ 
and that $\relone$ admits induction if 
 $$
 {\relone}{\searrow} p \subseteq p \implies \idvrel \subseteq p.
 $$
 
We now generalise this construction to the setting of $\Quantale$-relations. 
% The advantage of doing so is twofold: on one hand, we obtain a firm, algebraic 
% foundation for induction in a \emph{\quant{}} and \emph{metric} setting; on the other hand, 
% we obtain an elegant, algebraic, proof of the (\quant{)} Newman's Lemma.
First, we define the notion of a $\Quantale$-predicate. There are several ways to 
define predicates relationally. For instance, thinking about a $\Quantale$-relation 
$\vrelone: A \torel B$ as 
a $\Quantale$-valued matrix, we can see  
a predicate as a --- row or column --- vector~\cite{relational-mathematics}. 
Accordingly, we define a predicate over $A$ as a $\Quantale$-relation 
$p: A \torel 1$, with $1 \defeq \{*\}$ one-element 
set.\footnote{We thus view predicates as column vector. Equivalently, 
we may define predicates as row vectors, i.e. as $\Quantale$ relations $p: 1 \torel A$.} 
Notice that since $\reltop: 1 \torel 1$ coincides with $\idvrel$, any predicate $p$ satisfies 
$p; \reltop = p$.
Another way to define predicates is by means of \emph{coreflexive} $\Quantale$-relations~\cite{backshouse-calculational-approach-to-mathematical-induction} 
(also known as monotypes), whereby a predicate on $A$ is a $\Quantale$-relation 
$p: A \torel A$ such that $p \leq \idvrel$.
Vectors and coreflexives are equivalent notions, in the sense that there is an isomorphism 
between (column) vectors and coreflexives: any vector $p: A \torel 1$ 
gives the coreflexive\footnote{Notice that $\vrelone \tensor \idvrel$ = 
$\vrelone \wedge \idvrel$.} $(p;\reltop) \tensor \idvrel: A \torel A$; 
vice versa, any coreflexive $p: A \torel A$ gives the vector $p; \reltop$.

Given a column  $\Quantale$-vector $p$ and a $\Quantale$-relation $\vrelone$, 
we notice that 
$$(\vrelone \setminus p)(\ttwo,*) = \meet_\tone \vrelone(\tone,\ttwo) \mmap p(\tone,*)$$
gives exactly the formula we have used to define inductive predicates.
Moreover, the same formula can be obtained if $p$ is a coreflexive 
by considering $\vrelone \setminus (p;\reltop)$. Since we will extensively work with coreflexives, 
we introduce the notation ${\vrelone}{\searrow} p$ for 
$\vrelone \setminus (p;\reltop)$.

\begin{definition}
\label{def:inductive-2}
    Let $\vrelone: A \torel A$ be a $\Quantale$-relation.
    \begin{enumerate}
        \item A $\Quantale$-vector $p: A \torel 1$ is \emph{$\vrelone$-inductive} if 
            $\vrelone \setminus p \leq p$. We say that 
            $\vrelone$ \emph{admits vector induction} if, for any 
            $\Quantale$-vector $p: A \torel 1$, we have:
            $$
            \vrelone \setminus p \leq p \implies \reltop \leq p.
            $$
        \item A coreflexive $p: A \torel A$ is \emph{$\vrelone$-inductive} if 
            ${\vrelone}{\searrow} p \leq p$. We say that 
            $\vrelone$ \emph{admits coreflexive induction} if, for any 
            coreflexive $p: A \torel A$, we have:
            $$
            {\vrelone}{\searrow} p \leq p \implies \idvrel \leq p.
            $$
    \end{enumerate}
\end{definition}

\begin{remark}
\label{rem:condition-well-founded}
Thanks to the correspondence between vectors and coreflexives, it is easy to see that 
the notions of vector and coreflexive induction are equivalent, and 
that they indeed correspond to the pointwise notion of an inductive 
$\Quantale$-relation as given in \autoref{def:inductive}.
Morevoer, we can
abstract \autoref{def:inductive-2} from vectors and monotypes, 
and say, in full generality, that a $\Quantale$-relation $\vrelone$ \emph{admits induction} if 
$$
\vrelone \setminus \vreltwo \leq \vreltwo \implies \reltop \leq \vreltwo
$$
for any $\Quantale$-relation $\vreltwo$. Obviously, this definition subsumes 
those in \autoref{def:inductive-2}. One can also show that the vice versa holds too, 
and that the above three definitions of an inductive $\Quantale$-relation are equivalent. 
\end{remark}

To prove (the \quant{} refinement of) Newman's Lemma, we will not do indution on 
an arbitrary $\Quantale$-property, but on a \emph{Boolean} one. 
In previous section, we have modelled Boolean predicates as 
$\Quantale$-properties $p$ that are either equal to $\reltop$ or 
to $\relbot$. Even if correct, such a definition is operationally weak 
since it does not readily  come with useful algebraic laws and proof 
techniques. We overcome the problem by giving a modality-based 
definition of Boolean $\Quantale$-properties in the spirit of the 
exponential modality of linear logic \cite{DBLP:journals/tcs/Girard87}. 
To do so, we proceed as follows: first, 
we define a way to extract a Boolean property out of an $\Quantale$-enriched one. 
Since we can inject Boolean properties into $\Quantale$-enriched ones, 
we can then pick a $\Quantale$-property, extract a Boolean predicate out of it, 
and the (re)enriched it in $\Quantale$. We say that a property is Boolean if 
it is invariant under the above procedure.

% We conclude our analysis of \qars{} with a refinement of the well-known Newman's Lemma 
% to \qars s. Before proceedings any further, it is useful to internalise in the language 
% of $\Quantale$-relations the distinction 
% between Boolean and $\Quantale$-valued properties.
%, although the reader can 
%safely skip the next remark. 

% \autoref{def:inductive} defines induction principles 
% relying on $\Quantale$-valued properties. Sometimes, however, 
% what we want to prove by induction are actually Boolean, rather than $\Quantale$-valued, properties. 
% As an example of such a property, consider confluence. 
% In general, we say that $p$ is \emph{Boolean} if, 
% for any $\tone \in A$, $p(\tone)$ is either $\qunit$ or $\qbot$.\footnote{
% More generally, $\vrelone: A \torel B$ is Boolean if 
% $\vrelone(\tone, \ttwo)$ is either $\qbot$ or $\qunit$.} 
% The kind of Boolean 
% properties we are interested in are usually given by inequalities of the 
% form $\qone \leq \qtwo$, for $\qone,\qtwo \in \quantale$. To make such 
% inequality into $\Quantale$-property the reader may be tempted, following 
% the Boolean case, to internalise the order by
% considering the element $\qone \mmap \qtwo$ of $\quantale$. 
% This proposal is obviously wrong, as the 
% correct correspondence is: $\qone \leq \qtwo \iff \qunit \leq \qone \mmap \qtwo$. 

% Nonetheless, we can always extract a Boolean property out of a general one. 
% In fact,

Given a quantale $\Quantale$, there is a canonical adjucnction between 
$\Quantale$ and $\Two$ given by the maps $\varphi: \quantale \to \two$ and 
$\psi: \two \to \quantale$ defined thus: 
$$
\varphi(\qone) \defeq 
\begin{cases}
    \top & \text{ if } \qone = \qunit
    \\
    \bot & \text{ otherwise}
\end{cases}
\qquad 
\psi(x) \defeq
\begin{cases}
    \qunit & \text{ if } x = \top
    \\
    \qbot & \text{ otherwise}
\end{cases}
$$
Both $\varphi$ and $\psi$ form a so-called change of base functor 
\cite{Hoffman-Seal-Tholem/monoidal-topology/2014,Kelly/EnrichedCats}, and their composition 
$\psi \circ \varphi: \quantale \to \quantale$ is a change of base endofunctor.\footnote{
Change of base functors will play a crucial role in \autoref{sect:beyond-non-expansive-systems}}
Since we define Boolean $\Quantale$-relations (and thus $\Quantale$-properties) 
as those that are invariant under the map $\psi \circ \varphi$, we introduce a special 
notation for the latter.

\begin{definition} 
Define the (set-indexed family of) map(s) $\Box: \vrel(A,B) \to \vrel(A,B)$ 
by $\Box \vrelone \defeq \psi \circ \varphi \circ \vrelone$. 
We say that a $\Quantale$-relation is \emph{Boolean} if $\Box \vrelone =\vrelone$.
\end{definition}

% The map $\Box$ is deeply connected with the exponential modality of linear 
% logic~\cite{DBLP:journals/tcs/Girard87} and constitute an example of 
% a corelator~\cite{DBLP:journals/pacmpl/LagoG22a}. Indeed, 
The following result
(whose proof is straightforward) simply states that $\Box$ satisfies 
(some of) the axioms of a corelator \cite{DBLP:journals/pacmpl/LagoG22a,modal-reasoning-equal-metric-reasoning}. 
We will extensively use this fact in the proof of \autoref{prop:newmans-lemma}.

\begin{proposition}
The map $\Box$ obeys the following laws, where 
%$\mathtt{d}: A \to A \times A$ is the duplicator 
%function mapping $\tone$ to $(\tone,\tone)$ and
$\vrelone \otimes \vreltwo: A \times B \to A' \times B'$ 
is defined pointwise, for $\vrelone:A \to A'$ and $\vreltwo: B \to B'$.
 \begin{align*}
    \idvrel &\leq \Box \idvrel
    \label{eq:relator-id}
    \tag{rel-id}
    \\
    \Box\relone; \Box\reltwo &\leq \Box(\relone; \reltwo)
    \label{eq:relator-comp}
    \tag{rel-comp}
     \\
    \Box{\relone} &\leq \relone
    \label{eq:relator-dereliction}
    \tag{rel-der}
    \\
    \Box \relone \tensor \Box \reltwo &\leq \Box(\relone \tensor \reltwo)
    \label{eq:relator-tensor}
    \tag{rel-tensor}
    \\
     \Box \relone &\leq \Box \Box \relone
    \label{eq:relator-contraction}
    \tag{rel-contraction}
    \\
    \relone \leq \reltwo &\implies \Box\relone 
    \leq \Box\reltwo 
    \label{eq:relator-monotone}
    \tag{rel-mon}
    % \\
    % \Box \relone&\subseteq  \mathtt{d} ; (\Box{\relone} \otimes 
    % \Box{\relone}); \dual{\mathtt{d}}
\end{align*}
\end{proposition}

Using the map $\Box$ we can specialise the notion of coreflexive (and of a vector) to 
Boolean properties. 

\begin{definition} 
A Boolean property on $A$ is a corefliexive $p: A \torel A$ (i.e. $p \leq \idvrel$) 
such that $p = \Box p$.
\end{definition} 

Before stating our \quant{} version of Newman's Lemma, let us spell out some useful 
facts about Boolean properties.

\begin{lemma}
\label{lemma:newman-help-1}
Given $\Quantale$-relations $\vrelone, \vreltwo: A \torel A$ and a Boolean property $p: A \torel A$, we have:
\begin{enumerate}
    \item $\vrelone \searrow p$ is a Boolean property.
    \item $(\vrelone \searrow p);(\vreltwo \searrow p) = (\vrelone \vee \vreltwo) \searrow p$.
    \item $\vrelone \searrow p = p \swarrow \dual{\vrelone}$.
    \item $\vrelone;(\vrelone \searrow p) \leq p;\vrelone$ and 
        $(p \swarrow \vreltwo);\vreltwo \leq \vreltwo; p$.
\end{enumerate}
\end{lemma}

We are now ready to state and prove the \quant{} refinement 
of the abstract Newman's Lemma by \cite{backshouse-calculational-approach-to-mathematical-induction}.
% Our proof is entirely relational and generalises 
% the abstract Newman's Lemma by \cite{backshouse-calculational-approach-to-mathematical-induction} 
% but the reader 
% not familiar with relational methods can find an elementary pointwise 
% proofs in \autoref{appendix:newmanslemma}

\begin{proposition}
\label{prop:newmans-lemma}
 Let $\vrelone, \vreltwo: A \torel A$ be $\Quantale$-relations such that 
 $\vrelone \vee \dual{\vreltwo}$ admits induction. Then 
 $\vrelone; \vreltwo \leq \vreltwo^*; \vrelone^*$ implies  
$\vrelone^*; \vreltwo^* \leq \vreltwo^*; \vrelone^*$
\end{proposition}

Before proving \autoref{prop:newmans-lemma}, let us observe the following elementary fact.

\begin{lemma}
\label{lemma:newmans-helper-2}
$\vrelone^*; \vreltwo; \vrelthree^* \leq (\vrelone^*;\vreltwo) \vee (\vrelone^*;\vrelone;\vreltwo;\vrelthree;\vrelthree^*) \vee (\vreltwo; \vrelthree^*)$.
\end{lemma}

\begin{proof}[Proof of \autoref{prop:newmans-lemma}]
Since $\vrelone \vee \dual{\vreltwo}$ admits induction, for any 
coreflexive $p$ we have
$$
(\vrelone \vee \dual{\vreltwo}) \searrow p \leq p \implies \idvrel \leq p. 
$$
By \autoref{lemma:newman-help-1}, 
$(\vrelone \vee \dual{\vreltwo}) \searrow p = (\vrelone \searrow p);(p \swarrow \vreltwo)$, 
so that we obtain the following induction principle:
$$
(\vrelone \searrow p);(p \swarrow \vreltwo) \leq p \implies \idvrel \leq p. 
$$
We have to prove $\vrelone^*; \vreltwo^* \leq \vreltwo^*; \vrelone^*$ which, 
by adjunction,
is equivalent to 
$$
\idvrel \leq \vrelone^* \setminus (\vreltwo^*;\vrelone^*) / \vreltwo^*.
$$
We notice that for any $\Quantale$-relation $\vrelthree$, we have 
$\idvrel \leq \vrelthree$ if and only if $\idvrel \leq \Box(\vrelthree \wedge \idvrel)$. 
In fact, the left to right direction follows since $\Box \idvrel = \idvrel$, 
whereas the right to left direction follows from \eqref{eq:relator-dereliction}
($\idvrel \leq \Box(\vrelthree \wedge \idvrel) \leq \vrelthree \wedge \idvrel \leq \vrelthree$).
Therefore, to prove $\idvrel \leq \vrelone^* \setminus (\vreltwo^*;\vrelone^*) / \vreltwo^*$, 
it is enough to show 
$$
\idvrel \leq \underbrace{\Box((\vrelone^* \setminus (\vreltwo^*;\vrelone^*) / \vreltwo^*) \wedge \idvrel)}_{p}.
$$
Notice that $p$ is a Boolean coreflexive, and thus we can rely on inductiveness of 
$\vrelone \vee \dual{\vreltwo}$ and obtain the proof obligation
$(\vrelone \searrow p);(p \swarrow \vreltwo) \leq p$.
Since $p$ is Boolean, then so are $\vrelone \searrow p$ and $p \swarrow \vreltwo$, 
so that \eqref{eq:relator-comp} gives us:
\begin{align*}
    (\vrelone \searrow p);(p \swarrow \vreltwo) 
    = \Box(\vrelone \searrow p);\Box(p \swarrow \vreltwo)
    \leq \Box((\vrelone \searrow p);(p \swarrow \vreltwo)).
    %\tag{By \eqref{eq:relator-comp}}
\end{align*}
Therefore, our thesis becomes 
$$
\Box((\vrelone \searrow p);(p \swarrow \vreltwo)) \leq p  
= \Box((\vrelone^* \setminus (\vreltwo^*;\vrelone^*) / \vreltwo^*) \wedge \idvrel).
$$
By \eqref{eq:relator-monotone}, it is sufficient to prove 
$(\vrelone \searrow p);(p \swarrow \vreltwo) \leq 
(\vrelone^* \setminus (\vreltwo^*;\vrelone^*) / \vreltwo^*) \wedge \idvrel
$ which amounts to show
\begin{align*}
   (\vrelone \searrow p);(p \swarrow \vreltwo) &\leq \idvrel
    \\
    (\vrelone \searrow p);(p \swarrow \vreltwo) &\leq 
    \vrelone^* \setminus (\vreltwo^*;\vrelone^*) / \vreltwo^*.
\end{align*}
The former inequation is straightforward as both 
$\vrelone \searrow p$ and $p \swarrow \vreltwo$ are coreflexives 
(and thus $\vrelone \searrow p \leq \idvrel$ and $p \swarrow \vreltwo \leq \idvrel$), 
since $p$ is.
Let us now move to the second inequation. By adjunction, we have to show
$$
 \vrelone^*; (\vrelone \searrow p);(p \swarrow \vreltwo); \vreltwo^* \leq \vreltwo^*;\vrelone^*.
$$
By \autoref{lemma:newmans-helper-2}, we reduce the proof to the following three inequations:
\begin{align*}
    \vrelone^*; (\vrelone \searrow p);(p \swarrow \vreltwo) &\leq \vreltwo^*;\vrelone^*
    \\
    \vrelone^*;\vrelone; (\vrelone \searrow p);(p \swarrow \vreltwo);\vreltwo; \vreltwo^* 
    &\leq \vreltwo^*;\vrelone^*
    \\
    (\vrelone \searrow p);(p \swarrow \vreltwo); \vreltwo^* 
    &\leq \vreltwo^*;\vrelone^*.
\end{align*}
For the first one, since both 
$\vrelone \searrow p$ and $p \swarrow \vreltwo$ are coreflexives, 
we have:
$$
\vrelone^*; (\vrelone \searrow p);(p \swarrow \vreltwo) 
\leq \vrelone^*; \idvrel; \idvrel \leq \vrelone^*; \vreltwo^*.
$$
We prove the third inequation in a similar fashion. Let us now move to the 
second one. For readability, let $\vrelthree$ be $\vrelone^* \setminus (\vreltwo^*;\vrelone^*) / \vreltwo^*$,
so that $p = \Box(\vrelthree \wedge \idrel)$.
We have:
\begin{align*}
    \vrelone^*;\vrelone; (\vrelone \searrow p);(p \swarrow \vreltwo);\vreltwo; \vreltwo^* 
    &\leq 
    \vrelone^*; p; \vrelone ; \vreltwo; p; \vreltwo^* 
    \tag{\autoref{lemma:newman-help-1}, item 4}
    \\
    &\leq \vrelone^*; p; \vreltwo^* ; \vrelone^*; p; \vreltwo^* 
    \tag{Hypothesis}
    \\
    &= \vrelone^*; \Box(\vrelthree \wedge \idvrel); 
    \vreltwo^* ; \vrelone^*; \Box(\vrelthree \wedge \idvrel); \vreltwo^* 
    \\
    &\leq \vrelone^*; (\vrelthree \wedge \idvrel); 
    \vreltwo^* ; \vrelone^*; (\vrelthree \wedge \idvrel); \vreltwo^* 
    \tag{\ref{eq:relator-dereliction}}
    \\
    &\leq \vrelone^*; \vrelthree; 
    \vreltwo^* ; \vrelone^*; \vrelthree; \vreltwo^* 
    %\tag{\ref{eq:relator-dereliction}}
    \\
    &= \vrelone^*; \vrelone^* \setminus (\vreltwo^*;\vrelone^*) / \vreltwo^*; 
    \vreltwo^* ; \vrelone^*; \vrelone^* \setminus (\vreltwo^*;\vrelone^*) / \vreltwo^*; \vreltwo^* 
    %\tag{\ref{eq:relator-dereliction}}
     \\
    &\leq \vreltwo^*;\vrelone^*; \vrelone^*; \vrelone^* \setminus (\vreltwo^*;\vrelone^*) / \vreltwo^; \vreltwo^* 
    \\
    &\leq \vreltwo^*;\vrelone^*; \vrelone^* \setminus (\vreltwo^*;\vrelone^*) / \vreltwo^; \vreltwo^* 
    %\tag{\ref{eq:relator-dereliction}}
    \\
    &\leq \vreltwo^*;\vreltwo^*;\vrelone^*
    \\
    &\leq \vreltwo^*;\vrelone^*.
    %\tag{\ref{eq:relator-dereliction}}
\end{align*}
\end{proof}

\begin{corollary}[Newman's Lemma]
\label{corollary:newman-lemma}
Let $(A,\vrelone)$ be a \qars{.} If 
$\vrelone$ is SN, then $\vrelone$ is confluent if and only if it is locally confluent. 
\end{corollary}

% To recover the usual formulation of Newman's Lemma (properly refined to a 
% \quant{} setting), we instantiate $\vrelone$ and $\vreltwo$ in \autoref{prop:newmans-lemma} 
% as $\dual{\vrelthree}$ and $\vrelthree$, respectively, for $\vrelthree$ arbitrary 
% $\Quantale$-relation. Consequently, the hypothesis that $\vrelone \vee \dual{\vreltwo}$ 
% admits induction collapses to $\dual{\vrelthree}$ admitting induction, which is equivalent 
% to $\vrelthree$ being SN, and the thesis precisely gives confluence of $\vrelthree$ 
% (assuming its local confluence).

\begin{proof}
We instantiate $\vrelone$ and $\vreltwo$ in \autoref{prop:newmans-lemma} 
as $\dual{\vrelone}$ and $\vrelone$, respectively. 
Consequently, the hypothesis that $\vrelone \vee \dual{\vreltwo}$ 
admits induction collapses to
$\dual{\vrelone}$ admitting induction, which is equivalent 
to $\vrelone$ being SN. 
\autoref{prop:newmans-lemma} then precisely gives confluence of $\vrelone$ 
(assuming its local confluence).
\end{proof}

\begin{remark}
In the proof of \autoref{corollary:newman-lemma}, we have actually used 
the equivalence between $\Quantale$-relations admitting induction and terminating (well-founded)
ones, as proved in \autoref{prop:induction-iff-SN}. We can indeed safely do 
so as we have seen that the relational pointfree definition of an 
inductive relation (\autoref{def:inductive-2}) coincides with its 
pointwise counterpart (\autoref{def:inductive}). 
Nonetheless, a complete relational analysis of (\quant{)} Newman's Lemma 
requires a relational account of termination too. 
Doing that is beyond the scope of this paper, although it can be 
done with a reasonable effort. As a guideline, we simply say that 
a $\Quantale$-relation $\vrelone$ is \emph{well-founded} if 
$$
\vreltwo \leq \vreltwo; \vrelone \implies \vreltwo \leq \relbot
$$
for any $\vreltwo$ (similar definitions can be obtained restricting to 
vectors and monotypes, as in \autoref{rem:condition-well-founded}), 
and that $\vrelone$ is SN if $\dual{\vrelone}$ is well-founded.
\end{remark}

\section{QUANTITATIVE TERM REWRITING SYSTEMS: A SHORT PHENOMENOLOGY}
Having introduced the general theory of \quant{} abstract rewriting systems, 
in the remaining sections of this paper we shall introduce \emph{\quant{} term rewriting systems}  
and their connection with \quant{} algebras. Contrary to traditional term rewriting systems, 
there are several notions of a \quant{} term rewriting system (and of their associated notion of a 
\quant{} equational theory), 
each of which is associated with a 
suitable notion of non-expansiveness of functions. In the next section, 
we shall deal with \emph{non-expansive} 
\quant{} term rewriting systems, leaving to \autoref{sect:beyond-non-expansive-systems} 
the analysis of \emph{graded} \quant{} term rewriting systems, the most general 
class of \quant{} term-based systems we will study in this work. 
Before diving into the theory of non-expansive systems, however, it is instructive to 
anticipate a bit of term-systems phenomenology. 

\paragraph{Non-Expansive Systems}
Non-expansive term rewriting systems (\qtrs{s}, for short) are quantitative systems in which 
reducing terms inside contexts \emph{non-expansively} propagates distances. 
Therefore, if $\termone$ reduces to $\termtwo$ with distance $\qone$, then 
$C[\termone]$ reduces to $C[\termtwo]$ with distance $\qone$, too. 
That is, by thinking about the context $C$ as a function on terms, then 
$C$ is non-expansive with respect to the rewriting distance. 
To make this semantic choice coherent at a rewriting level, systems have to be \emph{linear}, as 
non-linearity of terms breaks non-expansiveness 
(cf. distance amplification in \autoref{sect:combinators-intro}).

\paragraph{Additive Systems}
Additive (term rewriting) systems constitute the subclass of \qtrs{s} whose quantale is idempotent. 
Even if \quant{,} the monoidal structure of additive systems collapses to a \emph{cartesian} one, as 
the tensor product 
of an idempotent quantale coincides with the meet of its underlying lattice. 
The main consequence of that is that non-expansiveness of rewriting is semantically 
coherent even with \emph{non}-linearity of systems, so that we can have non-linear additive systems 
that do not suffer neither confluence nor distance amplification issues. 
The theory of additive systems is essentially the same as the one of traditional rewriting systems, 
the latter being the prime examples of additive systems. 

\paragraph{Graded Systems}
Graded systems constitute the largest class of term-based \quant{} rewriting systems. 
Contrary to non-expansive systems, in a graded system the distance 
generated by a reduction  $\termone \qreduce{\qone} \termtwo$ can be
amplified (or reduced\footnote{In which case we may talk of \emph{contractive systems}.}) when performed in a context. Thus for instance, 
we may have that $\termone$ reduces to $\termtwo$ with distance $\qone$, but 
$C[\termone]$ reduces to $C[\termtwo]$ with distance $\baseone_{C}(\qone)$. 
The map $\baseone_C$ is known as the \emph{grade} or \emph{sensitivity} of the context $C$, 
and it gives the law determining how much distances are amplified by $C$. 
For instance, if we work with the Lawvere quantale, $\baseone_C$ is usually a 
multiplication by a constant map, the intended semantic meaning of 
such a map being a generalised Lipschitz constant associated 
to $C$ when regarded as a function. 
Graded term rewriting systems are an example of \emph{modal} and \emph{coeffectful} systems; 
and because of their modal nature, they allow us to drop the linearity constraint 
of non-expansive systems without incurring in (semantic and rewriting) (in)consistency 
issues. The price to pay for that is the need for a more sophisticated (meta)theory 
than the one of non-expansive systems. 
The latter, in fact, can be seen as trivial graded systems in which all contexts have grade 
given by the identity function (i.e. no amplification).

\section{Quantitative Term Rewriting: Non-Expansive Systems}
\label{sect:qtrs}

Let us now formally introduce non-expansive systems.
Through this section, let $\Quantale = (\quantale, \leq, \tensor, \qunit)$. 
be a fixed \emph{continuous} quantale. % with base $\qbase$. 
Before going any further, we shortly recall some of the (standard) notions and notation 
we will use in the rest of the paper.
\begin{description}
    \item[Terms]
    For a signature $\signature$ and a countable set of variables $\variables$, we write $\terms{\signature}{\variables}$ 
    for the collection of ($\signature$-)terms over $\variables$. 
    We use small Latin letters $\termone, \termtwo, \termthree, \hh$ to range over terms, 
    sometimes using letters $a, b, c, \hh$ too.
    \item[Positions]
    Recall that a \emph{position} $\posone$ is a finite string of positive integers. We denote by 
    $\posroot$ the empty string and by $\posone\postwo$ the concatenation of positions $\posone$, $\postwo$;
    we write 
    $\posone\leq\postwo$ if $\posone$ is a prefix of $\postwo$, i.e, if there is $\posthree$ such that $\postwo=\posone\posthree$. 
    We write $\posone\parallel \postwo$ if $\posone\not\leq\postwo$ and $\postwo\not\leq\posone$.
    Finally, we denote by $\subtermpos{\termone}{\posone}$ the subterm of $\termone$ at position $\posone$. 
    If $\subtermpos{\termone}{\posone}=\termtwo$, we will also write $\termone[\termtwo]_{\posone}$.
    \item[Context] 
    A context is a term over the signature $\signature \cup \{\Box\}$. 
    We write $\ctxone[\cdot]$ for a context containing a single occurrence of $\Box$ 
    and use the notation $\ctxone[\termone]$ to denote the term obtained by replacing 
    the (single) occurrence of $\Box$ with $\termone$ in $\ctxone[\cdot]$.
    % If the position is not relevant, we shall use the notation $\termone[\termtwo]$ and think about $\termone$ 
    % as a context (to stress that, the letter $C$ will also be employed to range over terms acting 
    % as contexts)
    \item[Substitution]
    We denote substitutions by $\sone, \stwo, \hh$ and write 
    $\termone^{\sone}$ in place of $\sone(\termone)$. Furthermore, given two substitutions $\sone,\stwo$, 
    we write $\sone \preceq \stwo$ if there exists $\sthree$ such that $\stwo=\sone\sthree$, where $(\sone\sthree)(\termone) \defeq \sone(\sthree(\termone))$.
    Given two term $\termone$ and $\termtwo$, if $\termone^{\sone}=\termtwo^{\sone}$, then $\sone$ is a unifier of $\termone$ and $\termtwo$, 
    while $\termone$ and $\termtwo$ are said to be \emph{unifiable}. Finally, 
    recall that the \emph{most general unifier} ($\mgu$) of two unifiable terms is 
    their minimal unifier with respect to $\preceq$. 
    \item[Linearity] 
    We say that a term $\termone$ is linear if it has no multiple occurrences of the same variable. 
    We say that a mathematical expression (such as a relation or a predicate) involving terms is linear if 
    all terms appearing in it are linear.
\end{description}

We are now ready to define non-expansive \quant{} term rewriting systems, 
which we simply refer to as $\Quantale$-term rewriting systems.
\begin{definition}
\label{def:qtrs}
    A \emph{$\Quantale$-term rewriting system} (\qtrs{,} for short) 
    is a pair $\qtrsone = (\signature, \stepto_{\vbaserelone})$ consisting of 
    a signature $\signature$ and
    a $\Quantale$-ternary relation\footnote{
    In \quant{} algebra,
    it is customary to consider ternary relations on a \emph{base} of the quantale, rather than 
    on the quantale itself~\cite{An-Internal-Language-for-Categories-Enriched-over-Generalised-Metric-Spaces}. 
    Thus for, instance, in the case of the Lawvere quantale, 
    we should take relations over non-negative rationals in place of those over $[0,\infty]$. 
    Although this choice makes definitions computationally lighter, for our results 
    working with ternary relations over elements of the quantale or over a base thereof 
    makes no difference. Consequently, we will continue working with full $\Quantale$-ternary relations. 
    Nonetheless, the reader can safely pretend such relations to be over a base of $\Quantale$ 
    (see any textbook on lattices and domains \cite{AbramskyJung/DomainTheory/1994,continuous-lattices-and-domains,DaveyPriestley/Book/1990} 
    for the definition of a base of a complete lattice, and the work by
    \citet{An-Internal-Language-for-Categories-Enriched-over-Generalised-Metric-Spaces} 
    for an example of ternary relations over a base of a quantale.
%     To make definitions computationally lighter, the literature on \quant{} algebraic
% theories usually considers ternary relations over a \emph{base} of $\Quantale$, 
% the carrier of such 
% a base usually being considerably smaller than $\Omega$. 
% For instance, \quant{} equational theories are often defined using ternary relations 
% over non-negative rationals, the latter being a base for $[0,\infty]$. 
% Even if our results are independent of working with bases or with full quantales, 
% we stick with the literature and define (in \autoref{sect:qtrs}) \quant{} term rewriting 
% systems relying on bases of quantales. 
% ▶ Example 11. The Boolean quantale (({0 ≤ 1}, ∨), ⊗ := ∧) is finite and thus continuous [14].
% Since it is continuous, {0, 1} itself is a basis for the quantale that satisfies the conditions above.
% For the Gödel t-norm [11] (([0, 1], ∨), ⊗ := ∧), the way-below relation is the strictly-less
% relation < with the exception that 0 < 0. A basis for the underlying lattice that satisfies
% the conditions above is the set Q ∩ [0, 1]. Note that, unlike real numbers, rationals numbers
% always have a finite representation. For the metric quantale (also known as Lawvere quantale)
% (([0, ∞], ∧), ⊗ := +), the way-below relation corresponds to the strictly greater relation with
% ∞ > ∞, and a basis for the underlying lattice that satisfies the conditions above is the set of
% extended non-negative rational numbers. The latter also serves as basis for the ultrametric
% quantale (([0, ∞], ∧), ⊗ := max).
    } $\stepto_{\vbaserelone}$ over $\signature$-terms.
    The (rewriting) $\Quantale$-ternary relation $\reduce_{\vrelone}$ generated by $\stepto_{\vrelone}$ is
    defined by the rules 
    in \autoref{figure:qtrs}. 
\end{definition}

    \begin{figure}
    {
    \centering
    \begin{tcolorbox}[boxrule=0.5pt,width=\linewidth,colframe=black,colback=black!0!white,arc=0mm]
    \vspace{-0.3cm}
      \[
    \infer{\qone \Vdash C[\subst{\redex}{\sone}] \to_{\vbaserelone} C[\subst{\contractum}{\sone}]}
        {\qone \Vdash \redex \stepto_{\vbaserelone} \contractum}
    \quad
    \infer{\qtwo \Vdash \termone \reduce_{\vbaserelone} \termtwo}
    {\qone \Vdash \termone \reduce_{\vbaserelone} \termtwo & \qtwo \leq \qone}
    \quad 
    \infer{\join \qone_i \Vdash \termone \reduce_{\vbaserelone} \termtwo}
    {\qone_1 \Vdash \termone \reduce_{\vbaserelone} \termtwo & \hdots & \qone_n \Vdash \termone \reduce_{\vbaserelone} \termtwo}
    \quad 
    \infer{\qone \Vdash \termone \reduce_{\vbaserelone} \termtwo}
    {\forall \qtwo \ll \qone.\ \qtwo \Vdash \termone \reduce_{\vbaserelone} \termtwo}
    \]
    \end{tcolorbox}
    }
    \caption{Definition of $\reduce_{\vrelone}$}
    \label{figure:qtrs}
    \end{figure}

\begin{notation}
We refer to a triple $(\redex, \qone, \contractum) \in {\stepto_{\vrelone}}$, i.e. such that 
$\redex \qstepto{\qone}_{\vrelone} \contractum$ as a (reduction) \emph{rule}. We call $\redex$ the \emph{redex}, 
and $\contractum$ the \emph{contractum}. 
Moreover, when $\vbaserelone$ is irrelevant or clear from the context, we shall write $\reduce$ in place of 
$\reduce_{\vbaserelone}$ and use the notation $\termone \qreduce{\qone} \termtwo$ in 
place of $\qone \Vdash \termone \reduce \termtwo$. 
% Moreover, since any \qtrs{} $(\signature, \stepto_{\vbaserelone})$ induces a \qars{}, 
% we can apply all definitions and 
% results seen in previous sections to \qtrs s. In doing so, we well oftentimes 
% refer to 
% $(\signature, \stepto_{\vbaserelone})$ (saying, e.g. that the latter is induces a \qars{} 
%     $(\terms{\signature}{\variables}, \vrelone)$
%Finally, we adopt classic notational conventions, 
%writing $\leftarrow_{\vbaserelone}$ and $\twoheadrightarrow_{\vbaserelone}$ 
%for $\dual{(\reduce_{\vbaserelone})}$ and $\reduce_{\vbaserelone}^*$, respectively.
\end{notation}

The first defining rule of the relation $\reduce_{\vrelone}$ in \autoref{def:qtrs} is the 
main rewriting rule: it states that rewriting can be performed inside any context and on any instance 
of reductions in $\stepto_{\vrelone}$. This rule --- which is standard in traditional 
rewriting --- reflects the (semantic) assumption that operation symbols in $\signature$ behave 
as \emph{non-expansive} functions: accordingly, contexts do not amplify rewriting distances.
The reaming rules encode structural properties of \quant{} rewriting: the first 
giving a form of \quant{} weakening, the second stating that rewriting is closed under \emph{finite} join,
and the third stating a generalised continuity property. 
On the Lawvere quantale, for instance, we can read $\termone \qreduce{\qone} \termtwo$ as stating that 
$\termone$ reduces to $\termtwo$ within an error of at most $\qone$. Equivalently, we can redeuce 
$\termone$ to the \emph{non}-semantically equivalent term $\termtwo$, which differs from 
$\termone$ of at most $\qone$. Accordingly, the structural rules in \autoref{def:qtrs} respectively state that 
if $\termone \qreduce{\qone} \termtwo$ and $\qone \leq \qtwo$, then we also have 
$\termone \qreduce{\qtwo} \termtwo$; that rewriting is closed under (necessarily \emph{finite}) minima; 
and that rewriting satisfies the Archimedean property~\cite{plotkin-quantitative-algebras-2016}: 
to prove $\termone \qreduce{\qone} \termtwo$, 
it is enough to prove $\termone \qreduce{\qtwo} \termtwo$ for any $\qtwo$ strictly bigger than $\qone$ 
(i.e. $\qtwo > \qone$). 

Any \qtrs{} $(\signature, \stepto_{\vbaserelone})$ induces a \qars{} 
whose objects are $\signature$-terms and whose 
rewriting $\Quantale$-relation 
$\vrelone: \terms{\signature}{\variables} \torel \terms{\signature}{\variables}$ 
    defined by
    $$
    \vrelone(\termone, \termtwo) \defeq \join \{\qone \mid \qone \Vdash \termone \reduce_{\vbaserelone} \termtwo\}.
    $$
Consequently, all definitions and results seen so far extend to \qtrs{s}. 
for that reason, we oftentimes say that a \qtrs{} $(\signature, \stepto_{\vbaserelone})$ 
has a given property when we actually mean that its associated \qars{} 
$(\terms{\signature}{\variables}, \vrelone)$ has it.

\begin{remark}
\label{rem:structural-rules}
\autoref{def:qtrs} stipulates that $\reduce_{\vrelone}$ must be closed under suitable 
structural rules, viz. weakening, closure under finite joins, and the so-called 
(\emph{infinitary}) Archimedean 
rule \cite{plotkin-quantitative-algebras-2016}. 
We have included such rules to stay as close as possible to the literature on \quant{} 
equational theories, where structural rules are used to ensure completeness of 
equational proof systems. From a rewriting perspective, however, such rules 
can be safely (and maybe naturally) avoided, arguably with the exception of weakening. 
In fact, not only having weakening as the only structural rule is a natural design choice 
at a semantic level (making, e.g., the defining rules of $\reduce_{\vrelone}$ 
\emph{finitary}), but it also strengthen the theory of \qtrs{} we are going to develop. 
In particular, the presence of the Archimedean rule forces us to formulate, e.g.,
confluence results at the level of the $\Quantale$-relation $\vrelone$, and one may wonder  
whether confluence holds also at the level of the ternary relation $\reduce_{\vrelone}$. 
The answer is, in general, in the negative. Nonetheless, an affirmative answer can 
be given if we drop all structural rules but weakening. This suggests that 
it is worth considering an alternative, structurally-free definition of 
\qtrs{s}. We will follow this path in \autoref{sect:beyond-non-expansive-systems}, where 
we shall define graded systems using weakening as the only structural rule.
\end{remark}

Let us now see some examples of \qtrs s, focusing, in particular, to the systems 
presented in \autoref{section:long-intro}.

\begin{example}
Traditional term rewriting systems are nothing but $\maketrs{\Two}\text{s}$.
\end{example}

\begin{example} 
All the examples seen in \autoref{section:long-intro} are $\maketrs{\Lawvere}\text{s}$. 
In particular, systems 
\begin{align*}
    \NATS &= (\signature_{\NATS}, \stepto_{\natrelone})
    &
    \BA &= (\signature_{\mathbf{\BA}}, \stepto_{\baryrelone})
    &
    \BCK &= (\signature_{\BCK}, \stepto_{\combrelone})
    &
    \TICK &= (\signature_{\TICK}, \stepto_{\tickrelone}) 
\end{align*}
as well combinations thereof (e.g. 
system $\BCKprob = (\signature_{\BCK} \cup \signature_{\BA}, \stepto_{\combrelone\baryrelone})$)
are all $\maketrs{\Lawvere}\text{s}$.
\end{example}

\begin{example}
Any \quant{} string rewriting system can be modelled as \qtrs{.}. 
In particular, all \quant{} string rewriting systems of \autoref{sect:string-rewriting} 
can be gives as $\maketrs{\Lawvere}$.
To do so, we consider modify the signature seen in \autoref{sect:string-rewriting} 
by taking $\signature_{\DNAone} \defeq \{\A, \C, \G, \T, \oo\}$, where 
$\A, \C, \G, \T$ are \emph{unary} function symbol and $\oo$ is a constant acting 
as the empty string. Thus, for instance, we model the string 
$\A\G\T\C$ as the term $\A(\G(\C(\T(\oo))))$.
Next, we adapt the rewriting relation previously introduced to 
act on terms (rather than strings). We thus obtain
the rewrite $\Lawvere$-relation $\stepto_{\dnarelone}$ defined as follows, 
where $b,c \in \{\A, \C, \G, \T\}$ and $b \neq c$ in the third rule.

{
\centering
\begin{tcolorbox}[boxrule=0.5pt,width=(\linewidth*2)/3,colframe=black,colback=black!0!white,arc=0mm]
$$
    x \qstepto{1}_{\dnarelone} b(x) 
    \qquad
    b(x) \qstepto{1}_{\dnarelone} x
    \qquad
    b(x) \qstepto{1}_{\dnarelone} c(x) 
$$
\end{tcolorbox}
}

\noindent As seen in \autoref{sect:string-rewriting}, (the \qtrs{} version of) system 
$\DNAone = (\signature_{\DNAone}, \stepto_{\dnarelone})$ operationally describes the Levenshtein 
distance~\cite{string-algorithms} between DNA sequences. 
Further edit distances on DNA molecules can be easily obtained 
modifying system $\DNAone$. For instance,
considering the third defining rule of $\stepto_{\dnarelone}$ (i.e.  $b(x) \qstepto{1}_{\dnarelone} c(x)$) 
only, we obtain an operational description of the Hamming distance~\cite{string-algorithms}, 
whereas the following system gives the Eigen–McCaskill–Schuster distance (one 
obtains the Watson–Crick distance similarly) \cite{encyclopedia-of-distances}.

{
\centering
\begin{tcolorbox}[boxrule=0.5pt,width=(\linewidth*2)/3,colframe=black,colback=black!0!white,arc=0mm]
\vspace{-0.3cm}
\begin{align*}
    \A(x)&\qstepto{1}_ \C(x)
    &
    \G(x) &\qstepto{1} \T(x)
    &
    \A(x) &\qstepto{1} \T(x)
    \\
    \A(x) &\qstepto{0} \G(x)
    &
     \G(x)&\qstepto{1} \C(x)
     &
     \C(x) &\qstepto{0} \T(x)
\end{align*}
\end{tcolorbox}
}
% Moreover, we can easily generalise this example to other edit-like distances. 
% For instance, the system 
% \begin{align*}
%     M(x) &\qstepto{1} N(x) & M, N \in \{\A, \C, \G, \T\}, M \neq N
% \end{align*}
% gives the Hamming distance \cite{string-algorithms} between DNA sequences --- so that the induced distance 
% measures the number of mutations between molecules --- whereas the 
% following system gives the Eigen–McCaskill–Schuster distance (one 
% obtains the Watson–Crick distance similarly) \cite{encyclopedia-of-distances}, 
% which is used to study virus and
% cancer proliferation under control of drugs or the immune system:
% \begin{align*}
%     \A(x) &\qstepto{0} \G(x) 
%     \\
%     \A(x) &\qstepto{1} \C(x) 
%     \\
%     \A(x) &\qstepto{1} \T(x) 
%     \\
%     \C(x) &\qstepto{0} \T(x)
%     \\
%     \C(x) &\qstepto{1} \G(x).
% \end{align*}
\end{example}

\begin{example}
\label{ex:quantitative-semilattices}
Consider the signature $\signature_{\QSL}$ containing a single binary 
operation $\cup$ for nondeterministic choice and let $\stepto_{\qslrelone}$ be
the following rewriting relation.

{
\centering
\begin{tcolorbox}[boxrule=0.5pt,width=(\linewidth*3)/4,colframe=black,colback=black!0!white,arc=0mm]
\vspace{-0.3cm}
\begin{align*}
   x \qstepto{0}_{\qslrelone} x \cup x
    \qquad
    (x \cup y) \cup z \qstepto{0}_{\qslrelone} x \cup (y \cup z)
    \qquad 
    (x \cup y) \qstepto{0}_{\qslrelone} (y \cup x)
\end{align*}
\end{tcolorbox}
}
As it is, this rewriting system is not that interesting. 
The key point here is the choice of the quantale used for distances. Contrary to previous examples, 
here we consider the \emph{strong} Lawvere quantale $\StrongLawvere$. 
The choice of this quantale largely impacts on the definition of $\reduce_{\qslrelone}$, which 
now gives a form of non-expansiveness of $\cup$ reflecting the 
\emph{ultrametric}~\cite{steen/CounterexamplesTopology/1995} structure of $\StrongLawvere$: 
\[
\infer{x \cup y \qreduce{\max(\qone,\qtwo)} x' \cup y'}
{x \qreduce{\qone} x' & y \qreduce{\qtwo} y'}
\]
The convertibility distance $\makedistance{\qslrelone}$ gives the so-called 
theory of \quant{} semilattices~\cite{plotkin-quantitative-algebras-2016} 
and axiomatises the Hausdorff distance between sets~\cite{Munkres/Topology/2000}. 
Ultrametricity of $\StrongLawvere$ ultimately relies on its tensor product 
being idempotent, i.e. satisfying the law $\qone \tensor \qone = \qone$. 
In the case of $\StrongLawvere$, this law trivially holds as the tensor coincides with 
the meet. When the underlying quantale is idempotent, then \quant{} and metric reasoning 
becomes similar to traditional, Boolean reasoning, up to the point that the linearity 
assumption mentioned in \autoref{sect:combinators-intro} (which we shall formally rely on 
it the next section) is not necessary to avoid distance trivialisation and to ensure 
confluence properties of systems. 
Finally, we can combine system $\QSL = (\signature_{\QSL}, \stepto_{\qslrelone})$ 
with, e.g., system $\BCK$, this way obtaining a \quant{} system for nondeterministic affine 
combinators. 
\end{example}

We summarise the examples of \qtrs{s} seen so far in 
\autoref{table:examples-qtrs-1} (we will see further examples 
of \qtrs{s} in \autoref{sect:beyond-non-expansive-systems}). 
The rest of this section is dedicated to the development of a general theory of \qtrs{s} 
and to instantiate it to infer nice computational 
properties of systems in \autoref{table:examples-qtrs-1}. 
In particular, we shall prove (by means of general techniques) confluence 
of all of them. Before diving into that, however, it is useful to spend 
few words on \quant{} equational theories.

\begin{table*}[htbp]
\centering
\begin{tabular}{|c|c|c|c|c|}
	\hline
  \textbf{System} & \textbf{Objects/Name} & \textbf{Distance Induced} 
  \\
  \hline 
    $\NATS = (\signature_{\NATS}, \stepto_{\natrelone})$
    & Natural Numbers
    & Euclidean Distance
    \\
    $\BA = (\signature_{\mathbf{\BA}}, \stepto_{\baryrelone})$
    & Multi-distributions/Barycentric algebras
    & Total Variation distance 
    \\
    $\BCKNATS = (\signature_{\BCKNATS}, \stepto_{\combnatrelone})$
    & Affine combinators with Arithmetic
    & Higher-order Euclidean Distance
  \\
  $\TICK = (\signature_{\TICK}, \stepto_{\tickrelone})$
  & Ticking
  & Cost distance
  \\
  $\DNAone = (\signature_{\DNAone}, \stepto_{\dnarelone})$
  & DNA molecules 
  & Edit distances%\footnote{E.g. Levenshtein, Hamming, Eigen–McCaskill–Schuster, Watson–Crick, etc} 
  \\
  $\QSL = (\signature_{\QSL}, \stepto_{\qslrelone})$
  & Quantitative (semi)lattices
  & Hausdorff distance
  \\
  \hline
  %$\mathsf{false}$	& $\infty$ 		& $\qbot$.
\end{tabular}
%\vspace{-0.2cm}
\caption{Main Examples of \emph{non-expansive} \qtrs{s}.}
\label{table:examples-qtrs-1}
\end{table*}

\subsection{Quantitative Equational Theories}
In this section, we formally introduce \quant{} equational theories 
and their connection with \qtrs s. Approaching the former in light of the latter 
allows us to highlights some operationally questionable design choices in the definition of a 
\quant{} equational theories

\begin{definition}
\label{def:quantitative-algebras}
A \quant{} equational theory is a pair $\eqtheoryone= (\signature, \approx_\veqone)$, where 
$\signature$ is a signature and $\approx_\veqone$ 
is a $\Quantale$-ternary relation over $\signature$-terms. The $\Quantale$-ternary (equality) relation
$\equal_{\veqone}$ generated by $\approx_\veqone$ is defined by the rules in 
\autoref{figure:qalgebra}.
\end{definition}

 \begin{figure}
{
\centering
\begin{tcolorbox}[boxrule=0.5pt,width=(\linewidth*5)/6,colframe=black,colback=black!0!white,arc=0mm]
%\vspace{-0.3cm}
     \[
    \infer{\qone \Vdash \termone \equal_{\vbaseeq} \termtwo}{\qone \Vdash \termone \approx_{\vbaseeq} \termtwo}
    \qquad
    \infer{\qunit \Vdash \termone \equal_{\vbaseeq} \termone}{}
    \qquad 
    \infer{\qone \Vdash \termtwo \equal_{\vbaseeq} \termone}{\qone \Vdash \termone \equal_{\vbaseeq} \termtwo}
    \qquad 
    \infer{\qone \tensor \qtwo \Vdash \termone \equal_{\vbaseeq} \termthree}
    {\qone \Vdash \termone \equal_{\vbaseeq} \termtwo 
    & \qtwo \Vdash \termtwo \equal_{\vbaseeq} \termthree}
    \]

    \[
    \infer{\bigotimes_i \qone_i \Vdash \op(\termone_1, \hh, \termone_n) \equal_{\vbaseeq} \op(\termtwo_1, \hh, \termtwo_n)}
    {\qone_1 \Vdash \termone_1 \equal_{\vbaseeq} \termtwo_1 & \cc & 
    \qone_n \Vdash \termone_n \equal_{\vbaseeq} \termtwo_n}
    \qquad 
     \infer{\qone \Vdash \subst{\termone}{\sone} \equal_{\vbaseeq} \subst{\termtwo}{\sone}}
    {\qone \Vdash \termone \equal_{\vbaseeq} \termtwo}
    \]
    \vspace{-0.05cm}
    \[
    \infer{\qtwo \Vdash \termone \equal_{\vbaseeq} \termtwo}
    {\qone \Vdash \termone \equal_{\vbaseeq} \termtwo & \qtwo \leq \qone}
    \qquad 
    \infer{\join \qone_i \Vdash \termone \equal_{\vbaseeq} \termtwo}
    {\qone_1 \Vdash \termone \equal_{\vbaseeq} \termtwo & \hdots & \qone_n \Vdash \termone \equal_{\vbaseeq} \termtwo}
    \qquad 
    \infer{\qone \Vdash \termone \equal_{\vbaseeq} \termtwo}
    {\forall \qtwo \ll \qone.\ \qtwo \Vdash \termone \equal_{\vbaseeq} \termtwo}
    \]
\end{tcolorbox}
}
\caption{Quantitative Equational Theory of $\equal_{\vbaseeq}$}
 \label{figure:qalgebra}
\end{figure}

The first block of rules in \autoref{figure:qalgebra} states that $\equal_{\vbaseeq}$ is a \quant{} 
\emph{equivalence} relation containing $\approx_{\vbaseeq}$, whereas the last block contains essentially the  
same structural rules defining a \qtrs. The second block of rules, instead, states that 
function symbols and substitution behave as \emph{non-expansive functions}. In fact, 
defining the $\Quantale$-relation $\veqone$ by
$$
\veqone(\termone, \termtwo) \defeq \join \{\qone \mid \qone \Vdash \termone \equal_{\vbaseeq} \termtwo\}
$$
we see that $\veqone$ is reflexive, symmetric, and transitive. 
By regarding any $n$-ary function symbol $f$ as a function 
$f: \terms{\signature}{\variables}^n \to \terms{\signature}{\variables}$, we also see 
that 
$$
\veqone(\termone_1, \termtwo_1) \tensor \cc \tensor \veqone(\termone_n, \termtwo_n) 
\leq \veqone(\op(\termone_1, \hh, \termone_n), \op(\termtwo_1, \hh, \termtwo_n)),
$$
meaning that function symbols indeed behave as non-expansive functions.

% so that we obtain a \francesco{$\quantale$-metric space} $(\terms{\signature}{\variables}/{\sim}, \veqone)$, 
% where $\sim$ is the equivalence relation relating terms at $\veqone$-distance $\qunit$. Each function symbol 
% $f$ induces, as usual, a map $f_{\sim}: \terms{\signature}{\variables}/{\sim} \to \terms{\signature}{\variables}/{\sim}$, 
% and the defining rules of $\equal_{\vbaseeq}$ ensure such a map to \emph{non-expansive}. Symbolically, 
% defining $\veqone_{\sim}([\termone]_{\sim}, [\termtwo]_{\sim}) \defeq \veqone(\termone, \termtwo)$, we have:
% $$
% \bigotimes_i \veqone_{\sim}([\termone_i]_{\sim}, [\termtwo_i]_{\sim})
% \leq 
% \veqone_{\sim}(f_{\sim}([\termone_1]_{\sim}, \hh, [\termone_n]_{\sim}), f_{\sim}([\termtwo_1]_{\sim}, \hh, [\termtwo_n]_{\sim}))
% $$

\begin{remark}
\label{rem:idempotent-equational-theories}
Sometimes~\cite{plotkin-quantitative-algebras-2016,plotkin-quantitative-algebras-2017}, 
\quant{} equational theories are defined using the following rule to deal with function symbols. 
 \[
    \infer{\bigwedge_i \qone_i \Vdash \op(\termone_1, \hh, \termone_n) \equal_{\vbaseeq} \op(\termtwo_1, \hh, \termtwo_n)}
    {\qone_1 \Vdash \termone_1 \equal_{\vbaseeq} \termtwo_1 & \cc & 
    \qone_n \Vdash \termone_n \equal_{\vbaseeq} \termtwo_n}
\]
Semantically, that means requiring function symbols to behave as \emph{strongly non-expansive} maps: 
% $$
% \bigwedge_i \veqone(\termone_i, \termtwo_i)
% \leq 
% \veqone(\op(\termone_1, \hh, \termone_n), \op(\termtwo_1, \hh, \termtwo_n)). 
% $$
$$
\veqone(\termone_1, \termtwo_1) \wedge \cc \wedge \veqone(\termone_n, \termtwo_n) 
\leq \veqone(\op(\termone_1, \hh, \termone_n), \op(\termtwo_1, \hh, \termtwo_n)).
$$

In the case of the Lawvere quantale, for instance, we require the distance between two function applications to 
bound the \emph{maximum} distance between their arguments, rather than by the \emph{sum} of such distances. 
Strong non-expansiveness, however, does not properly interact with transitivity, 
which is based on the tensor, rather than the meet, 
of the quantale. Consider terms $t_1, t_2, s_1, s_2$ with 
$\qone \Vdash t_1 \equal_{\vbaseeq} s_1$ and $\qtwo \Vdash t_2 \equal_{\vbaseeq} s_2$. 
For a binary function symbol $f$, we can consider the following two derivations:
\[
\infer{\qone \tensor \qtwo \Vdash f(\termone_1, \termone_2) \equal_{\vbaseeq} f(s_1, s_2)}
{
\infer{\qone \Vdash f(t_1, t_2) \equal_{\vbaseeq} f(s_1,t_2)}
    {\qone \Vdash t_1 \equal_{\vbaseeq} s_1 
    &
    \qunit \Vdash t_2 \equal_{\vbaseeq} t_2
    }
& 
\infer{\qtwo \Vdash f(s_1, t_2) \equal_{\vbaseeq} f(s_1,s_2)}
    {\qunit \Vdash s_1 \equal_{\vbaseeq} s_1
    &
    \qtwo \Vdash t_2 \equal_{\vbaseeq} s_2
    }
}
\qquad 
\quad 
\infer{\qone \wedge \qtwo \Vdash f(\termone_1, \termone_2) \equal_{\vbaseeq} f(s_1, s_2)}
{
\qone \Vdash t_1 \equal_{\vbaseeq} s_1 
&
\qtwo \Vdash t_2 \equal_{\vbaseeq} s_2
}
\]

From a rewriting perspective, these two derivations show that rewriting $t_1$ into $s_1$ and $t_2$ into $s_2$ 
inside $f$ \emph{sequentially} gives a different distance than performing the same rewriting 
\emph{in parallel}. This is not surprising: the non-expansiveness rule for function symbols is defined 
ultimately relying on the idempotent quantale $(\quantale, \leq, \wedge, \top)$, 
whereas transitivity relies on $(\quantale, \leq, \tensor, \qtop)$. 
Harmony is restored by taking the following transitivity rule, which amounts to instantiate 
\autoref{def:quantitative-algebras} with the idempotent quantale $(\quantale, \leq, \wedge, \qtop)$. 
\[
 \infer{\qone \wedge \qtwo \Vdash \termone \equal_{\vbaseeq} \termthree}
    {\qone \Vdash \termone \equal_{\vbaseeq} \termtwo 
    & \qtwo \Vdash \termtwo \equal_{\vbaseeq} \termthree}
\] 
% We refer equational theories obtained by instantiating \autoref{def:quantitative-algebras} with idempotent 
% quantales as \emph{additive} or \emph{\quant{} idempotent equational theories}.
\end{remark}

Following \autoref{rem:idempotent-equational-theories}, we introduce some further terminology and 
refer to equational theories over \emph{idempotent} 
quantales as \emph{additive} or \emph{idempotent} \quant{} equational theories. 
We employ a similar terminology for \quant{} rewriting systems.

\begin{example} 
\begin{enumerate}
    \item Since the Boolean quantale is obviously idempotent, 
        traditional rewriting systems and equational theories are additive (\quant) systems. 
    \item Since the strong Lawvere quantale is idempotent, system $\QSL$ of 
         \autoref{ex:quantitative-semilattices} is additive. 
         Its associated \quant{} equational theory is the one of \quant{} 
         semilattices \cite{plotkin-quantitative-algebras-2016,plotkin-quantitative-algebras-2017}, 
         which is additive too.  
    \item Consider the powerset quantale $\powerset(\{\A,\C,\G,\T\})$ and the system 
        $b \qstepto{\{b\}} c$, for $b, c \in \{\A,\C,\G,\T\}$ and $b \neq c$. 
        This way, we obtain a qualitative distance between molecules giving which 
        bases change between two molecules. 
    \item The open sets of a topological space form a frame~\cite{Vickers/Topology-via-logic} 
    and thus an idempotent quantale. Taking open sets as distances between objects, we can model 
    effectively measurable differences or approximated ones. This way, we stipulate that measuring
    can be done with a limited precision only. Such systems are indeed additive.
\end{enumerate}
\end{example}

Any \quant{} equational theory $(\signature, \approx_\vbaseeq)$ induces a \qtrs{} whose rewriting rules
are given by the equations of $\approx_{\vbaseeq}$ (actually, it is preferable 
to consider a subset thereof obtained by giving equations an appropriate orientation), 
so that we can use \qtrs s to study properties of \quant{} equational 
reasoning, the main ones being related to confluence, termination, and (therefore) metric word problems. 
In particular, given a \quant{} equational theory $(\signature, \approx_\vbaseeq)$, 
we can define a \qtrs{} $(\signature, \stepto_{\vrelone})$ such that $\vbaseeq = \makedistance{\vrelone}$. 
Consequently, by \autoref{prop:church-rosser}, if $\vrelone$ is confluent, then we recover the
equational distance between terms by looking at their common reducts. If, additionally, 
the system is terminating (i.e. SN), then we can approximate such a distance 
by looking at normal forms only: this way, we also obtain decidability of the reachability 
metric word problem for $(\signature, \approx_\vbaseeq)$. 
It is thus desirable to develop handy techniques to prove confluence and termination of \qtrs s.

\subsection{Confluence and Critical Pairs, Part I}
\label{sect:qtrs-confluence}
From (\quant) Newman's Lemma (\autoref{prop:newmans-lemma}), 
we know that to prove confluence of a terminating \qtrs{} 
%$(\signature, \stepto_{\vrelone})$, 
we only need to verify its \emph{local} confluence.
%, i.e. $\dual{\vrelone}; \vrelone \leq \vrelone^*; \dualstar{\vrelone}$. 
Proving local confluence of a \qtrs, however, can be difficult, as reductions may happen 
inside arbitrary contexts and on arbitrary instances of reduction rules. 
It is thus natural to ask whether we can 
prove local confluence \emph{locally}, i.e. by looking at \emph{ground rewriting} only. 

In this section, we show that local confluence of a \emph{linear} \qtrs{} 
follows directly from local confluence of its \emph{critical pairs}~\cite{Huet80,terese}. 
Linearity, as we have already discussed in \autoref{section:long-intro}, is a crucial property 
in \quant{} and metric reasoning: forcing non-expansiveness on non-linear systems 
often let distance 
trivialise~\cite{CrubilleDalLago/ESOP/2014,CrubilleDalLago/ESOP/2017,DBLP:phd/basesearch/Gavazzo19,Gavazzo/LICS/2018}, 
this way collapsing \quant{} equational deduction to traditional, Boolean reasoning. 
On rewriting systems, non-linearity leads to further undesired consequences, as shown 
by the following example.

\begin{example}
\label{ex:linearity-for-confluence}
Consider the signature $\signature \defeq \{f, e, i\}$ with 
$f$ a binary function symbol and $e, i$ constants. 
Let $\stepto_{\vrelone}$ be the reduction rule over the Lawvere quantale:
\begin{align*}
    f(x,x) &\qstepto{0} x 
    \\
    e &\qstepto{1} i. 
\end{align*}
It is easy to see that the system is confluent in the traditional, non-quantitative sense. 
Taking quantitative information into account, however, we have $e \stackrel{0}{\leftarrow} f(e, e) \qreduce{1} 
f(i, e)$, and thus $\vrelone(f(e, e),e) = 0$, 
$\vrelone(f(e, e), f(i, e)) = 1$. 
To close the diagram given by 
$e \stackrel{0}{\leftarrow} f(e, e) \qreduce{1} 
f(i, e)$, we need to reduce $e$ \emph{twice}: 
$$
e \qreduce{1} i \stackrel{0}{\leftarrow} f(i, i)
 \stackrel{1}{\leftarrow} f(i, e).
$$
This gives 
$\vrelone^*(e, i) = 1$ and 
$\vrelone^*(f(i, e), f(i, i)) = 1$,
this way breaking (local) confluence. 
%Notice that one way to restore confluence is 
%replacing $\Lawvere$ with an idempotent quantale. 
\end{example}

%$\termtwo_{1}, \termtwo_{2}$ are %\emph{joinable} with  \emph{decreasing distance} (i.e there is $\termthree$ such that $\termtwo_{1}\twoheadrightarrow \termthree \twoheadleftarrow \termtwo_{2}$  with  $\qone\tensor \qtwo \leq \vrelone^{*} (\termtwo_{1},\termthree)\tensor  \vrelone^{*} (\termtwo_{2},\termthree)$. In this section we prove that in order to state the confluence of a \qtrs{} we just need to verify the joinaibility with decreasing distance of specific pairs, the so-called critical pairs. 

Let us now recall the notion of a critical pair and refine the well-known critical pair 
lemma~\cite{Huet80} to a \quant{} setting.

{
\renewcommand{\redex}{c_1}
\renewcommand{\redextwo}{c_2}
\renewcommand{\contractum}{d_1}
\renewcommand{\contractumtwo}{d_2}

\begin{definition}
 Let $(\signature, \stepto_{\vrelone})$ be a \qtrs.
\begin{enumerate}
    \item Let 
        $\redex \qstepto{\qone}  \contractum$, $\redextwo \qstepto{\qtwo} \contractumtwo$ 
        be renamings of rewrite rules without common variables. Then 
        $\redex \qstepto{\qone}  \contractum$, $\redextwo \qstepto{\qtwo} \contractumtwo$  overlap at position $p$ if:
        \begin{itemize}
            \item $p$ is a function symbol position of $\redextwo$; 
            \item $\redex$ and $\subtermpos{\redextwo}{p}$ are unifiable;
            \item If $p=\posroot$, then the two rules are not variants 
            (i.e. they cannot be obtained one from the other by variables renaming)  . 
        \end{itemize}
    \item Let $\redex\qstepto{\qone}\contractum$, $\redextwo\qstepto{\qtwo}\contractumtwo$ overlapping at position $p$ and
         $\sone$ be the mgu of $\subtermpos{\redextwo}{p}$ and $\redex$. 
         Then the term $\subst{\redextwo}{\sone}$ can be rewritten in two ways: 
        %  \begin{align*}
        %      \subst{\redextwo}{\sone} &\qreduce{\qone} \subst{\redextwo}{\sone}[\subst{\contractum}{\sone}]_p 
        %      & \text{and} & &
        %     \subst{\redextwo}{\sone} &\qreduce{\qtwo} \subst{\contractumtwo}{\sone}
        %  \end{align*}
        \begin{align*}
             \subst{\contractumtwo}{\sone}
             \stackrel{\qtwo}{\leftarrow} 
             \subst{\redextwo}{\sone} 
             \qreduce{\qone} 
             \subst{\redextwo}{\sone}[\subst{\contractum}{\sone}]_{\posone}
            \end{align*}
         We call the triple $(\redex \qstepto{\qone}  \contractum, p, \redextwo \qstepto{\qtwo} \contractumtwo)$ 
         a \emph{critical overlap} and 
         the pair $(\subst{\redextwo}{\sone}[\subst{\contractum}{\sone}]_p,\subst{\contractumtwo}{\sone} )$ a \emph{critical pair}. 
        % $\tertwo_1 \qreduceleft{\qone_1} \termone  \qreduce{\qone_2} \termtwo_2$  
        %Furthemore, let $\mathtt{CP}\defeq \{(u,v)\mid (
        % u,v) \text{ is a  critical pair }\}$. %Given $(u,v)\in \mathtt{CP}$ we will say that $(u,v)$ are joinable, $u\downarrow v$ , iff $(\dual{\vrelone};\vrelone)(u,v) \leq( \vrelone^{*};\dual{\vrelone^{*}})(u,v)$ 
\end{enumerate}
\end{definition}
}

\begin{example}
\begin{enumerate}
    \item The \maketrs{$\Lawvere$} of \autoref{ex:linearity-for-confluence} has no critical pair 
    since there is no \emph{function symbol position} at which (renamings of) the rules 
    $f(x,x) \qstepto{0} x$ and $e \qstepto{1} i$ overlap.
\item Consider system $\BA$  
    of Barycentric algebras and the (no-common-variable renamings of) 
    commutativity and associativity rules:
    \begin{align*}
     x' \barplus{\probone_{1}} y' &\qstepto{0}_{\delta} y' \barplus{1 - \probone_{1}} x' 
     \\
     (x' \barplus{\probone_1} y') \barplus{\probone_2} z 
     &\qstepto{0}_{\delta}  
     x' \barplus{\probone_1 \probone_2} (y' \barplus{\frac{\probone_1 - \probone_1\probone_2}{1 - \probone_1\probone_2}} z)
     \end{align*}
 with $\probone_1, \probone_2 \in (0,1)$. Then, we see that
 the substitution $\sone$ mapping the variable $x'$ to $x$ and $y'$ to $y$
 is the $\mgu$ of $\subtermpos {\big((x \barplus{\probone_1} y) \barplus{\probone_2} z)\big)}{1}$ and 
 $x' \barplus{\probone} y'$. Consequently, the triple and the pair 
\begin{align*}
\big(x' \barplus{\probone_1} y' \qstepto{0}_{\delta} y' \barplus{1 - \probone} x',\ &1,\  
x \barplus{\probone_1 \probone_2} 
(y \barplus{\frac{\probone_1 - \probone_1\probone_2}{1 - \probone_1\probone_2}} z)\big)
\\
\big((y\barplus{1-\probone_{1}} x)\barplus{\probone_{2}} z,&\ x\barplus{\probone_{1}\probone_{2}} (y\barplus{\frac{\probone_{2}-\probone_{1}\probone_{2}}{1-\probone_{1}\probone_{2}}} z)\big)
\end{align*}
form a \emph{critical overlap} and 
a \emph{critical pair}, respectively. 
Diagramtically, we have a critical pair defined by the following peak:
\begin{figure}[H]
\centering
\begin{tikzpicture}[thick,scale=0.80, every node/.style={transform shape, sibling distance=8cm},level 1/.style={sibling distance=45mm},level 2/.style={transform shape,sibling distance=45mm}
]
\node (left) at (-4,0)    {$(x\barplus{\probone_{1}}y)\barplus{\probone_{2}}z$}
    child {node (19){$(y\barplus{1-\probone_{1}} x)\barplus{\probone_{2}} z$} 
    edge from parent [->] node [left] {0}}
    child { node (29)
    {$x\barplus{\probone_{1}\probone_{2}} (y\barplus{\frac{\probone_{2}-\probone_{1}\probone_{2}}{1-\probone_{1}\probone_{2}}} z)$}
    edge from parent [->] node [right] {0}};
\end{tikzpicture}
\end{figure}
\end{enumerate}
\end{example}

\begin{notation}
Given a \qtrs{} $\qtrsone = (\signature, \stepto_{\vrelone})$, we denote by $\CP(\qtrsone)$ 
the collection of its critical pairs. Moreover, we write 
$\termtwo_1 \qreduceleft{\qone_1} \termone  \qreduce{\qone_2} \termtwo_2$ if 
$\termone  \qreduce{\qone_1} \termtwo_1$ and 
$\termone  \qreduce{\qone_2} \termtwo_2$.
\end{notation}

% \begin{definition}
% Let $\redex\qreduce{\qone}\contractum$, $\redextwo\qreduce{\qtwo}\contractumtwo$ overlapping at position $p$.  Let $\sigma$ be the mgu of $\subtermpos{\redextwo}{p}$ and $\redex$. Then the term $\redextwo^{\sigma}$ can be rewritten in two ways: $\redextwo^\sigma[\contractum^\sigma]_p\qreduceleft{\qone}\redextwo^{\sigma} \qreduce{\qtwo} \contractumtwo^\sigma$.  We will call the pair $(\redextwo^\sigma[\contractum^\sigma]_p,\contractumtwo^\sigma )$ \emph{critical pair}. Furthemore, let $\mathtt{CP}\defeq \{(u,v)\mid (
% u,v) \text{ is a  critical pair }\}$. %Given $(u,v)\in \mathtt{CP}$ we will say that $(u,v)$ are joinable, $u\downarrow v$ , iff $(\dual{\vrelone};\vrelone)(u,v) \leq( \vrelone^{*};\dual{\vrelone^{*}})(u,v)$ 
% \end{definition}

We are now ready to prove the \quant{} refinement of the so-called Critical Pair Lemma~\cite{Huet80}. 
In its traditional version, such a lemma states that to prove local confluence 
of a rewriting relation it is enough to prove its local confluence on critical pairs only.
From Newman's Lemma it thus follows that if a rewriting relation is terminating and locally confluent 
on critical pairs, then it is confluent.
When we move to \quant{} rewriting, the Critical Pair Lemma needs a further assumption, namely 
\emph{linearity}.

{
\renewcommand{\redex}{a_1}
\renewcommand{\contractum}{b_1}
\renewcommand{\redextwo}{a_2}
\renewcommand{\contractumtwo}{b_2}

\begin{lemma}[Critical Pair]
\label{lemma:critical-pair}
Let $\qtrsone = (\signature, \stepto_{\vbaserelone})$ be a \emph{linear} \qtrs{}. 
If $\vrelone$ is locally confluent on all critical pairs of $\qtrsone$, then 
it is locally confluent.
%
%$\termtwo_{1},\termtwo_{2}$ are joinable with decreasing distance. %$\termtwo_{1}\twoheadrightarrow \cdot \twoheadleftarrow \termtwo_{2}$ with $(\dual{\vrelone};\vrelone) (\termtwo_{1},\termtwo_{2})\leq( \vrelone^{*};\dual{{\vrelone^{*}}})(\termtwo_{1},\termtwo_{2})$.
\end{lemma}
\begin{proof}
We have to show $\dual{\vrelone};\vrelone \leq \vrelone^{*};\dual{{\vrelone^{*}}}$
given that $(\dual{\vrelone};\vrelone )(t,s) \leq (\vrelone^{*};\dual{{\vrelone^{*}}})(t,s)$, 
for any $(t,s)\in \CP(\qtrsone)$. Pointwise, we need to prove  
\begin{align*}
\join_{\termone}\vrelone (\termone, \termtwo_{1})\tensor \vrelone (\termone, \termtwo_{2})\leq\join_{\termthree} \vrelone^{*}(\termtwo_{1}, \termthree)\tensor  \vrelone^{*}(\termtwo_{2}, \termthree)
\end{align*}
for arbitrary terms $\termtwo_{1}$ and $\termtwo_{2}$. 
Since $\vrelone(\termone,\termtwo) = \join \{\qone \mid \mrelto{\qone}{\termone}{\termtwo}\}$ 
and the join distributes over the tensor, it is enough to show that
% Then, we show that for every $\termone$
% \begin{align*}
% \vrelone (\termone, \termtwo_{1})\tensor \vrelone (\termone, \termtwo_{2})\leq\join_{\termthree}( \vrelone^{*}(\termtwo_{1}, \termthree)\tensor  \vrelone^{*}(\termtwo_{2}, \termthree)) 
% \end{align*}
% Recall that $\vrelone (\termone, \termtwo_{1})= \join \{\qone \mid  \termone \qreduce{\qone} \termtwo_{1}\}$ and  $\vrelone (\termone, \termtwo_{2})= \join \{\qtwo \mid  \termone \qreduce{\qtwo} \termtwo_{2}\}$. Consider then a 
for any local peak $\termtwo_{1}\qreduceleft{\qone_1}\termone \qreduce{\qone_2} \termtwo_{2}$, we have
\begin{align*}
\qone_1 \tensor \qone_2 \leq\join_{\termthree} \vrelone^{*}(\termtwo_{1}, \termthree)\tensor  \vrelone^{*}(\termtwo_{2}, \termthree).
\end{align*}
We proceed by structural induction on $\mrelto{\qone_1}{\termone}{\termtwo_{1}}$ and 
$\mrelto{\qone_2}{\termone}{\termtwo_{2}}$
 (see \autoref{figure:qtrs}). 
We begin with the structural rules, as those are easy. 
Suppose that one between $\mrelto{\qone_1}{\termone}{\termtwo_{1}}$ and $\mrelto{\qone_2}{\termone}{\termtwo_{2}}$ 
is 
obtained by one of the structural rules in \autoref{figure:qtrs}. 
Without loss of generality, we shall assume that this is the case for $\mrelto{\qone_1}{\termone}{\termtwo_{1}}$. 
\begin{itemize}
    \item Suppose that $\mrelto{\qone_1}{\termone}{\termtwo_{1}}$ 
    is obtained by quantitative weakening, so that we have:
    \[
        \infer{\mrelto{\qone_1}{\termone}{\termtwo_{1}}}
        {\mrelto{\qtwo}{\termone}{\termtwo_{1}} & \qone_1 \leq \qtwo}
    \]
    By induction hypothesis, we know that  
    $
    \qtwo \tensor \qone_2 \leq \join_{\termthree} \vrelone^{*}(\termtwo_{1}, \termthree)\tensor  \vrelone^{*}(\termtwo_{2}, \termthree),
    $ so that we conclude (recall that $\Quantale$ is integral)  
    $$\qone_1 \tensor \qone_2 \leq \qtwo \tensor \qone_2 \leq \join_{\termthree} \vrelone^{*}(\termtwo_{1}, \termthree)\tensor  \vrelone^{*}(\termtwo_{2}, \termthree)$$.
    \item Suppose that $\mrelto{\qone_1}{\termone}{\termtwo_{1}}$ is obtained by closure under finite join,
    so that we have $\qone_1 = \join_i \qtwo_i$ for some $\qtwo_1, \hh, \qtwo_n$ and
    \[
    \infer{\mrelto{\join_i \qtwo_i }{\termone}{\termtwo_{1}}}
    { \mrelto{\qtwo_1}{\termone}{\termtwo_1} & \hdots &   \mrelto{\qtwo_n}{\termone}{\termtwo_{1}}}.
    \]
    By induction hypothesis, we know that $\forall i \leq n.\ \qtwo_i \tensor \qone_2 \leq \join_{\termthree} \vrelone^{*}(\termtwo_{1}, \termthree)\tensor  \vrelone^{*}(\termtwo_{2}, \termthree)$ 
    which gives, by the universal property of joins, 
    \[
    \join_i(\qtwo_i \tensor \qone_2) \leq \join_{\termthree} \vrelone^{*}(\termtwo_{1}, \termthree)\tensor  \vrelone^{*}(\termtwo_{2}, \termthree)
    \]
    and thus the desired thesis, since $\join_i(\qtwo_i \tensor \qone_2) = (\join_i \qtwo_i) \tensor \qone_2$.
    \item Suppose that $\mrelto{\qone_1}{\termone}{\termtwo_{1}}$ is obtained by the Archimedean property, 
    so that we have:
    \[
        \infer{ \mrelto{\qone_1}{\termone}{\termtwo_{1}}}
        {\forall \qtwo_i\ll \qone_1.\ \mrelto{\qtwo_i}{\termone}{\termtwo_{1}}}
     \]
    By induction hypothesis, we have 
    $\qtwo_i \tensor \qone_2 \leq \join_{\termthree}\vrelone^{*}(\termtwo_{1}, \termthree) \tensor \vrelone^{*}(\termtwo_{2}, \termthree)$, for any $\qtwo_i \ll \qone_1$, 
    and thus 
    $$
    \join_{\qtwo_i \ll \qone_1} \qtwo_i \tensor \qone_2 
    \leq \join_{\termthree}\vrelone^{*}(\termtwo_{1}, \termthree) \tensor \vrelone^{*}(\termtwo_{2}, \termthree).
    $$
    We conclude the thesis since $\qone_1 = \join_{\qtwo_i \ll \qone_1} \qtwo_i$
    % That gives us  
    % $$
    % \join \{\qthree_i \tensor \qtwo \mid \qthree_i \ll \qone\} 
    % \leq \join_{\termthree}\vrelone^{*}(\termtwo_{1}, \termthree) \tensor \vrelone^{*}(\termtwo_{2}, \termthree).
    % $$
    % Since $\join \{\qthree_i \tensor \qtwo \mid c_i \ll \qone\} = \join \{\qthree_i \mid c_i \ll \qone\} \tensor \qtwo$ 
    % and $\qone = \join \{\qthree_i \mid c_i \ll \qone\}$, we obtain the desired thesis.
\end{itemize}
Let us now move to the main case, namely the one in which $\termtwo_{1}$ and $\termtwo_{2}$ are 
obtained from $\termone$ by closure under substitution and context of two rewriting rules. 
More precisely, suppose that  $\mrelto{\qone_1}{\termone}{\termtwo_{1}}$ is obtained by applying rule $\mrelstepto{\qone_1}{\redex}{\contractum}$ at position $\posone_{1}$ and 
$\mrelto{\qone_2}{\termone}{\termtwo_{2}}$ is obtained 
by applying  rule $\mrelstepto{\qone_2}{\redextwo}{\contractumtwo}$ at position $\posone_{2}$. 
We also assume that the two rules have disjoint variables. Thus, we have that $\subtermpos{\termone}{\posone_{1}}=\subst{\redex}{\sone}$,  $\subtermpos{\termone}{\posone_{2}}=\subst{\redextwo}{\sone}$, $\termtwo_{1}=\termone[\subst{\contractum}{\sone}]_{\posone_{1}}$ and $\termtwo_{2}=\termone[\subst{\contractumtwo}{\sone}]_{\posone_{2}}$. %$\termtwo_{i}=\termone[\contractum_{i}^{\sigma}]_{\posone_{i}}$. Now, recalling that $\vrelone(t,\termtwo_{1})=\join\{a\mid \termone \qreduce{\qone} \termtwo_{1}\}$ and $\vrelone(t,\termtwo_{2})=\join\{a\mid \termone \qreduce{\qtwo} \termtwo_{2}\}$, then we should verify that   $a\tensor b \leq \join_{u}{\vrelone^{*}(\termtwo_{1},u)\tensor {\vrelone^{*}(\termtwo_{2},u)}}$.  
We proceed by cases, depending on the relationship between $\posone_1$ and $\posone_2$. 
\begin{itemize}
\item Suppose $\posone_{1}\parallel \posone_{2}$. Then, $\termone=\termone[\subst{\redex}{\sone}]_{\posone_1}[\subst{\redextwo}{\sone}]_{\posone_2}$, $\termtwo_{1}=\termone[\subst{\contractum}{\sone}]_{\posone_1}[\subst{\redextwo}{\sone}]_{\posone_2}$ and $\termtwo_{2}=\termone[\subst{\redex}{\sone}]_{\posone_1}[\subst{\contractumtwo}{\sone}]_{\posone_2}$. 
Applying $\mrelstepto{\qone_2}{\redextwo}{\contractumtwo}$ to $\subtermpos{{\termtwo_{1}}}{{\posone_{2}}}$ and  
$\mrelstepto{\qone_1}{\redex}{\contractum}$ to $\subtermpos{{\termtwo_{2}}}{{\posone_{1}}}$, we have that 
$\termtwo_1\qreduce{\qone_2} \termone[\subst{\contractum}{\sone}]_{\posone_1}[\subst{\contractumtwo}{\sone}]_{\posone_2} \qreduceleft{\qone_1} \termtwo_{2}$. Consequently, 
$$
\qone_1 \tensor \qone_2 \leq \vrelone (\termtwo_{1},\termone[\subst{\contractum}{\sone}]_{\posone_1}[\subst{\contractumtwo}{\sone}]_{\posone_2})
\tensor \vrelone(\termtwo_{2},\termone[\subst{\contractum}{\sone}]_{\posone_1}[\subst{\contractumtwo}{\sone}]_{\posone_2})
\leq \join_{\termthree}\vrelone^{*}(\termtwo_{1}, \termthree) \tensor \vrelone^{*}(\termtwo_{2},\termthree).
$$
 \item Without loss of generality, suppose that 
 $\posone_{1}\leq \posone_{2}$. Then, there is $\postwo$ such that 
 $\posone_{2}=\posone_{1}\postwo$. We distinguish two cases:
 \begin{enumerate}
     \item If $\mrelstepto{\qone}{\redex}{\contractum}$ and $\mrelstepto{\qtwo}{\redextwo}{\contractumtwo}$ do not overlap at position $\postwo$, 
     then we have that either $\mrelstepto{\qone}{\redex}{\contractum}$ and 
     $\mrelstepto{\qtwo}{\redextwo}{\contractumtwo}$ are variants 
     and $p=\posroot$, or $\posone_{2}=\posone_{1}\postwo_{1}\postwo_{2}$ with 
     $\subtermpos{\redex}{{\postwo_{1}}}$ a variable, say $x$. 
     In the latter case, we have that $\termtwo_1=\termone [\subst{\contractum}{\sone}]_{{\posone_1}}=\termone [\subst{\contractum}{\sone}[\subst{\redextwo}{\sone}]_{{\postwo_1}{\postwo_2}}]_{{\posone_1}}$, 
     whereas $\termtwo_2= \termone[\subst{\redex}{\sone}[\subst{\contractumtwo}{\sone}]_{{\postwo_1}{\postwo_2}}]_{{\posone_1}}$. Consider the substitution 
     $\stwo$ mapping $x$ to $\subst{\contractum}{\sone}[\subst{x}{\sone}]_{\postwo_2}$. Since $\redex$ is linear, applying $\mrelstepto{\qtwo}{\redextwo}{\contractumtwo}$ to $\subtermpos{{\termtwo_{1}}}{{\posone_{2}}}$ and  $\mrelstepto{\qone}{\redex}{\contractum}$ to $\subtermpos{\termtwo_{2}}{{\posone_{1}}}$ with substitution $\stwo$, we have that $\mrelto{\qtwo}{\termtwo_1}{\termthree}$ and 
     $\mrelto{\qone}{\termtwo_2}{\termthree}$, with $\termthree=\termone[\subst{\contractum}{\sone}[\subst{\contractumtwo}{\sone}]_{{\postwo_1}{\postwo_2}}]_{\posone_1}$. Distance analysis proceeds as in the first case. \\
If instead, $\mrelstepto{\qone}{\redex}{\contractum}$ and 
$\mrelstepto{\qtwo}{\redextwo}{\contractumtwo}$ are variants with $\postwo=\posroot$, 
i.e. $\posone_{1}=\posone_{2}$, then necessarily $\termtwo_{1}=\termtwo_{2}$, that is, 
$\mrelto{\qunit}{\termtwo_{1}}{\termtwo_{1}=\termtwo_{2}}$ and 
$\mrelto{\qunit}{\termtwo_{2}}{\termtwo_2}$. 
In this case, we have that 
$$\qvalone\tensor\qvaltwo \leq \qunit \tensor \qunit  \leq \vrelone (\termtwo_{1},\termtwo_{2} )\tensor \vrelone(\termtwo_{2},\termtwo_{2}) \leq \join_{\termthree}\vrelone^{*}(\termtwo_{1}, \termthree)\tensor \vrelone^{*}(\termtwo_{2},\termthree).
$$
\item If $\mrelstepto{\qone}{\redex}{\contractum}$ and $\mrelstepto{\qtwo}{\redextwo}{\redextwo}$ 
overlap at position $\postwo$. Then, $\subtermpos{\subst{\redex}{\sone}}{\postwo}=\subtermpos{(\subtermpos{\termone}{\posone_1})}{\postwo}
=\subtermpos{\termone}{\posone_2}=\subst{\redextwo}{\sone}$. That is, $\sone$ is an unifier of $\subtermpos{\redex}{\postwo}$ and $\redextwo$. Let $\stwo$ be their most general unifier, so 
that $\sone= \sthree \circ \stwo $, for some substitution $\sthree$. Then $(\subst{\redex}{\stwo}[\subst{\contractumtwo}{\stwo}]_{\postwo},\subst{\contractum}{\stwo} )$ is a critical pair. 
By hypothesis, we know that $$(\dual{\vrelone};\vrelone)(\subst{\redex}{\stwo}[\subst{\contractumtwo}{\stwo}]_{\postwo},\subst{\contractum}{\stwo} )
\leq
\join_{\termthree} \vrelone^{*}(\subst{\redex}{\stwo}[\subst{\contractumtwo}{\stwo}]_{\postwo}, \termthree)\tensor  \vrelone^{*}(\subst{\contractum}{\stwo}, \termthree),$$
and thus $\qone\tensor\qtwo \leq 
\join_{\termthree}
\vrelone^{*}(\subst{\redex}{\stwo}[\subst{\contractumtwo}{\stwo}]_{\postwo}, \termthree)\tensor 
\vrelone^{*}(\subst{\contractum}{\stwo}, \termthree)$. 
To prove the thesis, it is sufficient to prove 
$$
\join_{\termthree}
\vrelone^{*}(\subst{\redex}{\stwo}[\subst{\contractumtwo}{\stwo}]_{\postwo}, \termthree)\tensor 
\vrelone^{*}(\subst{\contractum}{\stwo}, \termthree)
\leq 
\join_{\termfour} \vrelone^*(\termtwo_1, \termfour) \tensor \vrelone^*(\termtwo_2, \termfour).
$$
As usual, it is enough to show that for any $\termthree$ such that  $\mrelto{\qthree}{\subst{\redex}{\stwo}[\subst{\contractumtwo}{\stwo}]_\postwo}{\termthree}$ 
and $\mrelto{\qfour}{\subst{\contractum}{\stwo}}{\termthree}$ 
we have 
$\qthree \tensor \qfour \leq \join_{\termfour} \vrelone^*(\termtwo_1, \termfour) \tensor \vrelone^*(\termtwo_2, \termfour)$.
Fixed $\termthree$ as above, we have:
% and $\qone\tensor \qtwo \leq c \tensor d$. 
\begin{align*} 
    \termtwo_{1} &=\termone[\subst{\contractum}{\sone}]_{\posone_{1}}
=\termone[\contractum^{\stwo\sthree}]_{\posone_{1}}
\\
\termtwo_{2} 
&=\termone[
    \subst{\contractumtwo}{\sone}]_{\posone_{2}}
=\termone[
        \redex^{\sone}
            [\subst
                {\contractumtwo}
                {\sone}
            ]_{\postwo}
            ]_{\posone_{1}}
=\termone[\redex^{\stwo\sthree}
            [\contractumtwo^{\stwo\sthree}
            ]_{\postwo}
            ]_{\posone_{1}}
\end{align*}
%=\termone[\subst{(\subst{\redex}{\stwo}[\subst{\contractumtwo}{\stwo}]_{\postwo})}{\sthree}]_{\posone_{1}}.
Therefore, by very definition of $\reduce$, we have:
$$
\termtwo_{2}=\termone[\subst{(\subst{\redex}{\stwo}[\subst{\contractumtwo}{\stwo}]_{\postwo})}{\sthree}]_{\posone_{1}}
\qreduce{\qthree} \termone[\subst{\termthree}{\sthree}]_{\posone_1} 
\qreduceleft{\qfour} 
\termone[\contractum^{\stwo\sthree}]_{\posone_{1}}
=\termtwo_{1},
$$
and we are done.
%But then, we have that $\qone\tensor\qtwo\leq  \join_{u}(\vrelone^{*}(\termtwo_{1}, u)\tensor \vrelone^{*}(\termtwo_{2},u))$. % So, $(\redex^\sone[\contractumtwo^{\sone}]_{\postwo},\contractum^{\sone} )=((\redex^{\stwo})^{\vartheta}[(\contractumtwo^{\stwo})^{\vartheta}]_{\postwo},(\contractum^{\stwo})^{\vartheta} ) \in \mathtt{CP}$ %By hypothesis  $\redex^\stwo[\contractumtwo^{\stwo}]p  \to^{*} u \leftarrow^{=} \contractum^{\stwo} p$ and $\redex^\stwo[\contractumtwo^{\stwo}]p  \to^{=}  u \leftarrow^{*} \contractum^{\stwo} p$. Then, by closure under substitution (applying $\vartheta$) and noticing that distances are substitution-invariant,  we have the thesis.
 \end{enumerate}
\end{itemize}
\end{proof}
}

\begin{theorem}
\label{theorem:critical-pair-theorem}
Any \emph{linear} and terminating \qtrs{} locally confluent on its critical pairs is 
confluent.
\end{theorem}

\begin{proof}
It directly follows from \autoref{prop:newmans-lemma} and \autoref{lemma:critical-pair}.
\end{proof}

\begin{remark}
Example\autoref{ex:linearity-for-confluence} shows that, contrary to what happens in the 
traditional case, the linearity assumption is 
indeed necessary in \autoref{theorem:critical-pair-theorem}. In fact, it is an easy exercise 
to prove that if $\Quantale$ is idempotent, then \autoref{lemma:critical-pair} 
extends to non-linear \qtrs{s}.
\end{remark}

\begin{example}
It is easy to see that system $\NATS$ of natural numbers is terminating 
% It is easy to see that the system is terminating 
% (for instance, we have $\nsucc(\nzero) \qreduce{1} \nzero$ and 
% $\nsucc(\nsucc(\nzero)) \qreduce{1} \nsucc(\nzero) \qreduce{1} \nzero$, 
% so that $\nsucc(\nsucc(\nzero)) \stackrel{2}{\to^*} \nzero$) 
and locally confluent on all its critical pairs. By \autoref{theorem:critical-pair-theorem}, 
we conclude that $\mathcal{N}$ is confluent.  
\end{example}

\begin{example}
Most of the \quant{} string rewriting systems introduced so far are not 
terminating, and thus we cannot rely on \autoref{theorem:critical-pair-theorem} to
prove their confluence. One easy way to overcome the problem is to rephrase them as terminating systems. 
For instance, 
we can stipulate that molecules can only be deleted and that substitution of molecules 
is directed, in the sense that, e.g., $\A$ can become $\C$, $\G$, and $\T$, but not 
vice-versa (similarly, $\C$ can become $\G$ and $\T$, and $\G$ can only become a $\T$). 
This way, we indeed obtain a terminating system locally confluent on its critical pairs, and 
thus a confluent system. Although this approach works fine as long as we are interested in 
reachability problems (and alike), some care is needed when dealing with optimal distances, 
as forcing termination may lead to increasing minimal distances between molecules.\footnote{
This observation generalises to a collection of interesting research problems asking whether 
completion algorithms on traditional rewriting systems can be extended to \quant{} systems. 
Notice that this question may not have a Boolean answer: in fact, some completion procedures may 
be correct from the point of view of reachability problems, but not so when it comes to deal with 
optimal distances.
}
\end{example}

\begin{example}
System $\TICK_{\checkmark}$ is terminating and locally confluent, and thus confluent.
System $\TICK$, instead, is not terminating, due to the rule $\writeop{n}{x} \qstepto{\qone} \writeop{m}{x}$, 
with $\qone \geq E(n,m)$ (here, $E$ denotes the Euclidean distance). We can easily fix that by imposing $m<n$, 
this way obtaining a terminating and locally confluent --- and thus confluent --- system.
\end{example}
%\begin{proof}
%Let us notice that if $(u,v)\in\mathtt{CP}$, then $(u,v)=(s^{\sigma}, t^{\sigma})$, with $(s, t)$, critical pair and $\sigma$ substitution. Since distances are substituion-invariant \cecilia{magari da qualche parte provalo}, from hypothesis and Critical Lemma we obtain local confluence.
%\end{proof}
%From the last result and the Newmann lemma  we have then that every linear and terminating \qtrs{}  with joinable critical pairs is confluent. 
\subsection{Confluence and Critical Pairs, Part II}
\label{sect:qtrs-confluence-part-2}
\autoref{theorem:critical-pair-theorem} constitutes a powerful tool to prove 
confluence of \emph{terminating} \qtrs{s}. In a \quant{} setting, however, 
termination might be a too strong condition, and interesting 
\qtrs{s} may not satisfy it. As an example, consider system $\BA$ of Barycentric 
algebras. 
Due to  commutativity and left invariance, the system is obviously non-terminating. 
We may handle the former point by considering \quant{} notions of rewriting 
modulo\footnote{We leave the detailed development of such a theory for future work.}~\cite{rewriting-modulo-1,rewriting-modulo-2,Huet80}, but the former one is 
a genuine \quant{} reduction (and actually the \quant{} essence of the total variation 
distance!). That makes simply not possible to prove confluence of $\BA$ 
via critical pairs and Newmann's Lemma. And yet, 
by simply working out examples, we have the feeling that $\BA$ is indeed confluent. 
To solve this issue, we modify \autoref{lemma:critical-pair} replacing local confluence with 
a stronger condition, namely \emph{strong confluence}~\cite{Huet80}.

\begin{definition}
We say that a $\Quantale$-relation $\vrelone: A \torel A$ is \emph{strongly confluent} 
if it satisfies inequality \eqref{eq:strong-confluence-1} below
and that it is
\emph{strongly closed} if it satisfies inequalities 
\eqref{eq:strong-confluence-1} and \eqref{eq:strong-confluence-2}. 
As usual, we say that a \qars{} is strongly confluent (resp. strongly closed) if 
its rewriting $\Quantale$-relation is.
\begin{align}
    \dual{\vrelone};\vrelone &\leq  \reflex{\vrelone};\dualstar{\vrelone} 
    \label{eq:strong-confluence-1} \tag{strong-1}
    \\
    \dual{\vrelone};\vrelone &\leq  \vrelone^{*};\dual{{\reflex{\vrelone}}}.
    \label{eq:strong-confluence-2}
    \tag{strong-2}
\end{align}
\end{definition}

The next result states that if a \qtrs{}
is \emph{linear}, then to prove that it is strongly closed, it 
is enough to look at its critical pairs.

\begin{lemma}
\label{lemma:strongly-closed-critical-pair}
A \emph{linear} 
\qtrs{} $\qtrsone = (\signature, \stepto_{\vrelone})$ is strongly closed 
if and only if $\vrelone$ is 
strongly closed on all the critical pairs of $\qtrsone$. 
\end{lemma}
\begin{proof}
The proof is essentially the same of the one of \autoref{lemma:critical-pair}, 
the only difference being that in the main case we replace local confluence with strong closure. 

\end{proof}

Strong confluence (resp. closure) by itself is not immediately informative for our 
purposes. Its relevance is given by the following result stating that strong confluence 
(and thus strong closure) entails confluence.

\begin{lemma}
\label{lemma:strong-confluence-implies-confluence}
Strong confluence implies confluence.
\end{lemma}
\begin{proof}[Sketch]
Let $\vrelone: A \torel A$ be a $\Quantale$-relation and $\vreltwo = \dual{\vrelone}$ (our proof works for 
an arbitrary $\vreltwo: A \torel A$, actually). 
Assume \eqref{eq:strong-confluence-1}, i.e. $\vreltwo; \vrelone \leq \reflex{\vrelone}; \vreltwo^*$. 
We prove $\vreltwo^*;\vrelone^* \leq \vrelone^*; \vreltwo^*$. Pointiwse proofs rely on lexicographic induction. 
A lightweight, pointfree proof is obtained by observing that since
$\vreltwo^* = \mu X. \idvrel \vee \vreltwo;X$, we have
$$
\vreltwo^*;\vrelone^* = 
(\mu X. \idvrel \vee \vreltwo;X); \vrelone^* 
= \mu X. \vrelone^*  \vee \vreltwo;X.
$$
Consequently, to prove $\vreltwo^*;\vrelone^* \leq \vrelone^*; \vreltwo^*$ 
it is sufficient to prove 
$\mu X. \vrelone^*  \vee \vreltwo;X \leq \vrelone^*; \vreltwo^*$, which can be done 
using fixed point induction.

\end{proof}

\begin{theorem}
\label{theorem:strongly-closed-critical-pairs}
If a \emph{linear} \qtrs{} is strongly closed on all its critical pairs, then 
it is confluent.
\end{theorem}

\begin{proof}
Directly from \autoref{lemma:strongly-closed-critical-pair} and \autoref{lemma:strong-confluence-implies-confluence}.
\end{proof}

We can now rely on \autoref{theorem:strongly-closed-critical-pairs} to prove 
confluence of $\BA$. 

\begin{proposition}
\label{prop:BA-is-confluent}
    System $\BA$ is strongly closed, and thus confluent.
\end{proposition}

\begin{proof}
By \autoref{theorem:strongly-closed-critical-pairs},
it is enough to prove that $\BA$ is strongly closed 
on all critical pairs.
To do so, we first notice that the associativity rule is `reversible' in the following 
sense: given $\probone_1, \probone_2 \in (0,1)$, the reduction
$$
\mrelto{0}{(x \barplus{\probone_1} y) \barplus{\probone_2} z}
{x \barplus{\probone_1 \probone_2} (y \barplus{\frac{\probone_1 - \probone_1\probone_2}{1 - \probone_1\probone_2}} z)}.
$$ 
has an inverse reduction
$$
\mrel{0}
{x \barplus{\probone_1 \probone_2} (y \barplus{\frac{\probone_1 - \probone_1\probone_2}{1 - \probone_1\probone_2}} z)} 
{\reduce^*}
{( x \barplus{\probone_1} y) \barplus{\probone_2} z}
$$
obtained by alternatively applying commutativity and associativity.
We then verify that $\BA$ is strongly closed. This is 
a routine case analysis on
critical pairs. 
\end{proof}

As for \autoref{lemma:critical-pair} and \autoref{theorem:critical-pair-theorem}, 
even for \autoref{lemma:strongly-closed-critical-pair} and \autoref{theorem:strongly-closed-critical-pairs} 
we can drop the linearity assumption if the underlying quantale is idempotent. 
This gives confluence of the Hausdorff distance. 

\begin{example}
Mimicking \autoref{prop:BA-is-confluent} we see that system $\QSL$ is confluent too.
\end{example}

\subsection{Confluence and Critical Pairs, Part III} 
In previous sections, we have proved confluence of most of the  systems 
introduced in \autoref{section:long-intro}: systems 
$\NATS$, $\DNAone$, $\BA$, $\TICK$, $\QSL$ are all confluent. 
An important class of systems not present in this list is the one of 
\quant{} extensions\footnote{Affine combinators having no nontrivial \quant{} reductions, 
they are essentially identical to their traditional counterpart (which indeed gives 
a confluent and terminating system).}
of affine combinators. This class includes system $\BCKNATS$ (combinators plus arithmetic), 
$\BCKprob$ (probabilistic combinators), $\BCK_{\TICK}$ (combinators with cost), and similar systems. 
Even if different, all these systems are obtained in essentially the same way, 
namely by joining systems together.\footnote{Algebraically, this operation 
corresponds to the \emph{sum} of algebraic theories \cite{DBLP:journals/tcs/HylandPP06}.} 
For instance, system $\BCKprob$ is obtained by joining systems $\BCK$ of pure affine combinators 
and $\BA$.\footnote{Further systems can be obtained either by modifying the `effectful' 
layer (hence adding to $\BCK$, for instance, \quant{} output, nondeterminism, etc) 
or by replacing $\BCK$ itself with other rewriting systems modelling programming languages, such 
as concurrent ones \cite{Milner/Communication-and-concurrency/1989,handbook-process-algebra}.} 
It is then natural to ask whether confluence of such systems can be proved \emph{compositionally} 
in terms of confluence of their component subsystems. In our case, for instance, we know 
that both $\BA$ and $\BCK$ are confluent,\footnote{Since $\BCK$ alone does not have 
truly \quant{} behaviours, its confluence can be proved as for its traditional 
counterpart.} and we would like to infer confluence of $\BCKprob$. 
Indeed, we can do so relying on the \quant{} refinement of the Hindley-Rosen Lemma (\autoref{lemma:hindley-rosen}). 
Let us begin by formalising the idea of joining \qtrs{s}.

\begin{definition}
Given \qtrs{s} $\qtrsone = (\signature_{\qtrsone}, \stepto_{\vrelone})$ and 
$\qtrstwo = (\signature_{\qtrstwo}, \stepto_{\vreltwo})$ with disjoint signatures, 
we define their \emph{sum} 
as the \qtrs{} $\qtrsone + \qtrstwo = (\signature_{\qtrsone\qtrstwo}, \stepto_{\vrelone\vreltwo})$ 
defined thus: $\signature_{\qtrsone\qtrstwo} \defeq \signature_{\qtrsone} \cup \signature_{\qtrsone}$ 
and $\stepto_{\vrelone\vreltwo} \defeq {\stepto_{\vrelone}} \cup {\stepto_{\vreltwo}}$.
\end{definition}

\begin{example}
We immediately see that $\BCKprob = \BCK + \BA$. Extensions of $\BCK$ with ticking 
are obtained as $\BCK + \TICK$ and $\BCK + \TICK_{\checkmark}$, whereas 
$\BCK + \QSL$ gives nondeterministic affine combinators. Finally, notice that even 
formally different, system $\BCKNATS$ is essentially $\BCK + \NATS$.
\end{example}

To relate the sum of \qtrs{s} $\qtrsone$, $\qtrstwo$ as above 
with Newman's Lemma, we first observe 
that the $\Quantale$-relation $\vrelone\vreltwo$ associated to $\qtrsone + \qtrstwo$ 
coincides with $\vrelone \vee \vreltwo$.

\begin{lemma}
\label{lemma:sum-qtrs-equal-union}
For all \qtrs{s} $\qtrsone = (\signature_{\qtrsone}, \stepto_{\vrelone})$, 
$\qtrstwo = (\signature_{\qtrstwo}, \stepto_{\vreltwo})$, 
we have $\vrelone\vreltwo = \vrelone \vee \vreltwo$.
\end{lemma}

\begin{proof}
Straightforward.
\end{proof}

\autoref{lemma:sum-qtrs-equal-union} puts ourselves in the condition to 
rely on the \quant{} Hindley-Rosen Lemma to prove confluence of 
$\qtrsone + \qtrstwo$. Accordingly, to infer confluence of $\vrelone \vee \vreltwo$ ($= \vrelone\vreltwo$) 
we need to have confluence of $\vrelone$ and $\vreltwo$ as well as commutation of 
$\vrelone$ with $\vreltwo$. Whereas the former usually is our starting hypothesis, the latter 
requires a specific analysis. For the cases we are interested in, such an analysis is smooth, 
as systems $\qtrsone$ and $\qtrstwo$ are essentially independent, in the sense that 
$\qtrsone + \qtrstwo$ does not create new critical pairs. 

\begin{lemma}
\label{lemma:strcl}
Given two linear \qtrs{} $\qtrsone$, $\qtrstwo$ as above, 
if the collection of critical pairs obtained by overlapping of a rule of $\qtrsone$ 
and a rule of $\qtrstwo$ 
is empty, then $\vrelone$ strongly commutes with $\vreltwo$, and thus $\vrelone$ commutes with $\vreltwo$
%$\dual{\vrelone};\vreltwo\leq \vreltwo^{=};\dual{{\vrelone^{*}}}$ and $\dual{\vrelone};\vrelone \leq  \vreltwo^{*};\dual{(\vrelone^{=})}$  
\end{lemma}

\begin{proof}
The proof that $\vrelone$ strongly commutes with $\vreltwo$ 
is a simplified instance of the proof of \autoref{lemma:critical-pair} 
and \autoref{lemma:strongly-closed-critical-pair}. From that, commutation  
of $\vrelone$ with $\vreltwo$ follows by \autoref{lemma:strong-confluence-implies-confluence}.
\end{proof}

Using \autoref{lemma:strcl}, we obtain the necessary hypotheses to apply \autoref{lemma:hindley-rosen}
and conclude confluence of $\BCKprob = \BCK + \BA$.

\begin{proposition}
\label{prop:bck-plus-ba-is-confluent}
    System $\BCK + \BA$ is confluent.
\end{proposition}

\begin{proof}
Confluence of $\BCK + \BA$ immediately follows \autoref{lemma:hindley-rosen}, provided 
that $\combrelone$ commute with $\baryrelone$. That is indeed the case, since $\BCK$ and 
$\BA$ have no common critical pair, and thus they (strongly) commute, by \autoref{lemma:strcl}.
\end{proof}

In a similar fashion, one proves that, e.g., $\BCK + \NATS$ and $\BCK + \TICK$ are confluent, 
as well as combinations thereof. Moreover, as usual, if the underlying quantale is 
idempotent, we can drop the linearity 
assumption in \autoref{lemma:strcl}, so that by enriching $\BCK$ over $\StrongLawvere$ 
(rather than on $\Lawvere$), we see that $\BCK + \QSL$ is confluent too. 

These results complete the (confluence) analysis of the examples introduced in 
\autoref{section:long-intro} as well as our general results on linear (or non-expansive) 
\qtrs{s}. The next --- last --- section of this work outlines a possible way to go 
beyond linearity: we are going to move from non-expansive to \emph{Lipschitz continuous} 
\qtrs{s}.

\section{BEYOND NON-EXPANSIVENESS: GRADES, MODALITIES, AND LIPSCHITZ CONTINUITY}
\label{sect:beyond-non-expansive-systems}

In this section, we introduce a new class of \quant{} term rewriting systems 
--- namely \emph{graded} rewriting systems ---
that allows us to model non-linear systems avoiding, at the same time, 
distance trivialisation and lack of confluence issues. 
So far, in fact, we have focused on linear, non-expansive functions and term constructors. 
This is reflected almost everywhere in our definitions: for instance, 
the rule 
\[
\infer{\mrel{\qone}{\ctxone[\subst{\termone}{\sone}]}{\reduce_{\vrelone}}{\ctxone[\subst{\termtwo}{\sone}]}}
{\mrel{\qone}{\termone}{\stepto_{\vrelone}}{\termtwo}}
\]
states that differences produced by $\stepto_{\vrelone}$ are \emph{non-expansively} propagated 
by $\reduce_{\vrelone}$
through contexts (hence term constructors) and substitution. 
This can be seen even more clearly in \quant{} algebraic theories, where the 
rule
\[
\infer{
\mrel
{\qone_1 \tensor \cc \tensor \qone_n}
{\op(\termone_1, \hh, \termone_n)}
{\equal_{\veqone}}
{\op(\termtwo_1, \hh, \termtwo_n)}
}
{\mrel{\qone_1}{\termone_1}{\equal_{\veqone}}{\termtwo_1}
&
\cc
&
\mrel{\qone_1}{\termone_1}{\equal_{\veqone}}{\termtwo_1}
}
\]
precisely tells us that the function symbol $\op$ behaves as a non-expansive function.
We have also seen stronger forms of non-expansiveness, namely strong (or ultra) non-expansiveness:
\[
\infer{
\mrel
{\qone_1 \wedge \cc \wedge \qone_n}
{\op(\termone_1, \hh, \termone_n)}
{\equal_{\veqone}}
{\op(\termtwo_1, \hh, \termtwo_n)}
}
{\mrel{\qone_1}{\termone_1}{\equal_{\veqone}}{\termtwo_1}
&
\cc
&
\mrel{\qone_1}{\termone_1}{\equal_{\veqone}}{\termtwo_1}
}
\]
As already remarked, strong non-expansiveness is, in its essence, just ordinary non-expansiveness 
on an idempotent quantale. 
Even if we can view all of that from a more semantic point of view in terms of arrows and constructions in 
suitable categories of $\Quantale$-spaces \cite{Hoffman-Seal-Tholem/monoidal-topology/2014}, 
such a level of abstraction is not necessary for our goals: 
it is sufficient to notice that, given a $\Quantale$-relation 
$\vrelone: A \torel A$,  
$f: A^n \to A$ is non-expansive (with respect to $\vrelone$) if:
\begin{align*}
   % \big\meet_i \vrelone(\tone_i, \ttwo_i) &\leq \vrelone(\op(\tone_1, \hh, \tone_n), \op(\ttwo_1, \hh, \ttwo_n))
%    \\
    \bigotimes_i \vrelone(\tone_i, \ttwo_i) &\leq \vrelone(\op(\tone_1, \hh, \tone_n), \op(\ttwo_1, \hh, \ttwo_n))
\end{align*}
% \begin{align*}
%   \vrelone(\tone_1, \ttwo_1) \wedge \cc \wedge \vrelone(\tone_n, \ttwo_n) 
%   &\leq \vrelone(\op(\tone_1, \hh, \tone_n), \op(\ttwo_1, \hh, \ttwo_n))
%     \\
% \vrelone(\tone_1, \ttwo_1) \tensor \cc \tensor \vrelone(\tone_n, \ttwo_n) 
% &\leq \vrelone(\op(\tone_1, \hh, \tone_n), \op(\ttwo_1, \hh, \ttwo_n))
% \end{align*}

%The above inequalities indeed correspond to 
%non-expansiveness instantiated to ultra and ordinary metric spaces, respectively. 
Non-expansive maps, however, are not the only maps one is interested in when 
working with metric spaces. Another interesting class of transformation is the 
one of \emph{contractions} and, more generally, the one of \emph{Lipschitz continuous} 
functions \cite{metric-spaces}. 
Such maps have been extensively studied in the context of metric
(program) semantics, due to their link with differential privacy \cite{Pierce/DistanceMakesTypesGrowStronger/2010,GaboardiEtAl/POPL/2017} and 
(bounded) linear and coeffectful types \cite{Orchard-icfp-2019,modal-reasoning-equal-metric-reasoning,Gaboradi-et-al/ICFP/2016}. 

Moving from an original observation by \citet{Lawvere/GeneralizedMetricSpaces/1973}, 
generalisations of non-expansive maps to $\Quantale$-relations 
have been given in terms of change of base functors~\cite{Gavazzo/LICS/2018,DBLP:phd/basesearch/Gavazzo19}, 
and corelators (viz. comonadic lax extension) \cite{DBLP:journals/pacmpl/LagoG22a}.
In a nutshell, we allow functions $f: A \to A$ to amplify distances, but in a 
controlled way. Such a way is given by a (family of suitable) function(s) 
$\baseone: \quantale \to \quantale$, so that we require:
$$
\baseone(\vrelone(\tone, \ttwo)) \leq \vrelone(f(\tone),f(\ttwo)).
$$
The map $\baseone$ is sometimes called the \emph{sensitivity} of $f$ 
and gives the law describing
how much differences between outputs are affected by differences between inputs. Accordingly, 
we think about sensitivity as generalising Lipschitz constants; and indeed, 
multiplication by a constant is a typical example of a map $\baseone$ on the Lawvere 
quantale. 

Technically speaking, we shall define function sensitivity by means of change of 
base functors~\cite{Kelly/EnrichedCats} which, in our simplified setting, take 
the form of quantale homomorphisms \cite{Hoffman-Seal-Tholem/monoidal-topology/2014}. 
Any change of base functor $\baseone: \quantale \to \quantale$ induces a map on $\Quantale$-relations 
sending a $\Quantale$-relation $\vrelone$ to the $\Quantale$-relation 
$\bang{\baseone}\vrelone$ mapping $(\tone, \ttwo)$ to $\baseone(\vrelone(\tone, \ttwo))$. 
Maps $\bang{\baseone}$ are examples of (graded) relational modalities known as \emph{corelators} 
\cite{DBLP:journals/pacmpl/LagoG22a}.\footnote{They actually provided a canonical example of 
a corelator.} Even if the theory of graded rewriting systems we are going to define can be given in 
full generality in terms of corelators, we shall work with change of base functors only. 
The authors hope this will choice will help the reader understanding this last section. 

When we move from unary to $n$-ary functions, it does make sense to talk about \emph{the} 
sensitivity of $f$; instead, we should talk about the sensitivity of $f$ \emph{on a given argument}. 
Assuming $f$ to have sensitivity 
$\baseone_i$ on the $i$th argument, 
we then obtain the following new, finer notion of 
non-expansiveness:
\begin{align*}
    \bigotimes_i \bang{\baseone_i}\vrelone(\tone_i, \ttwo_i) &\leq \vrelone(\op(\tone_1, \hh, \tone_n), \op(\ttwo_1, \hh, \ttwo_n)).
\end{align*}

Armed with this new notion of non-expansiveness, let us see how to make rewriting systems 
non-expansive in this new sense. The resulting notion, the one of a graded rewriting system, 
is the main subject of this last section. As usual, before introducing such systems in full 
generality, let us warm up with a concrete example.

\paragraph{Graded Combinatory Logic} 
The main example of a graded system we will deal with is \emph{graded combinatory logic}
\cite{DBLP:conf/lics/Atkey18,dagnino-1}, a generalisation of Abramsky's 
bounded combinatory logic \cite{abramsky-combinatory-1,abramsky-combinatory-logic-2}.

Recall that system $\BCK$ (as well as its extensions) is based on \emph{affine} combinators only. 
In particular, we have seen how adding the (cartesian) combinator $\Wcomb$ leads to 
distance trivialisation and non-confluent behaviours. The reason is that the reduction 
rule $\Wcomb \cdot x \cdot y \qstepto{0} x \cdot y \cdot y$ duplicates
the variable $y$, and thus the distance between combinators
$\Wcomb \cdot \termone \cdot \termtwo$ and $\Wcomb \cdot \termone \cdot \termtwo'$ 
is \emph{duplicated} when reducing $\Wcomb$. One way to overcome this problem is to refine system $\BCK$ by 
introducing graded exponential modalities $\bangg{n}$ constraining the usage of terms. 
The function symbol $\bangg{n}$ is an example of a coeffectful modality \cite{Orchard-icfp-2019,DBLP:conf/esop/GhicaS14,Mycroft-et-al/ICFP/2014,Gaboradi-et-al/ICFP/2016} 
and can be thought as providing $n$ copies of its argument, so that we 
can break linearity up to usage $n$. From a metric point of view, 
$\bangg{n}$ is a function symbol whose sensitivity 
is given by the multiplication by $n$ function, meaning that whenever we have terms 
$\termone$, $\termtwo$ that are $\qone$-apart, $\bangg{n}\termone$ and $\bangg{n}\termtwo$ 
are stipulated to be $n\qone$ apart. 

According to this strategy, the reduction $\Wcomb \cdot x \cdot y \qstepto{0} x \cdot y \cdot y$ 
is replaced by 
$$
\Wcomb \cdot x \cdot \bangg{n+m} y \qstepto{0} x \cdot \bangg{n}y \cdot \bangg{m}y.
$$
But that is not the end of the story. The introduction of 
$\bangg{n}$ affects not only ground reductions; it also hugely impacts on the definition of 
$\stepto$, which now becomes \emph{modal}. 
In fact, suppose to have combinators $\termone$, $\termtwo$ that are $\qone$-apart. If we now 
want to reduce $\termone$ to $\termtwo$ under the scope of $\bangg{n}$, we \emph{cannot} 
non-expansively propagate $\qone$ through $\bangg{n}\termone$ and $\bangg{n}\termtwo$, 
as we usually do in linear system. Instead, we have to amplify $\qone$ by $n$, this way obtaining 
the rule:
\[
\infer{\bangg{n}\termone \qreduce{n\qone} \bangg{n}\termtwo}
{\termone \qstepto{\qone} \termtwo}
\]
All of that extends to reductions inside arbitrary combinator (context) $C$. 
When reducing $C[\termone]$ to $C[\termtwo]$, we have to amplify the distance $\qone$ 
between $t$ and $s$
according to \emph{how} (much) $C$ uses its argument, i.e. according to the sensitivity of 
$\ctxone$ regarded as a term function. 
Writing $\degctx{\ctxone}$ for such a sensitivity, we obtain the rule
\[
\infer{\ctxone[\termone] \qreduce{\degctx{\ctxone}(\qone)} \ctxone[\termtwo]}
{\termone \qstepto{\qone} \termtwo}
\]
showing us that contexts now act not only on terms, but also on distances between them 
(or, taking a modal perspective \cite{modal-reasoning-equal-metric-reasoning,DBLP:journals/pacmpl/LagoG22a}, 
contexts act also on possible worlds).

Before giving a complete definition of the system of graded combinators, 
there is one last point we need to clarify: how do we determine the sensitivity 
of a context? A natural solution often employed in the literature on modal and 
graded calculi is to rely on a type system tracking variable usage in terms. 
Following such a proposal, we would work with judgements of the form 
$x:\baseone_1, \hh, x:\baseone_n \imp \termone$ stating that $x_i$ has sensitivity 
$\baseone_i$ in $\termone$. Introducing type systems, however, would require
unnecessary work in large measure independent of rewriting. We shall avoid that by 
recursively computing the grade of a variable 
(and even of a variable position) in a term \cite{dagnino-1}. 
Nonetheless, the reader may find useful to think in terms of type systems at first since 
trying to design typing rules helps to isolate the compositional properties 
a good notion of sensitivity should satisfy. 

First, we need to be able to \emph{add} and \emph{multiply} sensitivity functions, so to 
model nested and parallel use of terms. For instance, if a variable $x$ 
is used $n$ times by $\termone$ and $m$ times by $\termtwo$, it will 
by used $n+m$ times by $\op(\termone, \termtwo)$, provided that $\op$ is 
non-expansive (e.g. $x$ is used
 $n+m$ times by $\Bcomb \cdot \termone \cdot \termtwo \cdot y$). 
Similarly, if $g$ is a function symbol with sensitivity $j$ (e.g. take $\bangg{j}$ as $g$), 
then $x$ will be used $jn$ times in $g(\termone)$. Such operations are 
obviously available in the concrete example of graded combinators, 
where sensitivity is given by multiplication by a constant: in the general 
setting of quantale homomorphisms, we shall model multiplication 
as function composition and addition as pointwise tensor product.  
Secondly, we need to have distinguished functions modelling linear and zero sensitivity, 
i.e. neutral elements for multiplication and addition, respectively. 
With no surprise, those will be the identity and the constant $\qunit$ quantale 
homomorphisms.\footnote{Another approach is to model term sensitivity axiomatically 
\cite{DBLP:conf/esop/GhicaS14,Gaboradi-et-al/ICFP/2016,Orchard-icfp-2019} by 
means of suitable semi-ring like structures $\mathcal{G}$ and then define 
$\mathcal{G}$-indexed relational extensions (i.e. corelators) to model 
the action of grades on $\Quantale$-relations \cite{DBLP:journals/pacmpl/LagoG22a,modal-reasoning-equal-metric-reasoning}.}

Let us now formally introduce system $\BCKbounded$ of bounded combinators. 
The signature of the system is defined thus
$$
\signature_{\BCKbounded} \defeq \{\Bcomb, \Ccomb, \Kcomb, \Wcomb_{n,m}, 
\Dcomb, \Deltacomb_{n,m}, \Fcomb_n, \bangg{n}, \cdot \mid 
n,m \in [0,\infty]\}.$$
$\signature_{\BCKbounded}$ contains basic combinators $\Bcomb$, $\Ccomb$, $\Kcomb$, as well 
as the (family of) combinator(s) $\Wcomb_{n,m}$ 
and (families of) combinators $\Fcomb_{n}$, $\Dcomb$, and $\Deltacomb_{n,m}$ 
manipulating the function symbol(s) $\bangg{n}$. In fact, 
in addition to the usual binary function symbol for application, 
we have a $[0,\infty]$-family of exponential modalities $\bangg{n}$ \cite{DBLP:journals/tcs/GirardSS92,DBLP:conf/esop/GhicaS14}. 
Each function $\bangg{n}$ has sensitivity (constant multiplication by) $n$, 
which allows 
us to leave application non-expansive (meaning that it has sensitivity one 
on each argument). 
Notice that the signature $\signature_{\BCKbounded}$ specifies for each function symbol 
not only its arity, but also the sensitivity of all its arguments. We shall refer to 
such signatures as \emph{graded signatures}.
In general, we will employ the notation $\op: (\baseone_1, \hh, \baseone_n)$ 
to state that $\op$ is an $n$-ary function symbol with sensitivity $\baseone_i$ 
on its $i$th argument. 
% $$
% \signature_{\BCKbounded} \defeq \{\Bcomb, \Ccomb, \Kcomb, \Wcomb_{n,m}, 
% \Dcomb, \Deltacomb_{n,m}, \Fcomb_n, \bangg{n}, \cdot \mid 
% n,m \in [0,\infty]\},$$ 
% where 
% $\Bcomb, \Ccomb, \Kcomb, \Wcomb_{n,m}, \Dcomb, \Deltacomb_{n,m}, \Fcomb_n$ 
% are constants (basic combinators), $\bangg{n}$ is a unary function symbol with sensitivity $n$, 
% and $\cdot$ is a binary non-expansive (i.e. it has sensitivity one on each argument) function symbol.

To define rewriting $\Quantale$-relations for $\BCKbounded$, we first need to 
define the sensitivity of a variable in a term. Actually, we look at the 
sensitivity of a variable position in a term. We define the grade 
$\degree{\posone}{\termone}$ of a variable position $\posone$ in $\termone$ 
(i.e. $\termone_{|\posone}$ is a variable) 
as follows, where $X$ ranges over basic combinators:
\begin{align*}
    \degree{\posroot}{\varone} &\defeq 1 
    \\
    \degree{\posone}{X} &\defeq 0
    \\
    \degree{i\posone}{\termone_1 \cdot \termone_2} &\defeq 
    \degree{\posone}{\termone_i} 
    \\
     \degree{1\posone}{\bangg{n}\termone} &\defeq 
    n \cdot \degree{\posone}{\termone}.
\end{align*}
We then define the grade of a variable $x$ in a term $t$ as 
$\degree{x}{t} \defeq \sum \{\degree{\posone}{\termone} \mid \termone_{|\posone} = x\}$. 
For instance, the variable $x$ has sensitivity $9 = 3 + (3 \cdot 2)$ (regarded 
as the multiplication-by-three function)
in 
$\termone \defeq \bangg{3}(x \cdot \bangg{2} (\Icomb \cdot x))$, 
as it is under the scope both of $\bangg{3}$ and of $\bangg{2}$, and the latter, in turn,  
is itself 
under the scope of $\bangg{3}$ (and morally, we can think about a nested 
$\bangg{n} \bangg{m}$ as a unique $\bangg{nm}$). 
Indeed, the sensitivity of $x$ at position $1$ is $3$, 
whereas at position $1212$ is $6$.
% (which means that $\termone$ uses $\bangg{2}(\Kcomb \cdot x \cdot \bangg{0} z)$ three times, 
% and $\bangg{2}(\Kcomb \cdot x \cdot \bangg{0} z)$ uses $x$ two times, so that, in total, 
% $x$ is used two times three, i.e. six, times in $\termone$).
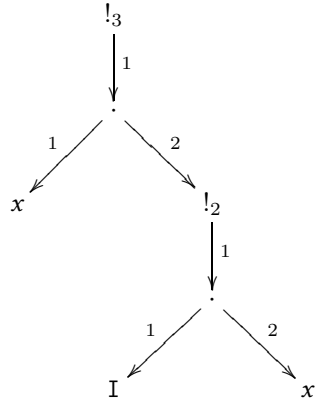
\[
\xymatrix{
& \bangg{3} \ar[d]^{1} & & 
\\
& \cdot \ar[ld]_{1} \ar[rd]^{2} & & 
\\
x & & \bangg{2} \ar[d]^{1} &
\\
& & \cdot \ar[ld]_{1} \ar[rd]^{2} &
\\
& \Icomb & & x
}
\]

\begin{notation}
Given a context $\ctxone$, we write $\degctx{\ctxone}$ for 
the sensitivity of the (unique occurrence of the) hole in $\ctxone$. 
\end{notation}

We now have all the ingredients to define system $\BCKbounded$, whose 
(ground) rewriting $\Quantale$-relation $\stepto_{\combboundrelone}$ and its extension 
$\reduce_{\combboundrelone}$ are defined in \autoref{figure:graded-combinators}. 
As usual, the $\Lawvere$-relation $\combboundrelone$ is defined by 
$\combboundrelone(t,s) \defeq \inf\{\qone \mid t \qreduce{\qone}_{\combboundrelone} s\}$.

\begin{figure}
{
\centering
\begin{tcolorbox}[boxrule=0.5pt,width=(\linewidth*5)/6,colframe=black,colback=black!0!white,arc=0mm]
%\vspace{-0.3cm}
 \[
     \Bcomb \cdot x \cdot y \cdot z \qstepto{0}_{\combboundrelone} x \cdot (y \cdot z)
     \qquad
    \Ccomb \cdot x \cdot y \cdot z \qstepto{0}_{\combboundrelone} x \cdot z \cdot y
    \qquad
    \Kcomb \cdot x \cdot \bangg{0}y  \qstepto{0}_{\combboundrelone} x
    \]
    \vspace{-0.2cm}
     \[
    \Dcomb \cdot \bangg{1} x \qstepto{0}_{\combboundrelone} x 
     \qquad
    \Deltacomb_{n,m} \cdot \bangg{nm} x \qstepto{0}_{\combboundrelone} \bangg{n} \bangg{m} x 
    \qquad
    \Fcomb_n \cdot \bangg{n} x \cdot \bangg{n} y  \qstepto{0}_{\combboundrelone} \bangg{n}(x \cdot y)
    \]
    \vspace{-0.2cm}
     \[
    \Wcomb_{n,m} \cdot x \cdot \bangg{n+m} y \qstepto{0}_{\combboundrelone} x \cdot \bangg{n} y \cdot \bangg{m}y 
    \]
    \vspace{-0.2cm}
    % \[
    % \infer
    % {\mrel
    % {\degree{p}{C}(\qone)}
    % {C[t^\sigma]_{\posone}}
    % {\reduce_{\combboundrelone}} 
    % {C[s^\sigma]_{\posone}}
    % }
    % {
    % \mrel{\qone}{t}{\stepto_{\combnatrelone}}{s}
    % }
    % \]
    %  \[
    % \infer
    % {C[t^\sigma]_{\posone} \qreduce{\baseone(\qone)}_{\combboundrelone} C[s^\sigma]_{\posone}}
    % {t \qstepto{\qone}_{\combboundrelone} s & \baseone = \degree{p}{C} }
    % \]
     \[
    \infer
    {C[t^\sigma] \qreduce{\degctx{\ctxone}(\qone)}_{\combboundrelone} C[s^\sigma]}
    {t \qstepto{\qone}_{\combboundrelone} s}
    \]
    % \[
    % \Acomb \cdot x \cdot \Zcomb \qstepto{0}_{\combnatrelone} x
    % \qquad
    % \Acomb \cdot x \cdot (\Scomb \cdot y) \qstepto{0}_{\combnatrelone} \Scomb \cdot (\Acomb \cdot x \cdot y)
    % \qquad
    % \Scomb \cdot x \qstepto{1}_{\combnatrelone} x
    % \]
    % \vspace{-0.2cm}
    % \[
    % \infer{C[t^\sigma] \qreduce{\qone}_{\combnatrelone} C[s^\sigma]}{t \qstepto{\qone}_{\combnatrelone} s}
    % \]
\end{tcolorbox}
}
\caption{System $\BCKbounded$ of graded combinators}
 \label{figure:graded-combinators}
\end{figure}

\subsection{Modal and Graded Rewriting: $(\Quantale, \CBE)$-Systems}
Now that the reader has familiarised with informal ideas behind 
modal and graded rewriting systems, we introduce such systems formally. 
To do so, we first recall the notion of a quantale homomorphism. 

\begin{definition}
Given quantales $\Quantale = (\quantale, \leq, \tensor, \qunit)$, 
$\mathbb{\Theta} = (\Theta, \cpoleq, \boxtimes, \qunittwo)$ a lax
quantale homomorphism is a \emph{monotone} map $h: \quantale \to \Theta$ such that 
\begin{align*}
    \qunittwo &\cpoleq h(\qunit)
    \\
    h(\qone) \boxtimes h(\qtwo) &\cpoleq h(\qone \tensor \qtwo).
\end{align*}
If we replace the above inequalities with full equalities and require
$h$ to be \emph{continuous} 
(i.e. $h(\join_i \qone_i) = \bigsqcup_i h(\qone_i)$),  
then we say that $h$ is a \emph{quantale homomorphism}.
\end{definition}

From now on, we shall work with quantale homomorphisms on the same 
quantale $\Quantale$. We denote such maps by $\baseone, \basetwo, \hh$ and refer 
to them as \emph{change of base (endo)functors} (CBEs, for short) 
\cite{Lawvere/GeneralizedMetricSpaces/1973, Hoffman-Seal-Tholem/monoidal-topology/2014}.  

\begin{remark}
CBEs have been successfully employed to study general program distances \cite{Gavazzo/LICS/2018,DBLP:phd/basesearch/Gavazzo19} 
and modal coeffectful reasoning \cite{modal-reasoning-equal-metric-reasoning,DBLP:journals/pacmpl/LagoG22a}. 
In such setting, actually, one works with \emph{lax} quantale homomorhpisms rather than with 
full homomorphisms. Although most (but not all) the results presented in this section can be 
given in terms of lax homomorphisms, important theorems such as confluence of orthogonal 
systems seem to require full homomorphisms. For that reason, we shall directly work with 
the latter maps. 
\end{remark}

\begin{example}
\begin{enumerate}
    \item 
The main example of CBEs we consider is multiplication by a constant on the Lawvere quantale 
(and variations thereof). Given $\kappa \in \mathbb{R}_{\geq 0}$, we regard $\kappa$ as mapping 
$\qone \in [0,\infty]$ to $\kappa\qone \in [0,\infty]$. Notice that we do not allow 
multiplication by infinity. That is because, from a rewriting perspective, 
multiplying by $\infty$ is semantically meaningless. Nonetheless, 
we could even include multiplication by
infinity in our analysis, provided that we restrict our definition to finitely continuous 
CBEs \cite{Gavazzo/LICS/2018,DBLP:phd/basesearch/Gavazzo19} and that we carefully define multiplication between zero 
and infinity (as first observed by \citet{GaboardiEtAl/POPL/2017}, algebra forces multiplication 
to become non-commutative, so that $0 \cdot \infty \neq \infty \cdot 0$).\footnote{
Extended multiplication is defined thus, for $y \neq 0$:
$ x \cdot \infty \defeq \infty$, 
$\infty \cdot 0 \defeq 0$, 
$\infty \cdot y \defeq \infty$.
}
\item Recall that in \autoref{section:qars} we have introduced a relational box modality 
    relying on the map
	 $\psi  : \two \to \quantale$ and its right adjoint 
	 $\varphi  : \quantale \to \two$.
% 	defined by:
% 	\begin{align*}
% 		\varphi(\qunit) &\defeq \true & \psi(\true) &\defeq 
% 		\\
% 		\varphi(\qvalone) &\defeq \false &\psi(\false) &\defeq \qbot.
% 	\end{align*}
	The map $\psi \circ \varphi$ is a CBE.
\item 
Other examples of CBEs, especially on quantales of modal predicates, can be found 
in the literature on relational reasoning about coeffects \cite{DBLP:journals/pacmpl/LagoG22a}.
\end{enumerate}
\end{example}

%More generally, given a quantale $\Quantale$ we can define CBEs
%   $n, \infty : \quantale \to \quantale$, for $n < \omega$ as follows:
%   \begin{align*} 
%   0(\qvalone) &\defeq \qunit 
%   \\
%   (n+1)(\qvalone) &\defeq 
%   \qvalone \tensor n(\qvalone)
%   \\ 
%   \infty(\qvalone) &\defeq \qbot.
%   \end{align*} 
%   Notice that $\baseid$ acts as the identity function, and that on the 
%   Lawvere and unit interval quantale we have 
%   $n(c) = n \cdot c$ and $\infty(c) = \infty$.

CBEs are closed under composition and the identity function 
$\baseid: \quantale \to \quantale$ is a CBE. 
Moreover, we can extend the order $\leq$ and 
the multiplication $\tensor$ of $\Quantale$ to CBEs pointwise.
Finally, 
we denote by $\qunitconst$ the constant 
$\qunit$ CBE. 

\begin{remark}
Any CBE $\baseone$ induces an action $\bang{\baseone}$ on 
$\Quantale$-relations defined by
$\bang{\baseone}\vrelone(\tone, \ttwo) \defeq \baseone(\vrelone(\tone,\ttwo)).
$ The map $\bang{\baseone}$ is an example of a corelator \cite{DBLP:journals/pacmpl/LagoG22a}.
\end{remark}

% Finally, we observe that the action of CBFs on a $\Quantale$-relation
% obeys the following laws:
% \begin{align*}
%   (h \comp h')(\vrelone)  
%   &= h \acts (h' \acts \vrelone), 
%   \\
%   (h \acts \vrelone) \comp (h \acts \vreltwo) 
%   &\leq h \acts (\vrelone \comp \vreltwo).
% \end{align*}

We now introduce a new class of rewriting systems, which we dub 
\emph{$(\Quantale, \CBE)$-systems} 
(or \emph{$\CBE$-systems} for short). 
Let us fix a quantale $\Quantale$ and a structure 
$\mathbb{\Phi} = (\Phi, \leq, \circ, \baseid, \tensor, \qunitconst)$, 
where $\Phi$ is a set of CBEs containing the identity and constant $\qunit$-functions,
and closed under function
composition and tensor.

\begin{definition}
\begin{enumerate}
    \item The \emph{modal arity} of an $n$-ary function symbol $\op$ 
        is a tuple $(\baseone_1, \hh, \baseone_n)$
        with $\baseone_i \in \Phi$. Given a function symbol $\op$ with 
        modal arity $(\baseone_1, \hh, \baseone_n)$ (notation $\op: (\baseone_1, \hh, \baseone_n)$), 
       we say that $\op$ has sensitivity (or modal grade) $\baseone_i$ on its $i$th argument. 
    \item A \emph{$\CBE$-graded signature} is a set $\signature$ containing function symbols 
        with their modal arity. Given a $\CBE$-graded signature $\signature$ and a set 
        $\variables$ of variables, the collection of $\terms{\signature}{\variables}$ is defined 
        as usual.
    \item Given a term $\termone$ and a position $\posone$ for a variable in $\termone$, 
        we define the grade $\degree{\posone}{\termone}$ of $\posone$ in $\termone$ as follows:
        \begin{align*}
            \degree{\posroot}{\termone} &\defeq \baseid
            \\
            \degree{i\posone}{\op(\termone_1, \hh, \termone_n)} 
            &\defeq \baseone_i \circ \degree{\posone}{\termone_i} 
            \qquad \quad (\op: (\baseone_1, \hh, \baseone_n)) \in \signature.
        \end{align*}
\end{enumerate}
\end{definition}

Given a term $\termone$ and a variable $x$, we can compute the grade $\degree{\varone}{\termone}$ 
of $x$ in $\termone$ 
by `summing' the grades of all position $\posone$ such that $\termone_{|\posone} = x$.
Formally,
$\degree{\varone}{\termone} \defeq \bigotimes\{\degree{\posone}{\termone} \mid \termone_{|\posone} = x\}$, 
where $\bigotimes \emptyset \defeq \qunitconst$. 
Notice that we can equivalently define $\degree{\varone}{\termone}$ recursively as follows:
%The grade of a variable $x$ in a term $\termone$ can be equivalently defined as follows:
    \begin{align*}
        \degree{x}{x} &\defeq \baseid
        \\
        \degree{x}{y} &\defeq \qunitconst
        \\
        \degree{x}{\op(\termone_1, \hh, \termone_n)} 
        &\defeq \bigotimes \baseone_i \circ \degree{x}{\termone_i} 
        \qquad \quad (\op: (\baseone_1, \hh, \baseone_n)) \in \signature.
    \end{align*}
We are now ready to define $(\CBE$-)graded term rewriting systems.
\begin{definition}
\label{def:cbetrs}
    A \emph{$(\Quantale,\CBE)$-term rewriting system} (\cbetrs{,} for short) 
    is a pair $\qtrsone = (\signature, \stepto_{\vbaserelone})$ consisting of 
    a $\CBE$-signature $\signature$ and
    a $\Quantale$-ternary relation.
    The (rewriting) $\Quantale$-ternary relation $\reduce_{\vrelone}$ generated by $\stepto_{\vrelone}$ is
    defined thus:
     \[
    \infer{\degctx{\ctxone}(\qone) \Vdash C[\subst{\redex}{\sone}] \to_{\vbaserelone} C[\subst{\contractum}{\sone}]}
        {\qone \Vdash \redex \stepto_{\vbaserelone} \contractum}
    \quad
    \infer{\qtwo \Vdash \termone \reduce_{\vbaserelone} \termtwo}
    {\qone \Vdash \termone \reduce_{\vbaserelone} \termtwo & \qtwo \leq \qone}
    \]
    We say that the system is \emph{balanced} if for any rule 
    $\mrel{\qone}{\redex}{\stepto}{\contractum}$ we have 
    $\degree{\varone}{\redex} = \degree{\varone}{\contractum}$, 
    for any variable $\varone$. From now on, we assume all \cbetrs{s} to be balanced. 
\end{definition}

Compared with the definition of (linear) \qtrs{s}, \autoref{def:cbetrs} has two main differences: 
first, the definition of the full rewriting relation $\reduce$ now takes into account 
the grade of the context; secondly, we omit all structural rules besides weakening. 
The omission of such rules (cf. \autoref{rem:structural-rules}) 
allows us to strengthen our results: in particular, whereas 
a critical pair lemma can be given for \cbetrs{s} extended with \emph{all} structural rules, 
our proof of confluence of orthogonal \cbetrs{s} seems not to scale to \cbetrs{s} 
extended with the Archimedean (infinitary) rule. 

\begin{notation}
We extend to \cbetrs{s} all notational conventions introduced for \qtrs{s}.
\end{notation}

Let us now have a closed look at the definition of $\reduce$.
First, it is instructive to characterise it inductively as follows:
\[
\infer{\mrel{\qone}{\redex}{\reduce}{\contractum}}
{\mrel{\qone}{\redex}{\stepto}{\contractum}}
\qquad
\infer{\mrel{\qone}{\subst{\termone}{\sone}}{\reduce}{\subst{\termtwo}{\sone}}}
{\mrel{\qone}{\termone}{\reduce}{\termtwo}}
\qquad 
\infer{\baseone_i(\qone) \Vdash \op(\termthree_1, \hh, \termone, \hh, \termthree_n) 
\reduce \op(\termthree_1, \hh, \termtwo, \hh, \termthree_n) }
{\qone \Vdash \termone \reduce \termtwo & \op: (\baseone_1, \hh, \baseone_n) \in \Phi}
\]
This characterisation clearly shows that performing reductions inside function symbols  
amplify distances, whereas applying substitutions does not. This is because in the 
definition of $\reduce$ we apply \emph{the same} substitution on terms. Intuitively, this 
reflects the fact that passing identical (i.e. at a null distance) arguments to a 
Lipschitz continuous function produces identical results, and thus there is no distance 
amplification. Indeed, rephrasing \autoref{def:cbetrs} to (balanced) \emph{equational theories}, 
we obtain the following substitution rule
\[
\infer{\qone \tensor \degree{\varone}{\termone}(\qtwo) \Vdash 
\termone[\termthree/\varone] =_{\veqone} \termtwo[\termfour/\varone]}
{\qone \Vdash \termone =_{\veqone} \termtwo & 
\qtwo \Vdash \termthree =_{\veqone} \termfour}
\]
When $\termthree$ and $\termfour$ coincide --- so that $\qtwo = \qunit$ --- we obtain 
$\qone \tensor \degree{\varone}{\termone}(\qtwo) = \qone \tensor \degree{\varone}{\termone}(\qunit) = 
\qone \tensor \qunit = \qone$, meaning that the distance $\qone$ is non-expansively propagated.

    % \begin{figure}
    % {
    % \centering
    % \begin{tcolorbox}[boxrule=0.5pt,width=\linewidth,colframe=black,colback=black!0!white,arc=0mm]
    % \vspace{-0.3cm}
    %   \[
    % \infer{\baseone_{\ctxone}(\qone) \Vdash C[\subst{\redex}{\sone}] \to_{\vbaserelone} C[\subst{\contractum}{\sone}]}
    %     {\qone \Vdash \redex \stepto_{\vbaserelone} \contractum}
    % \quad
    % \infer{\qtwo \Vdash \termone \reduce_{\vbaserelone} \termtwo}
    % {\qone \Vdash \termone \reduce_{\vbaserelone} \termtwo & \qtwo \leq \qone}
    % \quad 
    % \infer{\join \qone_i \Vdash \termone \reduce_{\vbaserelone} \termtwo}
    % {\qone_1 \Vdash \termone \reduce_{\vbaserelone} \termtwo & \hdots & \qone_n \Vdash \termone \reduce_{\vbaserelone} \termtwo}
    % \quad 
    % \infer{\qone \Vdash \termone \reduce_{\vbaserelone} \termtwo}
    % {\forall \qtwo \ll \qone.\ \qtwo \Vdash \termone \reduce_{\vbaserelone} \termtwo}
    % \]
    % \end{tcolorbox}
    % }
    % \caption{Definition of $\reduce_{\vrelone}$}
    % \label{figure:qtrs}
    % \end{figure}

As usual, any \cbetrs{} $(\signature, \stepto_{\vbaserelone})$ induces a \qars{} 
whose objects are $\signature$-terms and whose 
rewriting $\Quantale$-relation 
is defined by
    $$
    \vrelone(\termone, \termtwo) \defeq \join \{\qone \mid \qone \Vdash \termone \reduce_{\vbaserelone} \termtwo\}.
    $$
    
\begin{example}
{
\renewcommand{\code}[1]{\underline{#1}}
The main example of a \cbetrs{} we consider is system $\BCKbounded$, as introduced at
the beginning of this section. As for system $\BCK$ of affine combinators, 
we can also consider extensions of $\BCKbounded$. For instance, we can 
consider functions $f: \mathbb{N}^m \to \mathbb{N}$ with Lipschitz constants 
$(\baseone_1, \hh, \baseone_m)$ and 
add to $\signature_{\BCKbounded}$ (properly extended with constants $\code{0}$, $\code{1}, \hh$ 
for natural numbers) 
a combinator $\code{f}$ for any such a function, 
together with rules
$$
\code{f} \cdot \bangg{\baseone_1} \code{n}_1 \cc \bangg{\baseone_m} \code{n}_m 
\qstepto{0} \code{f(n_1, \hh, n_m)}.
$$
Another interesting example of a \cbetrs{} is obtained by grading the signature of
system $\BA$: any function symbol $+_{\probone}$ now takes signature 
$(\probone, 1-\probone)$, where by $\probone$ (resp. $1-\probone$) 
we mean multiplication by $\probone$ (resp. $1-\probone$). 
Notice that such a signature actually makes $+_\probone$ a \emph{contraction}~\cite{metric-spaces}.
\citet{plotkin-quantitative-algebras-2016} have shown that the system obtained from such a 
signature together with the (equational) rules of idempotency, commutativity, and associativity 
provides a (quantitative) equational axiomatisation of the (finitary) Wasserstein-Kantorovich 
distance \cite{Villani/optimal-transport/2008}.
}
\end{example}

\paragraph{Modal and Graded Equational Theories}
Before moving to the metatheory of \cbetrs{s}, 
we extend \quant{} equational theories to a graded setting \cite{dagnino-1}.

\begin{definition}
\label{def:graded-quantitative-algebras}
A graded (\quant{)} equational theory is a pair $\eqtheoryone= (\signature, \approx_\veqone)$, where 
$\signature$ is a $\CBE$-signature and $\approx_\veqone$ 
is a $\Quantale$-ternary relation over $\signature$-terms. The  $\Quantale$-ternary (equality) relation
$\equal_{\veqone}$ generated by $\approx_\veqone$ is defined by the rules in 
\autoref{figure:graded-algebra}.
We say that a graded equational theory is balanced if 
whenever $\qone \Vdash \termone \approx_\veqone \termtwo$, we have 
$\degree{\varone}{\termone} = \degree{\varone}{\termtwo}$, for any variable $x$. 
Notice that if equations in $\approx_{\veqone}$ are balanced, then so are equations in 
$=_{\veqone}$.
\end{definition}

 \begin{figure}
{
\centering
\begin{tcolorbox}[boxrule=0.5pt,width=(\linewidth*5)/6,colframe=black,colback=black!0!white,arc=0mm]
%\vspace{-0.3cm}
     \[
    \infer{\qone \Vdash \termone \equal_{\vbaseeq} \termtwo}{\qone \Vdash \termone \approx_{\vbaseeq} \termtwo}
    \qquad
    \infer{\qunit \Vdash \termone \equal_{\vbaseeq} \termone}{}
    \qquad 
    \infer{\qone \Vdash \termtwo \equal_{\vbaseeq} \termone}{\qone \Vdash \termone \equal_{\vbaseeq} \termtwo}
    \qquad 
    \infer{\qone \tensor \qtwo \Vdash \termone \equal_{\vbaseeq} \termthree}
    {\qone \Vdash \termone \equal_{\vbaseeq} \termtwo 
    & \qtwo \Vdash \termtwo \equal_{\vbaseeq} \termthree}
    \]

    \[
    \infer{\bigotimes_i \baseone_i(\qone_i) \Vdash \op(\termone_1, \hh, \termone_n) \equal_{\vbaseeq} \op(\termtwo_1, \hh, \termtwo_n)}
    {\qone_1 \Vdash \termone_1 \equal_{\vbaseeq} \termtwo_1 & \cc & 
    \qone_n \Vdash \termone_n \equal_{\vbaseeq} \termtwo_n &
    \op: (\baseone_1, \hh, \baseone_n) \in \signature}
    \qquad 
     \infer{\qone \Vdash \subst{\termone}{\sone} \equal_{\vbaseeq} \subst{\termtwo}{\sone}}
    {\qone \Vdash \termone \equal_{\vbaseeq} \termtwo}
    \]
    \vspace{-0.05cm}
    \[
    \infer{\qtwo \Vdash \termone \equal_{\vbaseeq} \termtwo}
    {\qone \Vdash \termone \equal_{\vbaseeq} \termtwo & \qtwo \leq \qone}
    \qquad 
    \infer{\join \qone_i \Vdash \termone \equal_{\vbaseeq} \termtwo}
    {\qone_1 \Vdash \termone \equal_{\vbaseeq} \termtwo & \hdots & \qone_n \Vdash \termone \equal_{\vbaseeq} \termtwo}
    \qquad 
    \infer{\qone \Vdash \termone \equal_{\vbaseeq} \termtwo}
    {\forall \qtwo \ll \qone.\ \qtwo \Vdash \termone \equal_{\vbaseeq} \termtwo}
    \]
\end{tcolorbox}
}
\caption{Graded Quantitative Equational Theory of $\equal_{\vbaseeq}$}
 \label{figure:graded-algebra}
\end{figure}

As for quantitative equational theories, we see that $\veqone$ is reflexive, symmetric, and transitive; and that
by regarding any $n$-ary function symbol $f$ as a function 
$f: \terms{\signature}{\variables}^n \to \terms{\signature}{\variables}$, we have
$$
\bang{\baseone_1}\veqone(\termone_1, \termtwo_1) \tensor \cc \tensor \bang{\baseone_n}\veqone(\termone_n, \termtwo_n) 
\leq \veqone(\op(\termone_1, \hh, \termone_n), \op(\termtwo_1, \hh, \termtwo_n)),
$$
meaning that function symbols behave as (generalised) Lipschitz continuous functions. 
Moreover, it is an easy exercise to prove that for any balanced graded theory 
the substitution rule 
\[
\infer{\qone \tensor \degree{\varone}{\termone}(\qtwo) \Vdash 
\termone[\termthree/\varone] =_{\veqone} \termtwo[\termfour/\varone]}
{\qone \Vdash \termone =_{\veqone} \termtwo & 
\qtwo \Vdash \termthree =_{\veqone} \termfour}
\]
is valid, from which we obtain the following substitution inequality:
$$
\veqone(\termone,\termtwo) \tensor \bang{\degree{\varone}{\termone}}\veqone(\termthree, \termfour) 
\leq \veqone(\termone[\termthree/\varone], \termtwo[\termfour/\varone]).
$$
Finally, at this point of the work it should be clear 
that the connection between \quant{} equational theories and \qtrs{s} extend 
\emph{mutatis mutandis} to graded equational theories and \cbetrs{{s}, modulo the 
addition of structural rules in the definition of the latter.

% so that we obtain a \francesco{$\quantale$-metric space} $(\terms{\signature}{\variables}/{\sim}, \veqone)$, 
% where $\sim$ is the equivalence relation relating terms at $\veqone$-distance $\qunit$. Each function symbol 
% $f$ induces, as usual, a map $f_{\sim}: \terms{\signature}{\variables}/{\sim} \to \terms{\signature}{\variables}/{\sim}$, 
% and the defining rules of $\equal_{\vbaseeq}$ ensure such a map to \emph{non-expansive}. Symbolically, 
% defining $\veqone_{\sim}([\termone]_{\sim}, [\termtwo]_{\sim}) \defeq \veqone(\termone, \termtwo)$, we have:
% $$
% \bigotimes_i \veqone_{\sim}([\termone_i]_{\sim}, [\termtwo_i]_{\sim})
% \leq 
% \veqone_{\sim}(f_{\sim}([\termone_1]_{\sim}, \hh, [\termone_n]_{\sim}), f_{\sim}([\termtwo_1]_{\sim}, \hh, [\termtwo_n]_{\sim}))
% $$

\subsection{Confluence and Critical Pairs, Part IV}

We now progressively extend the theory  of \qtrs{s} to 
\cbetrs{s}, beginning with \autoref{theorem:critical-pair-theorem}. 
The notion of an overlap and of a critical pair straightforwardly extend to \cbetrs{s}.
Notice, however, that if 
$\redex_1 \qstepto{\qone_1} \contractum_1$ and 
$\redex_2 \qstepto{\qone_2} \contractum_2$ overlap at position $\posone$, 
so that there is a substitution $\sone$ such that 
$\subst{\redex_2}{\sone} = \ctxone[\redex_1^{\sone}]$ (with $\ctxone = \redex_2^{\sone}[-]_{\posone}$), 
then the critical pick is 
        \begin{align*}
             \subst{\contractum_2}{\sone}
             \stackrel{\qone_2}{\leftarrow} 
             \subst{\redex_1}{\sone} 
            % \qreduce{\degree{\posone}{\redex_2^\sone}(\qone_2)} 
             \qreduce{\degctx{\ctxone}(\qone_2)} 
             %\subst{\redex_2}{\sone}[\subst{\contractum}{\sone}]_{\posone}
             \ctxone[\subst{\contractum_2}{\sone}].
            \end{align*}
We thus have all the ingredients to extend 
\autoref{lemma:critical-pair} to \cbetrs{s}. 
Compared to its \qtrs{} counterpart, however, the critical pair lemma for  \cbetrs{s} 
presents a major difference: we can relax the linearity assumption and require rewriting rules 
to be \emph{left-linear} only.

\begin{lemma}[Critical Pair, Graded]
\label{lemma:critical-pair-graded}
Let $\qtrsone = (\signature, \stepto_{\vbaserelone})$ be a \emph{left-linear} (balanced) \cbetrs{.} 
If $\vrelone$ is locally confluent on all critical pairs of $\qtrsone$, then 
it is locally confluent.
%
%$\termtwo_{1},\termtwo_{2}$ are joinable with decreasing distance. %$\termtwo_{1}\twoheadrightarrow \cdot \twoheadleftarrow \termtwo_{2}$ with $(\dual{\vrelone};\vrelone) (\termtwo_{1},\termtwo_{2})\leq( \vrelone^{*};\dual{{\vrelone^{*}}})(\termtwo_{1},\termtwo_{2})$.
\end{lemma}

\begin{remark}
\label{rem:critical-pair-graded}
Due to the absence of structural rules (besides weakening) in \autoref{def:cbetrs}, 
\autoref{lemma:critical-pair-graded} can be strengthen to to prove local confluence of 
$\reduce_{\vrelone}$ given its local confluence on critical pairs of $\qtrsone$.
\end{remark}

\begin{proof}[Proof of \autoref{lemma:critical-pair-graded}]
{
\renewcommand{\redone}{a_1}
\renewcommand{\redtwo}{a_2}
\renewcommand{\conone}{b_1}
\renewcommand{\contwo}{b_2}
Following \autoref{rem:critical-pair-graded}, we prove confluence of $\reduce$. 
The proof proceeds as for \autoref{lemma:critical-pair}, the main difference being the 
case of nested, non-critical redexes. We analyse this case in detail, and 
then extend it to the full case of a general peak. 
Suppose to have rules $\redone \qstepto{\qone_1} \conone$, 
$\redtwo \qstepto{\qone_2} \contwo$. Without loss of generality, we 
consider the case in which the second reduction happens inside (an instance) of the first one.
So there is a variable $x$ in $\redone$ and a substitution instance such that 
$\subst{x}{\sone}$ contains $\subst{\redtwo}{\sone}$.
Since $\qtrsone$ is left linear, we know that there is just one occurence of $x$ in 
$\redone$. Say it is at position $\posone$, so that we have $\redone[x]_{\posone}$. 
Say also that the relevant occurrence of $\subst{\redtwo}{\sone}$ in $\subst{x}{\sone}$ 
is at position $\postwo$, so that 
$\subst{x}{\sone}[\subst{\redtwo}{\sone}]_{\postwo}$ 
and, consequently, $\subst{\redone}{\sone}[\subst{x}{\sone}[\subst{\redtwo}{\sone}]_{\postwo}]_{\posone}$ 
and $\subst{\redone}{\sone}[\subst{\redtwo}{\sone}]_{\posone\postwo}$. 
The rule $\redone \qstepto{\qone_1} \conone$ may, in general, duplicate the single occurrence 
of $x$ in $\redone$. Say we have $\conone[x]_{\posone_1, \hh, \posone_n}$, meaning that 
$\conone$ has $n$ occurrences of $x$, each at position $\posone_i$. 
Therefore, reducing $\subst{\redone}{\sone}$ gives 
$\conone^{\sone}[x^{\sone}]_{\posone_1, \hh, \posone_n}$, and thus 
$\conone^{\sone}[\redtwo^{\sone}]_{\posone_1\postwo, \hh, \posone_n\postwo}$. 
We can now reduce each of the $n$ occurrences of $\redtwo^{\sone}$ in $\conone^{\sone}$. 
The distance obtained for each reduction, however, is not $\qone_2$ but 
$\degree{\posone_i\postwo}{\contwo^{\sone}}(\qone_2)$. Putting things together, we 
obtain the following 
reduction diagram:
\[
\xymatrix{
& \subst{\redone}{\sone}[\subst{\redtwo}{\sone}]_{\posone\postwo} 
\ar[ld]_{\qone_1}  \ar[rd]^{\degree{\posone\postwo}{\redone^{\sone}}(\qone_2)}&
\\
\conone^{\sone}[\redtwo^{\sone}]_{\posone_1\postwo, \hh, \posone_n\postwo} 
\ar@{->>}[rd]_{\bigotimes_i \degree{\posone_i\postwo}{\conone^\sone}(\qone_2) \quad} 
& & \redone^{\sone}[\contwo^{\sone}]_{\posone\postwo} \ar[ld]^{\qone_1}
\\
& \conone^{\sone}[\contwo^{\sone}]_{\posone_1\postwo, \hh, \posone_n\postwo} & 
}
\]
To obtain local confluence, we claim 
$$
\qone_1 \tensor \degree{\posone\postwo}{\redone^{\sone}}(\qone_2) 
\leq \bigotimes_i \degree{\posone_i\postwo}{\conone^\sone}(\qone_2) \tensor \qone_1.
$$
Since the system is (left-linear and) balanced, we have 
$$\degree{\posone}{\redone} = \degree{x}{\redone} = \degree{x}{\conone} = 
\bigotimes_i \degree{\posone_i}{\conone}.$$
Moreover, we notice that 
\begin{align*}
    \degree{\posone\postwo}{\redone^{\sone}} &= 
    \degree{\posone}{\redone} \circ \degree{\postwo}{x^\sone}
    \\ \bigotimes_i \degree{\posone_i\postwo}{\conone^\sone} 
    &= \bigotimes_i \degree{\posone_i}{\conone} \circ \degree{\postwo}{x^\sone}
\end{align*}
Writing $\qtwo$ for $\degree{\postwo}{x^\sone}$, we obtain the desired 
inequality as follows:
\begin{align*}
    \qone_1 \tensor \degree{\posone\postwo}{\redone^{\sone}}(\qone_2) 
    = \degree{\posone}{\redone}(\qtwo) \tensor \qone_1
    = \bigotimes_i \degree{\posone_i}{\conone}(\qtwo) \tensor \qone_1
   % = \bigotimes_i \degree{\posone_i}{\conone} \circ \degree{\postwo}{x^\sone}(\qone_2) \tensor \qone_1
    &= \bigotimes_i \degree{\posone_i\postwo}{\conone^\sone}(\qone_2) \tensor \qone_1.
\end{align*}
This shows how to deal with nested non-critical redexes in isolation.  
In the general case, we have to consider all of that happening inside a larger term 
$\termone$. This means nothing by considering cases of the form 
$\ctxone[\redone^\sone[\redtwo^\sone]_{\posone\postwo}]$. The proof proceeds exactly as in the isolated 
case, with the main difference that distances should now be scaled by $\degctx{\ctxone}$. 
But, due to the structural properties of CBEs, this creates no problem at all.
}
\end{proof}

\begin{theorem}
\label{theorem:critical-pair-theorem-graded}
Any \emph{left-linear} and terminating (balanced) \cbetrs{} locally confluent on its critical pairs is 
confluent.
\end{theorem}

\begin{proof}
It directly follows from \autoref{prop:newmans-lemma} and \autoref{lemma:critical-pair-graded}.
\end{proof}

\subsection{Orthogonality} 
Even if useful on many \cbetrs{s}, \autoref{theorem:critical-pair-theorem-graded} 
can only be used to infer local confluence of non-terminating systems, 
system $\BCKbounded$ being a prime example of such a system. Yet, examples 
seem to suggest $\BCKbounded$ to be indeed a confluent system. To make the latter intution into 
a proved mathematical result, 
we notice that the hypotheses of \autoref{theorem:critical-pair-theorem-graded}  
are trivially satisfied in the case of system $\BCKbounded$, as the latter simply has 
no critical pair. Taking advantage of this observation, we now generalise 
the well-known result \cite{rosen-70} that orthogonality implies confluence 
to a \quant{} and graded setting. 

\begin{definition}
A \cbetrs{} is \emph{orthogonal} if it is left-linear and has no critical pair. 
\end{definition}

As already remarked, our prime example of an orthogonal \cbetrs{} is system 
$\BCKbounded$ of graded combinators. To prove confluence of orthogonal systems we 
employ Tait and Martin-L\"of technique \cite{Barendregt/Book/1984}, properly instantiated to 
our rewrtiting setting
(see also Aczel's technique \cite{aczel-general-church-rosser}).
We extend $\reduce$ to a $\Quantale$-ternary relation $\multireduce$ allowing us to 
perform arbitrary (even nested) reductions in a term at once. 

\begin{definition}
Given a \cbetrs{} $\qtrsone = (\signature, \stepto_{\vrelone})$, 
we inductively define the multi-step reduction 
$\multireduce$ by the rules in \autoref{figure:multi-step-reduction}. 
We define the $\Quantale$-relation $\mathring{\vrelone}$ by 
$$
\mathring{\vrelone}(\termone, \termtwo) \defeq \join \{\qone \mid 
\qone \Vdash \termone \multireduce_{\vrelone} \termtwo\}.
$$
\end{definition}

 \begin{figure}
{
\centering
\begin{tcolorbox}[boxrule=0.5pt,width=(\linewidth*5)/6,colframe=black,colback=black!0!white,arc=0mm]
%\vspace{-0.3cm}
    \[
\infer{\qunit \Vdash \varone \multireduce_{\vrelone}  \varone}{}
\]
\vspace{-0.05cm}
\[
\infer{\bigotimes_i \baseone_i(\qone_i) \Vdash 
\op(\termone_1, \hh, \termone_n) \multireduce_{\vrelone}  \op(\termtwo_1, \hh, \termtwo_n)}
{\qone_1 \Vdash \termone_1 \multireduce_{\vrelone}  \termtwo_1 
& \cc &
\qone_n \Vdash \termone_n \multireduce_{\vrelone}  \termtwo_n
& \op: (\baseone_1, \hh, \baseone_n) \in \signature_{\qtrsone}
}
\]
\vspace{-0.05cm}
\[
\infer{\qone \tensor \bigotimes_i \degree{x_i}{\redone}(\qtwo_i) 
\Vdash \redone[v_1, \hh, v_n/x_1, \hh, x_n] \multireduce_{\vrelone} 
\conone[w_1, \hh, w_n/x_1, \hh, x_n]}
{\qtwo_1 \Vdash v_1 \multireduce_{\vrelone} w_1 
& \cc & 
\qtwo_n \Vdash v_n \multireduce_{\vrelone} w_n 
& \qone \Vdash \redone \stepto_{\vrelone} \conone
}
\]
\vspace{-0.05cm}
\[
\infer{\qtwo \Vdash \termone \multireduce_{\vrelone} \termtwo}
{\qone \Vdash \termone \multireduce_{\vrelone} \termtwo 
& \qtwo \leq \qone}
\]
\end{tcolorbox}
}
\caption{Multi-step reduction $\multireduce_{\qtrsone}$}
 \label{figure:multi-step-reduction}
\end{figure}

\begin{notation}
We extend the usual notational convention to $\multireduce$. 
Moreover, in what follows we often employ the vector notation $\vect{\varphi}$ 
for finite sequences $\varphi_1, \hh, \varphi_n$ of symbols.
\end{notation}

We immediately notice that since $\multireduce$ allows us to reduce several redexes in 
a term simultaneously, it gives a substitution property similar to the one 
of graded \quant{} equational theory.

\begin{lemma}[Substitution Lemma]
\label{lemma:graded-substituion-lemma}
The following inference is valid
% \[
% \infer{
% \qone \tensor \bigotimes \degree{\varone_i}{\termone}(\qtwo_i) \Vdash 
% \termone[v_1, \hh, v_n/x_1, \hh, x_n] \multireduce_{\vrelone} 
% \termtwo[w_1, \hh, w_n/x_1, \hh, x_n]}
% {\qone \Vdash \termone \multireduce_{\vrelone} \termtwo & 
% \qtwo_1 \Vdash v_1 \multireduce_{\vrelone} w_1 
% & \cc & 
% \qtwo_n \Vdash v_n \multireduce_{\vrelone} w_n 
% }
% \]
% Consequently, we also obtain the following substitution inequality:
% $$
% \mathring{\vrelone}(\termone,\termtwo) \tensor \bigotimes_i \bang{\degree{\varone_i}{\termone}}
% \mathring{\vrelone}(v_i, w_i) 
% \leq \mathring{\vrelone}(\termone[v_1, \hh, v_n/x_1, \hh, x_n],
% \termtwo[w_1, \hh, w_n/x_1, \hh, x_n]).
% $$
\[
\infer{
\qone \tensor \bigotimes \degree{\varone_i}{\termone}(\qtwo_i) \Vdash 
\termone[\vect{v}/\vect{x}] \multireduce_{\vrelone} 
\termtwo[\vect{w}/\vect{x}]}
{\qone \Vdash \termone \multireduce_{\vrelone} \termtwo & 
\qtwo_1 \Vdash v_1 \multireduce_{\vrelone} w_1 
& \cc & 
\qtwo_n \Vdash v_n \multireduce_{\vrelone} w_n 
}
\]
Consequently, we also obtain the following substitution inequality:
$$
\mathring{\vrelone}(\termone,\termtwo) \tensor \bigotimes_i \bang{\degree{\varone_i}{\termone}}
\mathring{\vrelone}(v_i, w_i) 
\leq \mathring{\vrelone}(\termone[\vect{v}/\vect{x}],
\termtwo[\vect{w}/\vect{x}]).
$$
\end{lemma}

\begin{proof}[Proof sketch]
By induction on the definition of 
$\multireduce_{\vrelone}$ and following the pattern of graded and quantitative substitution lemmas 
\cite{Gavazzo/LICS/2018,DBLP:phd/basesearch/Gavazzo19,DBLP:journals/pacmpl/LagoG22a}.
\end{proof}

Given a \cbetrs{} $\qtrsone = (\signature, \reduce_{\vrelone})$, 
we are going to prove confluence of $\vrelone$ by actually proving a stronger 
result, namely confluence of $\reduce_{\vrelone}$. This is possible thanks to the absence of 
(infinitary) structural rules in the definition of a \cbetrs{} (\autoref{def:cbetrs}). 
To achieve such a result, we shall prove that $\multireduce_{\vrelone}$ has the diamond 
property. Since ${\reduce_{\vrelone}} \subseteq {\multireduce_{\vrelone}} \subseteq {\reduce_{\vrelone}^*}$ 
(and thus $\vrelone \leq \mathring{\vrelone} \leq \vrelone^*$), confluence of
$\reduce_{\vrelone}$ follows. 
Before proving the diamond property for $\multireduce_{\vrelone}$, let us remark 
a useful property of orthogonal systems, namely that if we have a (necessarily unique) rule 
$\redone \qstepto{\qone} \conone$ and we reduce a term of the form 
$\redone[\vect{v}/\vect{x}]$, then either we reduce the (instance) of redex $\redone$, 
or the term obtained is itself an instance of the redex $\redone$, 
i.e. it is of the form $\redone[\vect{w}/\vect{x}]$, for some terms $\vect{w}$.

\begin{remark}
Given an orthogonal \cbetrs{,} 
suppose to have a reduction 
$\redone[\vect{v}/\vect{x}] \qmultireduce{\qone} \termone$ not an instance of weakening. 
Then, either:
\begin{enumerate}
    \item $\termone = \conone[\vect{w}/\vect{x}]$ with 
        $v_i \qmultireduce{\qtwo_i} w_i$, for any $i$;
        $\redone \qstepto{\qthree} \conone$; and
        $\qone = \qthree \tensor \bigotimes_i \degree{x_i}{\redone}(\qtwo_i)$. Or
    \item $\termone = \redone[\vect{w}/\vect{x}]$ with 
        $v_i \qmultireduce{\qtwo_i} w_i$, for any $i$, and
        $\qone = \bigotimes_i \degree{x_i}{\redone}(\qtwo_i)$.
\end{enumerate}
\end{remark}

\begin{proposition}
Let $\qtrsone = (\signature, \stepto_{\vrelone})$ be an orthogonal \cbetrs. 
Then, the relation $\multireduce_{\vrelone}$ has the diamond property. 
That is, if $\termtwo_1 \stackrel{\qone_1}{\circleleftarrow} \termone \qmultireduce{\qone_2} \termtwo_2$, 
there there exists a term $\termtwo$ such that 
$\termtwo_1 \qmultireduce{\qtwo_1} \termtwo \stackrel{\qtwo_2}{\circleleftarrow} \termtwo_2$ 
and $\qone_1 \tensor \qone_2 \leq \qtwo_1 \tensor \qtwo_2$.
\end{proposition}

\begin{proof}
The proof is by induction on $\termone$ with a case analysis on the defining clauses of 
$\multireduce$. 
The interesting case is for $\termone = \redone[\vect{v}/\vect{x}]$. By previous remark, 
there are two possibilities for the reduction $\redone[\vect{v}/\vect{x}] \qmultireduce{\qone_1} \termtwo_1$ 
(the case for weakening is straightforward).
\begin{enumerate}
    \item $\termtwo_1 = \conone[\vect{w}/\vect{x}]$ with 
        $v_i \qmultireduce{\qtwo_i} w_i$, 
        $\redone \qstepto{\qone} \conone$, and
        $\qone_1 = \qone \tensor \bigotimes_i \degree{x_i}{\redone}(\qtwo_i)$. 
        Since $\qtrsone$ is orthogonal, the rule $\redone \qstepto{\qone} \conone$ 
        is unique and $\termtwo_2$ must be of the form 
        $\redone[\vect{u}/\vect{x}]$ with $v_i \qmultireduce{\qthree_i} u_i$ and 
        $\qone_2 = \bigotimes_i \degree{x}{\redone}(\qthree_i)$ (otherwise, 
        we would have a critical pair). 
        That is, we have the peak:
         \[
        \xymatrix@=1.8pc{
        & \redone[\vect{v}/\vect{x}]\ar[ld]_{
        \qone \tensor \bigotimes_i \degree{x_i}{\redone}(\qtwo_i)}|\circ 
        \ar[rd]^{\bigotimes_i \degree{x_i}{\redone}(\qthree_i)}|\circ &
        \\
        \conone[\vect{w}/\vect{x}]  
        &  & \redone[\vect{u}/\vect{x}] 
        }
        \]
        By induction hypothesis on each $v_i$, we close the diagram
        \[
        \vcenter{
        \xymatrix@=0.9pc{
        & v_i\ar[ld]_{\qtwo_i}|\circ \ar[rd]^{\qthree_i}|\circ &
        \\
        w_i \ar[rd]_{\hat{\qthree}_i}|\circ &  & u_i \ar[ld]^{\hat{\qtwo}_i}|\circ
        \\
        & z_i &
        }
        }
        \qquad \quad \qtwo_i \tensor \qthree_i \leq \hat{\qthree}_i \tensor \hat{\qtwo}_i,
        \]
        so that we obtain
         \[
        \xymatrix@=1.8pc{
        & \redone[\vect{v}/\vect{x}]\ar[ld]_{
        \qone \tensor \bigotimes_i \degree{x_i}{\redone}(\qtwo_i)}|\circ 
        \ar[rd]^{\bigotimes_i \degree{x_i}{\redone}(\qthree_i)}|\circ &
        \\
        \conone[\vect{w}/\vect{x}] \ar[rd]_{
        \bigotimes_i \degree{x_i}{\conone}(\hat{\qthree}_i)}|\circ 
        &  & \redone[\vect{u}/\vect{x}] 
        \ar[ld]^{\qone \tensor \bigotimes_i \degree{x_i}{\redone}(\hat{\qtwo}_i)}|\circ
        \\
        & \conone[\vect{z}/\vect{x}] &
        }
        \]
        Since the system is balanced, $\qtwo_i \tensor \qthree_i \leq \hat{\qthree}_i \tensor \hat{\qtwo}_i$ 
        infer 
        $$
        \bigotimes_i \degree{x_i}{\redone}(\qtwo_i) \tensor 
        \bigotimes_i \degree{x_i}{\redone}(\qtwo_i)(\qthree_i) \leq 
        \bigotimes_i \degree{x_i}{\conone}(\hat{\qthree}_i) \tensor 
        \bigotimes_i \degree{x_i}{\conone}(\hat{\qtwo}_i)
        $$
        and thus
        $$
        \qone \tensor \bigotimes_i \degree{x_i}{\redone}(\qtwo_i) \tensor 
        \bigotimes_i \degree{x_i}{\redone}(\qtwo_i)(\qthree_i) \leq 
        \bigotimes_i \degree{x_i}{\conone}(\hat{\qthree}_i) \tensor 
        \qone \tensor
        \bigotimes_i \degree{x_i}{\conone}(\hat{\qtwo}_i).
        $$
        \item $\termtwo_1 = \redone[\vect{w}/\vect{x}]$ with 
        $v_i \qmultireduce{\qtwo_i} w_i$ and
        $\qone_1 = \bigotimes_i \degree{x_i}{\redone}(\qtwo_i)$. 
        Now,  if $\termone \qmultireduce{\qone_2} \termtwo_2$ is an instance 
        of a rule $\redone \qstepto{\qone} \conone$ (i.e. the `base' case in 
        the definition of $\multireduce$), then we proceed as in the previous case. 
        Otherwise, we must have 
        $\termtwo_2 = \redone[\vect{u}/\vect{x}]$ with $v_i \qmultireduce{\qthree_i} u_i$ and 
        $\qone_2 = \bigotimes_i \degree{x}{\redone}(\qthree_i)$. 
        That is, we have the peak:
         \[
        \xymatrix@=1.8pc{
        & \redone[\vect{v}/\vect{x}]\ar[ld]_{
         \bigotimes_i \degree{x_i}{\redone}(\qtwo_i)}|\circ 
        \ar[rd]^{\bigotimes_i \degree{x_i}{\redone}(\qthree_i)}|\circ &
        \\
        \redone[\vect{w}/\vect{x}]  
        &  & \redone[\vect{u}/\vect{x}] 
        }
        \]
        We proceed applying the induction hypothesis as in previous point and close 
        the diagram as follows, relying on the substitution lemma:
          
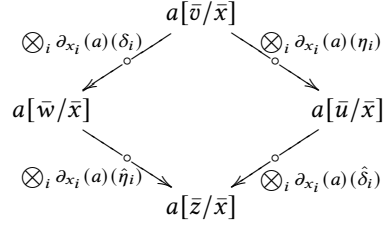
\[
        \xymatrix@=1.8pc{
        & \redone[\vect{v}/\vect{x}]\ar[ld]_{
         \bigotimes_i \degree{x_i}{\redone}(\qtwo_i)}|\circ 
        \ar[rd]^{\bigotimes_i \degree{x_i}{\redone}(\qthree_i)}|\circ &
        \\
        \redone[\vect{w}/\vect{x}] \ar[rd]_{
        \bigotimes_i \degree{x_i}{\redone}(\hat{\qthree}_i)}|\circ 
        &  & \redone[\vect{u}/\vect{x}] 
        \ar[ld]^{\bigotimes_i \degree{x_i}{\redone}(\hat{\qtwo}_i)}|\circ
        \\
        & \redone[\vect{z}/\vect{x}] &
        }
        \]
        Indeed, $\qtwo_i \tensor \qthree_i \leq \hat{\qthree}_i \tensor \hat{\qtwo}_i$ implies 
        $$
        \bigotimes_i \degree{x_i}{\redone}(\qtwo_i) \tensor \bigotimes_i \degree{x_i}{\redone}(\qthree_i) 
        \leq 
        \bigotimes_i \degree{x_i}{\redone}(\hat{\qthree}_i) \tensor \bigotimes_i \degree{x_i}{\redone}(\hat{\qtwo}_i).
        $$
\end{enumerate}
\end{proof}

\begin{corollary}
\label{corollary:multistep-is-diamond}
Let $\qtrsone = (\signature, \stepto_{\vrelone})$ be an orthogonal \cbetrs. 
Then $\mathring{\vrelone}$ has the diamond property.
\end{corollary}

We can finally prove confluence of orthogonal systems. 

\begin{theorem}
\label{thm:orthogonality-implies-confluence}
Let $\qtrsone = (\signature, \stepto_{\vrelone})$ be an orthogonal \cbetrs. 
Then, $\vrelone$ is confluent. 
\end{theorem}

\begin{proof}%[Proof of \autoref{thm:orthogonality-implies-confluence}]
Let $\vreltwo \defeq \dual{\vrelone}$. We prove 
$S^*;R^* \leq R^*;S^*$. By adjunction, it is sufficient to prove 
$S^* \leq (R^*;S^*) / R^*$. We proceed by fixed point induction, shwoing 
$\idvrel \leq (R^*;S^*) / R^*$ and 
$S; (R^*;S^*) / R^* \leq (R^*;S^*) / R^*$. The former is straightforward, 
whereas for the latter it is sufficient to prove 
$S; (R^*;S^*) / R^*;R^* \leq R^*;S^*$, i.e. 
$S;R^*;S^* \leq R^*;S^*$. Since $S \leq \mathring{S}$, it is enough to 
show 
$\mathring{S};R^*;S^* \leq R^*;S^*$
and thus
$R^* \leq \mathring{S} \setminus (R^*;S^*)/ S^*$. We do a second fixed point induction, 
hence proving $\idvrel \leq \mathring{S} \setminus (R^*;S^*)/ S^*$ and 
$R; \mathring{S} \setminus (R^*;S^*)/ S^* \leq \mathring{S} \setminus (R^*;S^*)/ S^*$. 
The former obviously holds since $\mathring{S} \leq S^*$, 
wheres for the latter we first use adjunction and reduce the proof obligation to 
$$
\mathring{S}; R; \mathring{S} \setminus (R^*;S^*)/ S^*; S^* \leq R^*;S^*,
$$
i.e. 
$\mathring{S}; R; \mathring{S} \setminus (R^*;S^*) \leq R^*;S^*$. 
Since $R \leq \mathring{R}$, it is sufficient to prove 
$\mathring{S}; \mathring{R}; \mathring{S} \setminus (R^*;S^*) \leq R^*;S^*$.  
Now, by \autoref{corollary:multistep-is-diamond}, we have 
$\mathring{S}; \mathring{R} \leq \mathring{R};\mathring{S}$, and thus:
\begin{align*}
\mathring{S}; \mathring{R}; \mathring{S} \setminus (R^*;S^*) 
\leq \mathring{R};\mathring{S}; \mathring{S} \setminus (R^*;S^*)
\leq \mathring{R}; R^*;S^*
\leq R^*;R^*;S^*
\leq R^*;S^*.
\end{align*}
\end{proof}

\begin{remark}
By replacing $\vrelone$ with $\reduce_{\vrelone}$ (and thus replacing 
algebraic operations on $\Quantale$-relations with their ternary relation 
counterparts) in the proof of 
\autoref{thm:orthogonality-implies-confluence}, we obtain confluence of 
$\reduce_{\vrelone}$.
\end{remark}

We conclude this section by observing that system $\BCKbounded$ is 
orthogonal, and thus by \autoref{thm:orthogonality-implies-confluence} 
it is confluent. 

\begin{theorem}
System $\BCKbounded$ of graded combinatory logic is confluent. 
\end{theorem}

To the best of the authors' knowledge, this is the first confluence 
result for a system of graded combinators endowed with a quantitative and 
modal operational (reduction) semantics. Such a result 
can seen as a first step towards a foundational study of operational 
properties of graded and coeffectful calculi.

% \section{RELATED WORK} 
%\section{Related Work} 

\section{CONCLUSION, RELATED, AND FUTURE WORK}
\label{sect:conclusion}

In this paper, we have started the development a systematic theory of 
metric and quantitative rewriting systems. 
The abstract nature of the notion of a distance employed makes our framework robust 
and allows for several conceptual interpretations of our rewriting systems. 
The latter, in fact, can be thought not only as metric and quantitative systems, 
but also as substructural (e.g. fuzzy or monoidal) and modal or coeffectful systems, this way 
suggesting possible applications of our theory to the development of quantitative and modal 
operational semantics of coffectful programming languages. We have shortly hinted at that 
at the end of 
previous section (the authors are currently working on applications of \quant{} rewriting to 
study operational properties of foundational graded $\lambda$-calculus). 

We have focused on 
fundamental definitions and confluence properties of abstract systems, 
as well as linear and graded term rewriting systems. 
Developing a general theory of \quant{} rewriting systems is an ambitious project 
that cannot be exhausted in a single paper. Among the many possible extensions of the theory 
presented in this paper, 
we mention the development of a theory of reduction strategies and their application to 
metric word problems; the design of completion algorithms for \quant{} term rewriting systems 
(both linear and graded), as well as their \quant{} correctness with respect to 
families of metric word problems; and the study of inductive and termination properties of 
quantitative systems. The latter, in particular, seem to suggest that new rewriting properties 
can be discovered by pushing the \quant{} enrichment one step forward, this way making 
the notions of termination, induction, confluence, etc \quant{} themselves.
Finally, we plan to investigate applications of \quant{} systems in the spirit of 
those outlined in \autoref{section:long-intro}. 

\paragraph{Related Work}
To the best of the authors' knowledge, this is the first systematic analysis of 
\quant{} and metric rewriting systems. This, of course, does not mean that 
isolated forms of \quant{} rewriting have not been proposed in the literature. 
For instance, specific forms of weighted reductions have been employed 
\cite{cost-analysis-term-rewriting-1,cost-analysis-term-rewriting-2} 
in the study of cost analysis of rewriting systems. Measured abstract rewriting systems, i.e. 
\emph{abstract} rewriting systems with a reduction relation 
enriched in a monoid,
have been introduced by
\citet{van-oostrom-2016} to study normalisation properties by random descent.  
In that context, a quantitative notion of confluence is introduced which, however, 
differs from ours in the way it compares distances between objects. In fact, given a peak 
$b_1 \stackrel{\qone_1}{\leftarrow} a \qreduce{\qone_2} b_2$ and a valley 
$b_1 \qreduce{\qtwo_1}  b \stackrel{\qtwo_1}{\leftarrow} b_2$, it is required 
$\qone_1 \tensor \qtwo_1 \leq \qone_2 \tensor \qtwo_2$. Even if this requirement has a natural 
reading when it comes to study normalisation properties of rewriting, it does not fit 
the algebra of \quant{} relations and 
seems ineffective 
when applied to the study of distances.
We also remark that measured rewriting systems have been studied in the context of 
abstract rewriting only, whereas our theory of \quant{} rewriting systems covers both 
abstract and (graded) term-based systems.

At the time of writing, the authors have discovered that abstract \emph{fuzzy}
rewriting systems have been studied by \citet{fuzzy-rewriting-1,fuzzy-rewriting-2} 
relying on the theory of fuzzy relations \cite{Fuzzy-relational-systems}. 
Even if the aforementioned theory of fuzzy rewriting systems does \emph{not} cover 
term-based systems (neither non-expansive nor graded), the development of fuzzy
abstract rewriting systems is in line with our \autoref{section:qars}. In particular, 
\citet{fuzzy-rewriting-1,fuzzy-rewriting-2} define fuzzy notions of confluence and 
prove a \quant{} Newman's lemma similar to (the pointwise version of) ours.  In fact, instantiating the theory of 
\autoref{section:qars} to Fuzzy quantales, we obtain an extension\footnote{\autoref{section:qars} 
contains results on abstract $\Quantale$-systems, such as the \quant{} Hindley-Rosen lemma, that 
are not given for Fuzzy systems.} of the theory of 
abstract Fuzzy rewriting systems. Remarkably, our pointwise analysis of 
\quant{} Newman's Lemma is close to the one by \citet{fuzzy-rewriting-1}.
Besides the absence of a theory term-based systems, 
major differences between our work and the one on Fuzzy rewriting can be found 
even at the level of abstract systems. First, as already remarked, our approach is more general 
and subsumes (and extends) Fuzzy rewriting. Moreover, our 
theory of abstract $\Quantale$-systems is largely pointfree and builds upon 
general relational techniques nontrivially extending the relational theory of abstract rewriting 
by \citet{backshouse-calculational-approach-to-mathematical-induction}, as well as 
 other pointfree theories of rewriting systems \cite{Struth-abstract-abstract-reduction,Struth-algebraic-notions-of-termination}. 
 In addition to all of that, we mention that, curiously, Fuzzy rewriting systems 
 have not been applied to metric reasoning. This is an interesting observation, 
 as it turns out that general \quant{} Fuzzy equational and algebraic theories \cite{fuzzy-equational-logic} 
 have been developed before the \quant{} algebraic theories by \citet{plotkin-quantitative-algebras-2016}. 
 However, even if mathematically sophisticated, such fuzzy theories have not been applied 
 (to the best of 
 the authors' knowledge)
 neither to metric reasoning nor to the semantics of 
 programming languages.

Contrary to the case of rewriting systems, general theories of quantitative equational reasoning 
have been developed. In addition to the aforementioned Fuzzy equational theories, 
we mention the rich research line on quantitative algebras and equational theories~\cite{plotkin-quantitative-algebras-2016,plotkin-quantitative-algebras-2017,plotkin-quantitative-algebras-2018,plotkin-quantitative-algebras-2018-bis,plotkin-quantitative-algebras-2021,plotkin-quantitative-algebras-2021-bis,DBLP:conf/lics/MioSV21}. With the exception of the recent work by 
\citet{dagnino-1}, such theories are usually not graded and, to the best of the authors' 
knowledge, are not capable to describe non-linear systems, such as system $\BCKbounded$ 
of graded combinators. From that point of view, our definition of a graded \quant{} equational theory 
can be seen as a first extension of \quant{} equational theories to graded systems. 

% In general, the authors would like to remark that, besides the aforementioned work on quantitative 
% equational theories 
% which do \emph{not} deal with the computational content of quantitative equality, 
% symbolic approaches to distances are not much considered leaving place to numerical 
% definitions and analytical methods only. This is quite unfortunate, as 
% several numerically-defined distances carry 
% simple symbolic (and oftentimes inductive) characterisations with a rich 
% computational content. 

\section*{Acknowledgements}
The authors would like to thank Melissa Antonelli, Francesco Dagnino, Ugo Dal Lago, 
and Claudia Faggian for their helpful suggestions and stimulating conversations 
on the subject.

\bibliography{main}

%% Appendix

\appendix

\section{Critical Pairs of $\BA$}
 
%  \begin{figure}
%     {
%     \centering
%     \begin{tcolorbox}[boxrule=0.5pt,width=\linewidth,colframe=black,colback=black!0!white,arc=0mm]

  \begin{minipage}{\linewidth}
      \begin{minipage}{0.40\linewidth}
          \begin{figure}[H]
          \captionsetup{labelformat=empty}
          \caption{}
          \centering
\begin{tikzpicture}[thick,scale=0.85, every node/.style={transform shape, sibling distance=6cm},level 1/.style={sibling distance=30mm},level 2/.style={transform shape,sibling distance=30mm}, 
]
    \node (1) {$x\barplus{1}y$}
    child { node (11) {$y\barplus{0} x$} edge from parent [->] node [left] {\tiny 0} }
    child { node (2) {$x$} 
        child { node  (21){$y\barplus{0} x$}  }
        edge from parent [->] node [right] {\tiny 0}
        %(2) edge [->]  node  (21) [left] {\tiny ($\sigma: x\mapsto y\barplus{0} x $)} (21)
        (2) edge [->]  node [right] {\tiny $0$} (21)
        (2) edge [->]  node [left] {\tiny ($\sigma: x\mapsto y\barplus{0} x $)} (21)
    }
    ;
\end{tikzpicture}

   \end{figure}
  \end{minipage}
      \hspace{0.05\linewidth}
      \begin{minipage}{0.40\linewidth} 
          \begin{figure}[H]
          \captionsetup{labelformat=empty}
          \caption{}
          \centering
\begin{tikzpicture}[thick,scale=0.85, every node/.style={transform shape, sibling distance=6cm},level 1/.style={sibling distance=30mm},level 2/.style={transform shape,sibling distance=30mm}, 
]
    \node (0) {$x\barplus{1}y$}
     child { node (1) {$x$} 
        child { node  (11){$y\barplus{1} x$}  }
        edge from parent [->] node [right] {\tiny $a$}
        %(2) edge [->]  node  (21) [left] {\tiny ($\sigma: x\mapsto y\barplus{0} x $)} (21)
          (1) edge [->]  node [left] {\tiny ($\sigma: x\mapsto y\barplus{1}x  $)} (11)
         (1) edge [-]  node [right] {\tiny $0$} (11)}
    child { node (2) {$z+_1 y$} 
        child { node  (21){$y\barplus{1} x$}  }
        edge from parent [->] node [right] {\tiny $a$}
        %(2) edge [->]  node  (21) [left] {\tiny ($\sigma: x\mapsto y\barplus{0} x $)} (21)
        (2) edge [->]  node [right] {\tiny $a$} (21)
    }
    ;
\end{tikzpicture}
          \end{figure}
      \end{minipage}
  \end{minipage}
\\

\begin{figure}[H]
      \captionsetup{labelformat=empty}
          \caption{}
\centering
    
\begin{tikzpicture}[thick,scale=0.80, every node/.style={transform shape, sibling distance=8cm},level 1/.style={sibling distance=45mm},level 2/.style={transform shape,sibling distance=45mm}
]
\node (right) at (4,0) {$(x\barplus{\probone_{1}}y)\barplus{\probone_{2}}z$}  
   child {node (1)  {$x\barplus{\probone_{1}\probone_{2}} (y\barplus{\frac{\probone_{2}-\probone_{1}\probone_{2}}{1-\probone_{1}\probone_{2}}} z)$}
          edge from parent [->] node [left] {\tiny 0} 
 }
    child { node (2) {$(u\barplus{\probone_{1}} y)\barplus{\probone_{2}})$} 
        child { node  (21){$(x\barplus{\probone_{1}} y)\barplus{\probone_{2}} z$}  }
         {child{node (22) {$x\barplus{\probone_{1}\probone_{2}} (y\barplus{\frac{\probone_{2}-\probone_{1}\probone_{2}}{1-\probone_{1}\probone_{2}}} z$} }}
        edge from parent [->] node [right] {\tiny $a$}
        %(2) edge [->]  node  (21) [left] {\tiny ($\sigma: x\mapsto y\barplus{0} x $)} (21)
        (2) edge [->]  node [right] {\tiny $a$} (21)
     (21) edge [->]  node [right] {\tiny $0$} (22)
    } ;
;
\node (left) at (-4,0)   {$(x\barplus{\probone_{1}}y)\barplus{\probone_{2}}z$}
    child {node (19)  {$x\barplus{\probone_{1}\probone_{2}} (y\barplus{\frac{\probone_{2}-\probone_{1}\probone_{2}}{1-\probone_{1}\probone_{2}}} z)$}
     child { node  (119){$(x\barplus{\probone_{1}} y)\barplus{\probone_{2}} z)$}  }
          edge from parent [->] node [left] {\tiny 0} 
           (19) edge [->>]  node [right] {\tiny $0$} (119)
 }
    child { node (29) {$(u\barplus{\probone_{1}} y)\barplus{\probone_{2}} z)$} 
        child { node  (219){$(x\barplus{\probone_{1}} y)\barplus{\probone_{2}} z)$}  }
          %{child{node (229) {$\;$} }}
        (left) edge [->]  node [right] {\tiny $a$} (29)
       % edge from parent [->] node [right] {\tiny $a$}
        %(2) edge [->]  node  (21) [left] {\tiny ($\sigma: x\mapsto y\barplus{0} x $)} (21)
        (2) edge [->]  node [right] {\tiny $a$} (21)
        %(21) edge [ draw=.]  node (22)
    }
    ;
\end{tikzpicture}

\begin{tikzpicture}[thick,scale=0.80, every node/.style={transform shape, sibling distance=8cm},level 1/.style={sibling distance=45mm},level 2/.style={transform shape,sibling distance=45mm}
]
\node (right) at (4,0) {$(x\barplus{\probone_{1}}y)\barplus{\probone_{2}}z$}
   child {node (1)  {$x\barplus{\probone_{1}\probone_{2}} (y\barplus{\frac{\probone_{2}-\probone_{1}\probone_{2}}{1-\probone_{1}\probone_{2}}} z)$}
     child { node  (11){$(x\barplus{\probone_{1}} y)\barplus{\probone_{2}} z)$}  }
          edge from parent [->] node [left] {\tiny 0} 
          (1) edge [->>]  node [right] {\tiny $0$} (11)
          }
 child { node (2) {$(u\barplus{\probone_{2}} z)$} 
        child { node  (21){$(x\barplus{\probone_{1}} y)\barplus{\probone_{2}} z$}  }
        edge from parent [->] node [right] {\tiny $a$}
        %(2) edge [->]  node  (21) [left] {\tiny ($\sigma: x\mapsto y\barplus{0} x $)} (21)
        (2) edge [->]  node [right] {\tiny $0$} (21) 
        };
;
\node (left) at (-4,0)   {$(x\barplus{\probone_{1}}y)\barplus{\probone_{2}}z$}
  child {node (19)  {$x\barplus{\probone_{1}\probone_{2}} (y\barplus{\frac{\probone_{2}-\probone_{1}\probone_{2}}{1-\probone_{1}\probone_{2}}} z)$}
          edge from parent [->] node [left] {\tiny 0} 
 }
    child { node (29) {$u\barplus{\probone_{2}} z$} 
        child { node  (219){$(x\barplus{\probone_{1}} y)\barplus{\probone_{2}} z$}  }
         {child {node (229)  {$x\barplus{\probone_{1}\probone_{2}} (y\barplus{\frac{\probone_{2}-\probone_{1}\probone_{2}}{1-\probone_{1}\probone_{2}}} z)$}}}
        edge from parent [->] node [right] {\tiny $a$}
        %(2) edge [->]  node  (21) [left] {\tiny ($\sigma: x\mapsto y\barplus{0} x $)} (21)
        (29) edge [->]  node [right] {\tiny $0$} (219)
         (219) edge [->]  node [right] {\tiny $0$} (229)
    }
    ;
\end{tikzpicture}
\end{figure}
\begin{figure}
\centering
        \captionsetup{labelformat=empty}
          \caption{}
\begin{tikzpicture}[thick,scale=0.80, every node/.style={transform shape, sibling distance=8cm},level 1/.style={sibling distance=45mm},level 2/.style={transform shape,sibling distance=45mm}
]
\node (right) at (4,0) {$((x\barplus{\probone_{1}}y)\barplus{\probone_{2}}z)\barplus{\probone_{3}}u$}
child {node (1)  {$(x\barplus{\probone_{1}\probone_{2}} (y\barplus{\frac{\probone_{2}-\probone_{1}\probone_{2}}{1-\probone_{1}\probone_{2}}} z))\barplus{\probone_3} u$}        
       child { node  (11){$((x\barplus{\probone_{1}}y)\barplus{\probone_{2}}z)\barplus{\probone_{3}}u$}  
        {child { node  (12){$(x\barplus{\probone_{1}}y)\barplus{\probone_{2}\probone_{3}}(z\barplus{{\frac{\probone_{3}-\probone_{2}\probone_{3}}{1-\probone_{3}\probone_{2}}}}u)$}
         edge from parent [->] node [right] {\tiny $0$}
        (1) edge [->>]  node [right] {\tiny $0$} (11)
        (right) edge [->]  node [right] {\tiny $0$} (1)}}}}
    child { node (2) {$(x\barplus{\probone_{1}}y)\barplus{\probone_{2}\probone_{3}}(z\barplus{{\frac{\probone_{3}-\probone_{2}\probone_{3}}{1-\probone_{3}\probone_{2}}}}u)$} 
        edge from parent [->] node [right] {\tiny $0$}
        %(2) edge [->]  node  (21) [left] {\tiny ($\sigma: x\mapsto y\barplus{0} x $)} (21)
    }
    ;
;
\node (left) at (-7,0)   {$((x\barplus{\probone_{1}}y)\barplus{\probone_{2}}z)\barplus{\probone_{3}}u$}
    child {node (1)  {$(x\barplus{\probone_{1}\probone_{2}} (y\barplus{\frac{\probone_{2}-\probone_{1}\probone_{2}}{1-\probone_{1}\probone_{2}}} z))\barplus{\probone_3} u$}        
         edge from parent [->] node [right] {\tiny $0$}
}
    child { node (29) {$(x\barplus{\probone_{1}}y)\barplus{\probone_{2}\probone_{3}}(z\barplus{{\frac{\probone_{3}-\probone_{2}\probone_{3}}{1-\probone_{3}\probone_{2}}}}u)$} 
        child { node  (219){$((x\barplus{\probone_{1}}y)\barplus{\probone_{2}}z)\barplus{\probone_{3}}u$}  
        {child { node  (22){$(x\barplus{\probone_{1}\probone_{2}} (y\barplus{\frac{\probone_{2}-\probone_{1}\probone_{2}}{1-\probone_{1}\probone_{2}}} z))\barplus{\probone_3} u$}
        % edge from parent [->] node [right] {\tiny $0$}
        (29) edge [->>]  node [right] {\tiny $0$} (219)
        (219) edge [->]  node [right] {\tiny $0$} (22)
        (left) edge [->]  node [right] {\tiny $0$} (29)}}}}
    ;
\end{tikzpicture}
\end{figure}

\begin{figure}[H]
\centering
          \captionsetup{labelformat=empty}
          \caption{}
\begin{tikzpicture}[thick,scale=0.80, every node/.style={transform shape, sibling distance=8cm},level 1/.style={sibling distance=45mm},level 2/.style={transform shape,sibling distance=45mm}
]
\node (right) at (5,0) {$(x\barplus{\probone_{1}}y)\barplus{\probone_{2}}z$}
    child {node (1)  {$x\barplus{\probone_{1}\probone_{2}} (y\barplus{\frac{\probone_{2}-\probone_{1}\probone_{2}}{1-\probone_{1}\probone_{2}}} z)$}
      child { node  (11){$(y\barplus{\frac{\probone_{2}-\probone_{1}\probone_{2}}{1-\probone_{1}\probone_{2}}} z)\barplus{1-\probone_{1}\probone_{2}} x$}  }
      { child { node  (12){$y\barplus{\probone_{2}-\probone_{1}\probone_{2}} (z\barplus{\frac{1-\probone_{2}}{1-\probone_{2}+\probone_{1}\probone_{2}}} x)$}  }}
      { child { node  (13){$(z\barplus{\frac{1-\probone_{2}}{1-\probone_{2}+\probone_{1}\probone_{2}}} x)\barplus{1-\probone_{2}-\probone_{1}\probone_{2}} y$}  }}
       { child { node  (14){$z\barplus{1-\probone_{2}} (x\barplus{\probone_{1}} y)$}  }}
        { child { node  (15){$z\barplus{1-\probone_{2}} (y\barplus{1-\probone_{1}} x)$}  }}
          edge from parent [->] node [left] {\tiny 0} 
         (1) edge [->]  node [right] {\tiny $0$} (11)
          (11) edge [->]  node [right] {\tiny $0$} (12)
           (12) edge [->]  node [right] {\tiny $0$} (13)
            (13) edge [->]  node [right] {\tiny $0$} (14)
             (14) edge [->]  node [right] {\tiny $0$} (15)
 }
    child { node (2) {$(y\barplus{1-\probone_{1}} x)\barplus{\probone_{2}} z$} 
        child { node  (21){$z\barplus{1-\probone_{2}} (y\barplus{1-\probone_{1}} x)$}  }
        edge from parent [->] node [right] {\tiny $a$}
        %(2) edge [->]  node  (21) [left] {\tiny ($\sigma: x\mapsto y\barplus{0} x $)} (21)
        (2) edge [->]  node [right] {\tiny $0$} (21)
    }
    ;
;
\node (left) at (-4,0)    {$(x\barplus{\probone_{1}}y)\barplus{\probone_{2}}z$}
    child {node (19)  {$x\barplus{\probone_{1}\probone_{2}} (y\barplus{\frac{\probone_{2}-\probone_{1}\probone_{2}}{1-\probone_{1}\probone_{2}}} z)$}
 }
    child { node (29) {$(y\barplus{1-\probone_{1}} x)\barplus{\probone_{2}} z$} 
        child { node  (219){$(x\barplus{\probone_{1}}y)\barplus{\probone_{2}} z$}  }
          {child { node  (229){$x\barplus{\probone_{1}\probone_{2}} (y\barplus{\frac{\probone_{2}-\probone_{1}\probone_{2}}{1-\probone_{1}\probone_{2}}} z)$}  }}
        edge from parent [->] node [right] {\tiny $a$}
        %(2) edge [->]  node  (21) [left] {\tiny ($\sigma: x\mapsto y\barplus{0} x $)} (21)
         edge from parent [->] node [right] {\tiny $0$}
        (29) edge [->]  node [right] {\tiny $0$} (219)
        (219) edge [->]  node [right] {\tiny $0$} (229)
    }
    ;
\end{tikzpicture}
\end{figure}

\end{document}

%% file: macros.tex
\usepackage[libertine]{newtxmath}

\RequirePackage[T1]{fontenc} 
\RequirePackage[varqu]{zi4} 
\RequirePackage[libertine]{newtxmath}

\DeclareMathAlphabet{\mathsfit}{T1}{\sfdefault}{m}{sl}
\SetMathAlphabet{\mathsfit}{bold}{T1}{\sfdefault}{bx}{sl}

\DeclareFontEncoding{LS1}{}{}
\DeclareFontSubstitution{LS1}{stix}{m}{n}

\usepackage[T1]{fontenc}
\usepackage{amsthm}
\usepackage{sansmath}
\usepackage{mathtools}
\usepackage{mathrsfs}
\usepackage{stackengine}
\usepackage[title]{appendix}

\usepackage[bbgreekl]{mathbbol}
\usepackage{multicol}
\usepackage{proof}
\usepackage[all]{xy}
\usepackage{xcolor}
\usepackage{epigraph}

\RequirePackage{geometry}
\usepackage{bm}
\DeclareMathAlphabet{\mathpzc}{OT1}{pzc}{m}{it}
\usepackage{url}
\usepackage{stmaryrd}
\usepackage{thm-restate}
\usepackage{float}
\usepackage{caption}
\usepackage{graphicx}
\usepackage{tikz}
\usepackage{tcolorbox}

\usepackage[square,numbers]{natbib}
\makeatletter
\newcommand{\circlearrow}{}% just in case
\DeclareRobustCommand{\circlearrow}{%
  \mathrel{\vphantom{\rightarrow}\mathpalette\circle@arrow\relax}%
}
\newcommand{\circle@arrow}[2]{%
  \m@th
  \ooalign{%
    \hidewidth$#1\circ\mkern1mu$\hidewidth\cr
    $#1\longrightarrow$\cr}%
}
\makeatother

\makeatletter
\DeclareRobustCommand{\circleleftarrow}{%
  \mathrel{\vphantom{\leftarrow}\mathpalette\circleleft@arrow\relax}%
}
\newcommand{\circleleft@arrow}[2]{%
  \m@th
  \ooalign{%
    \hidewidth$#1\circ\mkern1mu$\hidewidth\cr
    $#1\longleftarrow$\cr}%
}
\makeatother

\usepackage{hyperref}

\makeatletter

 \def\Autoref#1{%
   \begingroup
   \edef\reserved@a{\cpttrimspaces{#1}}%
   \ifcsndefTF{r@#1}{%
     \xaftercsname{\expandafter\testreftype\@fourthoffive}
       {r@\reserved@a}.\\{#1}%
   }{%
     \ref{#1}%
   }%
   \endgroup
 }
 \def\testreftype#1.#2\\#3{%
   \ifcsndefTF{#1autorefname}{%
     \def\reserved@a##1##2\@nil{%
       \uppercase{\def\ref@name{##1}}%
       \csn@edef{#1autorefname}{\ref@name##2}%
       \autoref{#3}%
     }%
     \reserved@a#1\@nil
   }{%
     \autoref{#3}%
   }%
 }
\makeatother

\definecolor[named]{ACMBlue}{cmyk}{1,0.1,0,0.1}
\definecolor[named]{ACMYellow}{cmyk}{0,0.16,1,0}
\definecolor[named]{ACMOrange}{cmyk}{0,0.42,1,0.01}
\definecolor[named]{ACMRed}{RGB}{235,83,60}
\definecolor[named]{ACMLightBlue}{cmyk}{0.49,0.01,0,0}
\definecolor[named]{ACMGreen}{cmyk}{0.20,0,1,0.19}
\definecolor[named]{ACMPurple}{RGB}{134,72,143}
\definecolor[named]{ACMDarkBlue}{RGB}{35,85,147}

\usepackage{titlesec}

\titleformat*{\section}{
  \scshape
  \sffamily
  \bfseries
  \color{black}
} 

\titleformat*{\subsection}{
  \sffamily
  \scshape
  \bfseries
  \color{black}
}

\titleformat*{\subsubsection}{
   \sffamily
  \normalsize
  \bfseries
  \color{black}
}

\titleformat*{\paragraph}{
  \scshape
  \sffamily
  \color{black}
}

\newcommand{\mathbbm}[1]{\text{\usefont{U}{bbm}{m}{n}#1}} % from mathbbm.sty

\newcounter{numberone}

\newtheorem{theorem}{Theorem}
\newtheorem{proposition}{Proposition}
\newtheorem{lemma}{Lemma}
\newtheorem{corollary}{Corollary}

\theoremstyle{definition}
\newtheorem{definition}{Definition}

\newtheorem{examplex}{Example}
\newenvironment{example}
  {\pushQED{\qed}\examplex}
  {\popQED\endexamplex}
\newtheorem{notation}{Notation}
\newtheorem{remark}{Remark}

\newcommand{\varone} {x}

\newcommand{\termone}{e}
\newcommand{\termtwo}{f}
\newcommand{\termthree}{g}
\newcommand{\termfour}{v}

\newcommand{\imp}{\vdash}

\newcommand{\vect}[1]{\bar{#1}}

\newcommand{\ctxone}{\mathcal{C}}

\newcommand{\signature}{\Sigma}

\newcommand{\op}{f}

\newcommand{\cc}{\cdots}
\newcommand{\hh}{\hdots}

\newcommand{\powerset}{\mathcal{P}}

\newcommand{\relone}{R}
\newcommand{\reltwo}{S}
\newcommand{\vrelone}{R}
\newcommand{\vreltwo}{S}
\newcommand{\vrelthree}{P}

\newcommand{\vdistone}{E}

\newcommand{\quantale}{\Omega}%{\mathsfit{V}}
\newcommand{\tensor}{\otimes}
\newcommand{\qunit}{\mathsfit{k}}
\newcommand{\qunittwo}{\mathsfit{j}}

\newcommand{\join}{\bigvee}
\newcommand{\meet}{\bigwedge}
\newcommand{\qvalone}{\qone}
\newcommand{\qvaltwo}{\qtwo}

\newcommand{\mmap}{\multimap}

\def\tobar{\mathrel{\mkern3mu  \vcenter{\hbox{$\scriptscriptstyle+$}}%
            \mkern-12mu{\to}}}
\newcommand{\torel}{\tobar}
\newcommand{\Vrel}[1]{#1\text{-}\mathit{Rel}}
\newcommand{\vrel}{\Vrel{\Quantale}}

\renewcommand{\powerset}{\mathcal{P}}

\newcommand{\idrel}{\Delta}
\newcommand{\idvrel}{\Delta}

\newcommand{\baseone}{\phi}
\newcommand{\basetwo}{\psi}

\newcommand{\baseid}{\bm{1}}
\newcommand{\bangg}[1]{{!}_{#1}}
\newcommand{\bang}[1]{[#1]}

\newcommand{\dual}[1]{#1^{\scriptstyle -}}

\newenvironment{claim}{\par\noindent\emph{Claim.}}{}
\newenvironment{proofoftheclaim}[1]{\par\noindent\emph{Proof of the claim.}\space#1}{\hfill $\diamond$}

\newcommand{\qtop}{\rotatebox[origin=c]{180}{$\Bot$}}
\newcommand{\qbot}{\bot}

\newcommand{\Two}{\mathbbm{2}}

\newcommand{\two}{\mathbb{2}}
\newcommand{\lawvere}{\mathbb{L}}
\newcommand{\Lawvere}{\mathbb{L}}
\newcommand{\StrongLawvere}{\lawvere^{\max}} %{\bm{\mathcal{S}}\bm{\mathcal{L}}}
\newcommand{\Quantale}{\mathbb{\quantale}}
\newcommand{\UnitInterval}{\mathbb{I}}

\newcommand{\posone}{p}
\newcommand{\postwo}{q}
\newcommand{\posthree}{r}
\newcommand{\posroot}{\lambda}
\newcommand{\subtermpos}[2]{{#1}_{\mid #2}}

\newcommand{\true}{\top}
\newcommand{\false}{\bot}
\newcommand{\defeq}{\triangleq}

\newcommand{\quant}{quantitative}
\newcommand{\qtrs}{$\Quantale$-TRS}
\newcommand{\qars}{$\Quantale$-ARS}
\newcommand{\cbetrs}{$(\Quantale,\CBE)$-TRS}

\newcommand{\subst}[2]{#1^{#2}}
\newcommand{\sone}{\sigma}
\newcommand{\stwo}{\tau}
\newcommand{\sthree}{\rho}

\newcommand{\qone}{\varepsilon}
\newcommand{\qtwo}{\delta}
\newcommand{\qthree}{\eta}
\newcommand{\qfour}{\iota}

\newcommand{\vbaserelone}{\vrelone}

\newcommand{\reduce}{\to}
\newcommand{\qreduce}[1]{\stackrel{#1}{\reduce}}
\newcommand{\reduceleft}{\leftarrow}
\newcommand{\qreduceleft}[1]{\stackrel{#1}{\reduceleft}}

\newcommand{\multireduce}{\circlearrow}
\newcommand{\qmultireduce}[1]{\stackrel{#1}{\circlearrow}}

\newcommand{\vbaseeq}{E}
\newcommand{\equal}{=}
\newcommand{\qequal}[1]{\stackrel{#1}{=}}

\newcommand{\veqone}{E}

\newcommand{\terms}[2]{#1(#2)}
\newcommand{\variables}{X}

\newcommand{\barplus}[1]{+_{#1}}
\newcommand{\probone}{\epsilon}

\newcommand{\stepto}{\mapsto}
\newcommand{\qstepto}[1]{\stackrel{#1}{\mapsto}}

\newcommand{\writeop}[2]{#1.#2}

\definecolor{cmykorange}{cmyk}{0,0.5,1,0}
\definecolor{cmykred}{cmyk}{0,1,1,0}
\definecolor{cmykyellow}{cmyk}{0,0,1,0}
\definecolor{cmykblue}{cmyk}{1,1,0,0}
\definecolor{cmykgreen}{cmyk}{1,0,1,0}
\definecolor{cmykviolet}{cmyk}{0.6,0.9,0,0.2}
\definecolor{cmykblack}{cmyk}{0,0,0,1}

\newcommand{\makedistance}[1]{#1^{\equiv}}
\newcommand{\reflex}[1]{#1^{=}}

\newcommand{\mrel}[4]{#1 \Vdash #2 \mathrel{#3} #4}
\newcommand{\mrelto}[3]{#1 \Vdash #2 \mathrel{\reduce} #3}
\newcommand{\mrelstepto}[3]{#1 \Vdash #2 \mathrel{\stepto} #3}

\newcommand{\A}{\texttt{A}}
\newcommand{\C}{\texttt{C}}
\newcommand{\G}{\texttt{G}}
\newcommand{\T}{\texttt{T}}
\newcommand{\moleculeone}{b}
\newcommand{\moleculetwo}{c}
\newcommand{\dnarelone}{M}
\newcommand{\dnareltwo}{}
\newcommand{\emptystring}{\lambda}
\newcommand{\DNAone}{\mathcal{M}}

\newcommand{\bck}{\mathcal{K}}
\newcommand{\BCK}{\bck}

\newcommand{\Icomb}{\texttt{I}}
\newcommand{\Bcomb}{\texttt{B}}
\newcommand{\Ccomb}{\texttt{C}}
\newcommand{\Kcomb}{\texttt{K}}
\newcommand{\Wcomb}{\texttt{W}}
\newcommand{\Zcomb}{\texttt{Z}}
\newcommand{\Scomb}{\texttt{S}}
\newcommand{\Acomb}{\texttt{A}}
\newcommand{\Dcomb}{\texttt{D}}
\newcommand{\Fcomb}{\texttt{F}}
\newcommand{\Deltacomb}{\delta}

\newcommand{\mgu}{\emph{mgu}}

\newcommand{\nzero}{\texttt{Z}}
\renewcommand{\nsucc}{\texttt{S}}
\newcommand{\nadd}{\texttt{A}}

\newcommand{\oo}{\mathtt{nil}}

\newcommand{\maketrs}[1]{#1\texttt{-TRS}}
\newcommand{\makears}[1]{#1\texttt{-ARS}}

\newcommand{\redex}{a}
\newcommand{\contractum}{b}
\newcommand{\redextwo}{\bm{s}}
\newcommand{\contractumtwo}{\bm{d}}

\newcommand{\redone}{a}
\newcommand{\redtwo}{a}
\newcommand{\conone}{b}
\newcommand{\contwo}{b}

\newcommand{\BA}{\mathcal{B}}
\newcommand{\baryrelone}{B}

\newcommand{\qtrsone}{\mathcal{R}}
\newcommand{\qarsone}{\mathcal{R}}

\newcommand{\NATS}{\mathcal{N}}
\newcommand{\natrelone}{N}

\newcommand{\tone}{a}
\newcommand{\ttwo}{b}
\newcommand{\tthree}{c}
\newcommand{\tfour}{d}

\newcommand{\combrelone}{K}

\newcommand{\code}[1]{c_{#1}}

\newcommand{\BCKprob}{\BCK_{\BA}}
\newcommand{\BCKNATS}{\BCK_{\NATS}}
\newcommand{\combnatrelone}{\combrelone_\natrelone}

\newcommand{\TICK}{\mathcal{T}}
\newcommand{\tickrelone}{T}

\newcommand{\relbot}{\Bot}
\newcommand{\reltop}{\qtop}

\newcommand{\QSL}{\mathcal{L}}
\newcommand{\qslrelone}{L}

\newcommand{\eqtheoryone}{\mathcal{E}}

\newcommand{\CP}{CP}

\newcommand{\qtrstwo}{\mathcal{S}}

\newcommand{\dualstar}[1]{#1^{* {\scriptstyle-}}}
\newcommand{\stardual}[1]{#1^{{\scriptstyle-}*}}

\newcommand{\combboundrelone}{W}
\newcommand{\BCKbounded}{\mathcal{W}}
\newcommand{\degree}[2]{\partial_{#1}(#2)}

\newcommand{\cpoleq}{\sqsubseteq}

\newcommand{\CBE}{\mathbb{\Phi}}

\newcommand{\degctx}[1]{\partial_{#1}}

\newcommand{\qunitconst}{\qunit^{\star}}